\documentclass[aps,prd,nofootinbib,onecolumn,superscriptaddress,preprintnumbers,balancelastpage,longbibliography,nobibnotes]{revtex4-1}

\usepackage[dvipsnames]{xcolor}
\usepackage{graphicx,color,amsmath,amssymb,flushend,bm,mathrsfs,comment}
\definecolor{linkcolor}{rgb}{0.0, 0.28, 0.67}

\usepackage[
   colorlinks=true,
    urlcolor=linkcolor,
   anchorcolor=linkcolor,
    citecolor=linkcolor,
    filecolor=linkcolor,
    linkcolor=linkcolor,
    menucolor=linkcolor,
    linktocpage=true,
    pdfproducer=medialab,
    pdfa=true
]{hyperref}

\usepackage[capitalise]{cleveref}
\usepackage{tikz}
\usepackage{ulem}
\usepackage{float}

\usepackage{amsmath}
\usepackage{physics}
\usepackage{simpler-wick}
\usepackage{amsfonts}
\usepackage{amssymb}
\usepackage[mathscr]{euscript}
\usepackage{setspace}
\usepackage{lipsum}
\usepackage{slashed}
\usepackage{cancel}
\usepackage{multirow}
\usepackage[utf8]{inputenc}
\usepackage{mathtools}
\usepackage{braket}
\usepackage{bbold}

\usepackage{soul}

\usepackage{fontawesome5}

\usetikzlibrary{calc}
\usepackage{siunitx}
\DeclareSIUnit{\year}{yr}
\DeclareSIUnit{\parsec}{pc}
\DeclareSIUnit{\eV}{e\kern-.05em V}
\DeclareSIUnit{\Jansky}{Jy}
\DeclareSIUnit{\sr}{sr}

\newcommand{\nocontentsline}[3]{}
\newcommand{\tocless}[2]{\bgroup\let\addcontentsline=\nocontentsline#1{#2}\egroup}

\def\dbar{{\mathchar'26\mkern-12mu d}}

\def\dbar{{\mathchar'26\mkern-12mu d}}

\newcommand{\bea}{\begin{eqnarray}\begin{aligned}}
\newcommand{\eea}{\end{aligned}\end{eqnarray}}

\newcommand{\const}{\text{const}}

\newcommand{\mpl}{m_\text{pl}}

\newcommand{\alphaem}{\alpha_\text{em}}

\newcommand{\eff}{\text{eff}}

\newcommand{\sm}{\text{SM}}
\newcommand{\cm}{\text{cm}}

\newcommand{\Ci}{\text{Ci}}

\newcommand{\gram}{\text{g}}

\newcommand{\erg}{\text{erg}}

\newcommand{\gammaE}{\gamma_\text{E}}

\newcommand{\eV}{\text{eV}}
\newcommand{\kev}{\text{keV}}
\newcommand{\mev}{\text{MeV}}
\newcommand{\gev}{\text{GeV}}
\newcommand{\tev}{\text{TeV}}

\newcommand{\Kelvin}{\text{K}}

\newcommand{\Eq}[1]{Eq.~(\ref{eq:#1})}

\newcommand{\formfactorreduce}{\widetilde{\mathcal{F}}}

\newcommand{\sct}{\text{sc}}

\newcommand{\psiinc}{\psi_\text{inc}}
\newcommand{\psisc}{\psi_\text{sc}}

\newcommand{\psiout}{\psi_\text{out}}

\newcommand{\psiint}{\psi_\text{int}}

\newcommand{\km}{\text{km}}

\newcommand{\bg}{\text{bg}}

\newcommand{\PQ}{\text{PQ}}

\newcommand{\Sec}[1]{Sec.~\ref{sec:#1}}
\newcommand{\Subsec}[1]{Sec.~\ref{subsec:#1}}
\newcommand{\Appx}[1]{Appendix~\ref{appx:#1}}

\newcommand{\Fig}[1]{Fig.~\ref{fig:#1}}

\newcommand{\osc}{\text{osc}}

\newcommand{\SN}{\text{SN}}
\newcommand{\HB}{\text{HB}}
\newcommand{\RG}{\text{RG}}

\newcommand{\CFresnel}{ \mathscr{C} }
\newcommand{\SFresnel}{ \mathscr{S} }

\newcommand{\sigmav}{\sigma_v}
\newcommand{\sigmak}{\sigma_k}

\newcommand{\mM}{m_\text{M}}
\newcommand{\mMtest}{m_{\text{M},\mathcal{T}}}
\newcommand{\mMsource}{m_{\text{M},\mathcal{S}}}
\newcommand{\mMearth}{m_{\text{M},\oplus}}

\newcommand{\vecE}{\mathbf{E}}
\newcommand{\vecr}{\mathbf{r}}

\newcommand{\veck}{\mathbf{k}}
\newcommand{\vecp}{\mathbf{p}}
\newcommand{\vecq}{\mathbf{q}}
\newcommand{\vecv}{\mathbf{v}}

\newcommand{\vecF}{\mathbf{F}}

\newcommand{\vecnabla}{\pmb{\nabla}}

\newcommand{\vectheta}{\boldsymbol{\theta}}
\newcommand{\vecphi}{\boldsymbol{\phi}}

\newcommand{\hc}{\text{h.c.}}

\newcommand{\deltaiso}{\delta_\text{iso}}

\newcommand{\testmass}{\mathcal{T}}
\newcommand{\sourcemass}{\mathcal{S}}

\newcommand{\fdecayphi}{\mathscr{F}_\phi}
\newcommand{\fdecayaxion}{\mathscr{F}_a}
\newcommand{\fdecaypiprime}{\mathscr{F}_{\pi'}}

\newcommand{\vac}{\text{vac}}

\newcommand{\second}{\text{s}}
\newcommand{\core}{\text{core}}

\newcommand{\kg}{\text{kg}}

\newcommand{\sphscr}{\text{sph}}

\newcommand{\formfactor}{\mathcal{F}}

\newcommand{\sph}{\text{sph}}

\newcommand{\erfc}{\text{erfc}}

\newcommand{\atom}{\text{atom}}
\newcommand{\nucleus}{\text{nucleus}}

\newcommand{\quality}{\mathcal{Q}}

\newcommand{\Fnoise}{F_\text{noise}}

\newcommand{\tialloy}{\text{Ti Alloy}}
\newcommand{\ptalloy}{\text{Pt Alloy}}

\newcommand{\sigmat}{\sigma_\text{T}}
\newcommand{\bmtx}{\begin{pmatrix}}
\newcommand{\emtx}{\end{pmatrix}}

\def\beq{\begin{equation}}
\def\eeq{\end{equation}}

\begin{document}

\preprint{DESY-25-060}

\title{Detecting Ultralight Dark Matter with Matter Effect}

\author{Xucheng Gan}
\altaffiliation{Corresponding author: Xucheng Gan}
\email{xucheng.gan@desy.de}

\affiliation{Deutsches Elektronen-Synchrotron DESY, Notkestr. 85, 22607 Hamburg, Germany}

\author{Da Liu}
\email{liudaphysics@gmail.com}
\affiliation{PITT PACC, University of Pittsburgh, Pittsburgh, PA, USA}

\author{Di Liu}
\email{liudisy@gmail.com}
\affiliation{Laboratoire d'Annecy-le-Vieux de Physique Th\'eorique, CNRS -- USMB, BP 110 Annecy-le-Vieux, F-74941 Annecy, France}
\affiliation{Centre for Mathematical Sciences, University of Plymouth, PL4 8AA, UK}

\author{Xuheng Luo}
\email{xluo26@jhu.edu}
\affiliation{The William H. Miller III Department of Physics and Astronomy, The Johns Hopkins University, Baltimore, Maryland, 21218, USA}

\author{Bingrong Yu}
\email{bingrong.yu@cornell.edu}
\affiliation{Department of Physics, LEPP, Cornell University, Ithaca, NY 14853, USA}

\begin{abstract}
Ultralight particles, with a mass below the electronvolt scale, exhibit wave-like behavior and have arisen as a compelling dark matter candidate. A particularly intriguing subclass is scalar dark matter, which induces variations in fundamental physical constants. However, detecting such particles becomes highly challenging in the mass range above $10^{-6}\,\text{eV}$, as traditional experiments face severe limitations in response time. In contrast, the matter effect becomes significant in a vast and unexplored parameter space. These effects include (i) a force arising from scattering between ordinary matter and the dark matter wind and (ii) a fifth force between ordinary matter induced by the dark matter background. Using the repulsive quadratic scalar-photon interaction as a case study, we develop a unified framework based on quantum mechanical scattering theory to systematically investigate these phenomena across both perturbative and non-perturbative regimes. Our approach not only reproduces prior results obtained through other methodologies but also covers novel regimes with nontrivial features, such as decoherence effects, screening effects, and their combinations. In particular, we highlight one finding related to both scattering and background-induced forces: the descreening effect observed in the non-perturbative region with large incident momentum, which alleviates the decoherence suppression. Furthermore, we discuss current and proposed experiments, including inverse-square-law tests, equivalence principle tests, and deep-space acceleration measurements. Notably, we go beyond the spherical approximation and revisit the MICROSCOPE constraint on the background-induced force in the large-momentum regime, where the decoherence and screening effects interplay. The ultraviolet models realizing the quadratic scalar-photon interaction are also discussed. 
\end{abstract}

\maketitle

\tableofcontents

\section{Introduction}

Observations related to gravitational interactions have confirmed the existence of dark matter. However, its particle physics properties remain unknown. One intriguing class of candidates is ultralight scalar dark matter, with its mass below the electronvolt scale. When coupled to Standard Model~(SM) fields, this dark matter induces variations in fundamental physical constants via coherent oscillations. This phenomenon has been extensively studied in various contexts~\cite{Damour:1994zq, Damour:1994ya, Olive:2001vz, Wetterich:2002ic, Armendariz-Picon:2002jez, Damour:2002mi,Damour:2002nv, Weiner:2005ac, Olive:2007aj, Piazza:2010ye, Damour:2010rp,Damour:2010rm,Uzan:2010pm,Stadnik:2015kia, Ghalsasi:2016pcj,Sibiryakov:2020eir,Bouley:2022eer, Hees:2018fpg,Ellis:2019flb,Brzeminski:2020uhm,Batell:2021ofv,Banerjee:2022sqg,Batell:2022qvr,Gan:2023wnp,Luo:2023cxo,Cyncynates:2024bxw,Cyncynates:2024ufu,Uzan:2024ded}. Various experimental approaches have been employed to detect dark matter in this mass range, including atomic clock comparisons~\cite{Arvanitaki:2014faa, VanTilburg:2015oza, Hees:2016gop, Stadnik:2016zkf, Kalaydzhyan:2017jtv, kennedy2020precision, Barontini:2021mvu, collaboration2021frequency,Kobayashi:2022vsf,Filzinger:2023zrs}, cold-atom interferometers~\cite{Arvanitaki:2016fyj,AEDGE:2019nxb, Badurina:2019hst, MAGIS-100:2021etm,Zhao:2021din}, and mechanical resonators~\cite{Arvanitaki:2015iga,Branca:2016rez,Manley:2019vxy}.  Additionally, natural fission reactions, such as Oklo~\cite{shlyakhter1976direct,Damour:1996zw,Gould:2006qxs,Flambaum:2008hu}, and astrophysical observations, such as the quasar absorption spectrum~\cite{Webb:1998cq,Petitjean:2009id,Srianand:2009gaa,Baryakhtar:2025uxs}, can also probe variations in physical constants induced by ultralight dark matter. Because the scalar dark matter has a large field value in the early universe, constraints can also be derived from cosmic microwave background~(CMB) observations~\cite{Hannestad:1998xp, Menegoni:2012tq, Stadnik:2015kia, Hart:2019dxi}, Lyman-alpha forest~\cite{Hamaide:2022rwi}, and Big Bang nucleosynthesis (BBN)~\cite{Stadnik:2015kia, Sibiryakov:2020eir, Bouley:2022eer}. However, the aforementioned methods are insensitive to the dark matter mass range above $10^{-6}\,\eV$, as the oscillation period of dark matter within this mass range is significantly shorter than the response time of current experimental apparatuses. Additionally, the scalar field amplitude is suppressed as its mass increases, thereby diminishing the variation of fundamental physical constants.

To detect ultralight dark matter with a mass above $10^{-6}\,\eV$, an effective approach is to utilize the matter effect arising from its coupling to Standard Model (SM) fields through shift-symmetry-breaking operators, which induce additional detectable forces. In this work, we specifically focus on a class of dark matter candidates that are quadratically coupled to SM fields through the operator:
\bea
\label{eq:quad_phi_sm_general}
\frac{1}{\Lambda^2} \phi^2 \mathcal{O}_\sm\,,\label{eq:coupling}
\eea
where $\phi$ is the scalar dark matter field, ${\cal O}_{\rm SM}$ is the renormalizable SM operator, and $\Lambda$ is the scale of new physics. This operator has been widely explored in Refs.~\cite{Olive:2007aj,Hees:2018fpg,Sibiryakov:2020eir,Bouley:2022eer,Banerjee:2022sqg,Gan:2023wnp,Beadle:2023flm,Kim:2023pvt,Kamionkowski:2024axz}. In this class of models, the linear scalar-SM coupling is strictly forbidden by imposing the $\mathbb{Z}_2$ symmetry under the transformation $\phi \rightarrow - \phi$. When dark matter propagates through ordinary matter, it can interact coherently with a bulk of ordinary matter if its de Broglie wavelength is larger than the atomic spacing in the matter, which modifies the dispersion relation of $\phi$ --- a phenomenon known as the matter effect. Under mean-field theory, this effect is typically described by replacing SM operators with their ensemble averages: ${\cal O}_\sm \to \langle {\cal O}_\sm \rangle$. As a result, $\phi$ acquires an effective mass in ordinary matter through the coupling in \Eq{coupling}, as
\bea
\mM^2\sim \frac{\langle {\cal O}_\sm \rangle}{\Lambda^2}\;.
\eea
The matter effect arises in a wide range of physical contexts, such as the Mikheyev-Smirnov-Wolfenstein~(MSW) effect in neutrino physics~\cite{Wolfenstein:1977ue,Mikheyev:1985zog,Mikheyev:1986wj,Botella:1986wy,Notzold:1987ik,Mirizzi:2009td,Huang:2023nqf}, the Meissner effect in superconductors~\cite{Meissner:1933ela}, and the reflection of visible light from metals~\cite{Jackson:1998nia}. In this work, we focus on the repulsive scalar-SM interaction, which induces a positive effective mass squared for the scalar within ordinary matter. However, our perturbative analysis is applicable to both repulsive and attractive interactions.

When the scalar interacts with ordinary matter, two types of forces arise: 
\begin{itemize}
\item {\bf Scattering force}, which results from the momentum transfer between the scalar field and a test mass;
\item {\bf Background-induced force}, which is experienced by the test mass due to the ripples in the scalar background generated by the source mass.
\end{itemize}
While both forces have been investigated in various contexts using different methodologies~\cite{Ferrer:2000hm,dePireySaintAlby:2017lwc,Fukuda:2018omk,Berezhiani:2018oxf,Hees:2018fpg,Fukuda:2021drn,Banerjee:2022sqg,Day:2023mkb,VanTilburg:2024xib,Barbosa:2024pkl,Luo:2024ocg,Grossman:2025cov,Cheng:2025fak}, they have not been systematically addressed within a unified framework, nor has the discussion covered all the phases within the parameter space. In particular, there is a substantial gap in understanding the behavior in the non-perturbative region, where the screening effect becomes significant, and the large-scale region, where decoherence is essential. Furthermore, it remains unclear how the forces behave in the regime where both effects become relevant.

\begin{figure}[t!]
\centering
\includegraphics[width=0.85\linewidth]{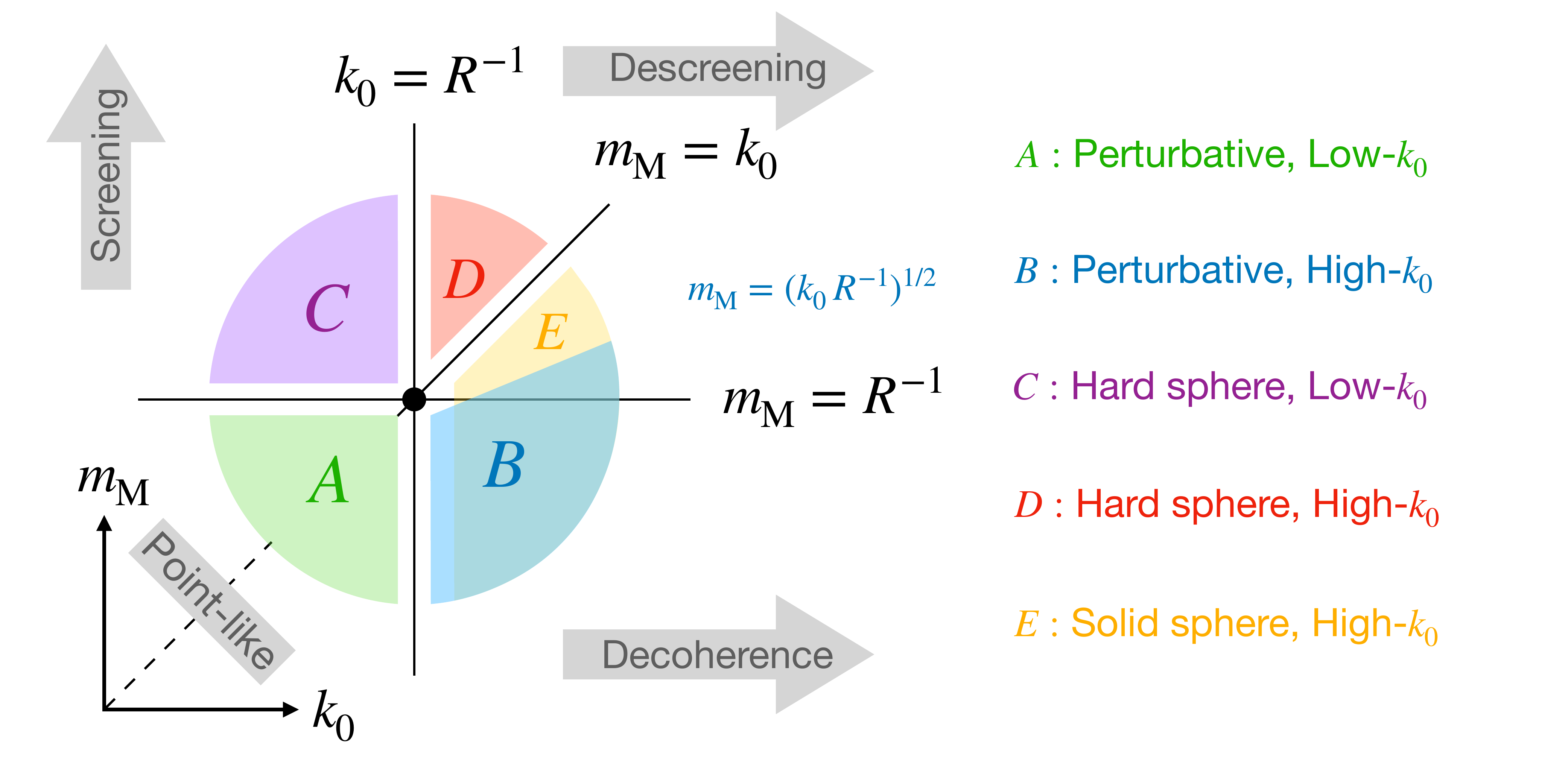}
\caption{Classification of scattering between ultralight dark matter and ordinary matter in the $k_0\text{-}\mM$ plane, where $k_0$ is the mean incident momentum of dark matter, $\mM$ is the effective mass of dark matter due to its coupling to ordinary matter, and $R$ is the size of the ordinary matter. Regions A and B are perturbative, and the remaining regions are non-perturbative. The screening effect becomes stronger as $\mM$ increases, while the decoherence effect becomes stronger as $k_0$ increases. In the strongly-coupled region, the descreening effect emerges at large $k_0$. {\bf Region A}: $R^{-1}>\mM,k_0$. The scattering occurs in the perturbative regime, where the scattered ordinary matter is treated as a point-like object. {\bf Region B}: $k_0 > R^{-1}$ and $(k_0 R^{-1})^{1/2}>\mM$. The scattering remains in the perturbative regime; however, it is necessary to take the finite size into account. {\bf Region C}: $\mM > R^{-1} > k_0$. This corresponds to low-momentum hard-sphere scattering. {\bf Region D}: $\mM > k_0 > R^{-1}$. This corresponds to high-momentum hard-sphere scattering. {\bf Region E}: $k_0 > \mM, \, R^{-1}$. This corresponds to high-momentum solid-sphere scattering, where the effective potential is treated as a finite well. In Region E$\setminus$B, the wave function enters the non-perturbative regime.}
\label{fig:phase_classify}
\end{figure}

By utilizing quantum mechanical scattering theory, we develop a unified framework to describe both the scattering force and the background-induced force arising from ultralight dark matter. Additionally, we provide a comprehensive classification of all the phases in the parameter space, labeled by the mean momentum-effective mass plane, as illustrated in \Fig{phase_classify}. As shown in the plot, in the strongly coupled region where the effective mass is large, the perturbative condition breaks down, and the \textbf{screening effect} becomes significant. This results in the suppression of both the scattering force and the background-induced force compared to the perturbative predictions. In the region where the mean momentum of the dark matter background is large, the \textbf{decoherence effect}, arising from the finite phase space and the finite size of the experimental objects, becomes significant. This leads to a suppression of the forces compared to those predicted using a point-like source in the near-field approximation. In the region where both $\mM$ and $k_0$ are large, the \textbf{descreening effect} emerges, competing with the suppression caused by the decoherence effect. In this work, we systematically explore all the regions identified in \Fig{phase_classify}, as well as the transitions between different regions. We also discuss the associated phenomenology and experimental tests using inverse-square-law~(ISL) tests~\cite{Spero:1980zz,Hoskins:1985tn,Tu:2007zz,Yang:2012zzb,Hoyle:2000cv,Hoyle:2004cw, Kapner:2006si,Murata:2014nra,Will:2014kxa}, equivalence principle~(EP) tests~\cite{Nobili:2012uj,Berge:2017ovy,Touboul:2017grn,MICROSCOPE:2019jix,MICROSCOPE:2022doy}, and deep-space acceleration measurements~\cite{Buscaino:2015fya}. Additionally, we analyze the theoretical sensitivities by comparing the derived results with thermal and quantum fluctuations~\cite{Kubo:1966fyg}, providing a more fundamental understanding of the experimental sensitivities.

\section{Summary}

For the scattering force, Refs.~\cite{Fukuda:2018omk,Fukuda:2021drn} discuss the acceleration sensitivities induced by the scattering force in the perturbative region using the Born approximation. However, they do not provide a rigorous analysis in the strongly coupled, non-perturbative regime, where perturbative methods are no longer valid. Ref.~\cite{Day:2023mkb} provides a dedicated discussion on the scattering force using partial wave analysis, which also covers the non-perturbative region. Although the partial wave analysis inherently includes some aspects of decoherence and descreening effects, this study does not comprehensively address these effects, nor does it thoroughly discuss the breakdown of the perturbative method in the non-perturbative regime. In our work, we conduct a systematic exploration of both the perturbative and non-perturbative regions by combining the Born approximation with partial wave analysis. We also address the decoherence effect in detail and investigate the optimal geometry of the test mass to maximize sensitivity. In addition, we perform a comparison between the Born approximation and partial wave analysis, identifying regions where the previously employed Born approximation breaks down. Moreover, we investigate the descreening effect in the scattering force. Furthermore, we compare the scattering and background-induced forces, identifying the regions where one effect dominates over the other.

For the background-induced force, Refs.~\cite{Ferrer:2000hm,VanTilburg:2024xib,Barbosa:2024pkl,Grossman:2025cov,Cheng:2025fak} use finite-density field theory, which has also been used to study the SM neutrino force in neutrino backgrounds~\cite{Horowitz:1993kw,Ferrer:1998ju,Ferrer:1999ad,Ghosh:2022nzo,Blas:2022ovz,Ghosh:2024qai}. This method is based on perturbative quantum field theory, which is well-suited for handling the perturbative region and can incorporate the decoherence effect in this region. However, it encounters technical challenges when extended to the strongly coupled non-perturbative regions. On the other hand, Refs.~\cite{dePireySaintAlby:2017lwc,Berezhiani:2018oxf,Hees:2018fpg,Banerjee:2022sqg} utilize the classical field theory approach, assuming a spherically symmetric field configuration. This approach is effective for describing the non-perturbative screening effect in the near-field region with low momentum. However, it breaks down in the high-momentum or far-field region and fails to account for the decoherence effect.  This limitation arises because the method is equivalent to including only the s-wave component in partial wave analysis and choosing the near-field limit. In this work, we systematically investigate all the phases identified in \Fig{phase_classify} and unify the previous approaches using quantum mechanical scattering theory. We derive the equivalent expressions as presented in Refs.~\cite{Ferrer:2000hm,VanTilburg:2024xib,Barbosa:2024pkl} 
using the Born approximation, and we extend the analysis of the perturbative decoherence effect to regimes where the finite size effect of the source mass becomes significant. Furthermore, we employ partial wave analysis to examine deviations from the spherical ansatz used in Refs.~\cite{dePireySaintAlby:2017lwc,Berezhiani:2018oxf,Hees:2018fpg,Banerjee:2022sqg}, particularly in regions where the decoherence effect is relevant. We also identify and analyze the previously unexplored descreening effect in the high-momentum, strongly coupled region, which counteracts the suppression caused by the decoherence effect. Additionally, we revisit the MICROSCOPE constraint on the repulsive quadratic scalar-SM interaction, providing an analysis beyond the spherically symmetric ansatz used in Refs.~\cite{Hees:2018fpg,Banerjee:2022sqg}.

For the effective theory of the quadratic scalar-photon interaction, which serves as the benchmark model in this paper, we present three possible ultraviolet~(UV) models. These include models with heavy $U(1)_Y$-charged fermions, heavy $U(1)_Y$-charged scalars, and dark QCD axion that couples to the SM photon via kinetic mixing. Using renormalization group analysis, we find that models with heavy $U(1)_Y$-charged fermions or heavy $U(1)_Y$-charged fermions or scalars can induce either repulsive or attractive interactions, depending on the model parameters. In contrast, the dark QCD axion induces exclusively attractive interactions.

The rest of this paper is organized as follows. In \Sec{matter_effect}, we use the repulsive quadratic scalar-SM coupling as a benchmark model to provide a brief introduction to the matter effect for the ultralight scalar. In \Sec{formalism}, we develop the formalism to treat both the scattering force and background-induced force in a unified framework. We also present the phase space integration, essential for analyzing the decoherence effect in the background-induced force. Furthermore, we compare the scattering and background-induced forces within the perturbative region. In \Sec{scforce}, we discuss the scattering force in the perturbative and non-perturbative regions, covering all the phases listed in \Fig{phase_classify}. We also address the decoherence effect, which plays a crucial role in optimizing the geometry of the test mass, and examine the breakdown of the perturbative result in the non-perturbative region. In addition, we analyze the experimental sensitivity, expressed in terms of acceleration, to assess the potential reach of future experiments. In \Sec{bg_force}, we discuss the background-induced force covering all the phases listed in \Fig{phase_classify}. We begin by discussing the perturbative decoherence effect for a point-like source and then extend the analysis to sources with finite size. Next, we review the non-perturbative treatment under the assumption that the scalar configuration is spherically symmetric. Then, we explore the high-momentum non-perturbative region, where the spherically symmetric ansatz breaks down, and highlight the emergence of the descreening effect. In addition, we discuss the experimental projections from recasting the inverse-square-law test and short-range equivalence-principle test and the experimental sensitivities in terms of acceleration. Finally, we revisit the constraint on the repulsive quadratic scalar interaction imposed by the MICROSCOPE satellite, extending the analysis beyond the previously adopted spherical ansatz. In \Sec{vary_alpha_UV}, we present the underlying models that generate our benchmark model, the quadratic scalar-photon interaction. Our main conclusions are summarized in \Sec{conclusion}.

We also provide a series of appendices containing the details of the calculation. \Appx{test_mass_bkg} discusses the motion of a test mass in an inhomogeneous scalar background. \Appx{force_qft} covers the perturbative computation of background-induced and vacuum forces within the framework of quantum field theory. \Appx{born_approx_appendix} presents detailed calculations using the Born approximation for computing the wave function, scattering amplitudes, and background-induced force. \Appx{phase_space_finite_size} includes the detailed derivations and explicit analytical expressions of the decoherence effect in the perturbative region, including the finite-phase-space effect and the monochromatic finite-size effect. \Appx{partial_wave_appx} provides the analysis of the partial wave method, including the computation of the wave function, phase shift, cross section, and s-wave approximations. \Appx{astro_constraint} discusses the stellar energy loss constraints on the quadratic scalar-photon interaction, including contributions from supernovae, horizontal branch stars, and red giants. In \Appx{uv_app}, we present the KSVZ-like dark QCD axion model.

\section{Matter Effect}\label{sec:matter_effect}

We start our discussion by introducing the benchmark model
\bea
\label{eq:lagrangian1}
\mathcal{L} = \frac{1}{2}\partial_\mu \phi \, \partial^\mu \phi - \frac{1}{2}m_0^2\,\phi^2 - \left( 1 - \frac{\phi^2}{2 \Lambda_\gamma^2} \right) \frac{1}{4} F_{\mu\nu}F^{\mu\nu}\,,
\eea
where $\phi$ is the scalar field, $F_{\mu \nu}$ is the electromagnetic field strength tensor, $m_0$ represents the bare mass of the scalar, and $\Lambda_\gamma$ is the scale suppressing the dimension-six effective operator of the scalar-photon interaction. In this theory, the linear scalar-photon interaction is forbidden because of the imposed $\mathbb{Z}_2$ symmetry with the transformation
\bea
\phi \rightarrow -\phi. 
\eea

In most parts of this work, we use \Eq{lagrangian1}, which includes the dimension-six scalar-photon interaction as the effective theory. We also systematically explore possible UV models, which are discussed in \Sec{vary_alpha_UV}. The quadratic scalar-photon interaction mentioned above is also parameterized as $\mathcal{L}_\text{int} = \frac{1}{4}\frac{d_{\gamma}}{2} \left(\frac{\sqrt{4\pi}}{\mpl}\right)^2 \phi^2 F_{\mu\nu} F^{\mu \nu}$ in Refs.~\cite{Bouley:2022eer, Hees:2018fpg,Banerjee:2022sqg,Gan:2023wnp,Arakawa:2024lqr}, where $\mpl = 1.22 \times 10^{19} \, \gev$ is the Planck mass. This parameterization treats $d_\gamma$ as a dimensionless quantity to describe the relative strength of the interaction compared with gravity in the hyper-weakly coupled regions. However, since our analysis focuses on interaction strengths that largely exceed gravitational effects, we use \Eq{lagrangian1} as the default parameterization. Throughout most parts of this paper, we consider the repulsive potential~($\mM^2>0$), as an attractive potential~($\mM^2<0$) leads to tachyonic instability during early-universe evolution~\cite{Sibiryakov:2020eir}. However, in the perturbative region, our discussion applies equally to an attractive potential of the same magnitude, differing only by a sign~\footnote{The complication arises in the non-perturbative regime: a repulsive potential leads to screening effects, as discussed in this work and previous studies~\cite{Hees:2018fpg,Banerjee:2022sqg,Day:2023mkb,Burrage:2024mxn}, while an attractive potential results in resonant enhancement, as explored recently in Refs.~\cite{Banerjee:2025dlo,delCastillo:2025rbr}. Screening also appears in the attractive potential away from the resonance peaks, as later shown explicitly in Refs.~\cite{Delaunay:2025pho,Burrage:2025grx,Gan:2025icr}. Refs.~\cite{Banerjee:2025dlo,delCastillo:2025rbr} further investigate the high-momentum regime using partial-wave methods and focus primarily on the scalar-field distribution near the Earth and its impact on ground-based experiments. This is complementary to our emphasis on unifying and extending the formalism of scalar-induced forces and on developing detection strategies based on the matter effect. We refer the reader to the aforementioned works for a broader overview of this landscape.}. For completeness, we also present representative non-perturbative scattering patterns for an attractive potential in \Fig{nonpeturb_attractive}, to build intuition for the differences between the two cases in the non-perturbative regime.

Although our discussion applies to general repulsive quadratic scalar-SM interactions, we focus on the theory in \Eq{lagrangian1} because the scalar-photon interaction induces a variation in the fine-structure constant, which has been extensively explored previously due to its fundamental significance~\cite{Bekenstein:1982eu,Damour:1996zw,Dvali:2001dd,Chacko:2002mf,Uzan:2002vq,Bekenstein:2002wz,Uzan:2010pm,Brzeminski:2020uhm,Beadle:2023flm,Kim:2023pvt,Uzan:2024ded}. Specifically, the variation in the fine-structure constant depends on the scalar field value and is given by
\bea
\frac{\Delta \alphaem(\phi)}{\alphaem} = \frac{\phi^2}{2 \Lambda_\gamma^2},
\eea
For scalar dark matter, we relate the field amplitude to its energy density and obtain $\phi = \sqrt{2 \rho_\phi}/m_0$, as derived in Refs.~\cite{Preskill:1982cy,Arias:2012az}. When scalar dark matter induces variations in the fine-structure constant, the density of ordinary matter also changes accordingly because electromagnetic energy depends on the fine-structure constant. Here, we provide a pedagogical approach to describe this density variation and the resulting matter effect of the scalar field following Refs.~\cite{Damour:1994zq,Damour:2010rp,Damour:2010rm,Uzan:2010pm,Uzan:2024ded}. Let $\rho$ denote the density of the bulk ordinary matter $(Z,A)$, where $Z$ and $A$ are the atomic number and mass number of the nucleus. Its variation is given by
\bea
\label{eq:delta_rho}
\Delta \rho(\phi)= \frac{\phi^2}{2\Lambda_\gamma^2} \times Q_\gamma \rho, \quad \quad \text{where $Q_\gamma \equiv \frac{\alphaem}{m_\atom} \frac{\partial \, m_\atom}{\partial \alphaem}$.}
\eea
The dimensionless quantity $Q_\gamma$ is defined as the dilaton charge, which depends on the properties of materials and quantifies how the density of ordinary matter responds to variations in the fine-structure constant. Here, $m_n$ and $m_p$ denote the neutron and proton masses, respectively, while $m_\atom$ represents the atomic mass. Based on the semi-empirical mass formula, the atomic mass can be approximated as $m_\atom \simeq (A-Z)\,m_n + Z m_p + Z m_e + E_\text{em} + \cdots$, where ``$\cdots$'' denotes the terms that are independent of $\alphaem$. The electrostatic energy term, $E_\text{em} = - \int d^3\vecr \, \abs{\vecE}^2/2$, describes the repulsion among protons, where $\vecE$ is the electrostatic field. Treating the nucleus as a liquid drop, we have $E_\text{em} = (3/5) \alphaem Z^2/R_\nucleus$~\cite{Weizsacker:1935bkz}. We then replace $Z^2 \rightarrow Z(Z-1)$ to subtract the electrostatic self-energy of the protons~\cite{Blatt:1952ije,kaplan1955nuclear}. We also replace $R_\nucleus \rightarrow A^{1/3} r_0$, where $r_0 \simeq 1.2\,\text{fm}$ is the empirical constant quantifying the proton size. Thus, the electrostatic energy among the protons is $E_\text{em} = a_0  Z(Z-1)/A^{1/3}$. Here $a_0 = (3/5)\,\alphaem/r_0 \simeq 0.72\,\mev$ is the constant quantifying the electrostatic energy of a single proton. Using the expressions above, we write the dilaton charge as
\bea
Q_\gamma = F_A \left[ \frac{A-Z}{A}\frac{1}{m_\text{amu}} \frac{\partial \, m_n}{\partial \log \alphaem} + \frac{Z}{A} \frac{1}{m_\text{amu}} \frac{\partial \, m_p}{\partial \log \alphaem} + \frac{a_0}{m_\text{amu}} \frac{Z(Z-1)}{A^{4/3}}\right],
\eea
where $F_A = A m_\text{amu}/m_\atom$ and $m_\text{amu} = 931 \, \mev$ is the atomic mass unit. The proton and neutron masses also depend on the fine-structure constant, with values $\partial m_p/\partial \log \alphaem = 0.63 \, \mev$ and $\partial m_n/\partial \log \alphaem = -0.13 \, \mev$~\cite{Gasser:1982ap}. The electron mass is independent of the fine-structure constant, so $\partial m_e/\partial \log\alphaem = 0$. We take $F_A \simeq 1$, introducing only a $\mathcal{O}(10^{-4})$ relative error due to the difference between the atomic mass unit and nucleon masses. Substituting the numerical values above, we obtain
\begin{equation}\label{eq:dilaton_charge_num}
    Q_\gamma \simeq \left[ -1.4 + 8.2 \frac{Z}{A} + 7.7 \frac{Z(Z-1)}{A^{4/3}}\right] \times 10^{-4}\,,
\end{equation}
which is consistent with previous results in Refs.~\cite{Damour:2010rp,Damour:2010rm}. From \Eq{delta_rho}, we see that the energy density of the ordinary matter depends on the scalar field value. Consequently, the scalar tends to relax to zero in a finite-density environment. Effectively, we do the replacement $\big\langle \frac{1}{4} F_{\mu\nu} F^{\mu\nu} \big\rangle \rightarrow - Q_\gamma \rho$ in \Eq{lagrangian1} and get the Klein-Gordon equation
\begin{align}
\label{eq:KG_eq}
\left(\square + m_0^2\right)\phi = - \mM^2(\vecr) \phi\,.
\end{align}
Here, $\mM$ represents the effective scalar mass induced by the finite-density environment and is given by
\begin{align}
\label{eq:effective_mass}
\mM(\vecr) \equiv \left( \frac{Q_\gamma\rho(\vecr)}{\Lambda_\gamma^2} \right)^{1/2},
\end{align}
which changes the dispersion relation of the scalar field. This phenomenon is known as the matter effect. Here, $\mM(\vecr)$ depends on the spatial coordinate $\vecr$ as it tracks the density of ordinary matter. Given the inhomogeneous distribution of ordinary matter, the scalar field experiences reflection and deflection. This process transfers momentum from the scalar field to the test mass, generating the scattering force. In the meantime, it induces inhomogeneities in the scalar background, inducing an extra potential that drives the motion of the test mass, known as the background-induced force. Both effects will be discussed in detail in the following sections. 

Before proceeding, we emphasize that our discussion applies generally to all repulsive quadratic scalar-SM interactions, including $\phi^2 F^2$, $\phi^2 G^2$, $\phi^2 m_q \bar{q}q$, and $\phi^2 m_e \bar{e}e$. For this reason, throughout most part of this paper, we use $\mM$ to represent the strength of the matter effect when presenting the formalism, independent of the specific choice of SM operator in \Eq{quad_phi_sm_general}. We restrict our discussion to the $\phi^2 F^2$ operator only when considering concrete experimental setups and comparisons with other existing constraints (e.g., from SN 1987A and BBN). The results for other quadratic operators can be obtained by directly recasting them in terms of the dilaton charges of the test and source masses. We recommend Refs.~\cite{Damour:2010rp,Hees:2018fpg} for comprehensive and pedagogical introductions to both dilaton charges and complete set of operators inducing varying fundamental constants.

\begin{figure}[h!]
\centering
\includegraphics[width=0.6\linewidth]{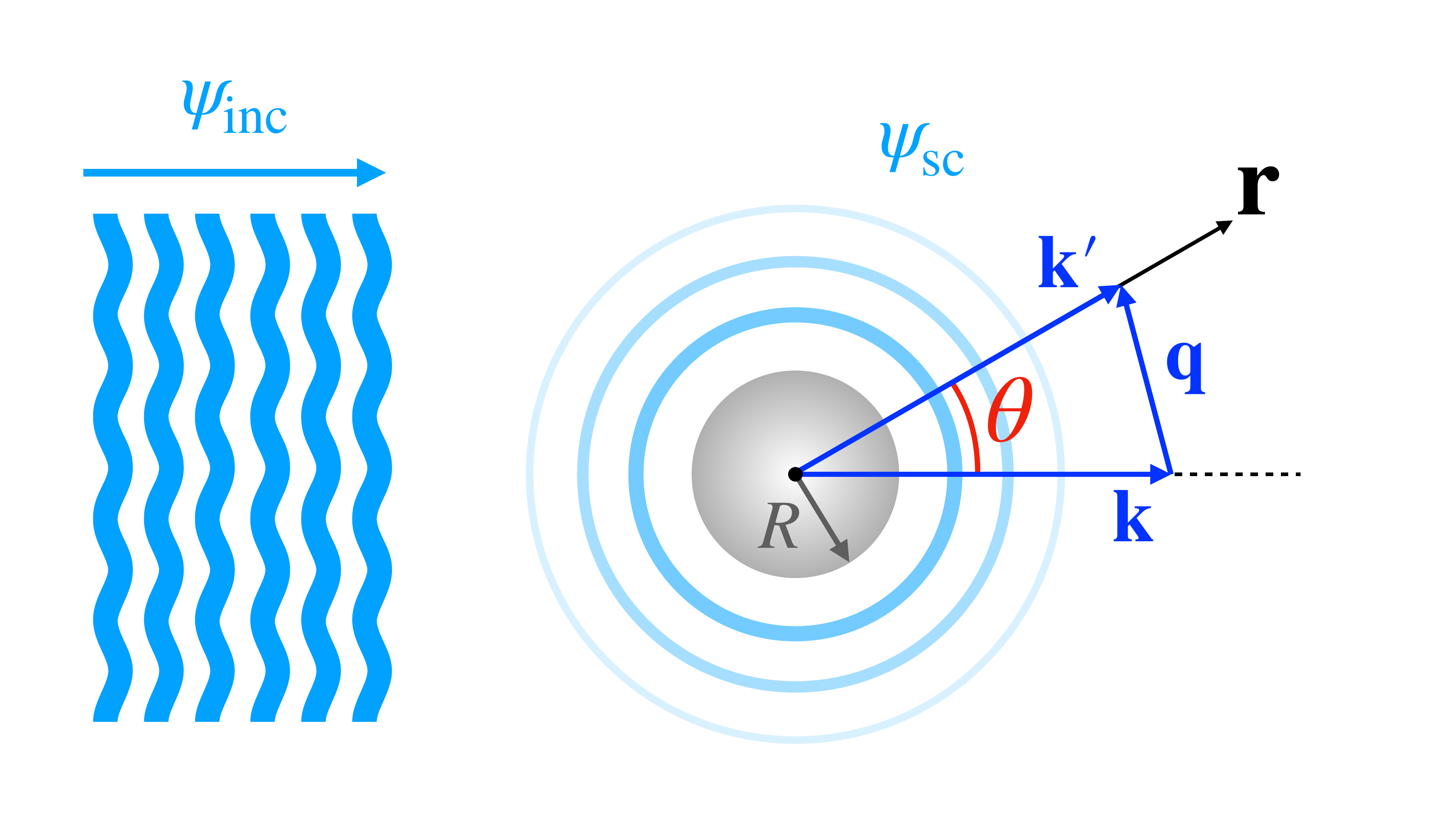}
\caption{Schematic representation of scalar wave scattering off a sphere of radius $R$ composed of ordinary matter. $\psiinc$ represents the incident plane wave propagating with the momentum $\veck$, while $\psisc$ represents the scattered wave. The observer at position $\vecr$ with an angle $\theta$~(referred to as the reflection angle) detects the wave with momentum $\veck'$, and the corresponding momentum transfer is represented by $\vecq = \veck' - \veck$. For the scattering force, the gray sphere represents the test mass $\testmass$ experiencing an additional acceleration due to dark matter collisions. For the background-induced force, the gray sphere represents the source mass $\sourcemass$. It generates a spatially inhomogeneous scalar potential, which in turn induces the acceleration of the test mass $\testmass$ in its vicinity~(not shown in the figure).}
\label{fig:qm_sca}
\end{figure}

\section{Formalism}\label{sec:formalism}

In this section, we develop a general formalism based on quantum mechanical scattering theory~\cite{Landau:1991wop,Sakurai:2011zz} to analyze how the matter effect affects the evolution of the ultralight scalar. Specifically, we analyze how the scattering force and background-induced force are derived within this framework. For non-relativistic (NR) cold dark matter, we reformulate \Eq{KG_eq} using the following ansatz:
\bea
\label{eq:phi_from_psi}
\phi(\vecr,t;\veck) = \Re\big[e^{-i \omega t} \psi(\vecr;\veck)\big]\,,
\eea
where $\omega \equiv  \sqrt{m_0^2+\veck^2}$ and $\psi$ represents the NR component of $\phi$. In the NR limit, $\psi$ depends only on spatial coordinates, since its time derivatives satisfy $m_0 \partial_t \psi, \partial^2_t \psi \ll m_0^2 \psi$~\footnote{A commonly used convention for the NR limit is $\phi(\vecr,t;\veck) = \Re[ e^{-i m_0 t} \widetilde{\psi}(\vecr,t;\veck) ]$~\cite{Bjorken:1965sts,Greiner:1997xwk}, which is equivalent to \Eq{phi_from_psi} under the transformation $\widetilde{\psi}(\vecr,t;\veck) = e^{-i k^2/2m_0 t}\psi(\vecr;\veck)$.}. Substituting \Eq{phi_from_psi} into \Eq{KG_eq}, we obtain the Schr\"{o}dinger equation
\bea
\label{eq:Schrodinger_Eq}
- \frac{1}{2 m_0} \vecnabla^2 \psi + V_\eff(\vecr)\psi = E_\eff(\veck) \psi\,.
\eea
In the above equation, $V_\eff$ is the effective potential, and $E_\eff$ is the effective energy, given by:
\bea
\label{eq:Veff_Eeff}
V_\eff(\vecr) = \frac{\mM^2(\vecr)}{2m_0}\,, 
\qquad E_\eff(\veck) = \frac{\veck^2}{2m_0}\,.
\eea

From these expressions, we see that the significance of the matter effect on the scalar field relies on the relative magnitudes of  $V_\eff$ and $E_\eff$, or equivalently speaking, the ratio of $\mM$ to $\veck$. Another key factor determining the significance of the matter effect is the geometry of scattered ordinary matter, characterized by $R$. However, this parameter does not appear explicitly in the Schr\"{o}dinger equation above but in matching the boundary conditions. Since we are studying dark matter scattering with ordinary matter, we write $\psi$ as the sum of an incident wave and a scattered wave given a specific momentum $\veck$:
\bea
\label{eq:psi_tot}
\psi(\vecr;\veck) = \psiinc(\vecr;\veck) + \psisc(\vecr;\veck).
\eea
Here, $\psi$ represents the total wave outside the scattered object. $\psiinc$ and $\psisc$ represent the incident and scattered wave functions, respectively, and are given by:
\bea
\label{eq:psi_in_sca}
\left\{
\begin{aligned}
& \text{Incident Wave:} && \quad \quad \psiinc(\vecr;\veck) = \abs{\psi_0} e^{i \veck \cdot \vecr} & \\
& \text{Scattered Wave:} && \quad \quad \psisc(\vecr;\veck) \simeq \abs{\psi_0} f(\theta;k) \frac{e^{ikr}}{r} \quad \quad \text{(Far-Field)}
\end{aligned}
\right. .
\eea
The scattered wave function $\psisc$ is presented in the asymptotic form, which is valid in the far-field limit. $\abs{\psi_0} = \abs{\phi_0}$ is the amplitude of the incident dark matter wave. $f(\theta;k)$ is the scattering amplitude. $\theta$ is the deflection angle of the scattered wave, defined as the angle between $\vecr$ and $\veck$. For simplicity, we consider only cylindrically symmetric collisions along the $\veck$-axis, making the calculations independent of the azimuthal angle. In $\psisc$, there is a factor $e^{ikr}/r$, which corresponds to an outgoing spherical wave. From the Lippmann-Schwinger equation~\cite{Lippmann:1950zz}, this is associated with the retarded Green’s function, reflecting the causal structure of the scattering process.

\subsection{Scattering Force}

The scattering amplitude $f(\theta;k)$ describes the asymptotic behavior of the scattered wave. Armed with this, we can get the scattering force on the ordinary object from the scalar dark matter with the momentum $\veck$. Specifically, we have
\bea
\label{eq:Fsca}
\vecF_\sct(\veck) = \rho_\phi v^2 \times \sigmat(k) \times \hat{\veck}.
\eea
$\sigmat$ is the momentum-transfer cross section, which is represented as
\bea
\label{eq:sigma_eff}
\sigmat(k) = \int d \Omega \, \abs{f(\theta;k)}^2 (1-\cos\theta). 
\eea
The factor $(1-\cos\theta)$ is the $\vecq$'s projection on the $\veck$-direction, and it suppresses the scattering amplitude
contribution from the forward direction. The force contributions being perpendicular to $\veck$ are canceled out due to the cylindrical symmetry of the system with respect to the $\veck$-axis. In the meantime, using $f(\theta;k)$ we could acquire the total scattering cross section
\bea
\label{eq:sigma}
\sigma(k) = \int d \Omega \, \abs{f(\theta;k)}^2.
\eea

\subsection{Background-Induced Force}

In the meantime, the scattering process induces the spatial inhomogeneity of the scalar distribution. When a test mass $\testmass$ moves inside such a scalar background, its mass varies with the spatial coordinate. Therefore, the test mass experiences an extra force
\bea
\label{eq:Fbg}
\vecF_\bg = - \vecnabla V_\bg.
\eea
In the above formula, $V_\bg$ represents the potential induced by the spatial variation of the ultralight scalar, and $\vecF_\bg$ is the corresponding background-induced force. $V_\bg$ is represented as
\bea
\label{eq:Vbg}
V_\bg(\vecr;\veck) & = \frac{\mMtest^2 \mathcal{V}_\testmass}{4} \big( \abs{\psi(\vecr;\veck)}^2 - \abs{\psi_0}^2\big) && \quad \text{(Exact)}\\
& \simeq \frac{\mMtest^2 \mathcal{V}_\testmass}{2} \Re\left[\psiinc^*(\vecr;\veck) \, \psisc(\vecr;\veck)\right]  && \quad (\abs{\psisc} < \abs{\psiinc}).
\eea
Here, $\mMtest$ denotes the effective mass induced by the test mass, and $\mathcal{V}_\testmass$ is its volume. In the parameter space we are interested in, the oscillation frequency of the ultralight scalar is much higher than the response frequencies of the experiments. For this reason, in \Eq{Vbg}, we do the time average over the background-induced potential. ``$\simeq$'' holds in the region satisfying $\abs{\psisc} < \abs{\psiinc}$, otherwise the contribution from $\abs{\psisc}^2$ should be included as shown in \Eq{V_bkg_timeave_appx_2}. For more detailed and quantitative discussions on the movement of the test mass in the scalar background, one can refer to \Appx{test_mass_bkg}. 

Substituting \Eq{psi_in_sca} into \Eq{Vbg}, we get the potential of $\testmass$ induced by the scalar scattering with $\sourcemass$ in the far-field limit, which is
\bea
\label{eq:Vbg_far}
V_\bg(\vecr;\veck) \simeq \frac{\rho_\phi}{m_0^2} \, \mMtest^2 \mathcal{V}_\testmass \abs{f(\theta;k)} \frac{\cos(k r - \veck\cdot \vecr + \arg f(\theta;k))}{r}\quad \quad \text{(Far-Field)}.
\eea
From the above equation, we see that the strength of the background-induced potential is determined by the factor $f(\theta;k)$ and $\mMtest^2 \mathcal{V}_\testmass$. The information about the source mass $\sourcemass$, including its size and the magnitude of the matter effect, is encoded in the scattering amplitude $f(\theta;k)$, while the information about the test mass is encoded in the factor $\mMtest^2 \mathcal{V}_\testmass$. The factor $\rho_\phi/m_0^2$ arises from replacing $\abs{\psi_0}$ with $\sqrt{2\rho_\phi}/m_0$. Given that $V_\bg \propto 1/m_0^2$, the background-induced force becomes more sensitive when the scalar mass becomes smaller. Notably, \Eq{Vbg_far} includes the term $\cos(kr - \veck \cdot \vecr + \arg f(\theta;k))$. Therefore, the background-induced potential experiences rapid oscillation when the large-distance or large-momentum condition $kr > 1$ is satisfied. More detailed discussion and comparisons with previous works will be provided in \Sec{bg_force}.

\begin{figure}[h!]
\centering
\includegraphics[width=0.45\linewidth]{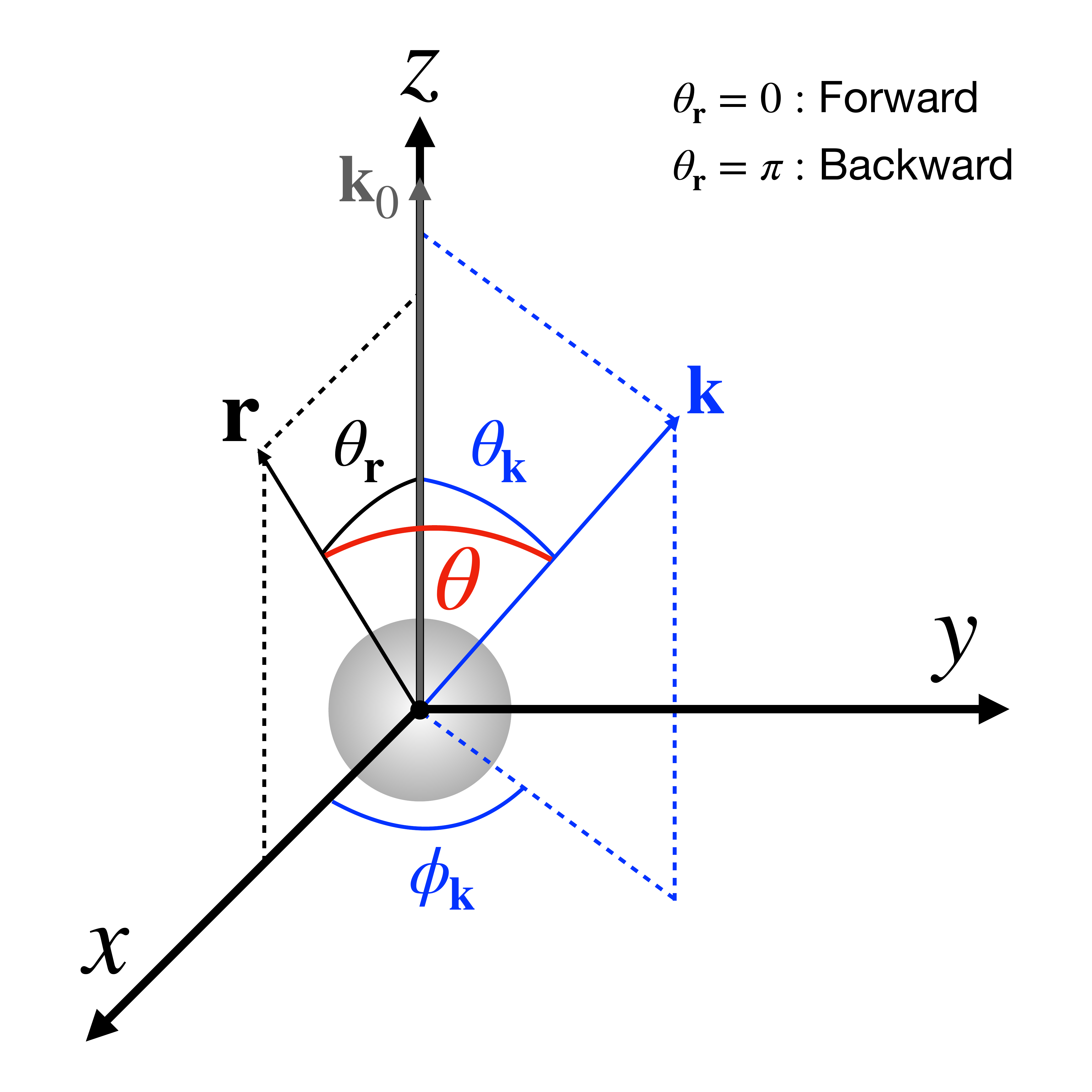}
\caption{Coordinate frame for phase space integration. The mean momentum $\veck_0$ of the dark matter is aligned with the z-axis, and the position vector $\vecr$ lies in the $x\text{-}z$ plane at an angle $\theta_\vecr$ relative to $\veck_0$. The forward direction corresponds to $\theta_\vecr = 0$, while the backward direction corresponds to $\theta_\vecr = \pi$. A monochromatic mode in the phase space distribution is represented by $\veck$, which has a polar angle $\theta_\veck$ and an azimuthal angle $\phi_\veck$. $\theta$, defined as the angle between the monochromatic momentum $\veck$ and the position vector $\vecr$, is shown in \Fig{qm_sca} and is referred to as the deflection angle in quantum mechanical scattering processes.}
\label{fig:frame}
\end{figure}

\subsection{Phase Space Distribution}\label{subsec:phase_space_dist}

In a realistic situation, dark matter in the Milky Way halo has a momentum distribution, which can be approximated as the Maxwell-Boltzmann distribution. Because of the relative movement between the solar system and the Milky Way halo, the dark matter phase space follows the boosted Maxwell-Boltzmann distribution\footnote{In the limit $\sigmak \rightarrow 0$, the phase-space distribution becomes $f_\phi(\veck) = (2\pi)^3 n_\phi \delta^{(3)}(\veck - \veck_0)$, commonly referred to as the monochromatic distribution characterized by momentum $\veck_0$. In this case, all incident waves share the same momentum. For notational convenience, we will also use $\veck$ to denote the monochromatic incident wave where appropriate.}
\bea
\label{eq:f_dm}
f_\phi(\veck) = n_\phi \left(\frac{2\pi}{\sigmak^2}\right)^{3/2} \exp\bigg[{-\frac{(\veck - \veck_0)^2}{2 \sigmak^2}}\bigg],
\eea
which satisfies the normalization condition $\int \dbar^3 \, \veck \, f_\phi(\veck) = n_\phi$. Here, $\int \dbar^3 \veck \equiv \int d^3 \veck/(2\pi)^3$. $n_\phi$ is the number density of the ultralight dark matter,  $\veck_0 = m_0 \vecv_0$ is the mean momentum, $\vecv_0 = - \vecv_\oplus$ is the mean velocity of the dark matter wind, $v_\oplus \simeq 238 \, \km/s$ is the velocity of the solar system in the Milky Way frame, $\sigmak = m_0 \, \sigmav$ is the momentum dispersion, and $\sigmav \simeq v_0/\sqrt{2}$ is the velocity dispersion~\cite{bland2016galaxy,Baxter:2021pqo}. To integrate over the phase space, we choose the coordinate system as shown in \Fig{frame}. We align $\veck_0$ with the z-axis. In this coordinate system, the momentum vector $\veck$ and the position vector $\vecr$ are
\bea
\left\{ 
\begin{aligned}
\veck & = (k \sin\theta_\veck \cos\phi_\veck, k \sin \theta_\veck \sin \phi_\veck, k\cos\theta_\veck)\\
\vecr & = (r \sin\theta_\vecr, 0, r \cos\theta_\vecr)
\end{aligned}
\right. .
\eea
The vector $\vecr$ is set on the $x\text{-}z$ plane by choosing the coordinate system accordingly. Given this, we know that the deflection angle $\theta$ is represented as
\bea
\label{eq:cos_theta}
\cos \theta =  \hat{\veck} \cdot \hat{\vecr} = \sin \theta_\veck \cos\phi_\veck \sin \theta_\vecr + \cos\theta_\veck \cos\theta_\vecr. 
\eea
In the above equation, $\hat{\veck}=\veck/|\veck|$ and $\hat{\vecr}=\vecr/|\vecr|$ represent the unit vectors in the directions of the momentum vector $\veck$ and the position vector $\vecr$, respectively

The scattering force given the phase space distribution in \Eq{f_dm} can be written as
\bea
\label{eq:Fsc_average}
\vecF_\sct = \frac{1}{n_\phi} \int \dbar^3 \, \veck \,  f_\phi(\veck) \, \vecF_\sct(\veck).
\eea
More specifically, we have
\bea
\label{eq:Fsc_average_2}
\vecF_\sct = \hat{\veck}_0 \times \frac{\rho_\phi v_0^2}{ (2\pi)^{1/2} \sigmak k_0^4 } \int^\infty_0 dk \ k^2 \sigmat(k) \left[ \left( k k_0 - \sigmak^2 \right) \exp\left(- \frac{ (k-k_0)^2 }{2 \sigmak^2}\right) + \left( k k_0 + \sigmak^2 \right)  \exp\left(- \frac{ (k+k_0)^2 }{2 \sigmak^2}\right) \right].
\eea
We have numerically verified that when compared with directly using the scattering force under the monochromatic phase space distribution, the exact integration over the phase space only adds an extra $\mathcal{O}(1)$ factor. For this reason, we use the monochromatic phase space distribution in \Sec{scforce} for a more intuitive and simplified illustration.

The background-induced force given the phase space distribution in \Eq{f_dm} is written as
\bea 
\label{eq:Fbg_average}
\vecF_\bg = - \nabla \langle V_\bg \rangle_\veck,
\eea
where the phase-space averaged background-induced potential is
\bea
\label{eq:Vbg_phaseint}
\langle V_\bg \rangle_\veck = \frac{1}{n_\phi}\int \dbar^3 \veck \, f_\phi(\veck) V_\bg(\vecr;\veck). 
\eea
We substitute \Eq{f_dm} and write \Eq{Vbg_phaseint} as
\bea
\label{eq:Vbg_phaseint_2}
\langle V_\bg \rangle_\veck = \left(\frac{1}{2 \pi \sigmak^2} \right)^{3/2} \int_0^{2\pi} d \phi_\veck \int_0^\pi d\theta_\veck \sin\theta_\veck \int_0^{\infty} dk \, k^2 \exp\left(- \frac{k^2+k_0^2}{2 \sigmak^2}\right) \exp\left(\frac{k k_0 \cos\theta_\veck}{\sigmak^2} \right) V_\bg(\vecr;\veck). 
\eea
Unlike the scattering force, the evaluation of the background-induced force requires integrating over the full phase-space distribution in \Eq{f_dm}, as shown in \Eq{Vbg_phaseint}. Because $V_\bg(\vecr;\veck)$ is oscillatory in both position and momentum, the average in \Eq{Vbg_phaseint} induces cancellations among different momentum components, which can substantially reduce the background-induced potential. See \Sec{bg_force} for a quantitative discussion. In this section, we merely give some overviews of the methodologies applied in the later discussion. To evaluate the \Eq{Vbg_phaseint_2}, one needs to do 3D integration. However, in the forward~($\theta_\vecr = 0$) and backward~($\theta_\vecr = \pi$) directions, the dependence on $\phi_\veck$ vanishes as one can see in \Eq{cos_theta}. In these two cases, one can reduce the 3D integration to 2D integration, which makes numerical and analytical evaluations easier. The background-induced potential in other directions~($0 < \theta_\vecr < \pi$) lies between the results in the forward and backward directions. Therefore, we can use these two benchmark directions to set the upper and lower limits on the physical effects of the background-induced potential.

Another possible simplification is averaging over $\hat{\veck} \cdot \hat{\vecr}$ when evaluating the background-induced potential. By doing this, one can get rid of the $\vecr$-dependence of $V_\bg(\vecr;\veck)$ and reduce the aforementioned 3D integration to 1D integration
\bea
\label{eq:Vbg_phaseint_ave}
\langle V_\bg \rangle_{\veck,\hat{\veck}\cdot\hat{\vecr}} = \frac{1}{(2 \pi)^{1/2}} \frac{1}{\sigmak k_0} \int_0^\infty dk \, k \left[ \exp\left(- \frac{(k-k_0)^2}{2\sigmak^2}\right) - \exp\left(- \frac{(k+k_0)^2}{2\sigmak^2}\right) \right] \langle V_\bg\rangle_{\hat{\veck}\cdot\hat{\vecr}} \,.
\eea
However, it is important to note that this simplification leads to significant suppression compared to results obtained using specific $\theta_\vecr$. This suppression shares similarities with the suppression caused by the isotropic distribution~\cite{Ghosh:2022nzo, VanTilburg:2024xib}. We confirm this effect through our independent evaluation in both perturbative and non-perturbative regions. Since most relevant experiments measure time-series data of accelerations or forces, relying on $\hat{\veck} \cdot \hat{\vecr}$-averaged background-induced force can lead to unphysically suppressed signals. Therefore, it is essential to compute the background-induced force for specific $\vecr$-directions, as discussed earlier.

\begin{figure}[h!]
\centering
\includegraphics[width=0.7\linewidth]{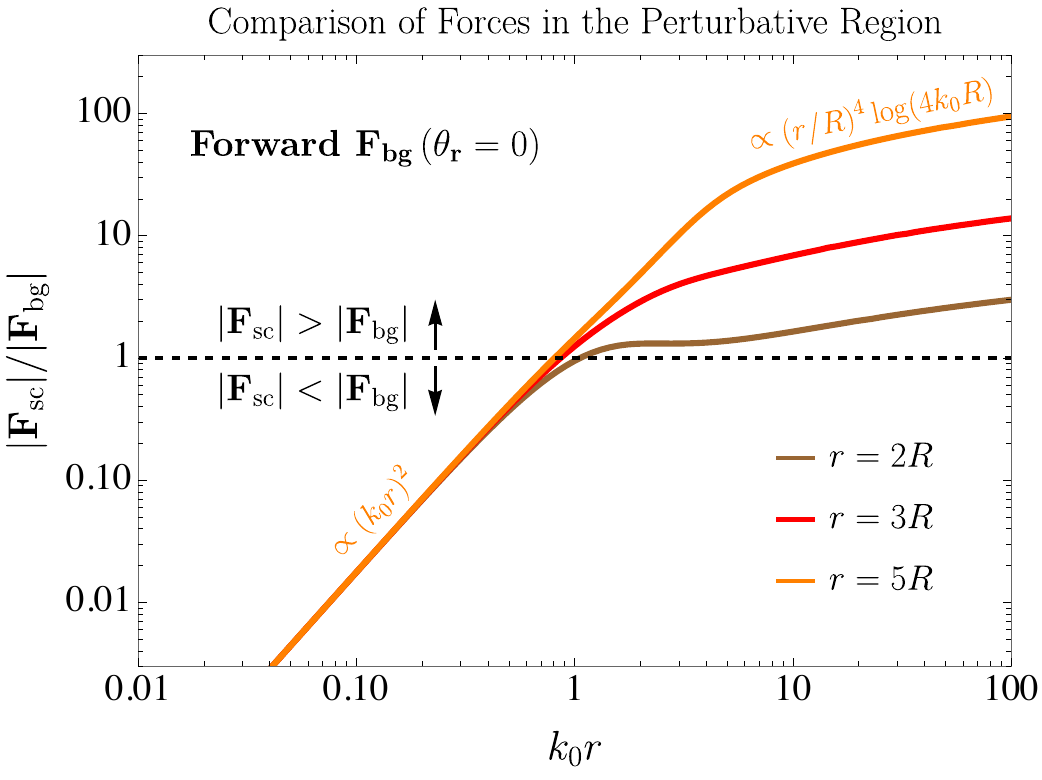}
\caption{The ratio of the scattering force over the background-induced force in the perturbative region for different $k_0$. The scattering force is measured using a sphere of radius $R$, while the background-induced force is measured using two identical spheres of the same radius and material, separated by a distance $r$. We plot the ratio $|\vecF_\sct|/|\vecF_\bg|$ for separations $r=2R$~(brown), $r=3R$~(red), and $r=5R$~(orange). When $k_0 r >1$, the scattering force dominates. When $k_0 r <1$, the background-induced force dominates. For the background-induced force, we assume the most optimistic sensitivity by considering the forward direction~($\theta_\vecr = 0$). In other directions~($0 < \theta_\vecr \leq \pi$), $\abs{\vecF_\bg}$ decreases by one to two orders of magnitude when $k_0 r > 1$.}
\label{fig:Fsc_vs_Fbg}
\end{figure}

\subsection{$\vecF_\sct$ vs $\vecF_\bg$}\label{sec:Fsc_vs_Fbg}

Before proceeding, we briefly compare the scattering force and the background-induced force in the perturbative regions, corresponding to Regions~A and~B in \Fig{phase_classify}, where the Born approximation is valid. Although these two effects can be tested through different experimental setups, exploring their relative magnitude is still valuable for better understanding. Here, we quote the results from \Subsec{perturbative_sc} and \Subsec{perturbative_bg} in the following discussions. For the scattering force, we consider a sphere with radius $R$. For the background-induced force, we choose a system with two identical spheres composed of the same material, having the same radius $R$, and separated by a distance of $r$. From \Eq{region_a_sigmat} and \Eq{region_b_sigmat} in \Subsec{perturbative_sc}, we have the scattering force
\bea
\label{eq:Fsc_compare}
\abs{\vecF_\sct} \simeq \rho_\phi v_0^2 \times \frac{(\mM^2 \mathcal{V})^2}{4\pi} \times \min\left[1, \frac{9}{8} 
\frac{\log(4k_0 R)}{(k_0 R)^4} \right],
\eea
where $\mathcal{V}$ is the volume of the spheres, and $\mM$ is the extra effective mass induced by the test spheres. From \Eq{Fsc_compare}, we find that when $kR \gtrsim 1$, there is an extra suppression factor $1/(kR)^4$. This is caused by the finite-size decoherence effect, which largely reduces the scattering amplitude. 

From \Subsec{perturbative_bg}, we have the background-induced force 
\bea
\label{eq:Fbg_compare}
\abs{\vecF_\bg} \simeq \rho_\phi v_0^2 \times \frac{(\mM^2 \mathcal{V})^2}{4\pi} \times \frac{1}{(k_0 r)^2} \times \min\left[1, \frac{12}{(k_0 r)^2}\right] \quad \quad (r \geq 2R). 
\eea
From \Eq{Fbg_compare}, we find that there is an extra suppression factor $1/(k_0 r)^2$ caused by the cancellation when integrating over the oscillatory $V_\bg$, which is also the decoherence effect caused by integrating over the finite experimental objects and finite phase space. In \Eq{Fbg_compare}, we choose the forward direction~($\theta_\vecr = 0$), which gives the most optimistic background-induced force. In other directions, especially the backward direction~($\theta_\vecr = \pi$), the background-induced force in the $k_0 r > 1$ region can be smaller by one or two orders of magnitude.

From \Eq{Fsc_compare} and \Eq{Fbg_compare}, we find that in the perturbative region, $\abs{\vecF_\sct}$ and $\abs{\vecF_\bg}$ have the same dependence on dark matter density $\rho_\phi$, the material of the experimental objects, the strength of the scalar-photon interaction, and the volumes of the experimental objects. Therefore, when we take the ratio of these two forces, all the aforementioned parameters cancel out, indicating that $\vecF_\sct$ and $\vecF_\bg$ are physical effects of the same perturbative order $\mathcal{O}(V_\eff^2)$. For $\abs{\vecF_\sct}/\abs{\vecF_\bg}$, only three parameters determine this ratio: dark matter mean momentum $k_0$, distance $r$~\footnote{Specifically, $r$ denotes the distance between the centers of the source mass $\sourcemass$ and the test mass $\testmass$. In this subsection, i.e. \Sec{Fsc_vs_Fbg}, both $\sourcemass$ and $\testmass$ have the same radius $R$, so $r = 2R$ corresponds to the case where the two spheres are in direct contact. In \Sec{bg_force}, the source mass $\sourcemass$ has radius $R$ whereas the test mass $\testmass$ is treated as point-like, and thus $r = R$ represents direct contact between $\sourcemass$ and $\testmass$.}, and size $R$. By choosing $r \sim \text{a few} \times R$, we find that\footnote{We use \Eq{Fbg_vs_Fsc_crit} rather than fixing a concrete numerical value, because this condition is independent of other factors, such as the geometric configuration of the setup (parameterized by $r/R$), the dark matter density $\rho_\phi$, the effective mass $\mM$, and the volume $\mathcal{V}$.}
\bea
\label{eq:Fbg_vs_Fsc_crit}
k_0 r =1
\eea
is the critical value that determines whether $\abs{\vecF_\bg}$ or $\abs{\vecF_\sct}$ dominates. Here, $k_0 r$ is a dimensionless quantity that can be recast in terms of the scalar bare mass $m_0$ for a given geometric configuration. When $k_0 r < 1$, $\abs{\vecF_\bg}$ dominates over $\abs{\vecF_\sct}$. In contrast, when $k_0 r > 1$, $\abs{\vecF_\sct}$ dominates over $\abs{\vecF_\bg}$. For a more quantitative illustration, see \Fig{Fsc_vs_Fbg}. This figure shows the ratio $\abs{\vecF_\sct}/\abs{\vecF_\bg}$ for separations $r=2R$~(brown), $r=3R$~(red), $r=5R$~(orange). To maximize the background-induced force, we consider the forward direction in the plot. The asymptotic behaviors of these three lines in the small and large $k_0 r$ limits agree well with the analytical expressions in \Eq{Fsc_compare} and \Eq{Fbg_compare}, as shown in the plot.

Based on the discussion in this subsection, we conclude that the physical effects induced by these two forces are sensitive to different mass ranges of dark matter and have different scaling laws in the parameter space labeled by the $(m_0, \Lambda_\gamma^{-1})$. Therefore, in the following sections, we treat these two forces as independent effects and separately discuss them in the next two sections.

\section{Scattering Force}\label{sec:scforce}

In this section, we discuss the effect of the ultralight dark matter scattering with the test object, which induces detectable anomalous acceleration for the test object. This topic has been previously explored in the literature, including the analyses in the perturbative Region A~(long-wavelength limit)~\cite{Fukuda:2018omk,Fukuda:2021drn,Day:2023mkb} and the non-perturbative Regions C and D~\cite{Day:2023mkb}. 

In this work, we provide a fine-grained scanning of the perturbative Regions A and B, and the non-perturbative Regions C, D, and E$\setminus$B. We derive a complete analytical expression for the Born approximation, covering both Regions A and B. Additionally, we find that the acceleration peaks at the boundary between these two regions, where the finite-size effect terminates the coherent enhancement in the long-wavelength limit. To further understand the transition from perturbative to non-perturbative regimes, we provide a detailed comparison between the Born approximation and partial wave analysis. In addition, we discuss the non-perturbative Region E$\setminus$B and point out its distinct characteristics. Furthermore, we investigate the descreening effect in Region D, which alleviates the decoherence suppression. One shall also note that the framework based on the quantum mechanical scattering theory is also suitable for discussing the background-induced force in \Sec{bg_force}.

In the following discussion, we analyze both perturbative and non-perturbative regions in order. We start with the perturbative Regions A and B, followed by a discussion on the perturbativity condition. We then explore the non-perturbative Regions C, D, and E$\setminus$B in detail. Finally, we discuss experimental sensitivities, including the sensitivities in current and proposed experiments, as well as the theoretical sensitivities derived by comparing thermal/quantum fluctuations. In this section discussing the scattering force, since we focus exclusively on the test mass, we use a simplified notation, and denote the effective mass, radius, and volume of the test mass by $\mM$, $R$, and $\mathcal{V}$, respectively, where $\mathcal{V} = 4\pi R^3/3$. In \Sec{bg_force}, where both the test and source masses are relevant to the discussion of the background-induced force, we assign the labels ``$\testmass$'' to the test mass and ``$\sourcemass$'' to the source mass.

\subsection{Perturbative Region}\label{subsec:perturbative_sc}

In this subsection, we discuss the scattering force in the perturbative regions, including Region A~($R^{-1} > \mM, k$) and Region B~($k > R^{-1}$ and $(kR^{-1})^{1/2} > \mM$). In these regions, the scattered wave function is perturbative, i.e., $\abs{\psi_\sct} < \abs{\psiinc}$. We summarize this subsection by showing the scaling law of the acceleration in the perturbative region as
\bea
a \propto \mM^4 \times
\left\{
\begin{aligned}
& R^3    & \quad \,\, (R<k^{-1})\\
& R^{-1} & \quad \,\, (R>k^{-1})
\end{aligned}
\right. ,
\eea
where $m_0$ is fixed, and the dimension of the acceleration is balanced by $k$. In Region A, acceleration is enhanced by coherent scattering, while in Region B, coherent enhancement is eliminated by the finite-size decoherence effect. We find that the test object size can be optimized to maximize sensitivity to ultralight dark matter of a specific mass. For instance, for dark matter with $m_0 \sim 10^{-2}\,\eV$, the optimal spherical radius is around $R \sim 3 \,\cm$. Since the exact mass of dark matter remains unknown, an experimental setup incorporating spheres of varying sizes can enhance sensitivity across a broader mass range. In the rest of the subsection, we will introduce the Born approximation, provide a detailed discussion of Regions A and B, and determine the perturbative condition.

Because the entire calculation in this section is performed in the perturbative regime, the use of the Born approximation is well justified. As we are only focusing on the momentum distribution in the asymptotically far region, we directly compute the scattering amplitude using
\bea \label{eq:f2}
f(\theta;k) = - \frac{\mM^2}{4 \pi} \int_{\mathcal{V}} d^3 \delta \vecr \,\,  e^{-i \vecq \cdot \delta \vecr}
\eea
throughout this subsection. A more detailed discussion of \Eq{f2} can be found in \Appx{wave_func_appendix}.

\begin{figure}[h]
\centering
\begin{tikzpicture} 
\node at(-8.8,0){\includegraphics[width=0.45\columnwidth]{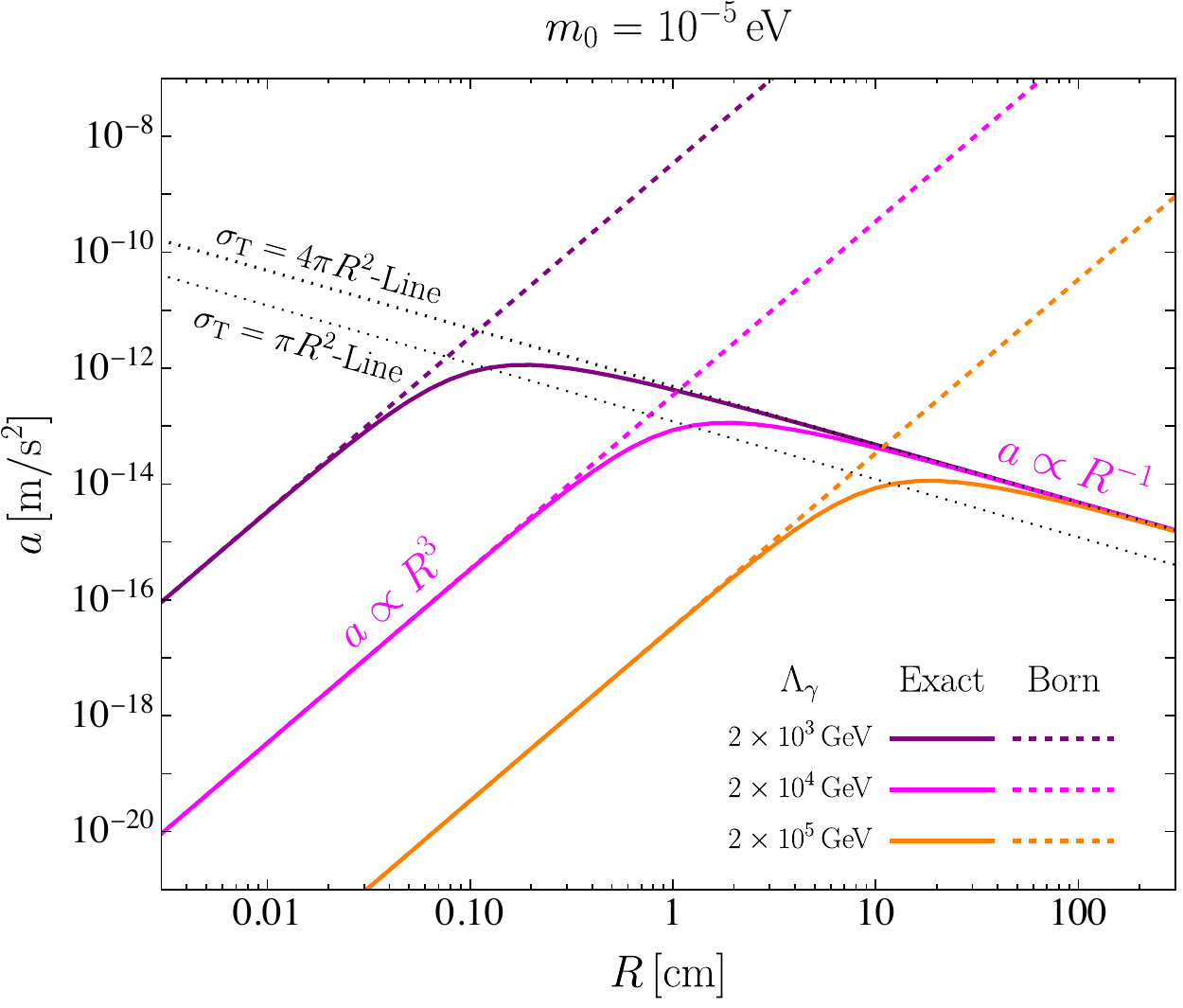}};
\node at (0.8,0){\includegraphics[width=0.45\columnwidth]{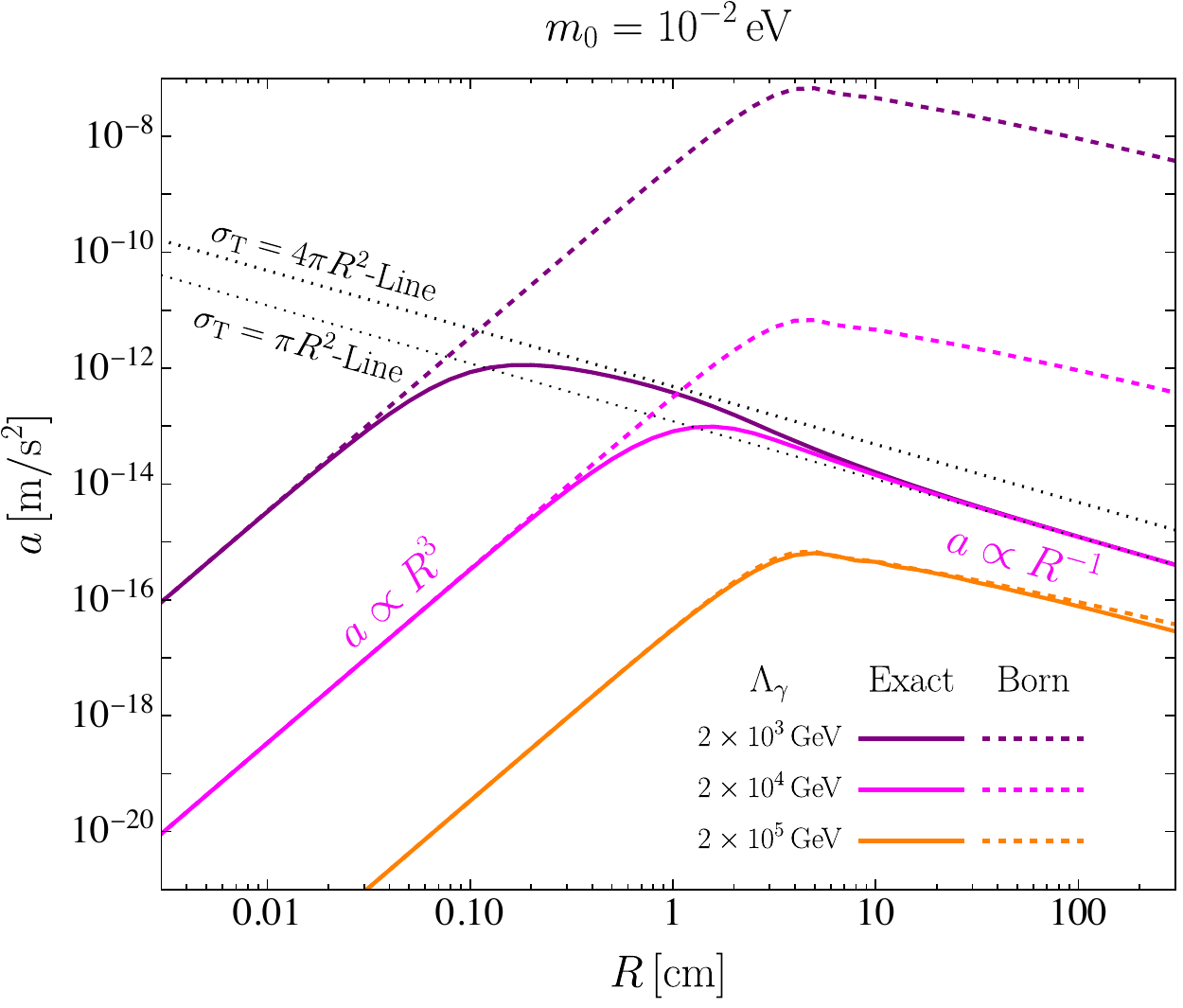}};
\end{tikzpicture}
\caption{The dark matter-induced acceleration as a function of the Platinum spherical radius. Solid lines represent the exact accelerations calculated using the partial wave analysis, while dashed lines represent the accelerations obtained via the Born approximation.  The $a$-$R$ curves for different values of $\Lambda_\gamma$ are shown in purple, magenta, and orange, respectively. In the small-$R$ region, where $\mM, k < R^{-1}$~(Region A), the solid lines closely match the dashed lines. {\bf Left}: $m_0 = 10^{-5}\,\eV$. The maximum accelerations occur at $R \sim \mM^{-1}$. As $R$ increases further, the acceleration approaches saturation, following the $\sigmat = 4 \pi R^2$ line. {\bf Right}: $m_0 = 10^{-2}\,\eV$. The purple and magenta lines exhibit maximum accelerations at $R \sim \mM^{-1}$. As $R$ increases, these lines initially follow the $\sigmat = 4 \pi R^2$ line before transitioning to the $\sigmat = \pi R^2$ line when $R > k^{-1}$. The orange solid line aligns closely with the orange dashed line, as it lies in the weakly coupled region. As $R$ increases, it reaches its maximum at $R \sim k^{-1}$ and subsequently decreases, following a power-law behavior of $a \propto R^{-1}$.}
\label{fig:R_a}
\end{figure}

In Region A, $e^{-i \vecq \cdot \delta \vecr}$ remains approximately one, meaning that the geometric shape of the scattered object is negligible. Utilizing \Eq{sigma_eff}, we derive
\bea
\label{eq:region_a_sigmat}
\text{Region A}: \quad \sigmat = \frac{(\mM^2 \mathcal{V})^2}{4 \pi}. \label{eq:sigmaA}
\eea
The above formula actually describes coherent enhancement, given by $f(\theta) = \mathcal{N}_\testmass f_\text{single}(\theta)$. $\mathcal{N}_\testmass$ is the test particle number, and $f_\text{single}$ is the scattering amplitude of a single test particle. In the long-wavelength perturbative limit, the scattering force is scaled as $F_{\sct} \propto \mathcal{V}^2 \propto R^6$. As the total test mass also increases as $\mathcal{M}_\testmass \propto R^3$, the acceleration induced by the scattering force obeys $a_{\sct} \propto \mathcal{V}^2/\mathcal{M}_\testmass \propto R^3$, explaining the scaling law in the small-$R$ limit in \Fig{R_a}.

In Region B, because $e^{-i \vecq\cdot \delta \vecr}$ is highly oscillatory when computing \Eq{f2}, we need to integrate over the concrete geometric shape of the test mass. Let us consider a uniform sphere with radius $R$, then \Eq{f2} can be analytically expressed as
\bea\label{eq:f3}
f(\theta;k) = - \mM^2 R^3 \times \frac{\sin(qR)- qR \cos(qR)}{(qR)^3}.
\eea
In the long-wavelength regime $kR < 1$~(Region A), the above expression returns to \Eq{f2} as expected. Although \Eq{f3} is derived for a uniform spherical potential, we expect an $\mathcal{O}(1)$ factor difference when the detailed shape of the bulk matter is considered. Substituting \Eq{f2} into \Eq{sigma_eff} and expanding $\sigmat$ in the large-$R$ limit, we have
\bea
\label{eq:region_b_sigmat}
\text{Region B}: \quad \sigmat = \pi R^2 \times \left(\frac{\mM}{k}\right)^4 \times \frac{\log(4kR)}{2},
\eea
which is written in a form suitable for comparison with the geometric cross section $\pi R^2$. Knowing that the total test mass scales as $\mathcal{M}_\testmass \propto R^3$, we have $a \propto \sigmat/\mathcal{M}_\testmass \propto R^{-1}$ scaling in the large-$R$ limit. The variation in the $\log(4kR)$ is ignored, as it introduces only an $\mathcal{O}(1)$ difference within the relevant parameter range. Equivalently, we can also write the momentum-transfer cross section as
\bea
\label{eq:region_b_sigmat_2}
\text{Region B}: \quad \sigmat = \frac{(\mM^2 \mathcal{V})^2}{4\pi} \times \frac{9}{8} \frac{\log(4kR)}{(kR)^4},
\eea
which is written in a form that enables comparison with  \Eq{region_a_sigmat}, the momentum-transfer cross section in Region A. In Region B, coherent enhancement is terminated as the dark matter wind begins to resolve the finite size of the test mass. This breakdown of coherent enhancement is referred to as the finite-size decoherence effect on the scattering force. For detailed derivations of the scattering cross section $\sigma$ and momentum-transfer cross section $\sigmat$ in the perturbative region (Regions A and B), one can refer to \Appx{sca_appendix}.

Before concluding the discussion on the perturbative region and transitioning to the non-perturbative regime, we must determine the conditions under which perturbativity breaks down. The key criterion is when the scattering cross-section in \Eq{sigma_appx} saturates $\sim \pi R^2$. We avoid using the momentum-transfer cross section for this comparison because, as seen in \Eq{sigma_eff}, the non-perturbative wave function contributions in the forward direction are suppressed, making the condition weaker. For $kR < 1$, Born approximation gives $\sigma \sim \pi R^2 \times (\mM R)^4$. Thus, perturbativity holds as long as $R^{-1} > \mM$. For $kR > 1$, Born approximation gives $\sigma \sim \pi R^2 \times (\mM^4 R^2/k^2)$, giving the perturbativity condition $(kR^{-1})^{1/2} > \mM$. To derive the aforementioned perturbativity condition directly from comparing the wave functions, one can refer to \Appx{wave_func_appendix}.

\subsection{Non-Perturbative Region}\label{subsec:non_perturb1}

As the interaction strength of the quadratic scalar-photon interaction increases, we eventually reach a non-perturbative regime where $\abs{\psisc} \sim \abs{\psiinc}$ in the vicinity of the scattered object. In this situation, the general solution in \Eq{born_approx_appx} is invalid. Therefore, the partial wave analysis becomes essential, enabling us to effectively explore the non-perturbative regions and provide us with physical intuition as well. In the following subsection, we will utilize the partial wave analysis to explore Region C~$(\mM > R^{-1} > k)$ and Region D~$(\mM > k > R^{-1})$, which is non-perturbative. In these regions, the scattered object can be approximated as a hard sphere with an infinite potential. We will also discuss the non-perturbative part of the Region E~($k > \mM > (k R^{-1})^{1/2}$), which is treated as the solid sphere that is penetrated by the incident scalar. We summarize that in the non-perturbative region there is an upper bound for the acceleration
\bea
\label{eq:sca_screen}
\sigmat \lesssim \xi \times \pi R^2,
\eea
where the momentum-transfer cross section saturates as the coupling strength increases. Here $\xi$ is a dimensionless factor that varies in the $kR\text{-}\mM$ plane. In Region C~(long-wavelength limit, hard sphere), $\xi = 4$ for the sphere, meaning that the whole surface of the sphere takes part in the scattering process when saturating the upper bound. In Region D~(short-wavelength limit, hard sphere), there is $\xi=1$, which means only the dark matter flux within the geometric cross section is reflected when saturating the upper bound. In Region E$\setminus$B~(short-wavelength limit, solid sphere), $\xi = \mathcal{O}(1) \times (\mM/k)^4$, which becomes similar to the perturbative result in \Eq{region_b_sigmat}. Based on the above discussion, we find that the upper bound of the acceleration of the scattered test mass $\testmass$ is
\bea
\label{eq:a_saturate}
a_\testmass \lesssim \frac{\rho_\phi v^2_0 \times \mathcal{O}(1) \times \pi R^2}{\rho_\testmass \times \mathcal{V}} \sim 10^{-13} \text{ m}/\text{s}^2 \times \left(\frac{\rho_\text{Pt}}{\rho_\testmass}\right) \left( \frac{1\cm}{R} \right),
\eea
where $\rho_\testmass$ is the density of the scattered test mass. To numerically evaluate the abovementioned maximal acceleration, we choose a $1\cm$ Platinum~(Pt) sphere to set the benchmark values. The density of Platinum is $\rho_\text{Pt} = 21.45\,\gram/\cm^3$. For materials with lower densities, such as Aluminum (Al), Titanium (Ti), or Beryllium (Be), the maximum acceleration can be an order of magnitude higher. From the above discussion, we can find that the non-perturbative region actually provides the upper cutoff for the induced acceleration in the strong coupling region. That is to say, when the induced acceleration saturates \Eq{a_saturate}, no matter how large the coupling is, the acceleration will never become larger. 

To further illustrate the screening effect causing the saturation of the acceleration, \Fig{R_a} shows how the acceleration of a Platinum sphere varies with $R$ for different values of $m_0$ and scalar-photon coupling strength $\Lambda_\gamma$. In this figure, solid lines represent the exact acceleration induced by the scattering force, calculated using the partial wave analysis, while dashed lines indicate the approximate acceleration computed using the Born approximation. In the left panel, we set $m_0 = 10^{-5}\,\eV$. As $R$ increases, the corresponding parameter points in the $m_0\text{-}\Lambda_\gamma$ plane sequentially transition through Regions A$\rightarrow$C$\rightarrow$D. In the small-$R$ region (Region A), the solid and dashed lines agree closely, following the power law $a \propto F_\sct/R^3 \propto R^3$. Such $a\text{-}R$ scaling law changes, and the exact results deviate from the results from the Born approximation at $R \sim \mM^{-1}$ because we enter a non-perturbative region as $R$ increases. In the large-$R$ region (Region C), which is highly non-perturbative, the exact momentum-transfer cross section saturates at $\sigmat = 4 \pi R^2$, making the acceleration to scale as $a \propto \sigmat/R^3 \propto 1/R$. Although Region D lies outside the plot range, numerical calculations confirm that the solid lines eventually approach the $\sigmat = \pi R^2$ line. In the right panel, we set $m_0 = 10^{-2}\,\eV$. For the purple and magenta lines, as $R$ increases, the parameter points swipe through Regions A$\rightarrow$C$\rightarrow$D, similar to the previous case. For the orange line, the parameter point transitions through Regions A$\rightarrow$B$\rightarrow$E as $R$ increases. In this case, the orange line bends downward at $R \sim k^{-1}$. The maximum acceleration described by \Eq{a_saturate} is not reached within reasonable $R$ values~(equivalently speaking, the parameter point associated with the orange line never enters Region E for moderate $R$). Consequently, the scattering process remains perturbative along the orange line, and the Born approximation provides a sufficiently accurate description of this region.

Even though the back-of-envelope estimation gives similar results on the maximum value of $\sigmat$, merely differing by $\mathcal{O}(0.1)$-$\mathcal{O}(1)$, the underlying physics varies in different regions. In the rest of this subsection, we will show the concrete calculation based on the partial wave analysis. To begin with, we list the momentum-transfer cross section under the framework of the partial wave analysis as
\bea
\label{eq:sigma_eff_partialwave}
\sigmat = \frac{4\pi}{k^2}\sum_{l=0}^{+\infty} (l+1) \left[ \abs{A_l}^2 + \abs{A_{l+1}}^2  - A_l^* A_{l+1} - A_l A_{l+1}^*\right]. 
\eea
$A_l$ is the amplitude of the $l$-th component of the scattered wave, and it can be represented as
\bea
\label{eq:Al_partialwave}
A_l = \frac{S_l - 1}{2} = \frac{e^{2 i \delta_l} - 1}{2}. 
\eea
$S_l$ is the S-matrix for the $l$-th component and $\delta_l$ is the phase shift. For the repulsive potential discussed in this paper, we generically have $\delta_l < 0$. Since we are dealing with elastic scattering, the S-matrix satisfies the unitarity condition $\abs{S_l}=1$, being equivalent to the probability current conservation. The momentum-transfer cross section listed in \Eq{sigma_eff_partialwave} has two parts. The first part contains the diagonal terms, and it is the contribution from the integral $\int d\Omega \, \abs{f(\theta)}^2$, which is the ordinary cross section. The second part is the one containing the off-diagonal terms, and it is the contribution from the integral $-\int d\Omega \, \cos\theta \abs{f(\theta)}^2$, which suppresses the forward direction's contribution. By matching the boundary condition at the spherical surface, we can exactly solve the phase shift of the uniform solid sphere as
\bea
\label{eq:tan_delta_l_solid_sphere}
\tan \delta_l = \frac{k j_l(k' R) j_{l+1}(k R) - k' j_l(k R) j_{l+1}(k' R)}{ k j_l(k' R) y_{l+1}(k R) - k' y_l(k R) j_{l+1}(k' R) }.
\eea
$j_l$ is $l$-th spherical Bessel function, $y_l$ is  $l$-th spherical Neumann function, $k$ is the momentum of the incident scalar dark matter, $k'=(k^2-\mM^2)^{1/2}$ is the scalar's momentum inside the solid sphere. $R$ is the radius of the sphere. For a more detailed discussion of scattering theory in the framework of the partial wave analysis, one can refer to \Appx{partial_wave_appx}. For the derivation of \Eq{sigma_eff_partialwave}, one can refer to \Appx{cross_sec_derivation_appx}. The derivations of the outgoing partial wave amplitude \Eq{Al_partialwave} and the phase shift \Eq{tan_delta_l_solid_sphere} can be found in \Appx{phase_shifts_appx}.

In Region C, we are discussing the low-momentum hard sphere scattering. Because the incident momentum $k$ is much smaller than $R^{-1}$, merely the s-wave component contributes to the wave function. Therefore, the scattering force is calculated as
\bea
\label{eq:sigmat_RegionC}
\text{Region C}: \quad \sigmat = \frac{4 \pi}{k^2} \abs{A_0}^2 =  4 \pi R^2. 
\eea
$A_0 = (e^{2 i \delta_0}-1)/2$ and $\delta_0$ is the phase-shift of the s-wave component. For the hard sphere low-$k$ scattering, $\delta_0 \simeq - k R$, therefore we can derive the above formula. To understand this formula more intuitively, the scattering process is in the long-wavelength limit. Therefore, the whole area of the scattered mass can participate in the scattering, and the momentum-transfer cross section is $4\pi R^2$. We can also equivalently express \Eq{sigmat_RegionC} as
\bea
\label{eq:sigmat_RegionC_2}
\text{Region C}: \quad \sigmat = \frac{(\mM^2 \mathcal{V})^2}{4\pi} \times \left[\frac{3}{(\mM R)^2}\right]^2,
\eea
where the factor $3/(\mM R)^2$ represents the characteristic suppression in Region C. Compared to the momentum-transfer cross section in Region A~(see \Eq{region_a_sigmat}), effectively only a fraction of the test mass participates in the scattering in Region C. Therefore, the suppressed cross section in Region C relative to Region A can be interpreted as the screening effect. This screening factor also appears in the discussion of the background-induced force in the far-field limit. The relevant discussion is provided in \Subsec{bg_nonp_lowk} and explicitly illustrated in \Eq{amplitudecalc_formfactor_sphscr_farfield_hard}. This similarity arises from their shared mathematical origin: the ratio of the scattering amplitude $f(\theta;k)$ between Regions C and A.

In Region D, we are dealing with the high-momentum hard sphere scattering. Because the incident momentum $k$ is much larger than $R^{-1}$, high-$l$ components of the wave function are excited. To acquire the correct result, we must sum over all the partial waves from $l=0$ to $l_{\max} \sim kR$. From the numerical calculation, we find that~\footnote{We notice that the scattering cross section is $\sigma = 2 \pi R^2$ in this region, which is different from $\sigmat = \pi R^2$. This is because the cross section $\sigma$ contains the classical reflection part~($\sigma_\text{refl} = \pi R^2$) and the shadow part~($\sigma_\text{shadow} = \pi R^2$). Only the classical part of the wave function contributes to the scattering force.}
\bea
\label{eq:sigmat_RegionD}
\text{Region D}: \quad \sigmat = \frac{4\pi}{k^2}\sum_{l=0}^{l_{\max}} (l+1) \left[ \abs{A_l}^2 + \abs{A_{l+1}}^2  - A_l^* A_{l+1} - A_l A_{l+1}^*\right] \simeq \pi R^2. 
\eea
The $l$-th partial wave phase shift satisfies $\tan \delta_l = j_l(k R)/y_l(k R)$, as derived in \Eq{tan_delta_l_hard_sphere_appx}, and can be used to compute $A_l$. To explain the above formula, the scalar dark matter behaves more particle-like in this region. Therefore, the scattering process can be treated as classical hard-sphere scattering. The classical treatment naturally gives the above formula, and it is interpreted as the geometric cross section of the scattered object. We can equivalently write \Eq{sigmat_RegionD} as
\bea
\label{eq:sigmat_RegionD_2}
\text{Region D}: \quad \sigmat \simeq \frac{(\mM^2 \mathcal{V})^2}{4\pi} \times \frac{1}{4}\left[\frac{3}{(\mM R)^2}\right]^2,
\eea
which can be used to compare with the result in Region A. From \Eq{sigmat_RegionD_2}, one can see that even in the region where the decoherence becomes significant~($kR>1$), the momentum-transfer cross section only reduces by a factor of $1/4$ compared to the result in Region C, as shown in \Eq{sigmat_RegionC_2}. This can be explained by the descreening enhancement in the large momentum region, which compensates the finite-size decoherence suppression shown in \Eq{region_b_sigmat_2}. Similar phenomena also occur in the background-induced force, as will be discussed in \Subsec{bg_nonp_highk}. The difference is that the scattering force is subject to finite-size decoherence suppression, whereas the background-induced force is subject to the decoherence suppression from the combination of finite-phase-space and finite-size effects.

\begin{figure}[t]
\centering
\begin{tikzpicture}
\node at(-8.8,0){\includegraphics[width=0.45\columnwidth]{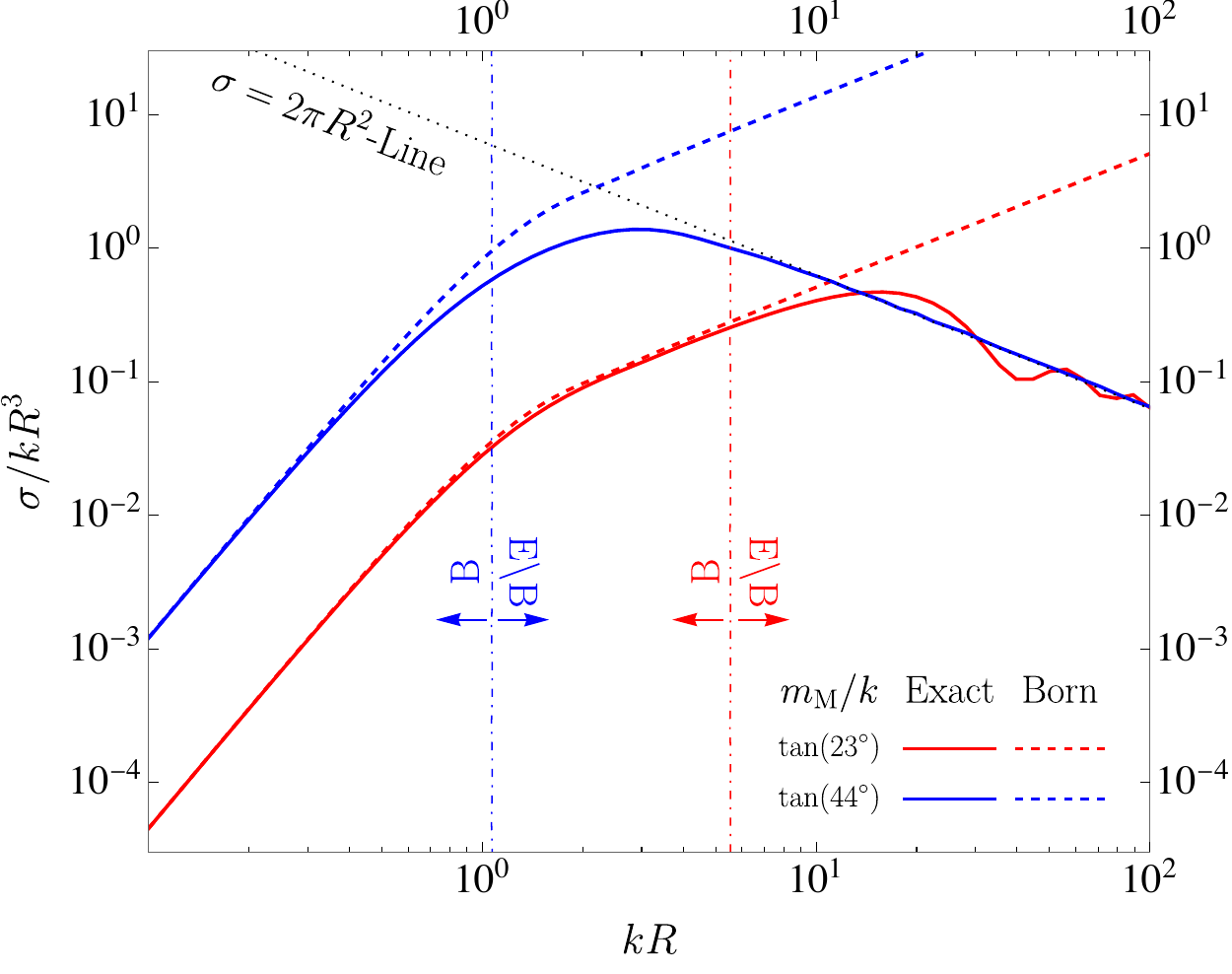}};
\node at (0.8,0){\includegraphics[width=0.45\columnwidth]{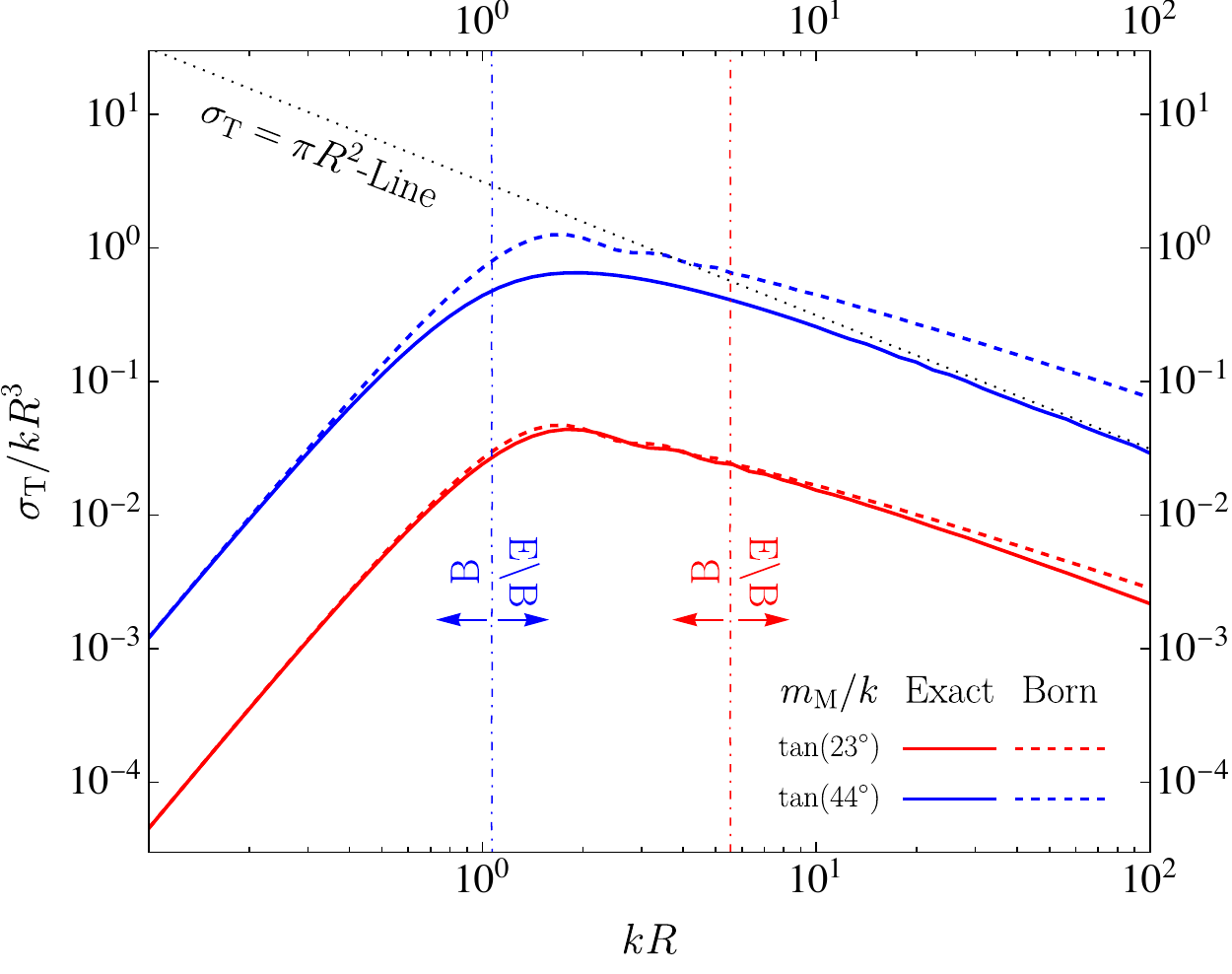}};
\end{tikzpicture}
\caption{We present the $kR$-$\sigma/k R^3$ and $kR$-$\sigmat/k R^3$ diagrams to illustrate the transition from the perturbative Region B to the non-perturbative Region E$\setminus$B. The solid lines represent the exact results calculated using the partial wave analysis, while the dashed lines correspond to the results obtained from the Born approximation. The red and blue lines denote marginal $(k,\mM)$ values that enter Region E$\setminus$B in large-$R$ limit. {\bf Left}: Cross section divided by the cube of the radius. The dashed line diverges from the solid line due to the forward-direction contribution when substituting \Eq{f3} into \Eq{sigma} to compute the cross section. {\bf Right}: Momentum-transfer cross section $\sigmat$ divided by the cube of the radius. The blue dashed line converges because the $(1-\cos\theta)$ term in \Eq{sigma_eff} eliminates the divergence in the forward direction when substituting \Eq{f3}.}
\label{fig:Region_E_minus_B}
\end{figure}

In Region E, because $k > \mM$, the finite potential needs to be considered, which indicates that we need to deal with the solid sphere scattering. In part of the parameter space belonging to Region E but not belonging to Region B~(labeled as Region E$\setminus$B), the non-perturbative effect should be taken into consideration.  Summing all the partial wave contributions up to $l_{\max} \sim kR$, we numerically find that
\bea\label{eq:sigmat_RegionE}
\text{Region E$\setminus$B}: \quad \sigmat = \frac{4\pi}{k^2}\sum_{l=0}^{l_{\max}} (l+1) \left[ \abs{A_l}^2 + \abs{A_{l+1}}^2  - A_l^* A_{l+1} - A_l A_{l+1}^*\right] \simeq \pi R^2 \times \mathcal{O}(1) \times \left(\frac{\mM}{k}\right)^4.
\eea
The phase shift $\delta_l$ used can be found in \Eq{tan_delta_l_solid_sphere_appx} and can be used to compute $A_l$. This result differs from the perturbative calculation utilizing the Born approximation by a $\mathcal{O}(1)$ factor. In the meantime, the precise cross section $\sigma$ from the partial wave analysis differs apparently from the Born approximation result. To understand such a difference, we can go back to the definitions in \Eq{sigma} and \Eq{sigma_eff}. For the momentum-transfer cross section quantifying the scattering force, there is an extra $(1-\cos \theta)$ factor, which cancels the divergence of $\abs{f(\theta)}^2$ in the forward direction. The aforementioned divergence in the integration behaves like $\int d \Omega \, \abs{f(\theta)}^2 \propto R^6$ is exactly the reason causing the perturbative cross section $\sigma$ to diverge in the large-$R$ limit. To allow the comparison with the result in Region A, we can equivalently write \Eq{sigmat_RegionE} as
\bea\label{eq:sigmat_RegionE_2}
\text{Region E$\setminus$B}: \quad \sigmat \simeq \frac{(\mM^2 \mathcal{V})^2}{4\pi} \times \mathcal{O}(1) \times \frac{1}{(kR)^4},
\eea
which has a similar form to \Eq{region_b_sigmat_2}. This implies that, once we enter the descreened region—where the kinetic energy of the incident scalar exceeds the barrier height~($k > \mM$), the momentum-transfer cross section is suppressed due to the finite-size decoherence effect. The comparison between the exact partial-wave results and the Born approximation, and their transition from the perturbative Region B to the non-perturbative Region E$\setminus$B, is illustrated in \Fig{Region_E_minus_B}.

\begin{figure}[t]
\centering
\includegraphics[width=0.49\linewidth]{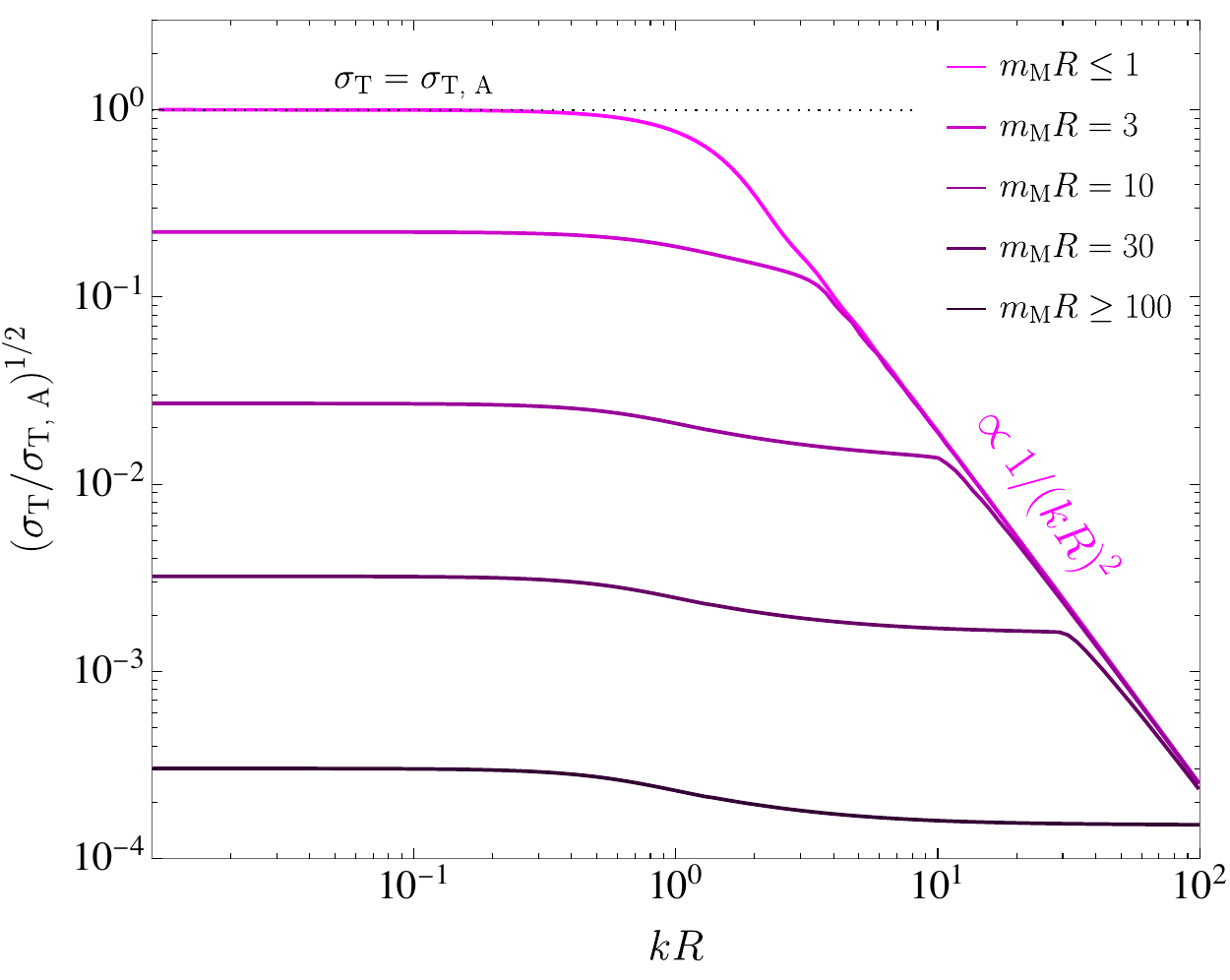}
\includegraphics[width=0.49\linewidth]{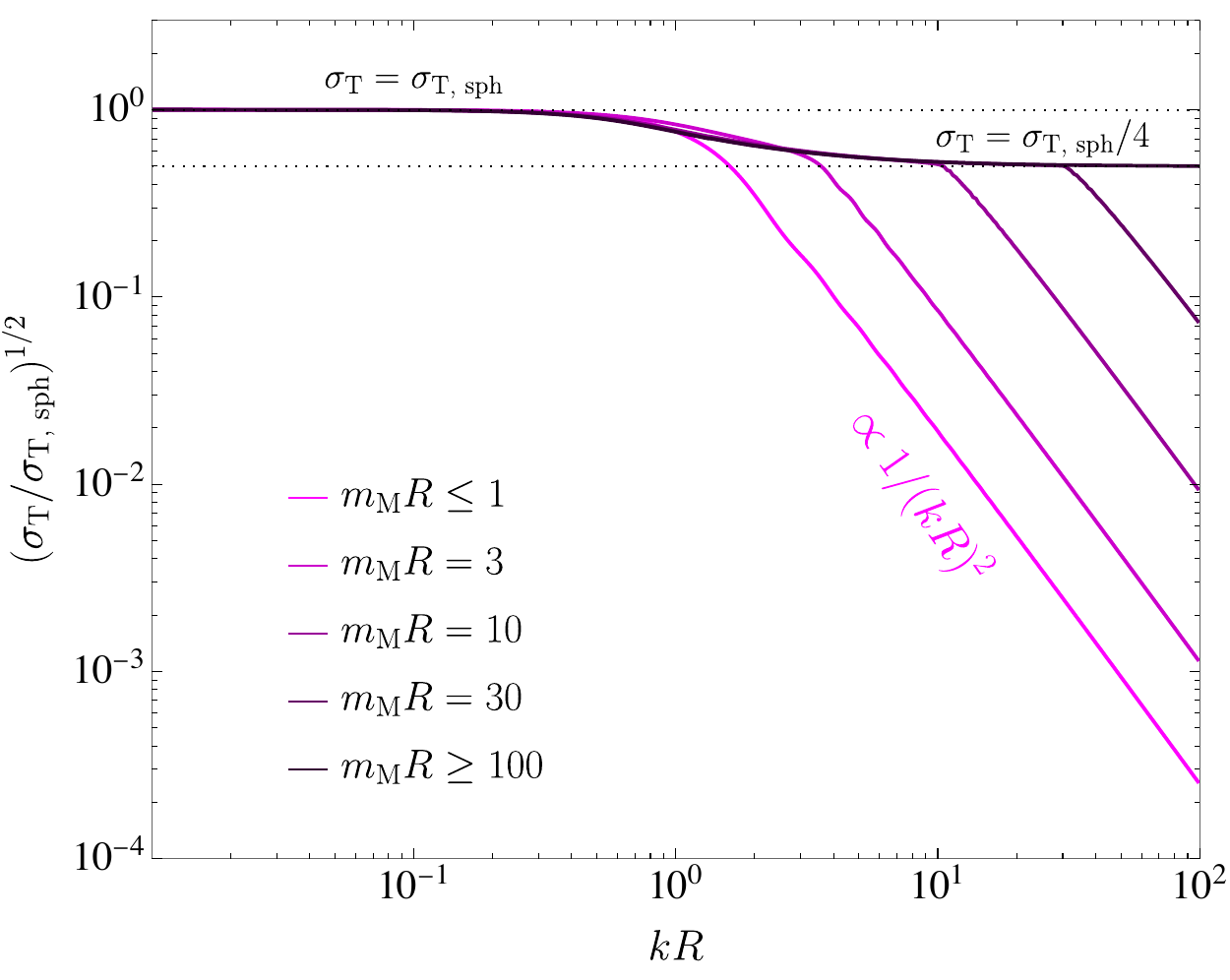}
\caption{{\bf Left:} The $kR$-$(\sigmat/\sigma_\text{T,\,A})^{1/2}$ plot. Different ratios for various $\mM R$ values are represented by the magenta lines with varying darkness. When $\mM R > 1$, the curves start below $\sigma_\text{T,\,A}$, illustrating the screening effect. For $k>\mM$, the lines overlap and follow $\sigmat \propto 1/(kR)^2$ due to finite-size decoherence suppression. When $R^{-1}<k<\mM$, the suppression from decoherence is compensated by descreening enhancement. {\bf Right:} The $kR$-$(\sigmat/\sigma_\text{T,\,sph})^{1/2}$ plot, where the ratio is normalized to $1$ for $kR < 1$. When $\mM R \gg 1$, corresponding to the hard-sphere scattering regime, the momentum-transfer cross section approaches $\sigma_\text{T,\,sph}/4$ for $kR \gg 1$.}
\label{fig:sca_force_nonp_formfactor}
\end{figure}

Based on the discussions in \Subsec{perturbative_sc} and \Subsec{non_perturb1}, we know how the momentum-transfer cross section (and the corresponding scattering force $\abs{\vecF_\sct}$) behaves in each phase listed in \Fig{phase_classify}. We now numerically illustrate the transitions between different phases as the incident momentum increases. In \Fig{sca_force_nonp_formfactor}, we show how $\sigmat$ changes with the dimensionless quantity $kR$ for given values of $\mM R$. In the left panel of \Fig{sca_force_nonp_formfactor}, we present $\sigmat$ in terms of the dimensionless quantity $(\sigmat/\sigma_\text{T,\,A})^{1/2}$, where $\sigma_\text{T,\,A}$ is the momentum-transfer cross section in Region A, as shown in \Eq{region_a_sigmat}. This dimensionless ratio describes how $\sigmat$ is affected by screening, decoherence, and descreening effects. We see that when $kR<1$, $\sigmat$ follows \Eq{sigmat_RegionC}, revealing the 
screening effect. For $k>\mM$, the momentum-transfer cross sections for different $\mM R$ values overlap and follow $\sigmat \propto 1/(kR)^2$, due to finite-size decoherence suppression. In the intermediate region where $R^{-1}<k<\mM$, the suppression from decoherence is compensated by descreening enhancement, leading to a much slower decline. In the right panel of \Fig{sca_force_nonp_formfactor}, we represent $\sigmat$ in terms of the dimensionless quantity $(\sigmat/\sigma_\text{T,\,sph})^{1/2}$, where $\sigma_\text{T,\,sph} = \frac{4\pi}{k^2}\abs{A_0}^2$ is the momentum-transfer cross section obtained by keeping only the s-wave component. Therefore, when $kR < 1$, the curves are normalized to one. We can see explicitly that for $\mM R \gg 1$, corresponding to the hard-sphere configuration, $\sigmat$ tends to $\sigma_\text{T,\,sph}/4 \simeq \pi R^2$ as shown in \Eq{sigmat_RegionD}. It is worth noting that the background-induced force exhibits similar behavior, which will be discussed in detail in \Subsec{bg_nonp_highk}.

\vspace{1.2cm}
\begin{table}[h!]
\centering
\small
\begin{tabular}{|c|c|c|c|c|c|}
\hline
Experiments &  Sensitivities to Acceleration & Test Masses  \\
\hline
MICROSCOPE Satellite ~\cite{Berge:2017ovy,Touboul:2017grn,MICROSCOPE:2019jix,MICROSCOPE:2022doy} &  $|\Delta a| \sim 10^{-14}\,\text{m/s}^2$ & Ti Alloy ((3.5,3.1) $\times$ 8.0 cm), Pt Alloy ((2.0,1.6) $\times$ 4.4 cm)  \\
\hline
E$\ddot{\rm o}$t-Wash Torsion Balance ~\cite{Schlamminger:2007ht} & $|\Delta a| \sim 10^{-15}\,\text{m/s}^2$  &  Be (0.85 cm), Ti (0.85 cm, 0.71 cm)\\
\hline
Galileo Galilei Satellite ~\cite{Nobili:2012uj} &  $|\Delta a| \sim 10^{-16}\,\text{m/s}^2$ &  \text{Not Specified}\\ 
\hline
Deep-Space Mission~\cite{Buscaino:2015fya} &   $\abs{a} \sim 10^{-18}\,\text{m/s}^2$  & Pt  (5 cm) \\
\hline
$T = 10\,\Kelvin$, $\quality=10^8$ &   $\abs{a} \sim 10^{-20} \left(1\,\kg/M_\testmass\right)^{1/2} \,\text{m/s}^2$  & \text{Not Specified} \\
\hline
SQL, $\quality=10^8$ &   $\abs{a} \sim 10^{-27} \left(1\,\kg/M_\testmass\right)^{1/2} \,\text{m/s}^2$  & \text{Not Specified} \\
\hline
\end{tabular}
\caption{Summary of current and future acceleration test experiments and the theoretical detection limit. The first four rows present the sensitivities and the test masses of the current and future acceleration tests, including MICROSCOPE satellite~\cite{MICROSCOPE:2019jix,MICROSCOPE:2022doy}, E$\ddot{\rm o}$t-Wash Torsion Balance~\cite{Schlamminger:2007ht}, Galileo Galilei satellite~\cite{Nobili:2012uj}, and the deep-space mission~\cite{Buscaino:2015fya}. The lowest two rows denote the theoretical sensitivities of the acceleration measurements derived by comparing the thermal/quantum noise, as shown in \Eq{acc_fluct}. The second-to-last row gives the sensitivity given $T = 10\,\Kelvin$ and $\quality = 10^8$. The last row gives the sensitivity saturating the standard quantum limit given $\quality = 10^8$. For both of the last two rows, we choose $t_\text{int} = 3\,\text{years}$, $\omega_\osc = 2\pi \times 1\, \text{mHz}$. $\abs{a}$ denotes the measurements of the absolute value of acceleration, while $\abs{\Delta a}$ denotes the measurements of the differential acceleration. In the test masses column, the two numbers separated by a comma indicate the outer and inner radii, respectively.}
\label{tab:acc_exp_table}
\end{table}

\subsection{Experimental Tests}\label{subsec:sc_force_exp}

\begin{figure}
\centering
\includegraphics[width=0.49\linewidth]
{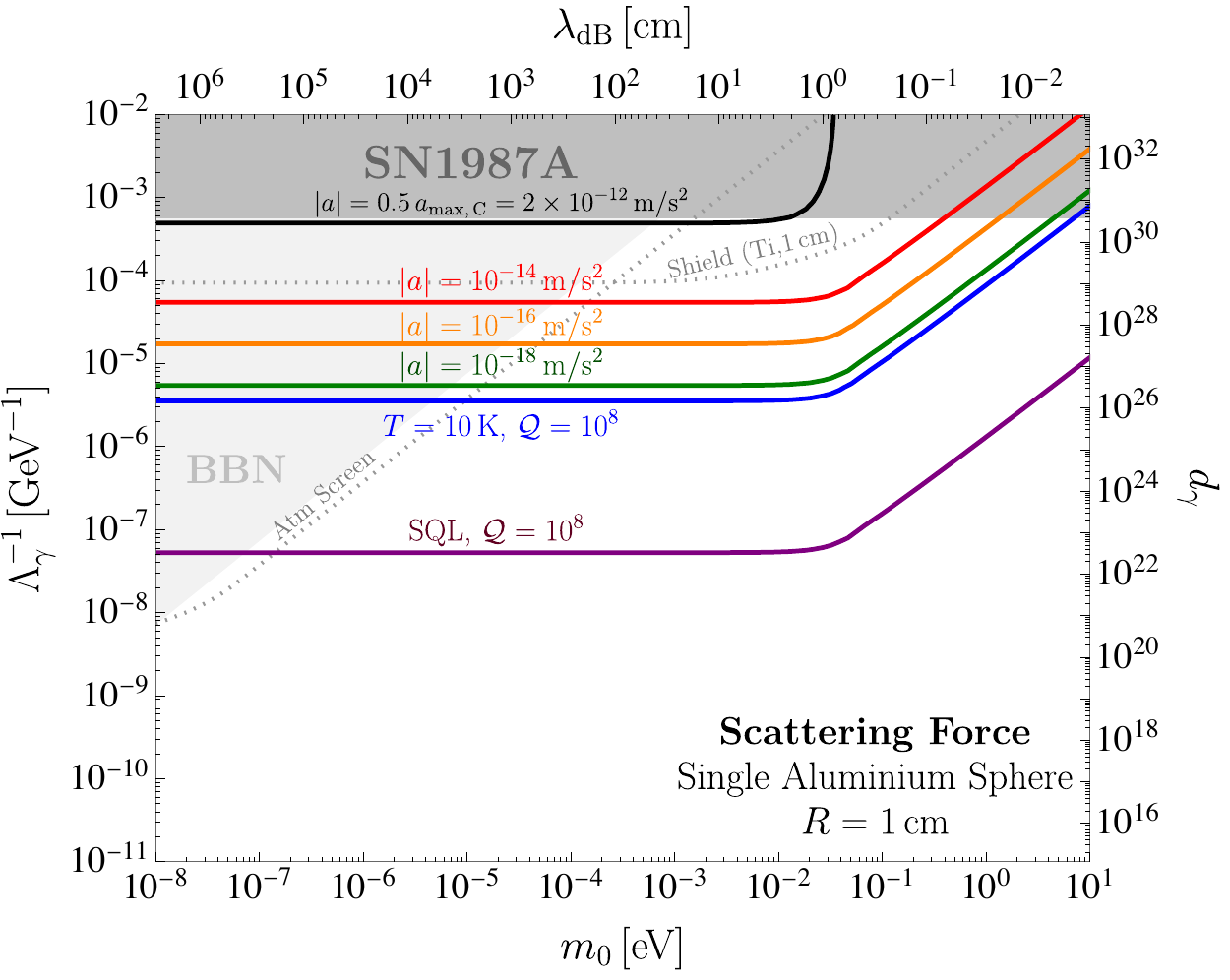}
\includegraphics[width=0.49\linewidth]
{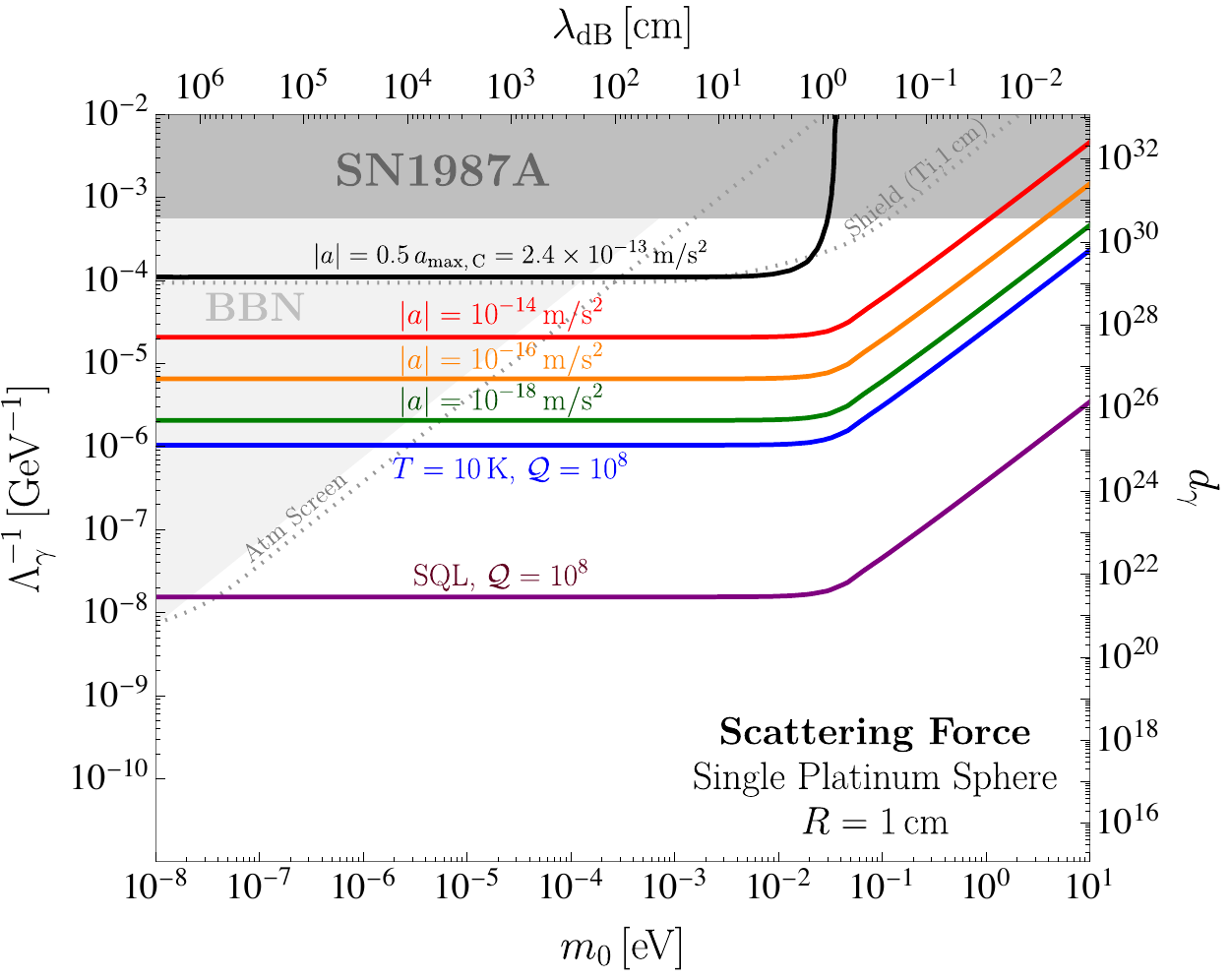}
\includegraphics[width=0.49\linewidth]{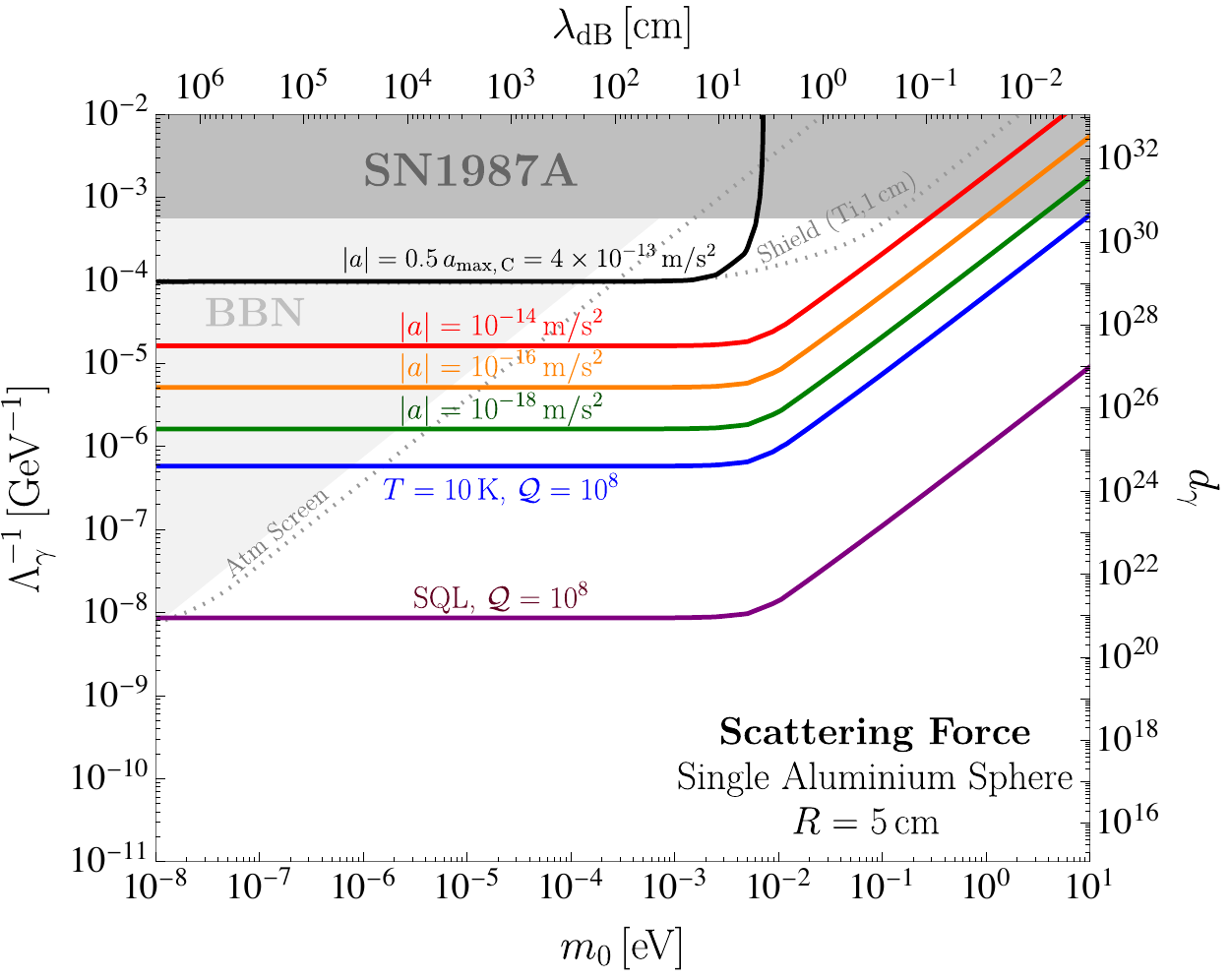}
\includegraphics[width=0.49\linewidth]{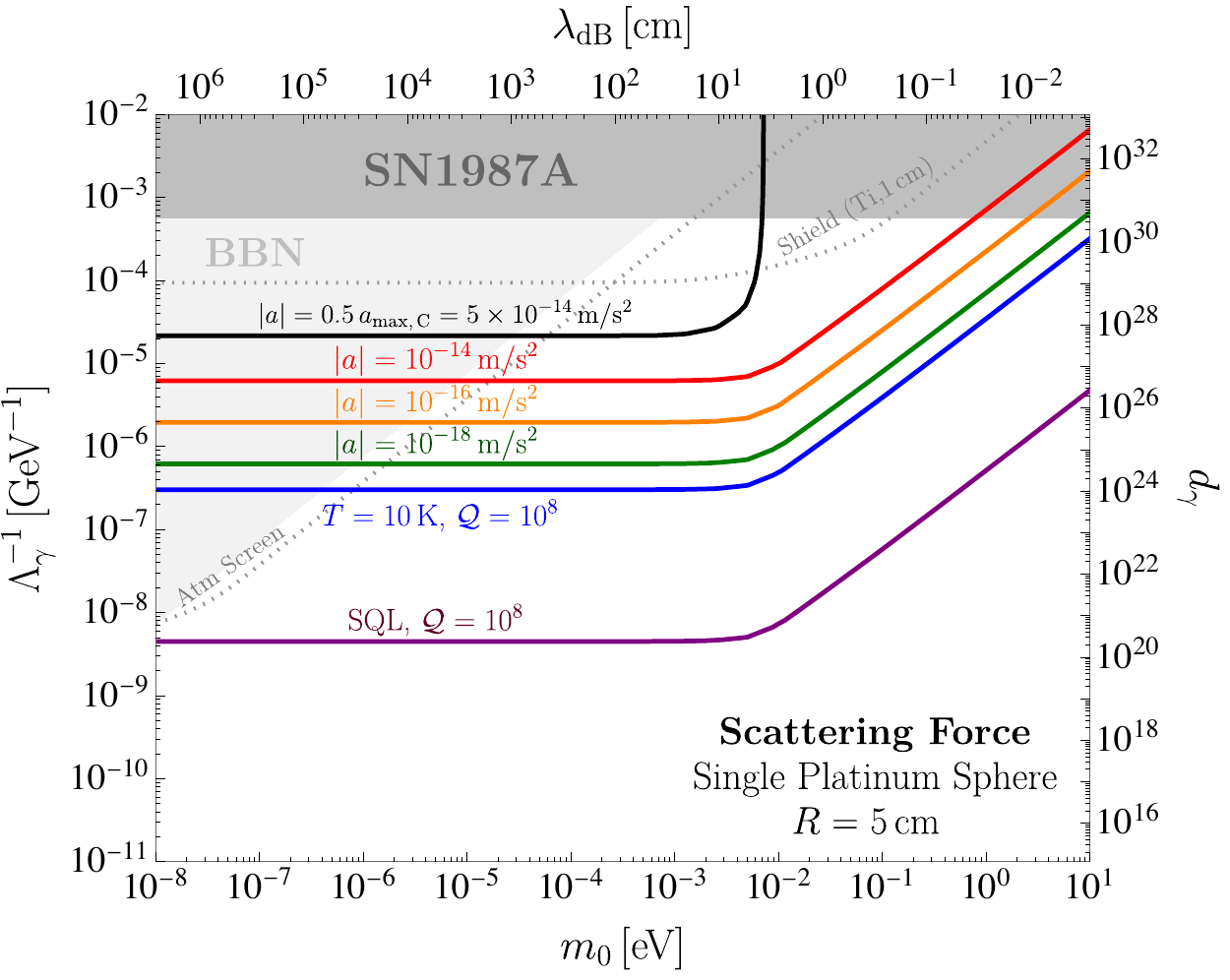}
\caption{Sensitivities from measuring the accelerations of test masses induced by the scattering force from the dark matter wind. We use uniform Aluminum and Platinum spheres with $R = 1\,\cm, 5\,\cm$ for illustration. The dark gray region represents the energy loss constraint from SN1987A~\cite{Raffelt:1990yz,Cox:2024oeb}. The light gray region represents the cosmological constraint from BBN~\cite{Stadnik:2015kia,Sibiryakov:2020eir,Bouley:2022eer}. The black line denotes the acceleration $\abs{a}=0.5\,a_{\max,\text{C}}$, where $a_{\max,\text{C}}$ is the maximum acceleration in Region C. The red, orange, and green lines denote the sensitivities with $\abs{a} = 10^{-14},\,\, 10^{-16},\,\,10^{-18}\,\rm m/s^2$, respectively. The blue line denotes the sensitivity for $T = 10\,\Kelvin$ and $\quality = 10^8$. The purple line denotes the sensitivity that saturates the standard quantum limit with $\quality = 10^8$. In each panel, the gray dotted lines labeled ``Atm Screen'' and ``Shield (Ti, $1\,\text{cm}$)'' indicate the boundaries above which the atmospheric and Titanium-shield~($1\,\text{cm}$ thickness) screening effects become significant. {\bf Upper Left}: Platinum sphere with $R=1\,\cm$. {\bf Upper Right}:  Aluminum sphere with $R=1\,\cm$. {\bf Lower Left}: 
Platinum sphere with $R=5\,\cm$.  {\bf Lower Right}: Aluminum sphere with $R=5\,\cm$.}
\label{fig:sensitivity1}
\end{figure}

In this subsection, we discuss experiments sensitive to the tiny acceleration of the test mass induced by the scattering force. These experiments can be categorized into two classes: terrestrial experiments \cite{Schlamminger:2007ht,Wagner:2012ui} and space-based experiments \cite{Nobili:2012uj,Buscaino:2015fya,MICROSCOPE:2019jix,MICROSCOPE:2022doy}. It is important to note that the current terrestrial experiments have relatively thick shielding and are mainly sensitive to the strongly coupled region, where the screening effects of the experimental shielding and atmosphere are important. In contrast, future proposed space-based acceleration tests have thin outer shells, avoiding the atmospheric screening and minimizing the screening effect from external shielding across a broad range of parameters~\cite{Buscaino:2015fya}. Future experiments with the improved quality factor and low temperature can detect more weakly coupled regions, where the screening effect becomes irrelevant. Since our work aims to provide a global perspective, we use acceleration as the key sensitivity metric, assuming the external shielding is fully reduced~\footnote{For completeness, we indicate the regions where atmospheric and experimental-shield screening effects become significant by gray dotted lines in \Fig{sensitivity1}~(scattering force) and \Fig{sensitivity_bg}~(background-induced force).}. In Table~\ref{tab:acc_exp_table}, we summarize the sensitivities of current/proposed experiments and the achievable sensitivities by comparing with the thermal and quantum fluctuations. We expect similar acceleration sensitivities can be achieved for future experiments specially designed for testing the dark matter scattering force. For clarity, we note that Table~\ref{tab:acc_exp_table} includes two types of experiments:
(1) relative acceleration measurements (tests of equivalence principle violation), denoted by $\abs{\Delta a}$; and
(2) absolute acceleration measurements, denoted by $\abs{a}$.
When recasting the sensitivities in terms of acceleration, we use the absolute acceleration for simplicity. Recasting them in terms of relative acceleration would only introduce an $\mathcal{O}(1)$ difference in typical cases, resulting in a negligible impact on the sensitivity curves.

We start from the terrestrial weak equivalence principle tests~\cite{Schlamminger:2007ht,Wagner:2012ui}, which are Cavendish-type torsion balance experiments being sensitive to the differential acceleration of the test bodies. Due to the atmosphere screening effect, these types of experiments can only probe the coupling strength below the condition in \Eq{screen_condition}. Here we take the E$\ddot{\rm o}$t-Wash experiment (Be-Ti) in Ref.~\cite{Schlamminger:2007ht} for an example. The test masses have an equal weight of 4.84 g, which can be modeled as a solid Beryllium sphere and a hollow Titanium sphere with an outer radius of 0.85 cm. The inner radius for the hollow Titanium sphere is $0.71 \,\cm$. The experimental sensitivity on the differential acceleration is $\abs{\Delta a} \sim 10^{-15} \, \rm m/s^2$~\cite{Schlamminger:2007ht}.

Then we discuss space-based weak equivalence principle tests, including the MICROSCOPE satellite~\cite{MICROSCOPE:2019jix,MICROSCOPE:2022doy}, the proposed Galileo Galilei satellite~\cite{Nobili:2012uj}, and the proposed deep-space mission~\cite{Buscaino:2015fya}. MICROSCOPE is a satellite conducting equivalence principle tests in low Earth orbit with a sensitivity to differential acceleration of $\abs{\Delta a} \sim 10^{-14} \  \rm m/s^2$. In \Subsec{bgforce_MICROSCOPE}, we will give a more detailed discussion on the MICROSCOPE experiment and revisit the constraint from the background-induced force. Galileo Galilei satellite, a next-generation space test of the weak equivalence principle, can reach the sensitivity $\abs{\Delta a} \sim 10^{-16}\,\rm m/s^2$ on the differential acceleration~\cite{Nobili:2012uj}.  Apart from experiments in near space, the proposal with a deep-space mission is also able to significantly increase the sensitivity of measuring an extra acceleration~\cite{Buscaino:2015fya}. In Ref.~\cite{Buscaino:2015fya}, the test mass is assumed to be made of a platinum sphere with a radius of 5 cm. Assuming a ranging accuracy of $10\,\cm$ over a 7-year lifetime of the mission, we estimate the corresponding sensitivity to be $\abs{a} \sim 10^{-18} \  \rm m/s^2$. 

After presenting examples of acceleration test sensitivities, we now discuss the achievable theoretical sensitivities through the fluctuation-dissipation theorem~\cite{Kubo:1966fyg}. Assuming the test mass is coupled to a damped harmonic oscillator, the force noise is given by $\Fnoise = (2 S_\text{FF} \Delta f)^{1/2}$. $\Delta f \sim 1/t_\text{int}$ is the bandwidth. $t_\text{int}$ is the integration time. $S_\text{FF} = \coth(\omega_\osc/2T) \mathcal{M}_\testmass \omega^2_\osc/\quality$ is the force noise spectrum, where $\quality$ is the quality factor, $\omega_\osc$ is the intrinsic frequency of the oscillator, $T$ is the temperature. For other types of noise, the experimental apparatus can be rotated at angular frequency $\omega_\text{rot} \simeq \omega_\osc$ to modulate the scattering force, thereby suppressing these noises at resonance. We use benchmark experimental parameters similar to those in Refs.~\cite{Graham:2015ifn,Arvanitaki:2023fij,VanTilburg:2024xib}. Specifically, we choose $\quality = 10^8$, which is higher than the $\quality \sim 10^3$ in the current torsion balance experiments but can be realized in the future experimental setups~\cite{Cagnoli:2000vr,Aspelmeyer:2013lha,Graham:2015ifn,Smetana:2024xwi}. We then choose $t_\text{int} = 3\,\text{years}$, $\omega_\osc = 2\pi \times 1\,\text{mHz}$. For sensitivity curves, we consider two benchmarks: 1. The finite temperature case with $T = 10\,\Kelvin$, which can be easily realized by the cooling of the liquid Helium~\footnote{One can also choose the benchmark temperature to be $T = 10\,\text{mK}$, which corresponds to the typical cooling temperature of a dilution refrigerator~\cite{wheatley1968principles,Pobell:2007egf}. In this case, the thermal noise can be further reduced compared with the $T=10\,\Kelvin$ scenario.}. Here the noise spectrum is $\Fnoise \simeq (4 T M_\testmass \, \omega_\osc/\quality \, t_\text{int})^{1/2}$. 2. The standard quantum limit~(SQL). At $T=0$ only the zero-point energy contributes to the noise power spectrum. The force noise is then given by $\Fnoise \simeq (2 M_\testmass \omega^2_\osc/ \quality \, t_\text{int})^{1/2}$. Now we can define the signal-to-noise ratio to be $\text{SNR} \equiv F_\text{ext}/\Fnoise$, where $F_\text{ext}$ is the currently discussed scattering force $F_\sct$ or the later discussed background-induced force $F_\bg$. Choosing $\text{SNR} = 1$, we derive the following sensitivities in the finite temperature~($T \simeq 10 \, \Kelvin$)
\bea
\label{eq:acc_fluct}
\text{Finite-$T$}: \quad  \abs{a} \sim 10^{-20}\, \text{m}/\text{s}^2 \times \left(\frac{T}{10 \, \Kelvin}\right)^{1/2} \left(\frac{\omega_\osc}{2\pi \times 1\,\text{mHz}}\right)^{1/2} \left(\frac{1\,\kg}{M_\testmass}\right)^{1/2} \left(\frac{10^8}{\quality}\right)^{1/2}  \left(\frac{3\,\text{years}}{t_\text{int}}\right)^{1/2},
\eea
and standard quantum limit~($T=0$)
\bea
\text{SQL}: \quad  \quad \,\,\,\, \abs{a} \sim 10^{-27}\, \text{m}/\text{s}^2 \times \left(\frac{\omega_\osc}{2\pi \times 1\,\text{mHz}}\right) \left(\frac{1\,\kg}{M_\testmass}\right)^{1/2} \left(\frac{10^8}{\quality}\right)^{1/2}  \left(\frac{3\,\text{years}}{t_\text{int}}\right)^{1/2}.
\eea

To illustrate the parameter space that can be reached using similar setups, we plot the sensitivity curves induced by the scattering force on a solid sphere. In \Fig{sensitivity1}, we choose the solid Platinum~($Q_{\gamma,\text{Pt}} \simeq 4.3 \times 10^{-3}$, $\rho_\text{Pt} \simeq 21.45\,\gram/\cm^3$) and Aluminum~($Q_{\gamma,\text{Al}} \simeq 1.7 \times 10^{-3}$, $\rho_\text{Al} \simeq 2.7\,\gram/\cm^3$) spheres with radii of $1\,\cm$ and $5\,\cm$. The scaling of the sensitivity contours in the weak coupling region~(colored lines in \Fig{sensitivity1}) as a function of model parameters is
\bea
\text{$m_0\text{-}\Lambda_\gamma^{-1}$ Plane:} \quad \quad \Lambda_\gamma^{-1} \propto \left\{ 
\begin{aligned}
& \const &  \quad k_0 R < 1\\
& m_0 & \quad k_0 R \gtrsim 1
\end{aligned}
\right. .
\eea
Such a scaling law can be understood as follows: Given the scattered test object $\testmass$, below the upper bound of the acceleration in \Eq{a_saturate}, the model parameter space is in either Regime A or Regimes B and E. For Region A where $k_0 R \lesssim 1$, we find that the acceleration is independent of $m_0$ based on \Eq{sigmaA}. Therefore, the fixed acceleration curve is flat in this regime. Otherwise, if $k_0 R \gtrsim 1$, the model parameters are in Regimes B and E, where the finite-size decoherence effect suppresses the acceleration caused by the scattering force as discussed in \Subsec{perturbative_sc}. In these regions, based on \Eq{region_b_sigmat} and \Eq{sigmat_RegionE}, we know that the acceleration obeys $a \propto \mM^4 k^{-4}_0 \propto \Lambda_\gamma^{-4} m_0^{-4}$. This results in the $\Lambda_\gamma^{-1}\propto m_0$ scaling law. One may notice that the black curves in \Fig{sensitivity1} have different scaling laws. It is because these curves are related to
\bea
\text{Black Lines:}\quad a = 0.5 \, a_{\max,\text{C}} = \frac{\rho_\phi v_0^2 \times 2 \pi R^2}{\rho_\testmass \times \frac{4\pi R^3}{3}}.
\eea
The acceleration sensitivities represented by the black lines fall between the maximum acceleration in the long-wavelength limit~($\sigmat \simeq 4 \pi R^2$) and the maximum acceleration in the short-wavelength limit~($\sigmat \simeq \pi R^2$). Consequently, as $m_0$ increases, black sensitivity lines bend sharply upward. This occurs because, in the $k_0 R \gtrsim 1$ region, the acceleration associated with $\sigmat > \pi R^2$ cannot be reached, regardless of the scalar-photon coupling strength. Based on \Fig{sensitivity1}, we can also summarize the guidelines for selecting test mass materials to optimize sensitivity in acceleration-based scattering force tests. In the weakly-coupled perturbative region, the acceleration follows: $a = \frac{\rho_\phi v_0^2 \,\, \sigmat}{\rho_\testmass \mathcal{V}} \propto Q_{\gamma,\testmass} \rho_\testmass$,
where $\sigmat \propto \mM^4$. This implies that maximizing sensitivity in this region requires a material with both a high dilaton charge and high density. Therefore, Platinum and Tungsten are suitable choices for enhancing sensitivity in the perturbative region. In the strong-coupling region, which is highly non-perturbative, the acceleration behaves as: $a = \frac{\rho_\phi v_0^2 \, \sigmat}{\rho_\testmass \mathcal{V}} \propto \frac{1}{\rho_\testmass}$, where the momentum-transfer cross-section saturates at $\sigmat \sim \mathcal{O}(1) \times \pi R^2$ and becomes independent of the test mass material. Therefore, a low-density material, such as Aluminum, is suitable for exploring the strong-coupling region.

In \Fig{sensitivity1}, we also show the constraints from astrophysics and the early universe. The strongest astrophysical constraint comes from SN1987A~\cite{Raffelt:1990yz,Cox:2024oeb}, which is shown as the dark gray shaded region. Specifically, the additional energy loss from $\gamma \gamma \rightarrow \phi \phi$ channel alters the neutrino emission of SN1987A, providing a stringent constraint on the quadratic scalar-photon interaction within the parameter space of interest. Other astrophysical objects, such as horizontal branch stars (HB) and red giants (RG), also impose energy loss constraints. Since the SN1987A bound is the strongest among them, we choose it as the benchmark and show it in the plots. Further details can be found in \Appx{astro_constraint}. Another relevant constraint comes from the BBN, as discussed in Refs.~\cite{Stadnik:2015kia,Sibiryakov:2020eir,Bouley:2022eer}. This constraint is based on the fact that the scalar field acquires a large field value in the early universe, leading to a modification of the fine-structure constant during BBN. Therefore, the primordial abundances of helium-4 and deuterium are changed. We use the light gray shaded region to represent the BBN constraint.

In reality, the scattering between the test mass $\testmass$ and dark matter may not be ideally isolated from the rest of the environment. One of the effects to be considered is the screening from various surrounding matter~\cite{Day:2023mkb}. When the interaction between dark matter and matter enters the non-perturbative regime as discussed in \Subsec{non_perturb1}, the dark matter is blocked from passing through the ordinary matter. We take the Earth's atmosphere as an example. Since the atmosphere thickness is much larger than the wavelengths considered here, Regime D is reached if the interaction strength satisfies $\mM > m_0 v_0$. As a result, the Earth's atmosphere blocks the dark matter before it reaches the ground-based laboratories. Considering the atmosphere~($78\%$ Nitrogen and $22\%$ Oxygen) above the ground, which has the dilaton charge $Q_{\gamma,\text{atm}} \simeq 1.3 \times 10^{-3}$ and the density $\rho_{\text{atm}} \simeq 1.2 \times 10^{-3} \, \text{g}/\cm^3$ near the earth surface, we estimate that the screening effect is significant in the region
\begin{equation}
\label{eq:screen_condition}
m_{\text{M}, \text{atm}}= \left(\frac{ Q_{\gamma,\text{atm}}  \, \rho_{\text{atm}}}{\Lambda_\gamma^2}\right)^{1/2} \gtrsim m_0 v_0 \quad \Longrightarrow \quad \frac{\Lambda_\gamma^{-1}}{\text{TeV}^{-1}} \gtrsim   4 \times 10^{-1}  \frac{m_0}{\text{meV}}.
\end{equation}
In \Fig{sensitivity1}, the gray dotted lines labeled ``Atm Screen'' mark the boundaries above which atmospheric screening becomes significant, as estimated in \Eq{screen_condition}. In this regime, the sensitivity of ground-based experiments is expected to be lost due to the screening of scalar dark matter within the atmosphere. However, conducting the acceleration test in space effectively eliminates atmospheric screening and preserves sensitivity.

In addition to the screening caused by the atmosphere, there is also the screening caused by the shield of the experimental setup. Before hitting the test mass, the dark matter passes through the shield containing the test mass. Compared with the atmosphere, these obstacles have much smaller thicknesses but much higher densities. To illustrate, we take a Titanium shield with width $L = 1\text{ cm}$ as a benchmark example. The screening effect is significant when~(Here $Q_{\gamma,\text{Ti}} \simeq 2.0 \times 10^{-3}$, $\rho_\text{Ti} \simeq 4.5 \, \text{g}/\cm^3$)
\begin{equation}
\label{eq:screen_condition2}
m_{\text{M},\text{Ti}} = \left(\frac{Q_{\gamma,\text{Ti}} \, \rho_\text{Ti}}{\Lambda_\gamma^2}\right)^{1/2} \gtrsim \max \left( m_0 v_0, L^{-1}  \right) \,\, \Longrightarrow \,\, \frac{\Lambda_\gamma^{-1}}{\text{TeV}^{-1}} \gtrsim  10^{-1} \left( \frac{L}{\text{1 cm}} \right)^{-1} \max \left( 1,\frac{m_0}{20\text{ meV}} \frac{L}{\text{1 cm}} \right ).
\end{equation}
In \Fig{sensitivity1}, the gray dotted lines labeled ``Shield (Ti, $1\,\text{cm}$)'' denote the boundaries above which screening from the $1\,\text{cm}$-thick Titanium shield becomes significant, as estimated in \Eq{screen_condition2}. Given this, a primary goal of future proposed experiments to detect the scattering force induced by ultralight dark matter in the strongly coupled region is to minimize shield thickness, allowing for sufficient penetration of the dark matter wind. In contrast, within the extensive weakly coupled parameter space accessible to future experiments, dark matter can easily pass through the shielding of experimental setups. Since the specific configurations of these experiments may vary, we do not incorporate such details in our analysis for simplicity.

\section{Background-Induced Force}\label{sec:bg_force}

In this section, we discuss the background-induced force. As shown in \Fig{qm_sca}, when the scalar dark matter encounters the source mass $\sourcemass$, a bulk of ordinary matter, it is scattered due to a change in its dispersion relation. Since the scattered wave amplitude decays over a large distance, a spatially inhomogeneous scalar background is generated. When a test mass $\testmass$ is placed near the source, it feels the scalar gradient induced by the source, as described by \Eq{Fbg} and \Eq{Vbg}.

This effect has been discussed in various contexts, including ultralight dark matter~\cite{Hees:2018fpg,Berezhiani:2018oxf,Ferrer:2000hm,Banerjee:2022sqg,VanTilburg:2024xib,Barbosa:2024pkl}, neutrino physics~\cite{Horowitz:1993kw,Ferrer:1998ju,Ferrer:1999ad,Ghosh:2022nzo,Blas:2022ovz,Ghosh:2024qai}, and modified gravity~\cite{dePireySaintAlby:2017lwc}. In addition, it has an interesting analogy in condensed matter physics, known as Friedel oscillations~\cite{friedel1952xiv,blandin1959knight,blandin1960effets,villain2016jacques},  where a long-range oscillatory potential emerges from a repulsive contact interaction. This effect can lead to electron pairing in the absence of phonon mediation, which is known as Kohn-Luttinger superconductivity~\cite{Kohn:1965zz}. The background-induced force also has an interesting optical analogy known as optical binding, where two spheres experience an oscillatory potential between each other when illuminated vertically by light from above~\cite{burns1989optical,chen2015lateral,Rudolph:2024cey}. Various approaches have been employed to study this effect, ranging from finite-density field theory~\cite{Horowitz:1993kw,Ferrer:1998ju,Ferrer:1999ad,Ferrer:2000hm,Ghosh:2022nzo,Blas:2022ovz,Ghosh:2024qai,VanTilburg:2024xib,Barbosa:2024pkl,Grossman:2025cov,Cheng:2025fak} to classical field theory~\cite{dePireySaintAlby:2017lwc,Berezhiani:2018oxf,Hees:2018fpg,Banerjee:2022sqg,VanTilburg:2024xib}. Notably, Ref.~\cite{VanTilburg:2024xib} provides a dedicated study with both methods and offers an intuitive interpretation of this effect, introducing the term ``wake force''.

However, we point out that the previous discussions on the background-induced force have limitations and are valid within specific regions. For example, Refs.~\cite{Horowitz:1993kw,Ferrer:1998ju,Ferrer:1999ad,Ferrer:2000hm,Ghosh:2022nzo,Blas:2022ovz,Ghosh:2024qai,VanTilburg:2024xib,Barbosa:2024pkl} treat the source as point-like and assume its matter effect to be perturbative, corresponding to Region A. In contrast, Refs.~\cite{dePireySaintAlby:2017lwc,Berezhiani:2018oxf,Hees:2018fpg,Banerjee:2022sqg} extend to non-perturbative regions where the screening effect becomes relevant. However, these studies rely on a spherically symmetric ansatz, which corresponds to taking the zero-momentum limit of the incident wave and retaining only the s-wave component of the wave function. This spherical ansatz is applicable only in Regions A and C in the short-range limit, where $kr<1$.

Utilizing the quantum mechanical scattering theory introduced in  \Sec{formalism}, we now discuss all phases of the parameter space outlined in \Fig{phase_classify}. This section is structured as follows. We begin our discussion in the perturbative region and then extend our discussion to the non-perturbative regions. First, we analyze the perturbative point-like region~(Region A), reproducing the results in Refs.~\cite{Ferrer:2000hm,VanTilburg:2024xib,Barbosa:2024pkl} using \Eq{Vbg_far}. We review the finite-phase-space effect on the background-induced potential due to the oscillatory potential and present previously unreported analytical results. Extending the discussion to Region B, we account for the finite size of the source mass in the perturbative region. Second, we investigate the non-perturbative regions where the screening effect becomes important. We begin by reviewing the previous treatment using the spherical ansatz. We then show a general method and the deviation from the spherical ansatz in Regions C, D, E. Third, we impose projections from ISL tests and EP tests. We also explore potential future sensitivities based on measurements of the induced acceleration. Finally, we revisit the MICROSCOPE satellite constraints, which were previously explored in Refs.~\cite{Hees:2018fpg,Banerjee:2022sqg}. We reproduce the results in the low-mass region~($m_0 \lesssim 10^{-10}\,\eV$, $k_0 R_\oplus \lesssim 1$) and extend the analysis to the high-mass region~($m_0 \gtrsim 10^{-10}\,\eV$, $k_0 R_\oplus \gtrsim 1$) where the spherical ansatz becomes invalid.

\begin{figure}[t!]
\centering
\includegraphics[width=0.49\linewidth]{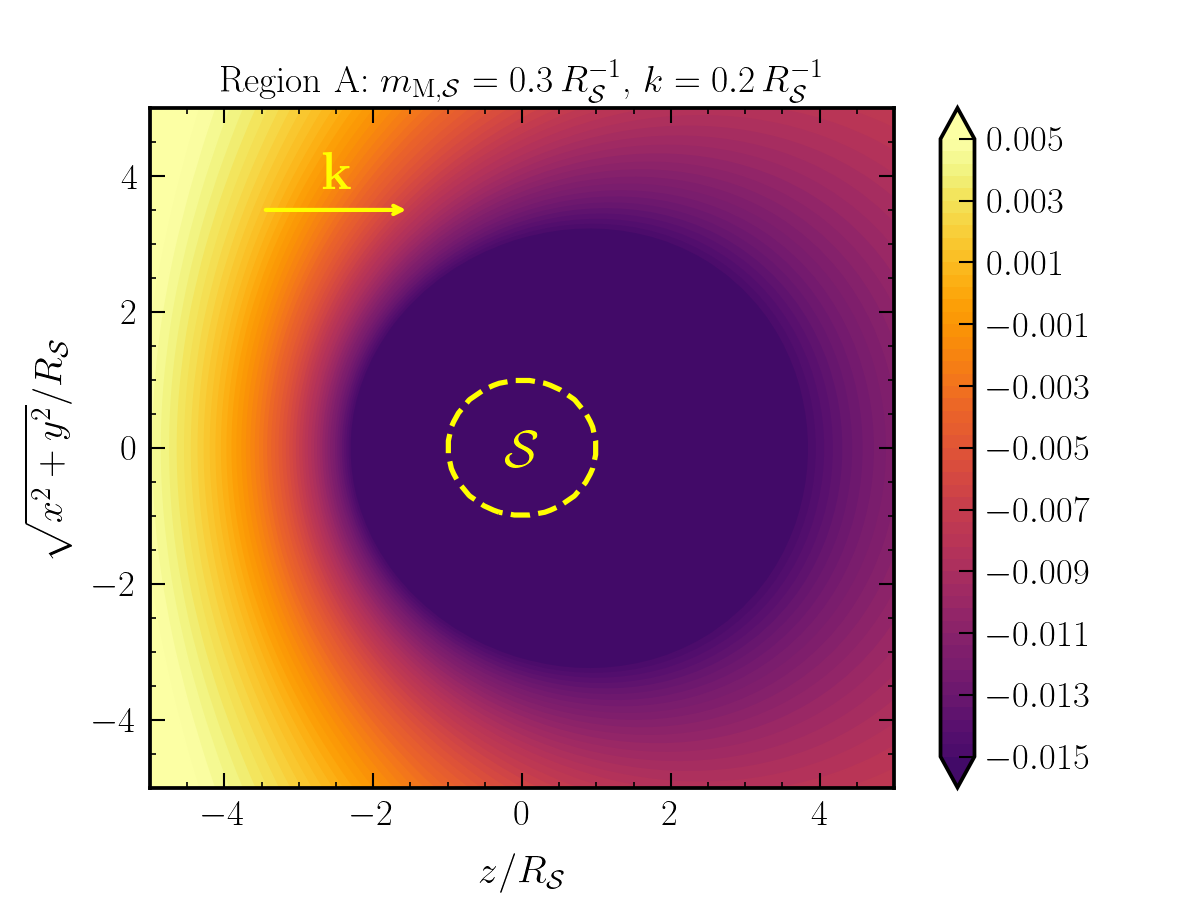}
\includegraphics[width=0.49\linewidth]{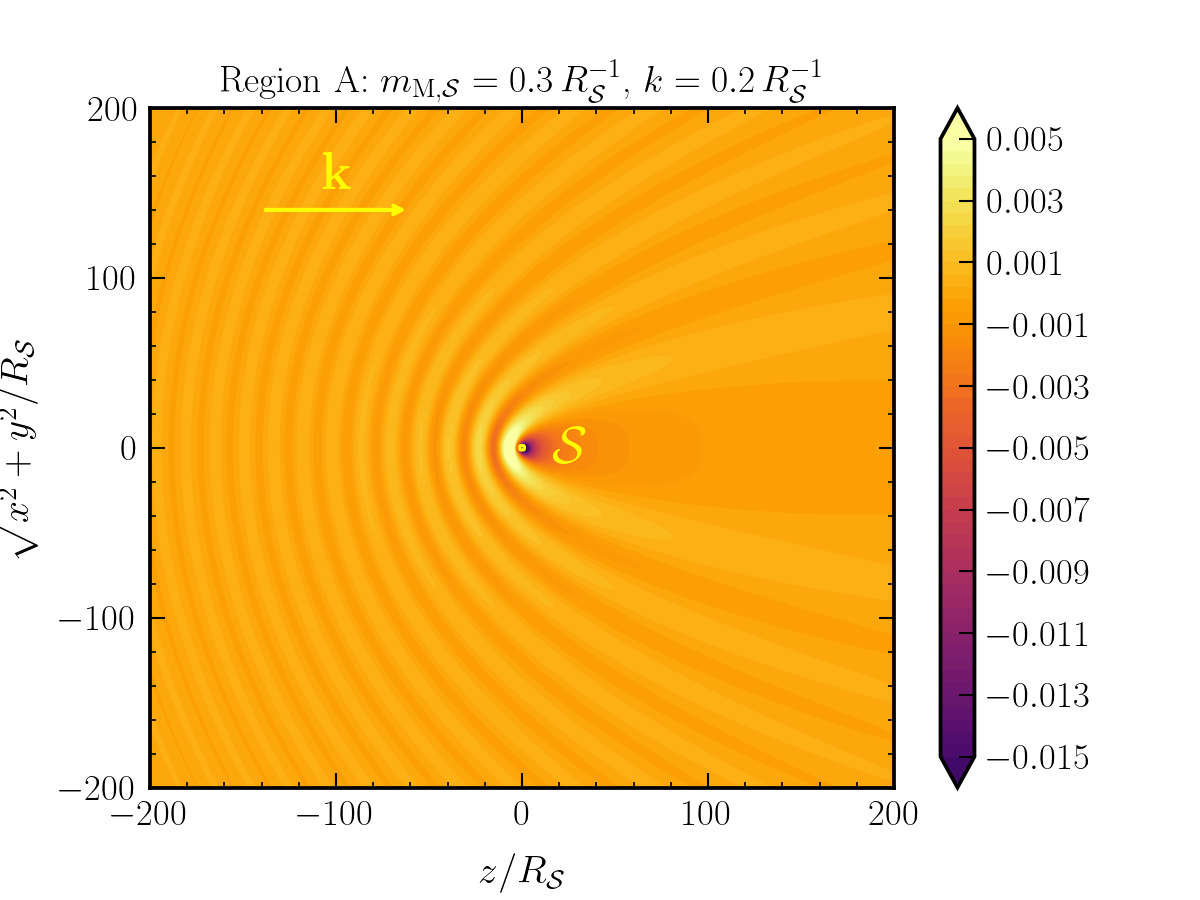}\\
\includegraphics[width=0.49\linewidth]{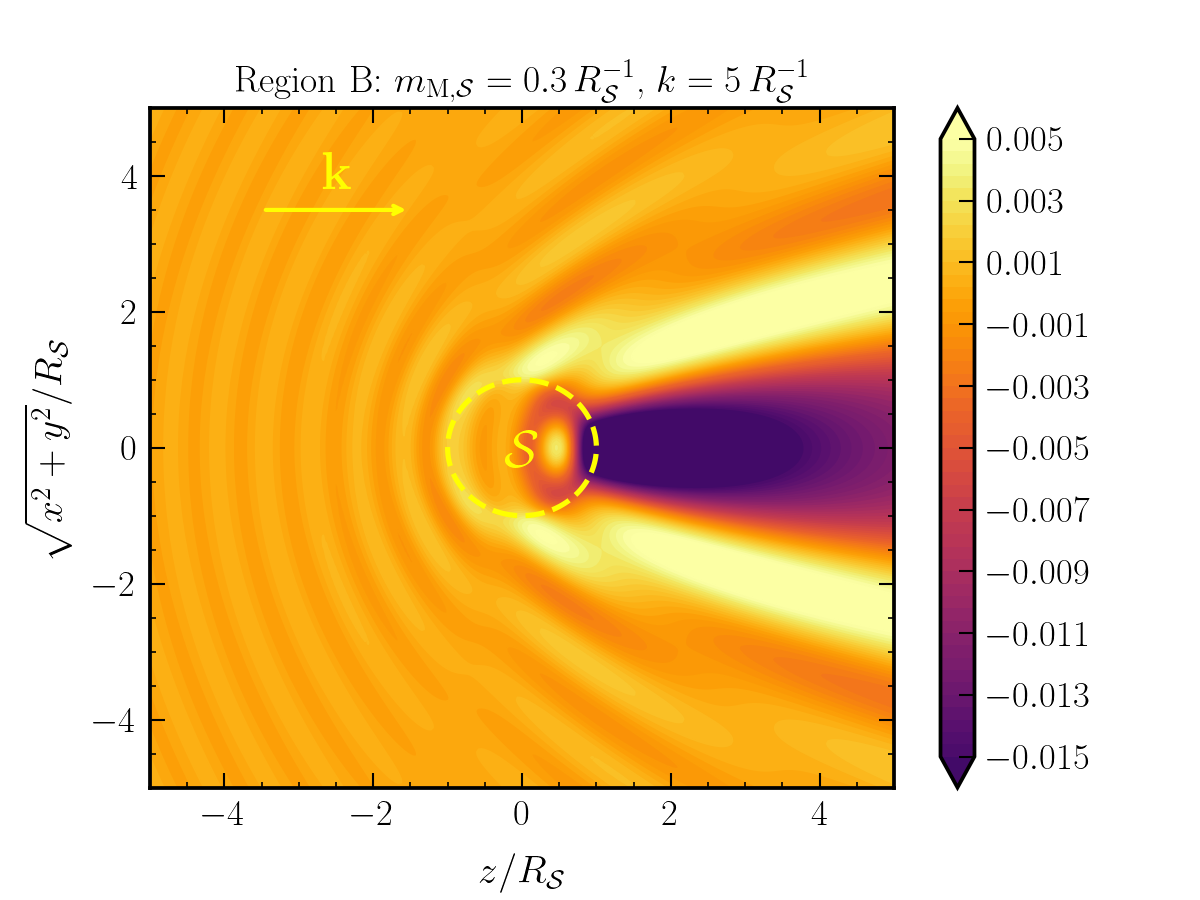}\\
\caption{The plots show $\abs{\psi}^2/\abs{\psi_0}^2 - 1$ in perturbative Regions~A and~B, as defined in Fig.~\ref{fig:phase_classify}. In both
regions one has $\abs{\psisc} < \abs{\psi_0}$, so that the Born approximation is valid. The red dashed lines indicate the boundary of the source (scattering) object $\sourcemass$. Here we use a narrower color scale, $[-0.015, 0.005]$, for $\abs{\psi}^2/\abs{\psi_0}^2 - 1$ so that the small scalar-field deviations are clearly visible, whereas Figs.~\ref{fig:nonpeturb_sca_plt} and~\ref{fig:nonpeturb_penetrate} use the range $[-1, 1]$. {\bf Upper Left:} Region~A. The scalar field is dominated by the s-wave ($l=0$). Near the source surface, $r \sim R_\sourcemass$, the scalar distribution is nearly spherically symmetric. {\bf Upper Right:} Region~A (zoomed out). Even though the field is s-wave dominated, the spherical symmetry is still broken once $k_0 \neq 0$. More specifically, at distances where $k_0 r \gtrsim 1$ (corresponding to $r \gtrsim 5 R_\sourcemass$ for this plot), the field becomes oscillatory and the departure from spherical symmetry becomes obvious. {\bf Lower:} Region~B. The scalar field receives contributions from multiple partial waves, summed up to $l_\text{max} \sim k_0 R_\sourcemass$. In this regime the field is oscillatory throughout the entire space. {\bf Common:} In all panels, for $k_0 r \gtrsim 1$ the scalar field becomes oscillatory, leading to a suppression from decoherence effect, as quantified later in Figs.~\ref{fig:FA_plt} and~\ref{fig:pert_finitesize_formfactor}.}
\label{fig:peturb_sca_plt}
\end{figure}

Before proceeding and splitting the discussion into separate regions, we write the generic background-induced potential as
\bea
\label{eq:Vbg_formfactor}
\langle V_\bg\rangle_\veck = - \frac{\rho_\phi}{m_0^2} \frac{ (\mMsource^2 \mathcal{V}_\sourcemass) (\mMtest^2 \mathcal{V}_\testmass) }{4\pi r} \times \formfactor(\vecr),
\eea
where we assign the labels ``$\testmass$'' and ``$\sourcemass$'' to the test mass and the source mass, respectively. In the above equation, $\formfactor$ represents the form factor in different regions listed in \Fig{phase_classify}. When $\formfactor = 1$, \Eq{Vbg_formfactor} becomes the background-induced force in the perturbative region, taking the short distance limit~($k_0 r < 1$). The deviations of $\formfactor$ from $1$ help to describe the characteristics of different regions, such as the finite-phase-space effect, finite-size effects, screening effects, and descreening effects. Substituting \Eq{Vbg_formfactor} into \Eq{Fbg}, we get the background-induced force
\bea
\label{eq:Fbg_formfactor}
\vecF_\bg = - \frac{\rho_\phi}{m_0^2} \frac{(\mMsource^2 \mathcal{V}_\sourcemass) (\mMtest^2 \mathcal{V}_\testmass)}{4 \pi r^2} \times \left( \formfactor - k_0 r \frac{\partial \formfactor}{\partial (k_0 r)} \right) \hat{\vecr} + (\text{$\hat{\vectheta},\hat{\vecphi}$ components}). 
\eea
Since $\vecF_\bg$ only has the radial components in the forward and backward directions, we focus solely on the radial direction throughout the discussion, omitting the components in the $\hat{\vectheta}$ and $\hat{\vecphi}$ directions.

\begin{figure}[h!]
\centering
\includegraphics[width=0.49\linewidth]{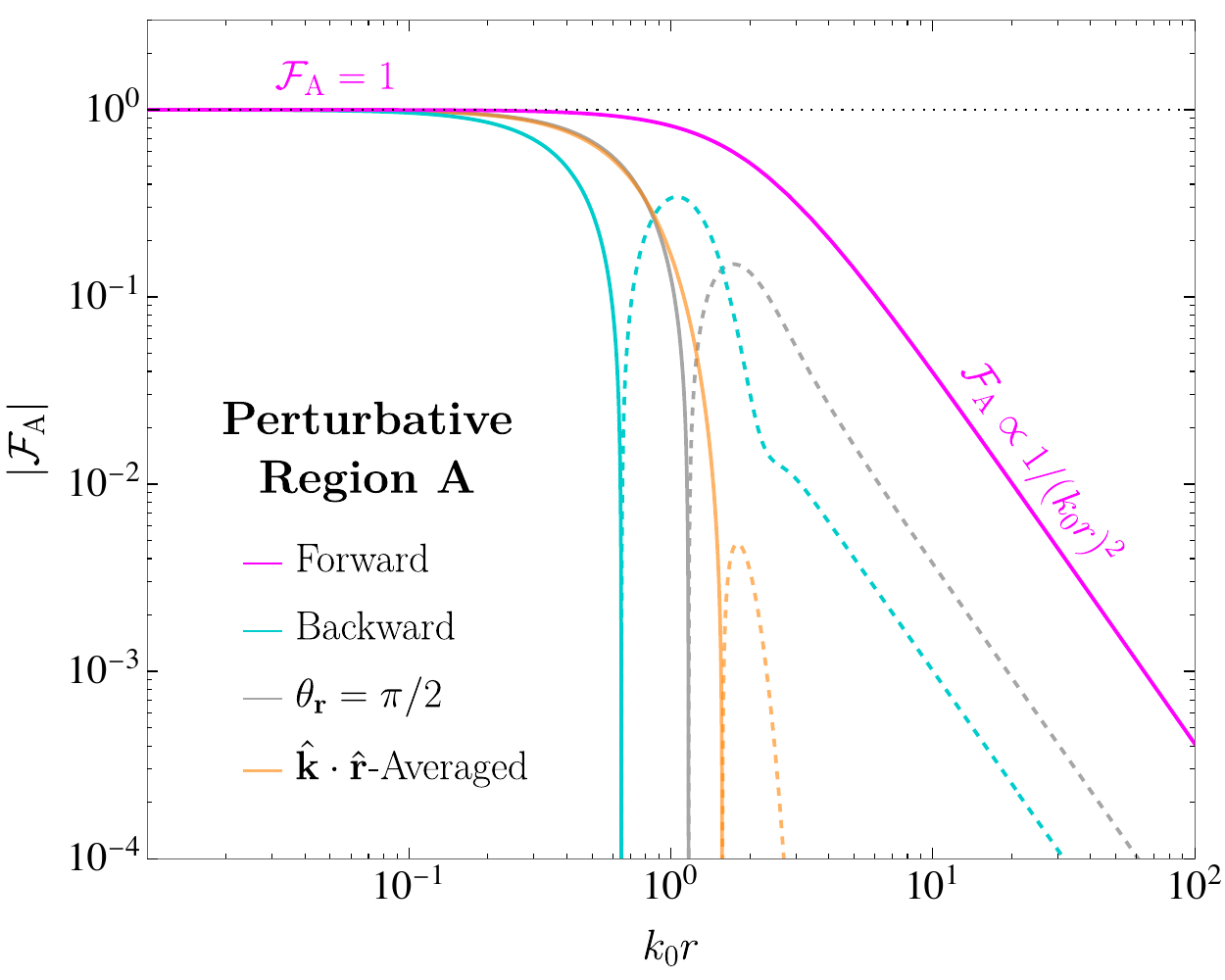}
\caption{The form factor due to the finite-phase-space decoherence effect in Region A, where the source mass $\sourcemass$ is point-like, and the matter effect is perturbative. The magenta, cyan, and gray lines represent the form factor in the forward, backward, and $\theta_\vecr = \pi/2$ directions, respectively. The orange line denotes the form factor averaged over $\hat{\veck} \cdot \hat{\vecr}$. The solid line represents $\abs{\formfactor_\text{A}}$ when $\formfactor_\text{A} > 0$, while the dashed line represents $\abs{\formfactor_\text{A}}$ when $\formfactor_\text{A}<0$. In the small-distance regime~($k_0 r < 1$), all the lines tend toward $\abs{\formfactor_\text{A}} = 1$. In the large-distance region~($k_0 r > 1$), the magenta, cyan, and gray lines obey the power law $\abs{\formfactor_\text{A}} \propto 1/(k_0 r)^2$, while the orange line exhibits the exponential oscillatory damping.}
\label{fig:FA_plt}
\end{figure}

\subsection{Perturbative Region}\label{subsec:perturbative_bg}

Let us start with the simplest case as a warm-up, where the parameter space is located in Region A. In this region, the source mass is treated as point-like in the long wavelength limit, and its distortion to the scalar background is perturbative. Recalling the scattering amplitude computed in \Eq{f2}, we have $ f(\theta;k) = - \mMsource^2\mathcal{V}_\sourcemass/4\pi$. Note that now the scattered object becomes the source mass rather than the test mass, so we replace the index $\testmass$ in \Eq{f2} with $\sourcemass$. Therefore, the background-induced potential under the monochromatic dark matter wind characterized by momentum $\veck$ is
\bea
\label{eq:Vbg_A}
\text{Region A}: \quad V_\bg(\vecr;\veck) = - \frac{\rho_\phi}{m_0^2} \frac{(\mMtest^2 \mathcal{V}_\testmass)\,(\mMsource^2 \mathcal{V}_\sourcemass)}{4 \pi r} \times \cos(kr - \veck \cdot \vecr).
\eea
In the upper two panels of \Fig{peturb_sca_plt}, we show the full spatial configuration of the dimensionless background-induced potential in the perturbative Region A for a monochromatic incident wave with momentum $\veck$, namely the factor $\abs{\psi}^2/\abs{\psi_0}^2-1$ appearing in \Eq{Vbg}~\footnote{We emphasize that the analytic expression in \Eq{Vbg_A} is primarily applicable to the zoomed-out view (right panel of \Fig{peturb_sca_plt}), where the source $\sourcemass$ can be approximated as point-like. Resolving the full spatial configuration, especially the zoomed-in structure shown in the left panel of \Fig{peturb_sca_plt}, requires either the partial-wave analysis presented in \Appx{partial_wave_appx} or the direct Born-level integration given in \Eq{Vbg_finitesize}. In perturbative Regions~A and~B, these two approaches are equivalent.}.

Integrating $V_\bg$ in \Eq{Vbg_A} over the phase space utilizing \Eq{Vbg_phaseint}, we have
\bea
\label{eq:phase_space_factor}
\text{Region A}: \quad \formfactor_\text{A}(\vecr)  = \int \dbar^3 \veck \, \frac{f_\phi(\veck)}{n_\phi}\cos(k r-\veck\cdot \vecr). 
\eea
After we substitute the phase space distribution in \Eq{f_dm} into \Eq{phase_space_factor}, the form factor can be computed. Detailed calculations are provided in  \Appx{finite_phase_space_appx}, and here we briefly summarize the main results. 

The form factors in the forward and backward directions can be computed analytically. We have
\bea
\label{eq:FA_forward}
\text{Forward:}\quad \formfactor_\text{A}(\theta_\vecr = 0) = \frac{2\,(1+\erf(1))}{4 + (k_0 r)^2} + \frac{e^{-(k_0 r)^2}}{ 4 + (k_0 r)^2 } \Re\left\{ e^{-2ik_0 r} [2 - i k_0 r + (k_0 r)^2] \, \erfc(1 - i k_0 r) \right\}
\eea
and
\bea
\label{eq:FA_backward}
\text{Backward:} \quad \formfactor_\text{A}(\theta_\vecr = \pi) & = \frac{2 \, \erfc(1)}{ 4 + (k_0 r)^2 } + \frac{e^{-(k_0 r)^2}}{4 + (k_0 r)^2} \Re \left\{ e^{-2i k_0 r} [2-i k_0 r +(k_0 r)^2 ] \erf(1-i k_0 r) \right\}\\
&  \quad + \frac{e^{-(k_0 r)^2}}{ 4 + (k_0 r)^2 } \left\{ [2+(k_0 r)^2] \cos(2 k_0 r) - (k_0 r) \sin(2 k_0 r) \right\},
\eea
in which we substitute the velocity dispersion $\sigmak = k_0/\sqrt{2}$. We list the concrete computation process in \Appx{finite_phase_space_appx}, which gives the analytical results in the forward and backward directions, as shown in \Eq{FA_forward} and \Eq{FA_backward}. We compare our analytical results from \Eq{FA_forward} and \Eq{FA_backward} with the numerical results from Ref.~\cite{VanTilburg:2024xib}, and we find that these results agree well. The form factors in other directions are computed through the 3D integration as shown in \Eq{F_phase_point_1_appx}. In \Fig{FA_plt}, we show the absolute values of the form factors in the forward, backward, and $\theta_\vecr = \pi/2$ directions in magenta, cyan, and gray lines, respectively. The $\hat{\veck} \cdot \hat{\vecr}$-averaged form factor $\langle \formfactor_\text{A} \rangle_{\hat{\veck}\cdot\hat{\vecr}}$ is plotted in the orange color. The solid lines represent the cases where $\formfactor_\text{A} > 0$, whereas the dashed lines represent the cases where $\formfactor_\text{A} < 0$. 

In the small-distance limit, the forward and backward form factors are
\bea
\label{eq:form_small_r}
k_0 r < 1: \quad  \left\{
\begin{aligned}
& \text{Forward} && \formfactor_\text{A}(\theta_\vecr = 0) \simeq 1-0.2 \, (k_0 r)^2\\
& \text{Backward} && \formfactor_\text{A}(\theta_\vecr = \pi) \simeq 1-3.8 \,(k_0 r)^2\\
\end{aligned}
\right.,
\eea
respectively. These are derived by substituting $\sigmak = k_0/\sqrt{2}$ into \Eq{FA_forward_small_r} and \Eq{FA_backward_small_r}. We find that the form factor in both directions tends to be one when the distance tends to zero, which can be explained by the fact that $\cos(kr - \veck \cdot \vecr)$ becomes non-oscillatory in the long-wavelength limit. When $\formfactor_\text{A} \simeq 1$, the background-induced force has no phase space suppression and behaves like an attractive force obeying the inverse square law $F_\bg \propto 1/r^2$, meaning that the background-induced force induces a shift in the Newtonian constant. From \Fig{FA_plt}, we find that $\formfactor_\text{A}$ in the backward direction turns down at $k_0 r \simeq 0.65$ and flips the signs, whereas $\formfactor_\text{A}$ in the forward direction keeps the positive sign in all the ranges. 

In the large-distance limit, the forward and backward form factors are
\bea
\label{eq:FA_forward_large_r_num}
k_0 r > 1: \quad 
\left\{
\begin{aligned}
& \text{Forward} && \formfactor_\text{A}(\theta_\vecr = 0) \simeq \frac{4}{(k_0 r)^2}\\
& \text{Backward} && \formfactor_\text{A}(\theta_\vecr = \pi) \simeq -\frac{0.1}{(k_0 r)^2}\\
\end{aligned}
\right. ,
\eea
separately. The above two equations are obtained by substituting $\sigmak = k_0/\sqrt{2}$ into \Eq{FA_forward_large_r} and \Eq{FA_backward_large_r}. We find that the form factors in both the forward and backward directions exhibit quadratic suppression in the large-distance limit. Additionally, the form factor in the forward direction is $\sim 40$ times larger than that in the backward direction. This hierarchy arises because, when integrating over phase space, the momentum $\veck$ has more weight around $\veck_0$ and the $\cos(kr-\veck \cdot \vecr)$ term is more oscillatory when $\veck$ is anti-parallel to $\vecr$ than when $\veck$ is parallel to $\vecr$. This leads to stronger cancellation when integrating over the finite phase space if the test mass is located in the backward direction, as described by \Eq{phase_space_factor}. As a result, the phase space distribution introduces a bias, making the forward form factor larger than the backward form factor. Numerical integration confirms that the form factors in all other directions~($ 0 < \theta_\vecr < \pi$) follow the same power-law suppression at large distances. We also find that the form factors in most other directions lie between the forward and backward form factor lines. Thus, the forward and backward form factors serve as good benchmarks for experimental sensitivities. In \Fig{FA_plt}, we illustrate this by plotting the form factor in the direction perpendicular to the mean dark matter momentum~($\theta_\vecr = \pi/2$) using a gray line as an example.

Before proceeding, we analytically compute the background-induced potential averaged over $\hat{\veck} \cdot \hat{\vecr}$, which is qualitatively discussed in \Subsec{phase_space_dist}. Taking the angular average of $\cos(kr - \mathbf{k}\cdot\mathbf{r})$, we have $\langle \cos(kr - \mathbf{k}\cdot\mathbf{r}) \rangle_{\hat{\veck}\cdot \hat{\vecr}} = \sin(2kr)/2kr$, which already exhibits power-law suppression when $kr > 1$. Substituting this averaged quantity into \Eq{Vbg_phaseint_ave} and comparing the result with \Eq{Vbg_formfactor}, we have
\bea
\langle \formfactor_\text{A} \rangle_{\hat{\veck} \cdot \hat{\vecr}} =   e^{-2 \sigmak^2 r^2}  \times \frac{\sin(2k_0 r)}{2 k_0 r}. 
\eea
We show the averaged form factor in \Fig{FA_plt} using orange lines. From the above formula, we find that the $\hat{\veck} \cdot \hat{\vecr}$-averaged form factor experiences exponential suppression in the large-distance limit, whereas $\formfactor_\text{A}$ along the 
$\vecr$-direction only follows a $1/(k_0 r)^2$ suppression law. This indicates strong cancellation among background-induced potentials in different directions due to $\hat{\veck} \cdot \hat{\vecr}$-averaging. This exponential suppression arises because $\hat{\veck}\cdot\hat{\vecr}$-averaging is analogous to rotating the observer in the time domain. In doing so, the observer effectively experiences the dark matter wind from all directions. Thus, we expect the averaged form factor to exhibit similarities with the form factor in an isotropic distribution. Indeed, in the zero-mean-momentum limit, we find $\lim_{k_0 \rightarrow 0} \formfactor_\text{A} = \lim_{k_0 \rightarrow 0} \langle \formfactor_\text{A} \rangle_{\hat{\veck} \cdot \hat{\vecr}} = e^{-2 \sigmak^2 r^2}$, which exactly reproduces the form factor for an isotropic distribution.

\begin{figure}[h]
\centering
\includegraphics[width=0.49\linewidth]{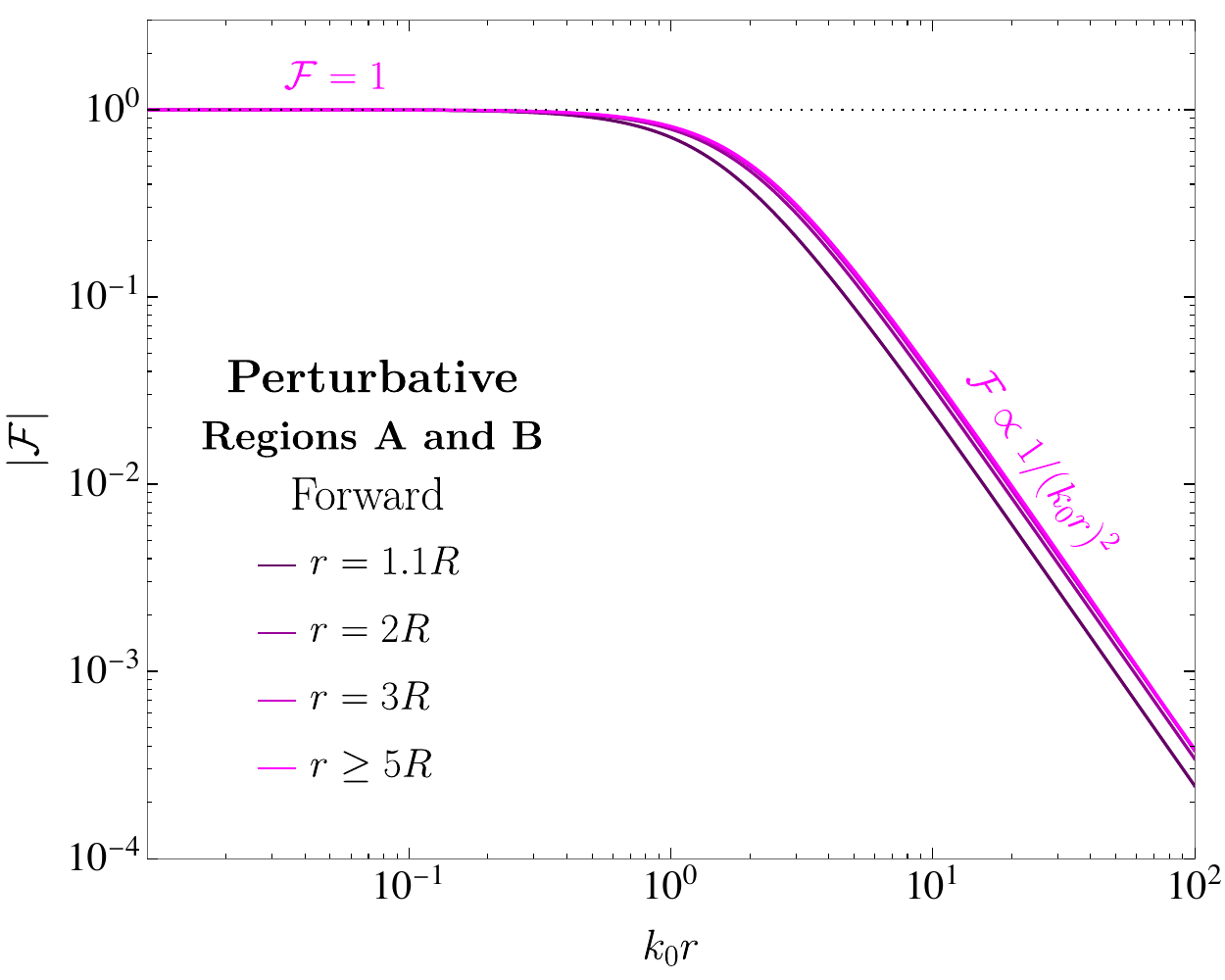}
\includegraphics[width=0.49\linewidth]{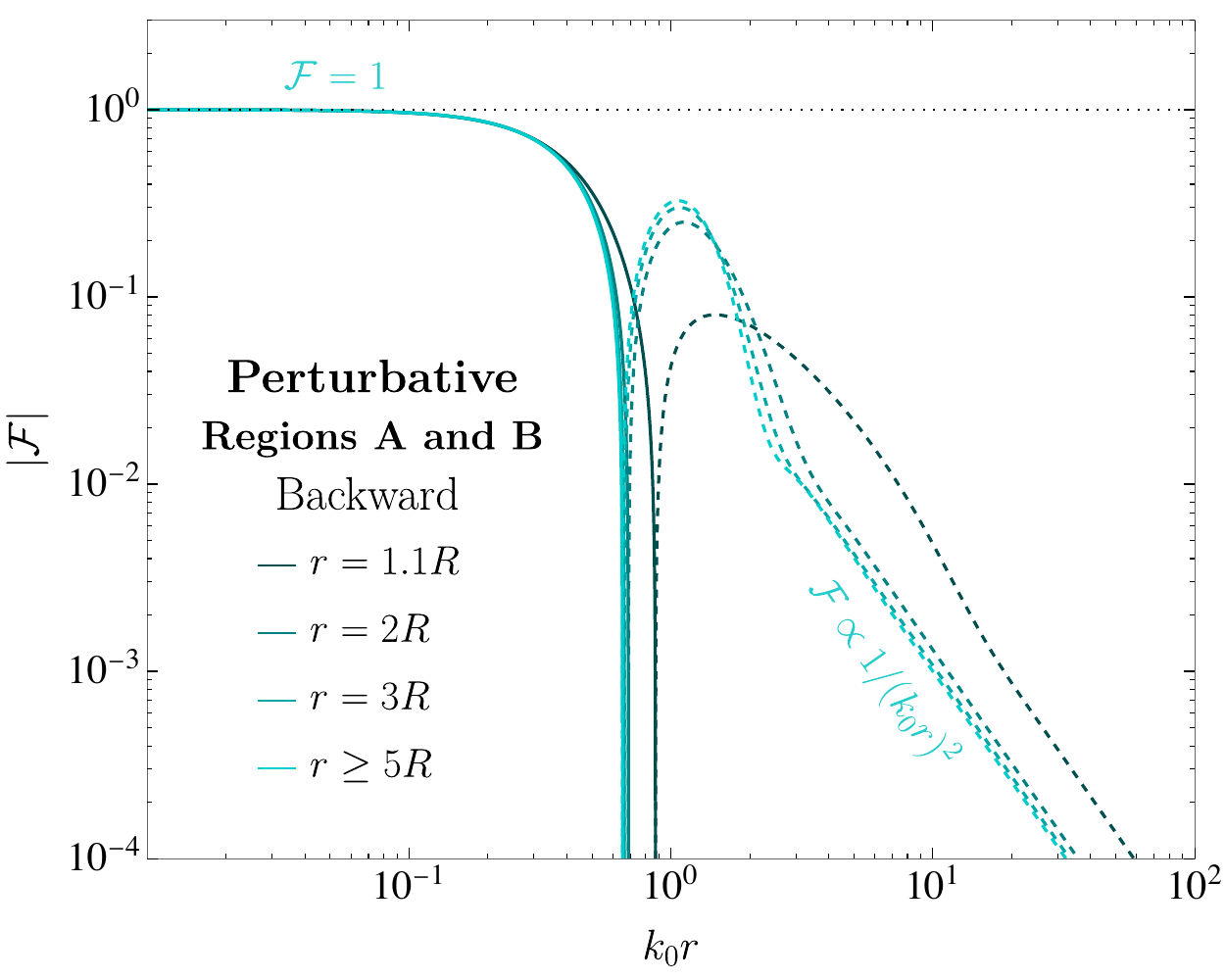}
\caption{Form factors of the background-induced force in the perturbative region, incorporating the finite-size effect of a source sphere with radius $R$. In the perturbative region~(Regions A and B), the form factors are independent of $\mM$ but depend on the geometric factor $r/R$. The finite-size effect becomes significant only when $\abs{r-R}/R < 1$. In other cases, we can directly use the point-source form factor shown in \Fig{FA_plt}. {\bf Left:} Form factors in the forward direction, with magenta lines of varying darkness representing different $r/R$ ratios. {\bf Right:} Form factors in the backward direction, with cyan lines of varying darkness indicating different $r/R$ ratios.}
\label{fig:pert_finitesize_formfactor}
\end{figure}

Now we consider a more general case where the source mass  $\sourcemass$ is not point-like but has a finite geometric shape. In principle, this finite-size effect needs to be taken into consideration in Region B, as illustrated in \Fig{phase_classify}, where the size of the source mass is large compared to the de Broglie wavelength. Since we focus on the perturbative regime, we can apply the Born approximation to obtain the scattered wave function
\bea
\label{eq:born_approx_wave_func}
\text{Regions A and B}:\quad\quad \psisc(\vecr;\veck) =  - \abs{\psi_0} \frac{\mMsource^2}{4 \pi} \int_{\mathcal{V}_\sourcemass} d^3 \delta \vecr_\sourcemass \frac{\exp(i \abs{\veck} \abs{\vecr - \delta \vecr_\sourcemass})}{\abs{\vecr - \delta \vecr_\sourcemass}} \exp(i \veck \cdot \delta \vecr_\sourcemass).
\eea
In the above formula, $\vecr$ is the position vector pointing from the reference point $O_\sourcemass$ on the source mass to the test mass $\testmass$ and $\delta \vecr_\sourcemass$ is the position vector pointing from the reference point $O_\sourcemass$ to points in the source mass. One can choose $O_\sourcemass$ flexibly as long as it is attached to the source. For a spherical source, it is naturally positioned at the center of the sphere. In the far-field limit, we can expand the integrand in terms of $\delta \vecr_\sourcemass$. Then, the wave function can be factorized into the product of the scattering amplitude $f(\theta;k)$ and the outgoing spherical wave $e^{ikr}/r$, as shown in \Eq{psi_sc_born_far_appx}. The derivation of this expression and its far-field approximation is included in \Appx{wave_func_appendix} for completeness. By substituting \Eq{born_approx_wave_func} into \Eq{Vbg} and then performing the phase space averaging using \Eq{Vbg_phaseint}, we obtain the monochromatic background-induced potential incorporating the finite-size effect, which is
\bea
\label{eq:Vbg_finitesize}
\text{Regions A and B}:\quad \, V_\bg(\veck) = - \frac{\rho_\phi}{m_0^2} \, \frac{(\mMtest^2 \mathcal{V}_\testmass) \, (\mMsource^2 \mathcal{V}_\sourcemass)}{4\pi r} \times \frac{r}{\mathcal{V}_\sourcemass} \int_{\mathcal{V}_\sourcemass} d^3 \delta \vecr_\sourcemass \frac{\cos(\abs{\veck}\abs{\vecr - \delta \vecr_\sourcemass}-\veck \cdot (\vecr - \delta \vecr_\sourcemass))}{\abs{\vecr - \delta \vecr_\sourcemass}}.
\eea
Here, we treat the test mass to be point-like to simplify the discussion. For the situation where the test mass also has a nonnegligible geometric size, we can use the general formula listed in \Eq{Vbg_finitesize_full_appx}. In \Fig{peturb_sca_plt}, we show the full spatial configuration of the monochromatic dimensionless background-induced potential in the perturbative Regions A and B, namely the factor $\abs{\psi}^2/\abs{\psi_0}^2-1$ appearing in \Eq{Vbg}.

Comparing the background-induced potential listed in \Eq{Vbg_finitesize} with \Eq{Vbg_formfactor}, we have
\bea
\label{eq:pert_finite_size_factor}
\text{Regions A and B}:\quad \quad \mathcal{F} = \frac{r}{\mathcal{V}_\sourcemass} \int_{\mathcal{V}_\sourcemass} \frac{d^3 \delta \vecr_\sourcemass}{\abs{\vecr - \delta \vecr_\sourcemass}} \int \dbar^3 \veck \, \frac{f_\phi(\veck)}{n_\phi}  \cos(\abs{\veck} \abs{\vecr-\delta\vecr_\sourcemass}-\veck\cdot\left(\vecr-\delta\vecr_\sourcemass\right)).
\eea
From \Eq{pert_finite_size_factor}, we find that, because of the linearity of the phase-space and spatial integrals, the total background-induced potential in the perturbative region can be treated as a linear superposition of the background-induced potentials~(averaged over phase space) arising from interactions between infinitesimal pieces of the source mass and the test mass. \Eq{pert_finite_size_factor} is the form factor incorporating both the finite-size and finite-phase-space effects. When the source also has a negligible geometric shape in the long wavelength limit, \Eq{pert_finite_size_factor} goes back to \Eq{phase_space_factor}.

For monochromatic phase space distribution where $f_\phi(\vecp) = (2\pi)^3 n_\phi \delta^{(3)}(\vecp- \veck)$, $\mathcal{F}$ is merely the integration over the spatial coordinate. In this situation, because the cosine term in \Eq{pert_finite_size_factor} is oscillatory, there is strong suppression of the form factor, especially in the non-forward direction where the background-induced force is more oscillatory. A more detailed discussion of the finite-size effect given the monochromatic phase space distribution can be found in \Appx{finite_size_appx}. Here, we merely list the analytical approximation in the far field limit, which is $\mathcal{F} \simeq 3(\sin(qR)-qR\cos(qR))/(qR)^3$. $q = 2k \sin(\theta/2)$ is the magnitude of aforementioned momentum transfer. In the forward direction, we have $\mathcal{F}(\theta_\vecr = 0) \simeq 1$. In the backward directions, we have $\mathcal{F}(\theta_\vecr = \pi) \propto 1/(qR)^2$. The quadratic suppression of the form factor exists in the angular region satisfying $\theta \gtrsim (kR)^{-1}$.

In realistic scenarios, the dark matter phase space follows the boosted Maxwell-Boltzmann distribution, as shown in \Eq{f_dm}. In this situation, both the source mass and the phase space have a finite width. Therefore, we need to compute the form factors numerically. In this work, we employ two methods to compute the form factor, which yields equivalent results in the perturbative regime. The first method involves directly evaluating the integral in \Eq{pert_finite_size_factor}. The second method involves computing the scattered wave function $\psi_\sct$ through the partial wave analysis, substituting the acquired wave function into \Eq{Vbg}, averaging the background-induced potential over the phase space, and directly comparing the potential with \Eq{Vbg_formfactor}. We have verified numerically that the form factors obtained from both methods agree very well in the perturbative regions~(Regions A and B). Importantly, in the non-perturbative regions~(Regions C, D, E$\setminus$B), the Born approximation breaks down, and only the partial wave analysis or full numerical simulation can reliably describe the system. A more detailed discussion of the partial wave analysis is provided in \Subsec{bg_nonp_highk}. In \Fig{pert_finitesize_formfactor}, we present the form factors for different values of the geometric ratio $r/R$, where $r/R=1.1,2,3,\geq 5$. The left panel shows the forward-direction form factors using magenta lines with varying darkness, while the right panel displays the backward-direction form factors using cyan lines with similar variations of darkness. From these plots, we observe that when $\abs{r-R}/R \gtrsim \mathcal{O}(1)$, the finite-size effect of the source becomes negligible, and the form factor closely matches that of a point source, as shown in \Fig{FA_plt}. However, when  $\abs{r-R}/R < 1$, the form factor deviates significantly from the point-source result. Notably, the $r=1.1R$ line in \Fig{pert_finitesize_formfactor} reveals that in this regime, the backward form factor is enhanced while the forward form factor is suppressed. This behavior can be understood by considering the geometry: as the test mass approaches the source, the portion of the source contributing to the potential from the non-backward direction increases for the test mass located at $\theta_\vecr = \pi$, while the contribution from the forward direction decreases for the test mass located at $\theta_\vecr = 0$. We have numerically verified that in the perturbative region, making the test mass even nearer~(such as choosing $r=1.01R$) does not further modify the form factor because the geometric configuration does not change as long as the $\abs{r-R}/R < 1$ condition is already satisfied.

\subsection{Non-Perturbative Region: Low Momentum}\label{subsec:bg_nonp_lowk}

Now, we move to the non-perturbative region where the screening effect becomes important. The non-perturbative background-induced force has previously been explored by Refs.~\cite{dePireySaintAlby:2017lwc,Berezhiani:2018oxf,Hees:2018fpg,Banerjee:2022sqg} using the spherical symmetric ansatz. However, these studies only cover the low-momentum and near-field region~(Regions A and C with $kr < 1$) rather than other regions with relatively high momentum of the incident dark matter. In this subsection, we review their treatment but adapt to the notation and formalism used in this paper. 

When the ultralight scalar forms a spherically symmetric configuration around the source, which is a uniform sphere, it has zero momentum in \Eq{Schrodinger_Eq}, which gives
\bea
\vecnabla^2 \psi_\sph = \mMsource^2 \, \theta(R_\sourcemass-r) \, \psi_\sph.
\eea
Using the spherical symmetric ansatz and imposing the boundary conditions that $\psi_\sph = \abs{\psi_0}$ when $r \rightarrow \infty$ and $d \psi_\sph/dr = 0$ at $r = 0$, we know the wave function inside and outside the sphere are
\bea
\label{eq:psi_bound}
\psi_\sph(r) = 
\abs{\psi_0} \times \left\{\begin{aligned}
& C_1  \frac{\sinh(\mMsource r)}{r}  \quad &(r \leq R_\sourcemass)\\
& 1 + \frac{C_2}{r}  \quad &(r \geq R_\sourcemass)
\end{aligned}
\right. .
\eea
Matching the boundary conditions of the wavefunctions inside and outside the spheres, we have
\bea
C_1 = \frac{1}{\mMsource \cosh(\mMsource R_\sourcemass)}, \quad \quad C_2 = - \frac{\mMsource^2 \mathcal{V}_\sourcemass}{4\pi}  J_+(\mMsource R_\sourcemass).
\eea
Therefore, the function outside the sphere is represented as
\bea
\label{eq:psi_out_bound}
r\geq R_\sourcemass: \quad \quad \psi_\sph(r) = \abs{\psi_0} \bigg[ 1- \frac{\mMsource^2 \mathcal{V}_\sourcemass}{4\pi r}  J_+(\mMsource R_\sourcemass)\bigg].
\eea
Here, $\mathcal{V}_\sourcemass = 4 \pi R_\sourcemass^3/3$ is the volume of the source. The function $J_+$ is defined as
\bea
J_+(x) = \frac{3(x-\tanh(x))}{x^3}.
\eea
In the region where $x \ll 1$ limit, there is $J_+ \rightarrow 1$, while when $x \gg 1$, there is $J_+ \rightarrow 3/x^2$ and $1 - (x^2/3)J_+ \rightarrow 1/x$. Therefore, in the perturbative region, $C_2 \rightarrow \mMsource^2 \mathcal{V}_\sourcemass/4 \pi$, while in the highly non-perturbative region, $C_2 \rightarrow 1$.

Substituting the wave function outside the sphere in \Eq{psi_out_bound} into \Eq{Vbg}, we have
\bea
V_{\bg,\,\sph} = - \frac{\rho_\phi}{m_0^2} \frac{(\mMsource^2 \mathcal{V}_\sourcemass)(\mMtest^2 \mathcal{V}_\testmass)}{4 \pi r} \times \formfactor_\sphscr(r).
\eea
$\formfactor_\sphscr$ is the form factor incorporating the screening effect of the spherically symmetric configuration, and it is represented as
\bea
\label{eq:formfactor_sphscr}
\formfactor_\sphscr(r) = J_+(\mMsource R_\sourcemass) \times \left[ 1 - \frac{1}{2} \frac{\mMsource^2 \mathcal{V}_\sourcemass}{4 \pi r} J_+(\mMsource R_\sourcemass)\right].
\eea
When $\mMsource R_\sourcemass < 1$, the matter effect of the source is perturbative. In this region, we have $\formfactor_\sphscr \simeq 1$. When deeply going inside the strong-coupling region, where $\mMsource R_\sourcemass > 1$, we have 
\bea
\label{eq:formfactor_sphscr_farfield_hard}
\formfactor_\sphscr \simeq \frac{3}{(\mMsource R_\sourcemass)^2} \quad \quad \quad \text{(Strongly-Coupled + Far Field)}
\eea
in the far-field region, which is understood as the screening effect. The suppression factor in \Eq{formfactor_sphscr_farfield_hard} has two contributions:
1.~When the potential barrier is sufficiently high, the scalar field cannot penetrate the solid sphere $\sourcemass$ and instead decays exponentially inside it. As a result, only a thin spherical shell with thickness set by the penetration depth, $\Delta R_\sourcemass \simeq 1/\mMsource$, contributes effectively to the background-induced force, leading to a geometric suppression factor $\frac{4\pi R_\sourcemass^2 \Delta R_\sourcemass}{\mathcal{V}_\sourcemass}\simeq \frac{3}{\mMsource R_\sourcemass}$. 2.~According to \Eq{psi_out_bound}, the scalar field at the surface of $\sourcemass$ is further suppressed by an additional factor $\frac{1}{\mMsource R_\sourcemass}$. This reduces the scalar background near the source $\sourcemass$ and hence the background-induced potential. Combining these two effects gives the form factor $\formfactor_\sphscr$ in \Eq{formfactor_sphscr_farfield_hard}.

Apart from the method described above, which utilizes the spherically symmetric ansatz to derive the form factor of the background-induced force, we introduce an alternative approach based on partial wave analysis in the s-wave limit. The detailed calculation for arbitrary distances from the source can be found in \Appx{swave_approx_appx}. Here, we choose the far-field limit to simplify the discussion. In the strong coupling regime (Region C), the source can be treated as a hard sphere. Because we focus on the region where the incident momentum is negligible~($k_0 R_\sourcemass < 1$), we only need to keep the s-wave component. Based on this, we substitute the s-wave phase shift 
$\delta_0 = - k R_\sourcemass$  into the scattering amplitude given in \Eq{ftheta_partial}, and have
\bea
\label{eq:fC_amp}
f_\text{C}(\theta;k) = \frac{\delta_0}{k} = - R_\sourcemass. 
\eea
Substituting \Eq{fC_amp} into \Eq{Vbg_far}, we obtain
\bea
\label{eq:amplitudecalc_formfactor_sphscr_farfield_hard}
\formfactor_\sphscr  \simeq \frac{f_\text{C}(\theta;k)}{f_\text{A}(\theta;k)} = \frac{-R_\sourcemass}{ - \frac{\mMsource^2 \mathcal{V}_{\sourcemass}}{4\pi} } =  \frac{3}{(\mMsource R_\sourcemass)^2} \quad \quad \quad \text{(Strongly-Coupled + Far Field)},
\eea
which gives the same form factor for the screening effect as \Eq{formfactor_sphscr} in the strongly coupled and far-field limit, but with a new physical interpretation. From \Eq{fC_amp}, we see that the scattering amplitude in the strongly coupled Region C is independent of $\mMsource$ or, equivalently, independent of the strength of the matter effect induced by the source. In contrast, the scattering amplitude in Region A follows the relation $f_\text{A} \propto \mMsource^2$. This indicates that when the source becomes sufficiently ``hard'', the scattering of the incident scalar reaches saturation, regardless of further increasing the coupling constant. However, when computing the form factor, one might still expect that increasing the coupling constant would enhance scalar scattering, as inferred from the perturbative Region A. Therefore, the screening effect can also be interpreted as the gap between the expected scattering behavior derived from perturbative calculations in Region A and the reality of the non-perturbative saturation in Region C. Notably, increasing the incident momentum can enhance the scattering and consequently increase the background-induced force. This momentum-dependent enhancement of the background-induced force is the basic characteristic of the descreening effect discussed in \Subsec{bg_nonp_highk}.

\begin{figure}[t!]
\centering
\includegraphics[width=0.49\linewidth]{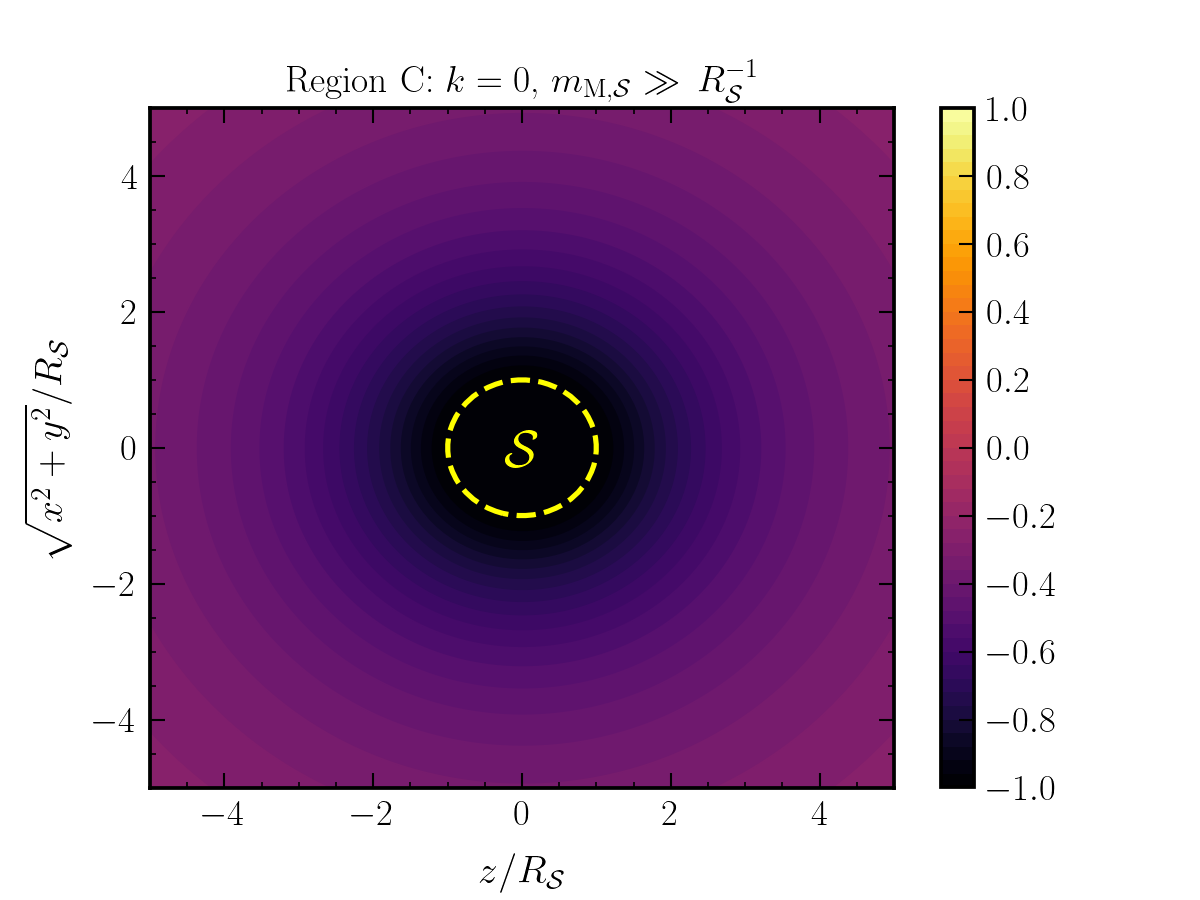}
\includegraphics[width=0.49\linewidth]{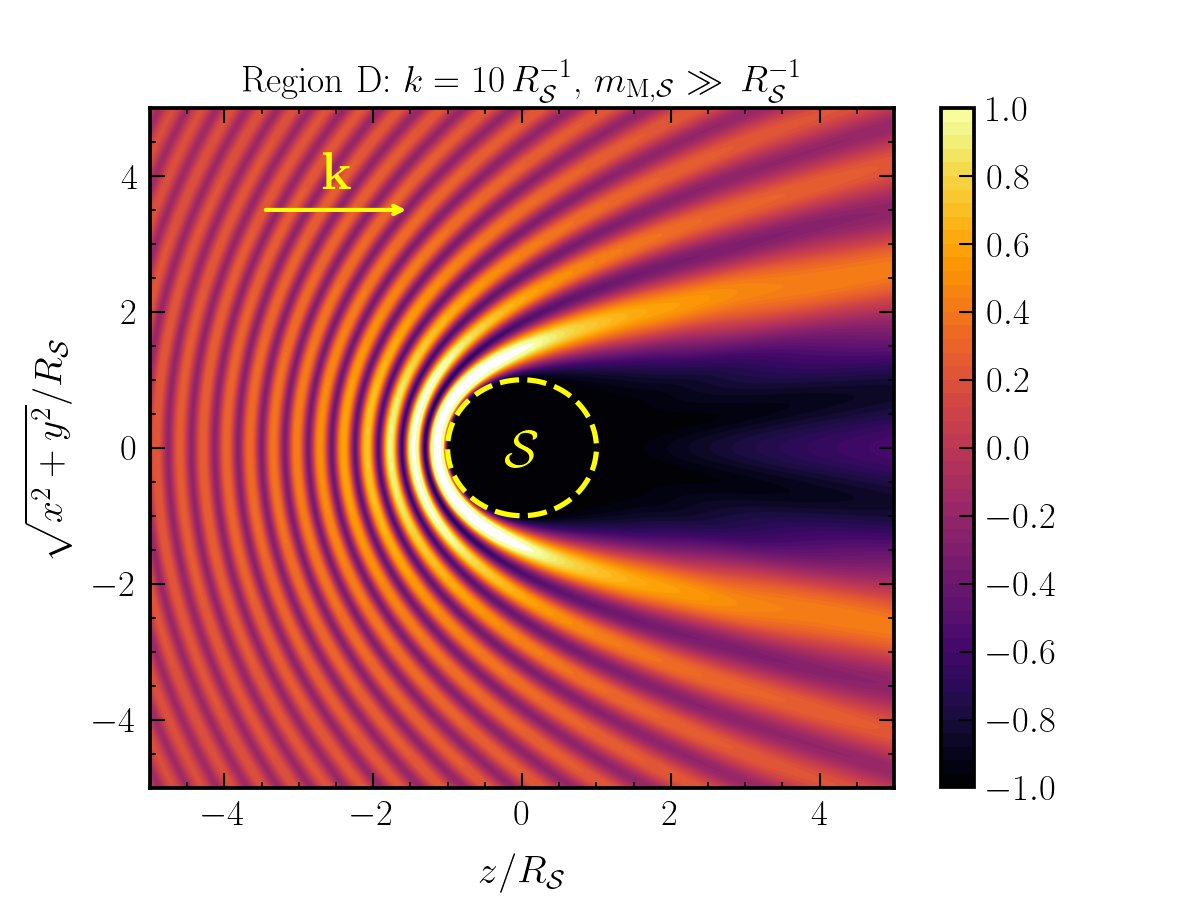}\\
\caption{The plots show $\abs{\psi}^2/\abs{\psi_0}^2-1$ in the non-perturbative
regime $\mMsource \gg k, R^{-1}_\sourcemass$, where the screening effect becomes significant, and the source is treated as a hard sphere. The yellow dashed lines indicate the boundary of the source~(scattered) object. {\bf Left:} Region C. This corresponds to low-momentum hard sphere scattering, where only the s-wave\,($l=0$) contributes to the wave function. Notably, the scalar field is suppressed at the sphere's surface, giving $\abs{\psi(r=R_\sourcemass)}/\abs{\psi_0} = 1/\mMsource R_\sourcemass$ as a consequence of matching the boundary conditions. {\bf Right:} Region D. This corresponds to high-momentum hard sphere scattering, where multiple partial waves contribute to the wave function. One can see that the scalar is pushed toward the surface of the source as its kinetic energy overcomes the gradient energy.}
\label{fig:nonpeturb_sca_plt}
\end{figure}

For an intuitive illustration of the screening effect, one can refer to the left panel of \Fig{nonpeturb_sca_plt}, which is computed through the partial wave analysis and consistent with the field configuration shown in \Eq{psi_bound}. This figure shows the spherically symmetric configuration of $\abs{\psi}^2/\abs{\psi_0}^2 - 1$, the dimensionless and test-mass-independent part of the background-induced potential. We can find that near or inside the sphere, the field value of the scalar field is highly suppressed, while in the large distance, the scalar field tends to $\abs{\psi} \simeq \abs{\psi_0}$, which is consistent with the imposed boundary condition for the spherically symmetric solution covered in this section.

However, as one can see in the right panel of \Fig{nonpeturb_sca_plt}, such a spherical ansatz breaks down when $kR > 1$. In this region, two additional effects arise compared with the spherical symmetric ansatz. The first effect is that the high momentum of the dark matter wind squeezes the dark matter to the near-surface of the source mass, alleviating the screening effect mentioned above. The second effect is the decoherence suppression from the finite-phase-space and finite-size effects when $kr > 1$, which is discussed in \Subsec{perturbative_bg} for the perturbative case. We stress that even in Regions A and C, where the s-wave approximation is legitimate, the spherical symmetric ansatz still cannot capture the complete information, as long as $k_0 \neq 0$. The reason is that we can always find the large distance region satisfying $k_0 r \gtrsim 1$ as long as there is a hierarchy between $r$ and $R$, where the finite-phase-space decoherence suppression shown in \Fig{FA_plt} or \Fig{pert_finitesize_formfactor} becomes significant. We attach more detailed discussions in \Appx{swave_approx_appx}.

\subsection{Non-Perturbative Region: High Momentum}\label{subsec:bg_nonp_highk}

In this section, we discuss the non-perturbative region characterized by high incident mean momentum $k_0$ of the dark matter. At the beginning of this section, we decompose the form factor in \Eq{Vbg_formfactor} as
\bea
\label{eq:FormFactor_Nonp_Decompose}
\formfactor(\vecr, \veck_0) = \formfactor_\sphscr(r)  \times \formfactorreduce(\vecr, \veck_0).
\eea
Here, $\formfactor_\sphscr$ is the spherically symmetric form factor described in \Eq{formfactor_sphscr}, incorporating the screening effect in the zero-momentum limit. In \Eq{FormFactor_Nonp_Decompose}, we define the reduced factor as $\widetilde{\formfactor} \equiv \formfactor/\formfactor_\sphscr$, which captures the interplay between the decoherence and screening effects in the strongly-coupled regime when the incident scalar momentum is significant. The deviation of the reduced form factor $\formfactorreduce$ from $1$ quantifies the deviation of the real background-induced force from the one assuming the spherically symmetric ansatz, as shown in \Eq{formfactor_sphscr}. In the strongly-coupled region where $\formfactor$ is highly suppressed by the screening effect, the definition of $\formfactorreduce$ normalizes the value to one when $k_0 r < 1$, which helps the visualization. 

Notably, we discover the descreening (or deshielding) effect in the background-induced force, which substantially alleviates the screening effect and compensates for decoherence suppression when the incident momentum increases. Moreover, in contrast to the perturbative finite-size effect discussed in \Fig{pert_finitesize_formfactor}, which at most changes the form factor by an order of magnitude, the finite size of the source mass becomes crucial in the non-perturbative region, leading to changes in the form factor by several orders of magnitude.

To numerically compute the values of $\formfactor$ and $\formfactorreduce$ crossing the perturbative and non-perturbative regions, we use the partial wave analysis introduced in \Appx{partial_wave_appx}. First, we calculate the incident wave function $\psiinc$ using \Eq{inc_wave_appx} and the scattered wave function $\psisc$ using \Eq{sca_wave_appx}. We utilize the complete expression of the incident plane wave $\psiinc$ and keep the partial wave components of the scattered wave $\psisc$ up to $l_{\max}\sim kR_\sourcemass$. The specific choice of $l_{\max}$ is listed in \Appx{phase_shifts_appx}. Next, we compute the background-induced potential using the full expression in \Eq{Vbg}. Finally, we perform the phase space integration using \Eq{Vbg_phaseint_2}. To optimize computational efficiency on the cluster, we evaluate the form factors in the forward and backward directions as a benchmark. This approach reduces the 3D integral in \Eq{Vbg_phaseint_2} to a 2D integral over $k$ and $\theta_\veck$. We set $\sigmak = k_0/\sqrt{2}$, which is consistent with the rest of this work. It is essential to ensure that the grid resolution for momentum and angular integrals is sufficiently high to capture the oscillatory features of the background-induced potential. We set the grid number for momentum integration to approximately $N_k = 50 \, k_0 r$. We choose the grid number of the $\theta_\veck$ integration to be $N_{\theta_\veck} = 200$ when $kr\leq 100$, and $N_{\theta_\veck} =1000$ when $100<kr\leq 500$. The numerical results remain stable when varying the grid size around these values. Before performing the non-perturbative calculations, we compute the form factor in the perturbative regions~(Regions A and B) and confirm that the results acquired from the partial wave analysis are consistent with the results based on the Born approximation listed in \Eq{pert_finite_size_factor}

\begin{figure}[t!]
\centering
\includegraphics[width=0.49\linewidth]{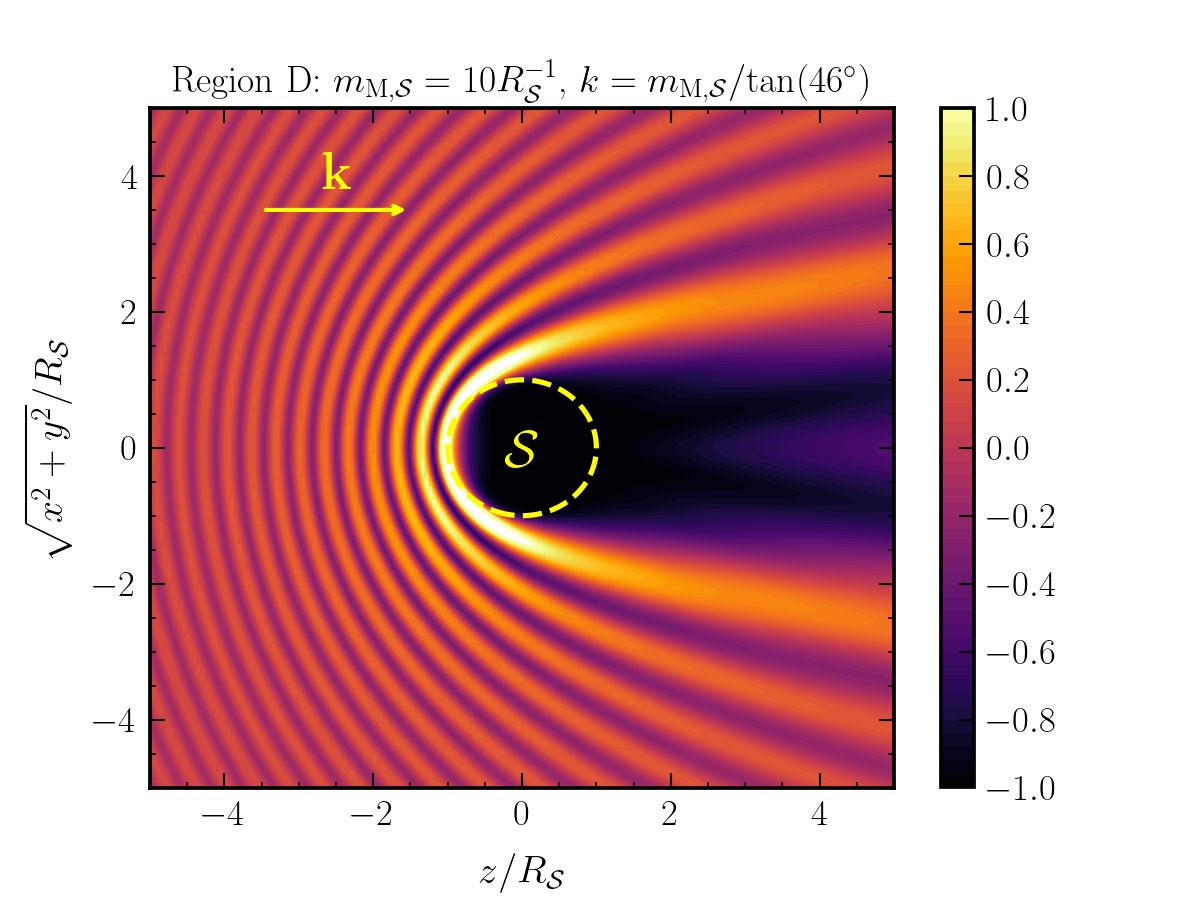}
\includegraphics[width=0.49\linewidth]{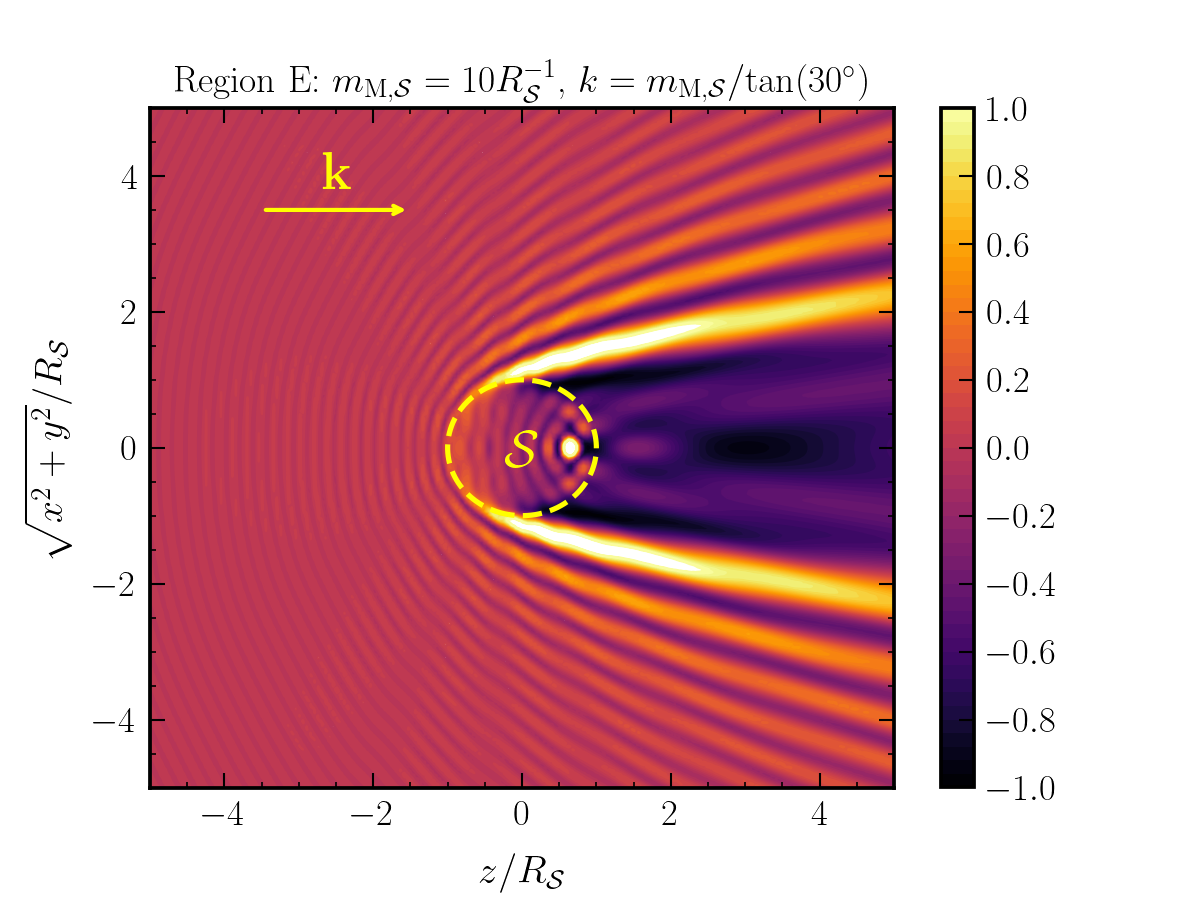}
\caption{The plots show $\abs{\psi}^2/\abs{\psi_0}^2-1$ with the same effective mass $\mMsource = 10\, R^{-1}_\sourcemass$ but with different incident momentum to illustrate the descreening effect. {\bf Left:} The scattering pattern for $k = \mMsource/\tan(46^\circ)$. The chosen momentum is slightly below the source barrier threshold, resulting in partial shielding of the incident wave. However, since the scalar field is already compressed near the source, the screening effect in the background-induced force is already significantly alleviated. {\bf Right:} The scattering pattern for $k = \mMsource/\tan(30^\circ)$. We can see that with the increased momentum, the entire barrier is fully penetrated, completely eliminating the screening effect. In this regime, the entire source contributes to the background-induced force.}
\label{fig:nonpeturb_penetrate}
\end{figure}

\begin{figure}[t]
\centering
\includegraphics[width=0.49\linewidth]{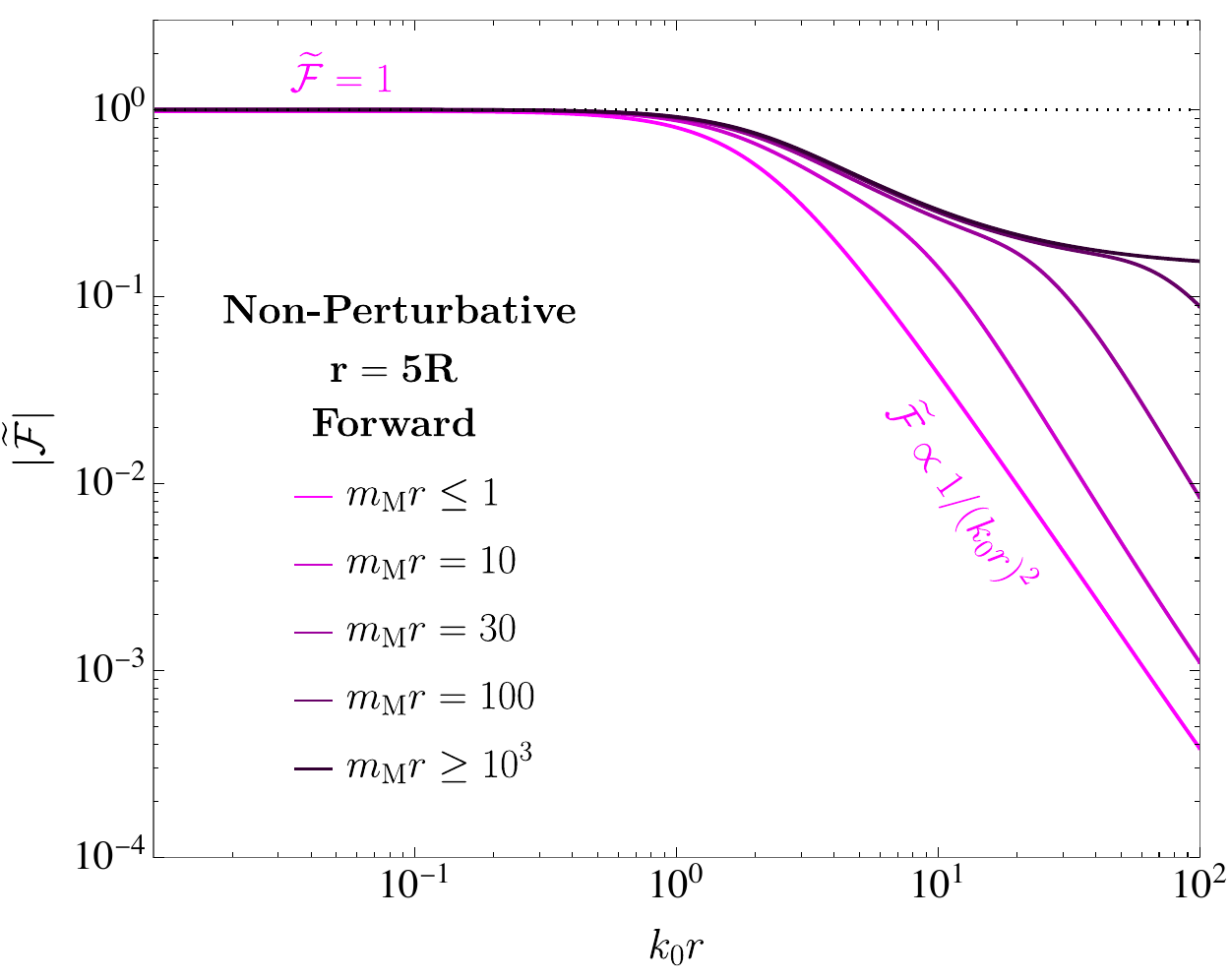}
\includegraphics[width=0.49\linewidth]{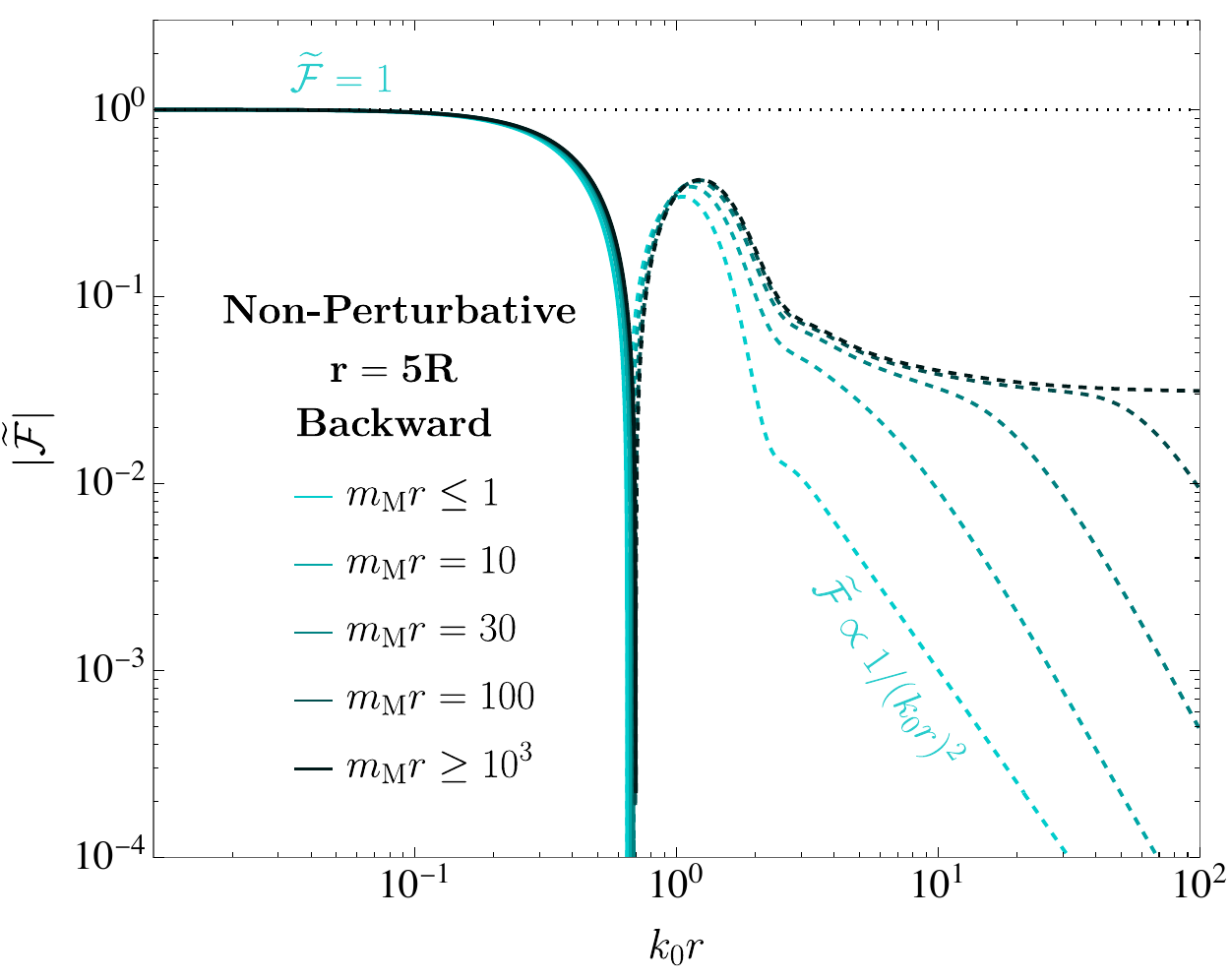}
\caption{The reduced form factor with $r=5R$ for different values of $\mM$. When $k_0 r < 1$, the reduced form factors approach $\formfactorreduce = 1$. In the intermediate regime where $1 < k_0 r < \mM r$, the reduced form factors become flatter, indicating the descreening effect, which reduces decoherence suppression. For $k_0 r > \mM r$, the reduced form factors follow the power law $\formfactorreduce \propto 1/(k_0 r)^2$, consistent with the perturbative form factor shown in \Fig{pert_finitesize_formfactor}. {\bf Left: } Reduced form factors in the forward direction~$(\theta_\vecr = 0)$, represented by magenta lines. {\bf Right:} Reduced form factors in the backward direction~$(\theta_\vecr = \pi)$, represented by cyan lines.}
\label{fig:nonp_formfactor}
\end{figure}

In the low momentum region with small distances, as discussed in \Subsec{bg_nonp_lowk}, the form factor is
\bea
\text{Regions A and C~($k_0 r<1$)}: \quad  \quad  \quad \formfactor \simeq \formfactor_\sphscr,
\eea
indicating that the spherically symmetric ansatz introduced in \Subsec{bg_nonp_lowk} is valid. In this regime, the reduced form factor is approximately $\formfactorreduce \simeq 1$. One can refer to \Fig{nonp_formfactor} for references. In the plots, we present the reduced form factors for various effective masses with a fixed geometric factor of $r/R=5$. The magenta lines in the left panel represent the form factors in the forward direction, while the cyan lines in the right panel show the reduced form factors in the backward direction. As shown in the plots, all lines converge to $1$ when $k_0 r < 1$, indicating that the spherically symmetric ansatz is a reliable approximation in this region.

In the region where the incident momentum exceeds the height of the barrier, we have
\bea
\label{eq:descr_regionb}
\text{Region E~$(k_0 > \mM)$:} \quad \quad \quad \formfactor \simeq \formfactor_\text{pert},
\eea
where $\formfactor_\text{pert}$ denotes the perturbative form factor calculated in \Subsec{perturbative_bg}. In this regime, the incident scalar penetrates the source barrier.  Consequently, the entire volume of the source contributes to the background-induced force, and the background-induced potential reverts to the scenario discussed in \Subsec{perturbative_bg}. It is important to note that, even within this region, a portion of the parameter space may still satisfy $\mM R > 1$,  indicating that $\formfactor_\sphscr$ is smaller than $1$. However, this suppression is balanced by the enhancement of $\formfactorreduce$ compared to the decoherence form factor $\formfactor_\text{pert}$ in the perturbative region.

For an intuitive illustration of the descreening effect, one can refer to \Fig{nonpeturb_penetrate},  where the effective mass is determined while varying the incident momentum. In the left panel of \Fig{nonpeturb_penetrate}, when the incident momentum $k$ is slightly smaller than $\mM$, the scalar is squeezed near the source barrier, resulting in an increased field strength compared to the zero momentum scenario shown in the left panel of \Fig{nonpeturb_sca_plt}. As the incident momentum approaches the barrier height, the barrier becomes partially penetrated. The right panel of \Fig{nonpeturb_sca_plt} depicts the scenario where the incident momentum exceeds the barrier height, allowing the scalar wave to fully penetrate the source. In this case, the entire volume of the source contributes to the background-induced potential. The coherence enhancement of the scalar within the source is also observed in the right panel. Similar phenomena occur in 1D quantum mechanical scattering for the low potential barrier or potential well. Furthermore, as depicted in \Fig{nonpeturb_sca_plt}, when the scalar penetrates the source, the forward direction amplitude is enhanced, while the backward direction amplitude is suppressed, consistent with the analysis presented in \Appx{wave_func_appendix}. This scalar distribution is qualitatively distinct from the scenario in which the source behaves as a hard sphere, as shown in the right panel of \Fig{nonpeturb_sca_plt}. In the latter case, the source obstructs the forward-moving scalar when $\abs{r-R}/R < 1$  and reflects the incident scalar backward.

We now provide a more quantitative illustration of the descreening effect, as shown in \Fig{nonp_formfactor}. Here, we fix the geometric factor to $r/R = 5$ and vary the values of the effective mass $\mM$. The left panel shows the reduced form factor in the forward direction~$(\theta_\vecr = 0)$, represented by magenta lines with varying darkness. The right panel shows the reduced form factor in the backward direction~$(\theta_\vecr = \pi)$, represented by cyan lines with varying darkness. In both panels, darker lines correspond to larger values of $\mM$. For $\mM r \leq 1$, the reduced form factors come back to the perturbative form factor, as shown in \Fig{pert_finitesize_formfactor}. The reduced form factors for $\mM r = 10^3$ are related to the hard sphere within the plot range $k_0 r \leq 10^2$. From the figures, we can see that as $\mM$ increases, the lines deviate further from the shape of the perturbative form factor. In the regime where $k_0 r < 1$, the reduced form factor tends to $\formfactorreduce \simeq 1$, indicating that the form factor $\formfactor$ returns to the spherical symmetric ansatz as shown in \Eq{formfactor_sphscr}. When $k_0 r > \mM r$, the reduced form factors follow the power law $\formfactorreduce \propto 1/(k_0 r)^2$. Moreover, we find that the form factors $\formfactor = \formfactorreduce \times \formfactor_\sphscr$ come back to the perturbative form factors as shown in \Fig{pert_finitesize_formfactor} because the source barrier cannot effectively screen the incident scalar. In the intermediate region $1 < k_0 r < \mM r$, the reduced form factors in both forward and backward directions become flatter than the power law $\formfactorreduce \propto 1/(k_0 r)^2$, indicating that the descreening effect compensates for decoherence suppression in the high-momentum region. Referring to \Fig{phase_classify}, we know that the turning of the reduced form factors at $\mM \simeq k_0$ marks the transition from the screened region (Region D) to the descreened region (Region E).

\begin{figure}[t]
\centering
\includegraphics[width=0.49\linewidth]{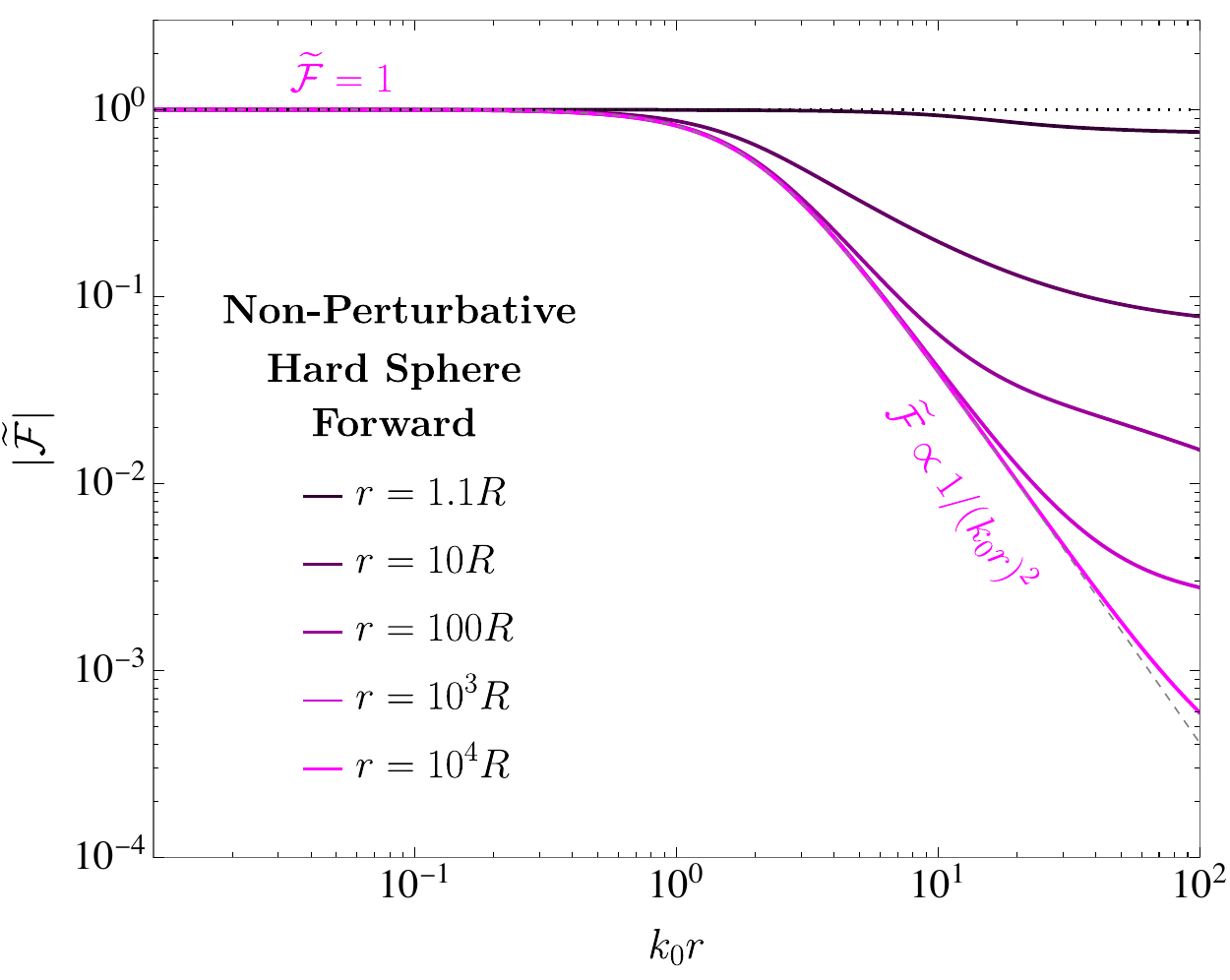}
\includegraphics[width=0.49\linewidth]{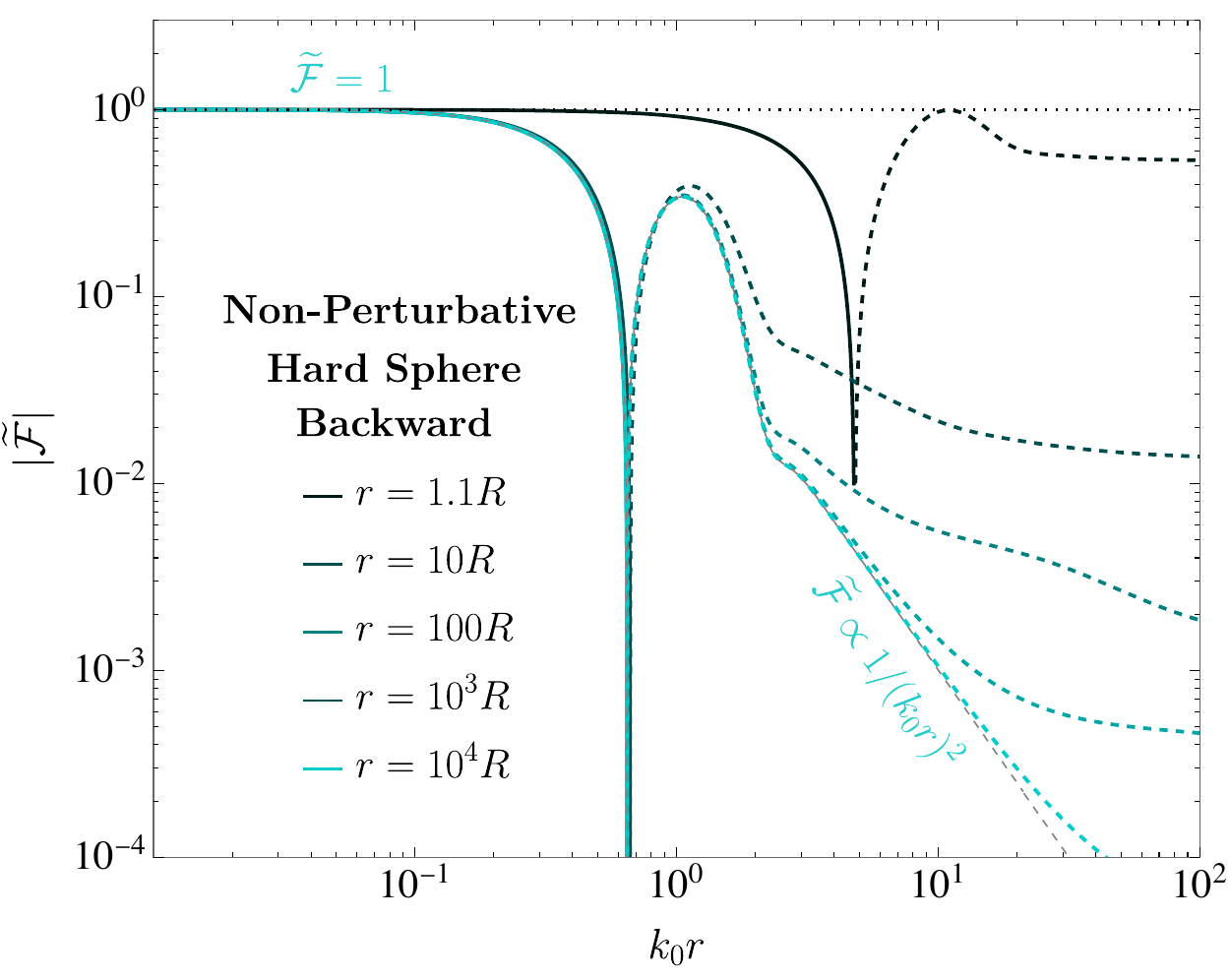}
\caption{Reduced form factors for the hard sphere~($\mM R \gg 1$) for different values of $r/R$. As $\abs{r-R}/R$ decreases, the reduced form factors deviate more significantly from the decoherence form factor of a point-like source~(shown in \Fig{FA_plt}) and are represented by the gray dashed lines in these figures. {\bf Left:} Reduced form factors in the forward direction~($\theta_\vecr = 0$), represented by magenta lines. {\bf Right:} Reduced form factors in the backward direction~($\theta_\vecr = \pi$), represented by cyan lines.}
\label{fig:nonp_formfactor_hardsph_finitesize}
\end{figure}

Before concluding the discussion on high-momentum behavior in the non-perturbative region, we discuss how the finite size of the source affects the form factors in the strongly coupled region. We find that the finite size of the source mass significantly changes the form factor compared to the perturbative finite-size effect shown in \Fig{pert_finitesize_formfactor}, where the form factors change by less than one order of magnitude. As shown in \Fig{nonp_formfactor}, we observe that as the coupling strength increases, the form factor lines tend to stabilize. For this reason, we choose the hard sphere configuration~(equivalent to $\mM R \gg 1$) to compute the reduced form factors in both the forward direction~($\theta_\vecr = 0$) and the backward direction~($\theta_\vecr = \pi$). The left panel shows the reduced form factors in the forward direction, represented by magenta lines of varying darkness. The right panel shows the reduced form factors in the backward direction, represented by cyan lines with varying darkness. In both panels, darker colors correspond to smaller geometric factors $r/R$, indicating a stronger impact from the finite size of the source mass. As shown in the figures, when $r/R = 10^4$~(the source is far away from the test mass), the reduced form factor matches the decoherence form factor $\formfactor_\text{A}$ for a point-like source, as shown in \Fig{FA_plt}. In \Fig{nonp_formfactor_hardsph_finitesize}, we use the gray lines to represent $\formfactor_\text{A}$ as a benchmark. We find that as the geometric factor $\abs{r-R}/R$ decreases, meaning the finite size effect of the source mass becomes more significant, the reduced form factors $\formfactorreduce$ deviate more noticeably from $\mathcal{F}_\text{A}$. It is worth noting that when $\abs{r-R}/R \lesssim 1$, the difference between the forward and backward form factors becomes much smaller compared to the perturbative case, as shown in \Fig{pert_finitesize_formfactor}. This effect can be explained by the fact that when $\abs{r-R}/R \lesssim 1$, the test mass in the forward direction moves within the shadow of the scalar wind. This reduces the forward contribution relative to the backward direction despite the forward direction exhibiting the weakest decoherence suppression.

\begin{figure}[h]
\centering
\includegraphics[width=0.49\linewidth]{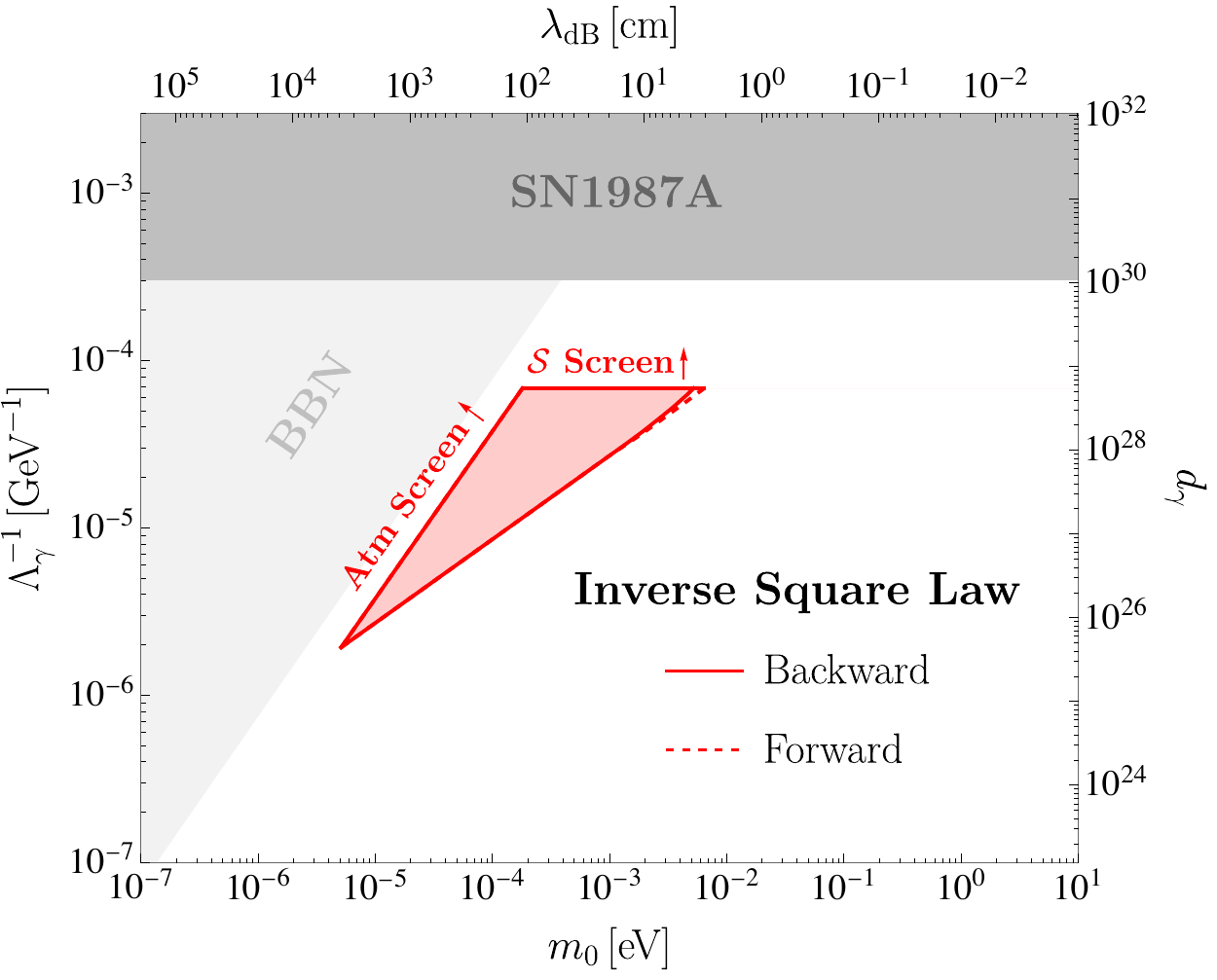}
\includegraphics[width=0.49\linewidth]
{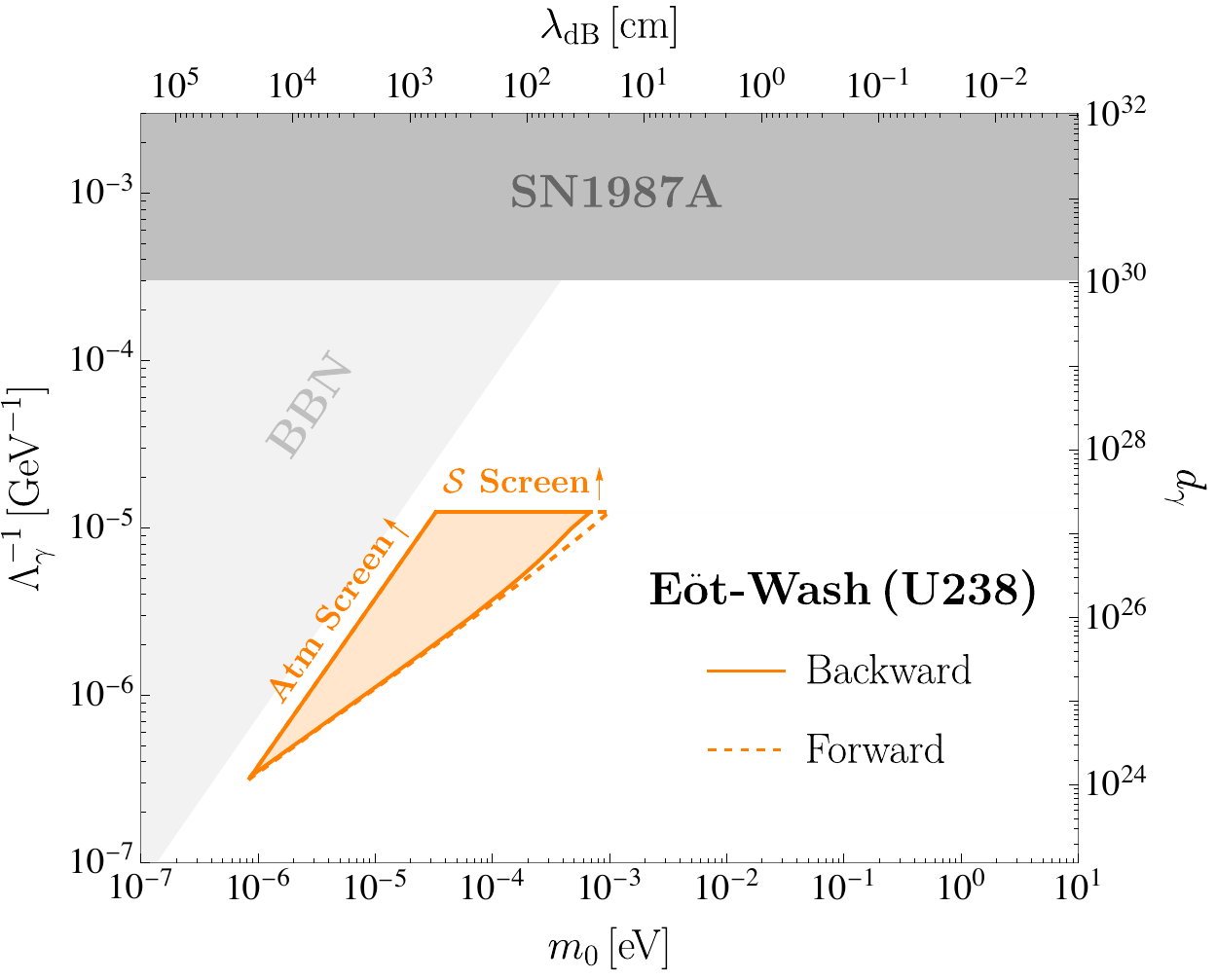}
\caption{Projections on the repulsive quadratic scalar-photon interaction based on the background-induced force. The dark gray region represents the constraint from SN1987A~\cite{Raffelt:1990yz,Cox:2024oeb}. The light gray region represents the constraint from BBN~\cite{Stadnik:2015kia,Sibiryakov:2020eir,Bouley:2022eer}. The solid lines indicate the conservative projection imposed using the backward form factor, while the dashed lines indicate the optimistic projection imposed using the forward form factor. Both projections vanish in the strongly coupled region, where the scalar field is screened either by the atmosphere or the source. We denote the boundary where atmospheric screening becomes significant as ``Atm Screen'', and the boundary where source screening becomes significant as ``$\sourcemass$ Screen''. {\bf Left}: Projection obtained from ISL tests~\cite{Spero:1980zz,Hoskins:1985tn,Hoyle:2000cv,Hoyle:2004cw, Kapner:2006si,Tu:2007zz,Yang:2012zzb,Murata:2014nra,Will:2014kxa}. {\bf Right}: Projection obtained from short-range E$\ddot{\rm o}$t-Wash EP experiments~\cite{Smith:1999cr}. Here the attractor is $\sim 3 \, \text{tons}$ of U238.}
\label{fig:plt_projection2}
\end{figure}

\subsection{Experimental Tests}\label{sec:lab_bif}

The background-induced force between two objects leads to deviations from the inverse square law of gravity between two objects, which have been searched extensively in various types of experimental setups and on many different physical scales (see \cite{Murata:2014nra,Will:2014kxa} for a general review of the search). Usually, experimental constraints on the new forces are parameterized by a Yukawa-like force with interaction strength $\alpha$ and interaction range $\lambda$, where the potential between two objects (with the test mass $M_\testmass$ and the source mass $M_\sourcemass$) is described by
\begin{equation}
V(r) = -G_N\frac{M_\testmass M_\sourcemass}{r}(1+ \alpha e^{-r/\lambda}),
\end{equation}
where $G_N$ is the gravitational constant. For the range of parameter space we are interested in ($\lambda \sim (m_0 v_0)^{-1} \gtrsim 1 \, \cm$), the dominant constraints come from Cavendish-type torsion balance experiments \cite{Murata:2014nra}, where the interaction strength is constrained to be much smaller than the gravitational interaction ($\alpha \lesssim 10^{-4}$).

Although these experiments will also probe the background-induced forces, the exact constraints on the model parameters $(m_0, \Lambda_{\gamma})$ are challenging to obtain. This is because, instead of measuring the absolute potential, the experiments are usually set up to measure the relative ratio of the force (or acceleration) between test and source masses. Therefore, if the new force has the same scaling as gravity $\propto r^{-2}$, the induced signals become degenerate in experiments. Conversely, if the new force is too short-ranged relative to the typical separation between source and test masses, the interaction is suppressed, thereby weakening the signal. Therefore, these experiments are most sensitive to the force that has an interaction range that is comparable to the separation between test and source masses. In this regime, the detailed configurations of experiments could have $\mathcal{O}(1)$ impact on the induced background-induced forces between test and source masses~\cite{Murata:2014nra,Will:2014kxa}. We recast the constraints on the $(m_0, \Lambda_{\gamma})$ parameter space as projections, following a strategy similar to that proposed in Ref.~\cite{VanTilburg:2024xib}. In particular, we require that the background-induced potential $V_\bg(r)$ has to be smaller than the Yukawa potential with existing strongest constraints on parameter $(\lambda_{\text{lim}},\alpha_{\text{lim}})$ at $r = \lambda$. The choice $r = \lambda$ corresponds to the radius where the Yukawa potential starts to be exponentially suppressed. We consider two types of experiments: fifth force experiments testing the ISL deviation~\cite{Spero:1980zz,Hoskins:1985tn,Tu:2007zz,Yang:2012zzb,Hoyle:2000cv,Hoyle:2004cw,Kapner:2006si,Murata:2014nra,Will:2014kxa, Tan:2016vwu} and ground-based short-range EP test~\cite{Smith:1999cr}. Their projections on the background-induced force are presented in \Fig{plt_projection2}. The projection from the ISL test is shown in the left panel of \Fig{plt_projection2} as the red-shaded region. We impose cutoffs at $m_\text{M, atm} = k_0$ and $m_{\text{M}, \sourcemass} = R^{-1}_\sourcemass$, with $R_\sourcemass$ taken to be the thickness of the source mass. The first cutoff arises from atmospheric screening, as discussed in \Eq{screen_condition}. The second cutoff is due to screening by the source mass. To evaluate $m_{\text{M}, \sourcemass}$, we consider a stainless steel source with thickness $1 \, \text{\cm}$. This is the thickness of the source in Refs.~\cite{Spero:1980zz, Hoskins:1985tn}. Other ISL experiments~\cite{Hoyle:2000cv,Hoyle:2004cw, Kapner:2006si,Tu:2007zz,Yang:2012zzb,Tan:2016vwu} have sources with smaller thicknesses and similar effective mass, so our cutoff gives a conservative projection. The projection from the ground-based short-range EP test is shown in the same panel as the orange-shaded region. Here, the source comprises approximately $ 3\, \text{tons}$ of U238, consisting of blocks with thickness $2.6 \,\cm$. Similar to before, we impose cutoffs at $m_\text{M, atm} = k_0$ and $m_{\text{M}, \sourcemass} = R^{-1}_\sourcemass$, corresponding to screening by the atmosphere and the source mass, respectively.

\begin{figure}[t]
\centering
\includegraphics[width=0.49\linewidth]
{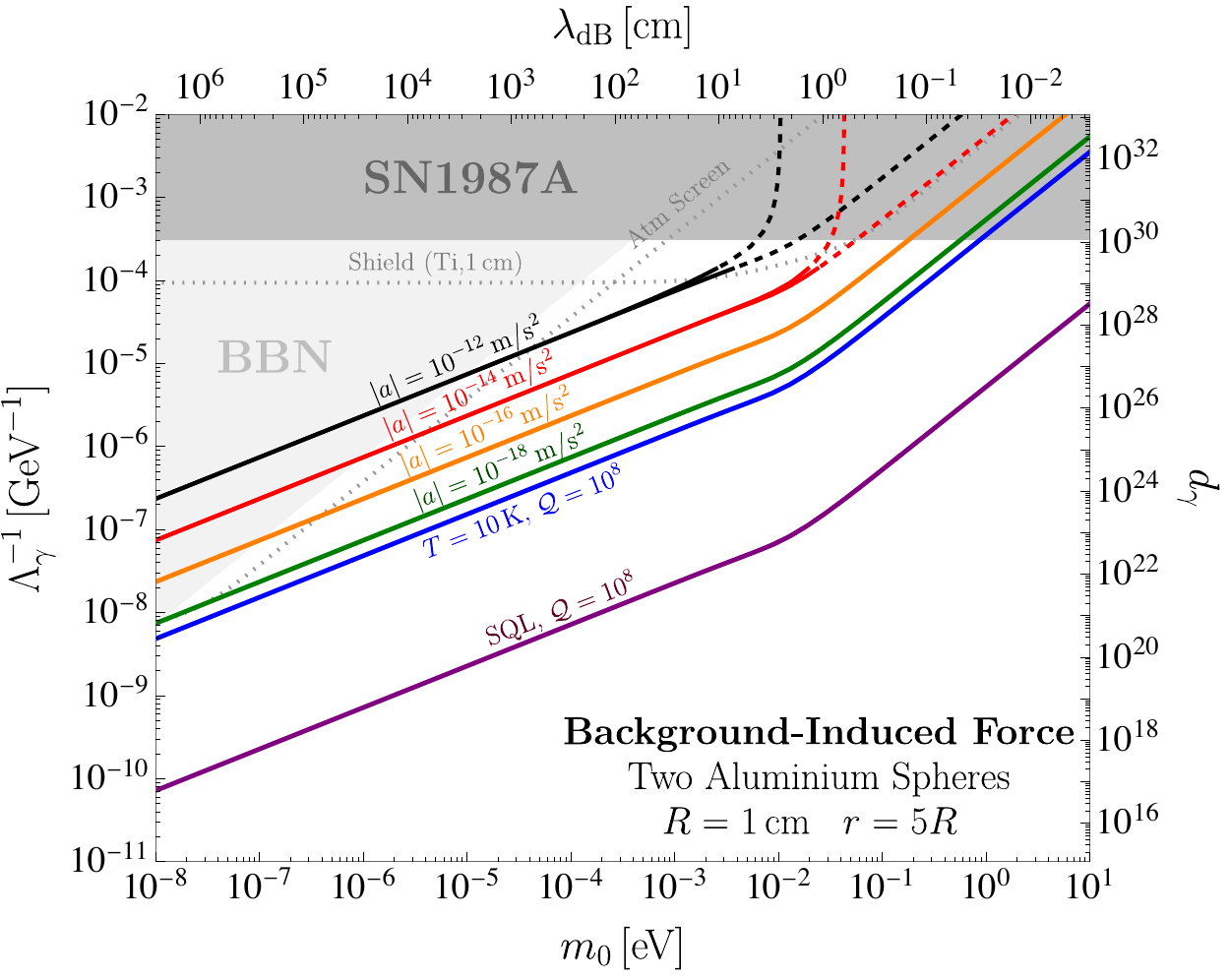}
\includegraphics[width=0.49\linewidth]
{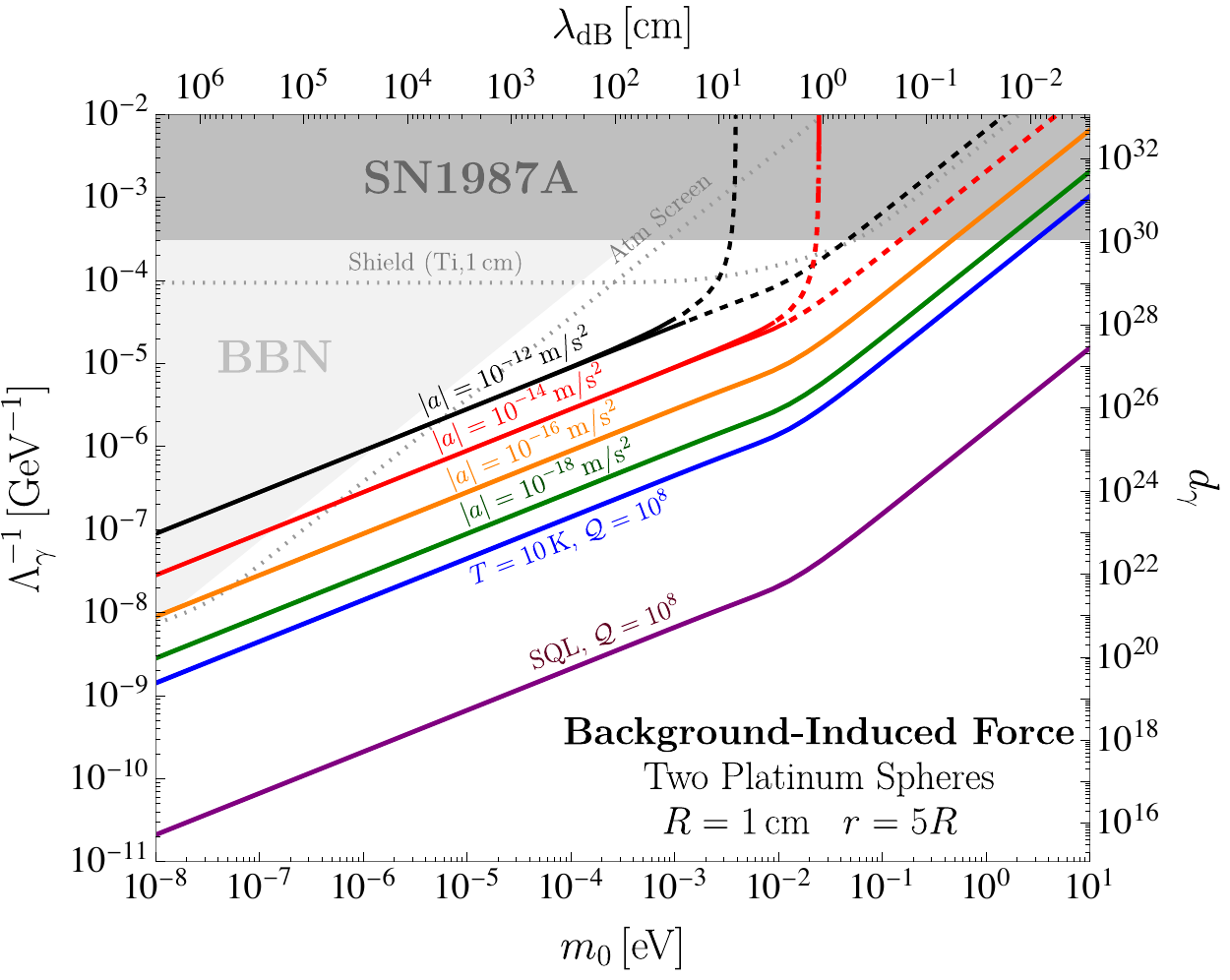}
\includegraphics[width=0.49\linewidth]
{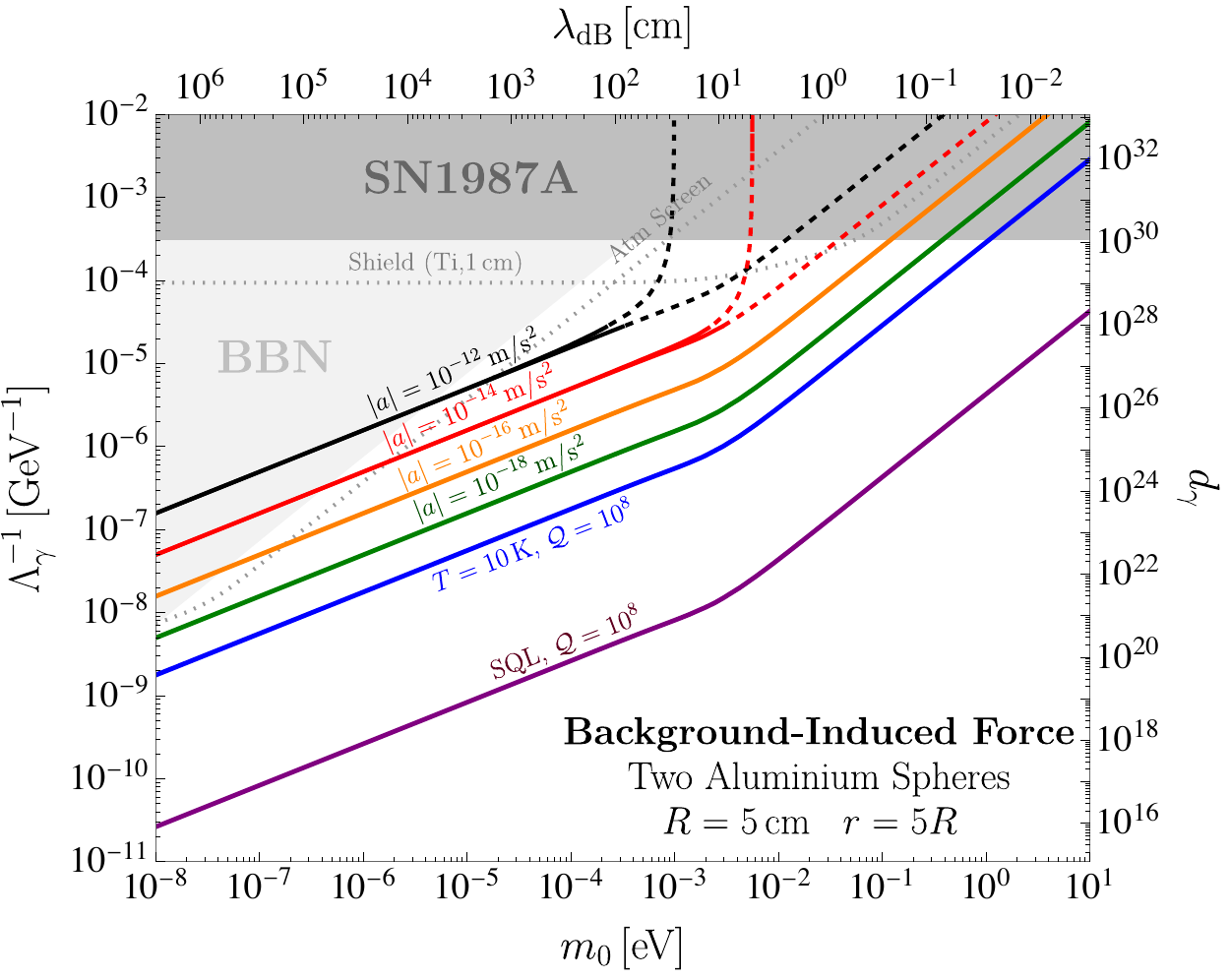}
\includegraphics[width=0.49\linewidth]
{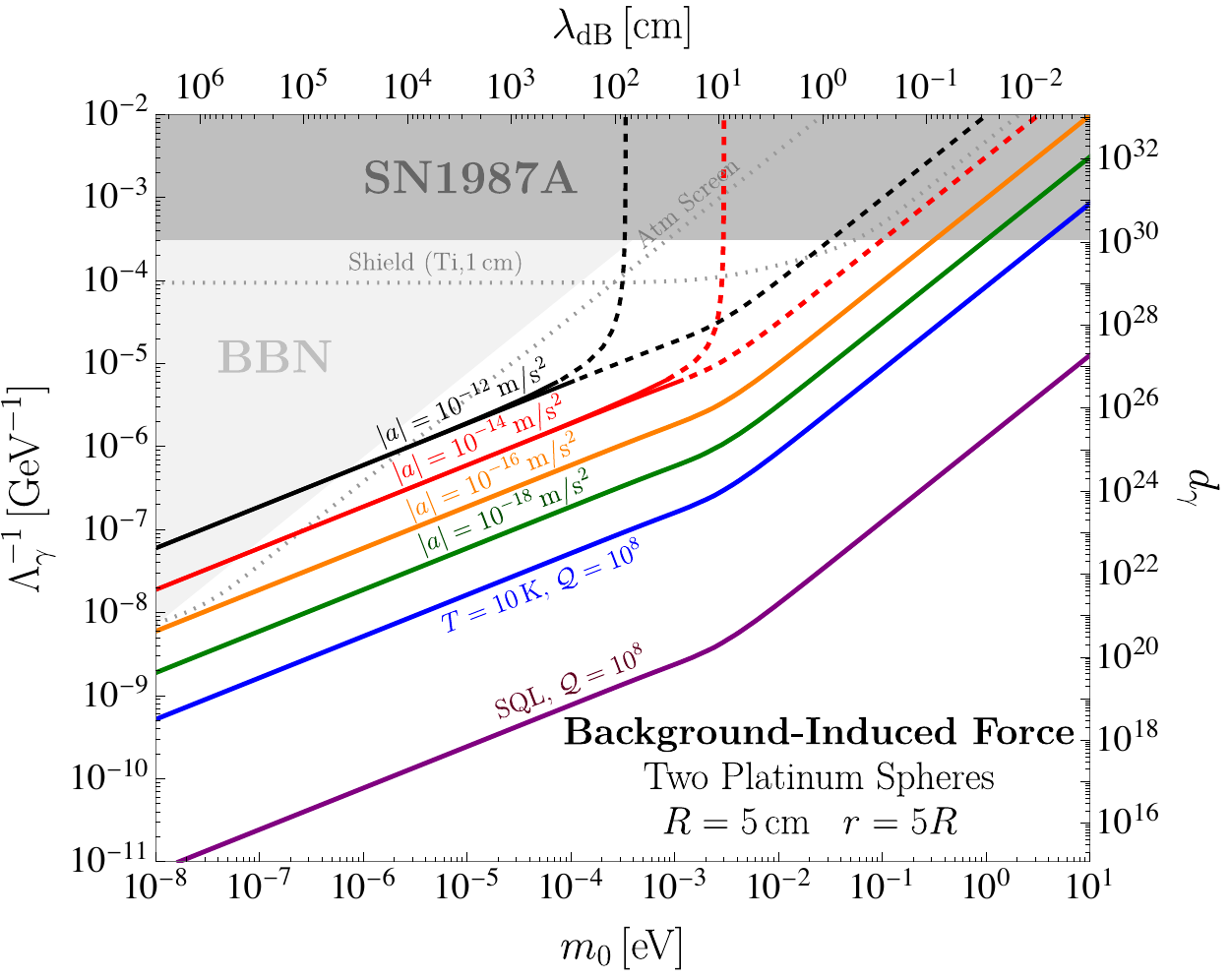}
\caption{Sensitivities from measuring the accelerations of test masses induced by the background-induced force from the source masses identical to the test masses. Here we choose the forward direction~($\theta_\vecr = 0$) to maximize sensitivity. We use pairs of uniform Aluminum and Platinum spheres with radii of $R = 1\,\cm, 5\,\cm$ to illustrate the sensitivities. The dark gray region represents the constraint from SN1987A~\cite{Raffelt:1990yz,Cox:2024oeb}. The light gray region represents the constraint from BBN~\cite{Stadnik:2015kia,Sibiryakov:2020eir,Bouley:2022eer}. The black, red, orange, and green lines indicate the sensitivities corresponding to accelerations of $\abs{a} = 10^{-12},\,\,10^{-14},\,\, 10^{-16},\,\,10^{-18}\,\rm m/s^2$, respectively. The blue line indicates the sensitivity for $T = 10\,\Kelvin$ and $\quality = 10^8$. The purple line indicates the sensitivity in the SQL with $\quality = 10^8$. For sensitivities in the unscreened region, we use solid curves, while in the screened region, we use dashed curves to represent the probable sensitivity ranges. In each panel, the gray dotted lines labeled ``Atm Screen'' and ``Shield (Ti, $1\,\text{cm}$)'' mark the boundaries above which the atmospheric and Titanium-shield~($1\,\text{cm}$ thickness) screening effects become significant.
{\bf Upper Right}: Aluminum sphere with $R=1\,\cm$. {\bf Upper Left}: Platinum sphere with $R=1\,\cm$. {\bf Lower Right}:  Aluminum sphere with $R=5\,\cm$. {\bf Lower Left}: Platinum sphere with $R=5\,\cm$.}
\label{fig:sensitivity_bg}
\end{figure}

Similar to \Subsec{sc_force_exp}, which discusses the experimental sensitivities to the scattering force, we discuss the sensitivities to the quadratic scalar-photon interaction induced by the background-induced force, as shown in \Fig{sensitivity_bg}. In the plot, we choose the forward direction~($\theta_\vecr = 0$) to maximize the sensitivity to the background-induced force. In this setup, we use a pair of identical Aluminum or Platinum spheres, where one sphere serves as the source mass and the other as the test mass. For illustration, we choose sphere radii of $R=1\,\cm$ and $R=5\,\cm$. The black, red, orange, and green lines represent the sensitivities for accelerations of $a = 10^{-12},\,\,10^{-14},\,\,10^{-16},\,\,10^{-18}\,\,\rm m/s^2$. The blue lines represent the sensitivities for $T = 10\,\Kelvin$ and $\quality = 10^8$, while the purple lines represent the sensitivities in the SQL with $\quality = 10^8$. Similar to \Subsec{sc_force_exp}, we set $t_\text{int} = 3\,\text{years}$ and $\omega_\osc = 2\pi \times 1\,\text{mHz}$ when plotting the blue and purple lines~\footnote{For simplicity, we present all sensitivities in terms of the absolute acceleration $\abs{a}$; rewriting them in relative acceleration $\abs{\Delta a}$ would shift the curves only at the $\mathcal{O}(1)$ level and does not affect our conclusions.}. We use solid curves to represent sensitivities in the unscreened region. In the unscreened region, the scaling behavior in \Fig{sensitivity_bg} is
\bea
\text{$m_0\text{-}\Lambda_\gamma^{-1}$ Plane:} \quad \quad \Lambda_\gamma^{-1} \propto \left\{ 
\begin{aligned}
& m_0^{1/2} &  \quad k_0 r < 1       \\
& m_0       &  \quad k_0 r \gtrsim 1
\end{aligned}
\right. .
\eea
In the region where $k_0 r<1$, there is no decoherence suppression. As shown in \Eq{effective_mass} and \Eq{Vbg_formfactor}, the background-induced potential scales as $V_\bg \propto 1/\Lambda_\gamma^4 m_0^2$. Therefore, we get the power-law relation $\Lambda_\gamma^{-1} \propto m_0^{1/2}$. In the region where $k_0 r \gtrsim 1$, decoherence suppression becomes significant and follows \Eq{FA_forward_large_r_num}. In this region, the background-induced potential scales as $V_\bg \propto 1/\Lambda_\gamma^4 m_0^4$, which gives the power-law relation $\Lambda_\gamma^{-1} \propto m_0$. 

Now we discuss the screened region in \Fig{sensitivity_bg}. Although our formalism explicitly treats screening only for the source mass, in estimating sensitivities we introduce two limiting benchmark scenarios that effectively parameterize the possible degree of screening for both masses. 1.\,Conservative Benchmark: Both source and test masses experience low-momentum screening, as discussed in \Subsec{bg_nonp_lowk}. Since we do not account for the alleviation due to the descreening effect, this represents the most conservative sensitivity estimate. 2.\,Optimistic Benchmark: Neither the source nor the test masses experience any screening effects, and they are treated as being entirely in the unscreened region. This gives the most optimistic sensitivity curves. Determining the final sensitivity curves when both the source and test masses experience the screening effect is beyond the scope of this work and is left for future studies.

When presenting the projections in \Fig{plt_projection2} and the sensitivities in \Fig{sensitivity_bg}, we model the incident field as a plane wave, \Eq{psi_in_sca}, to keep the analysis tractable and to provide a shielding-agnostic baseline. Following the same treatment as in \Subsec{sc_force_exp}, we present sensitivities assuming that external and atmospheric screening can be mitigated, for example in the weakly coupled region, in space-based implementations, or in setups with sufficiently thin outer shells. For ground-based experiments, atmosphere screening plays a decisive role in suppressing the background-induced force in the strongly coupled region: if the scalar field is repelled by the air surrounding ground-based experiments, the background-induced force is effectively eliminated. For completeness, the parameter regions where screening would become relevant for atmosphere and experimental shielding are denoted by gray dotted lines in \Fig{sensitivity_bg}. Even if experimental shielding is present, it cannot, in general, fully eliminate the background-induced force. The signal requires only a nonzero ambient scalar field, and scalar fields can still leak into the experimental setup before the test and source masses are sealed from the environment. In contrast, the scattering force can be completely removed if the incoming flux is blocked.

In addition to the possible effects of experimental and atmospheric shielding, the Earth itself can in principle distort the incident wave through a matter effect and source an additional background-induced force. In typical geometries, however, the Earth-sourced contribution points vertically and is therefore approximately orthogonal to the force sourced by $\sourcemass$, so it can be cleanly separated. Moreover, we have $k_0 R_\oplus \gtrsim 1$~($m_0 \gtrsim \text{few} \times 10^{-11}\,\eV$) in the experimentally relevant region. In this limit the Earth's surface is effectively locally planar and can be treated as a reflecting boundary/mirror. Even with strong screening, the near-surface abundance is modified only at the $\mathcal{O}(1)$ level. Consequently, our background-induced force estimates and the associated sensitivities are not qualitatively affected. For an intuitive illustration, one can refer to the right panel of \Fig{nonpeturb_sca_plt}, in which $k_0 R_\sourcemass \gtrsim 1$. If we identify the source mass ``$\sourcemass$'' with the Earth, i.e., $\sourcemass = \oplus$, we see that in this regime the kinetic energy of the field overcomes the effective surface tension, so the scalar field is pushed toward the Earth's surface.

While background-induced, scattering, vacuum, and Yukawa forces are all capable of generating anomalous accelerations, the background-induced force is distinguishable from the other forces under appropriate experimental conditions. First, we discuss the distinguishability of the background-induced force from static vacuum~(see \Appx{force_qft}) or Yukawa force~(from linear scalar-SM coupling). Because of screening effects from the Earth, the density of the scalar may vary over time. Since the background-induced force is proportional to the dark matter energy density, its signal would exhibit periodic modulation, making it distinguishable from static forces. Second, we discuss the distinction between the background-induced force and the scattering force. The background-induced force acts between two objects, allowing its direction to be manipulated by adjusting the experimental setup. In contrast, the scattering force always points along the dark matter velocity vector. This directional difference leads to distinct modulation patterns, enabling experimental distinction between the two forces. Moreover, the detection of quadratic scalar-SM couplings is easier when both background-induced and scattering forces are present. Recasting experimental sensitivities separately for each force is therefore a conservative approach.

We also show that different types of quadratic couplings between the scalar field $\phi$ and SM fields, as listed in Ref.~\cite{Hees:2018fpg}, can be distinguished based on their associated dilaton charges for specific elements. For example, in the case of the effective interaction $\phi^2\bar{N}N$, the nucleon dilaton charge is $1$. In contrast, in our scenario, the dilaton charge depends on $(Z,A)$ in a specific functional form, as shown in \Eq{dilaton_charge_num}. Therefore, by selecting test masses composed of different elements, the signal responses will differ among various types of quadratic scalar-SM couplings, enabling experimental discrimination between them. The method of distinguishing scalar-SM interactions also applies to the experimental tests of the scattering force discussed in \Subsec{sc_force_exp}.

\begin{figure}[h!]
\centering
\includegraphics[width=0.5\linewidth]{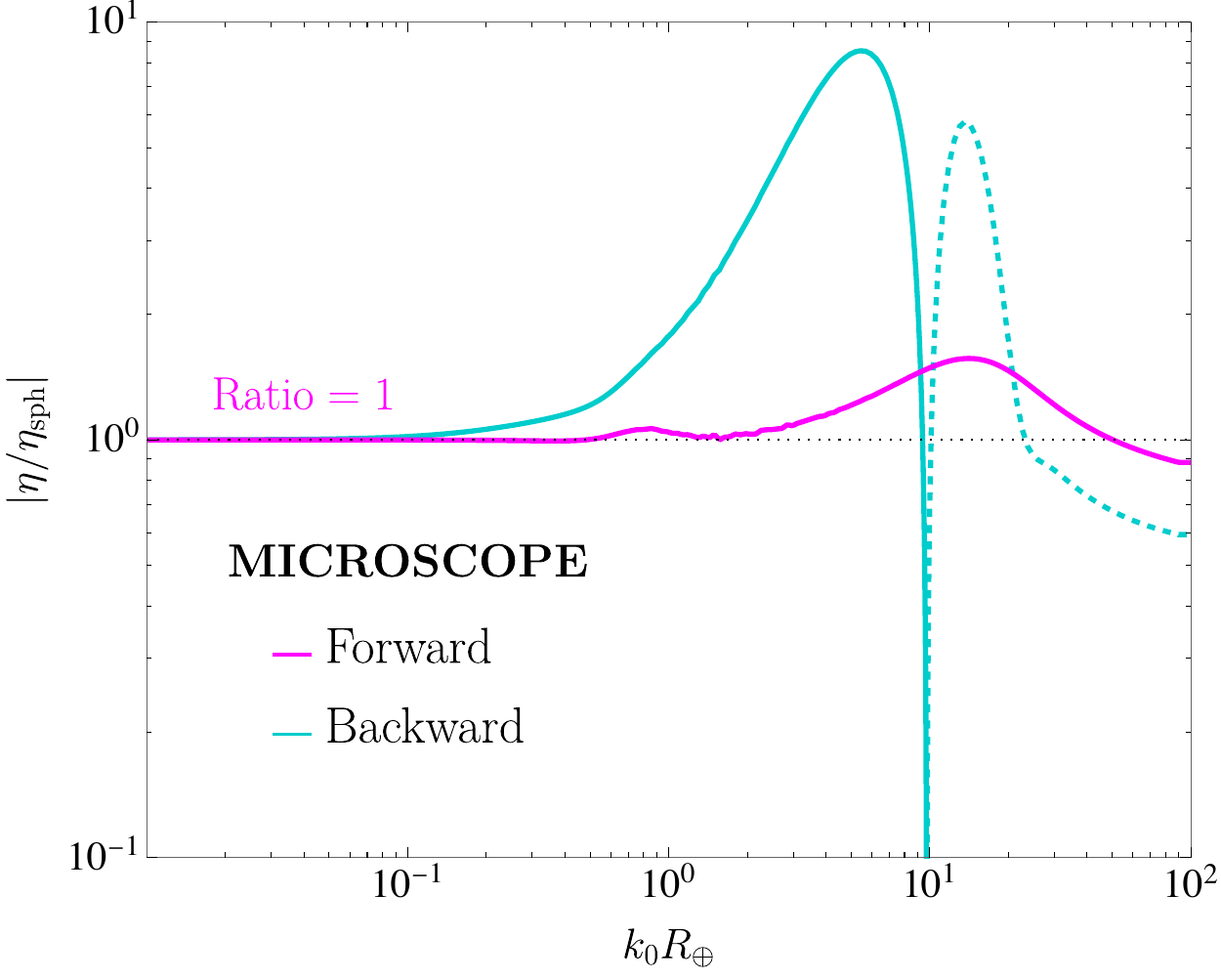}
\caption{The ratio of the E$\ddot{\rm o}$tv$\ddot{\rm o}$s parameter to the value computed from the spherically symmetric ansatz~(as given in \Eq{etovos_microscope_sph}). The magenta line represents the ratio of E$\ddot{\rm o}$tv$\ddot{\rm o}$s parameters in the forward direction, while the cyan line represents the ratio in the backward direction. Solid lines correspond to the value of $\abs{\eta/\eta_\sph}$ when $\eta/\eta_\sph>0$, representing the attractive force. Dashed lines correspond to the value of $\abs{\eta/\eta_\sph}$ when $\eta/\eta_\sph<0$, representing the repulsive force.}
\label{fig:MICROSCOPE_force_ratio}
\end{figure}

\subsection{MICROSCOPE Satellite}\label{subsec:bgforce_MICROSCOPE}

In this section, we revisit the constraint from the MICROSCOPE satellite, one of the most stringent tests of equivalence principle (EP) violation, as an illustrative example for the treatment in the non-perturbative region. Moreover, the MICROSCOPE experiment~\cite{Berge:2017ovy,Touboul:2017grn,MICROSCOPE:2019jix} imposes the most stringent terrestrial constraint on the quadratic scalar-standard-model coupling when $m_0 \gtrsim 10^{-18}\,\eV$~\cite{Hees:2018fpg,Banerjee:2022sqg}, which is independent of the assumption of the cosmological evolution. Similar discussions can also be applied to other ground-based equivalence principle tests, such as E$\ddot{\rm o}$t-Wash, which uses the Earth as the source~\cite{Schlamminger:2007ht}. The previous discussions place constraints on the repulsive quadratic scalar interaction utilizing the spherical symmetric ansatz, as shown in Refs.~\cite{Hees:2018fpg,Banerjee:2022sqg}. However, as we have discussed in \Subsec{bg_nonp_lowk} and \Subsec{bg_nonp_highk}, we find that the behavior of the background-induced potential in the high-momentum regions behaves qualitatively differently from the low-momentum region. The critical scalar mass at which the spherical ansatz for the scalar distribution around the Earth is
\bea
\text{Spherical Symmetry Violation:} \quad \quad m_0 \sim \frac{1}{R_\oplus v_0} \sim \text{few} \times 10^{-11}\,\eV,
\eea
where the scalar dark matter's de Broglie wavelength matches the earth's radius $R_\oplus = 6378 \, \km$~\cite{moritz2000geodetic,IAUInter-DivisionA-GWorkingGrouponNominalUnitsforStellarPlanetaryAstronomy:2015fjh}.
For this reason, it is necessary to go beyond the spherically symmetric ansatz, as shown in \Eq{psi_out_bound}, and revisit the MICROSCOPE constraint~\footnote{While we were finalizing the first version of this manuscript, Ref.~\cite{Gue:2025nxq} appeared. Their study focuses on attractive scalar interactions in the absence of self-interactions, where resonance occurs, whereas our work focuses on repulsive scalar interactions, which lead to screening effects. In addition, Ref.~\cite{Gue:2025nxq} considers the regime $kR_\oplus < 1$ using the s-wave approximation~($l=0$), while our analysis targets the $kR_\oplus > 1$ regime and includes all partial-wave components up to $l_{\max} \sim kR_\oplus$. The MICROSCOPE constraint on the light QCD axion, taking into account the Earth's phase transition triggered by a negative $\mM^2$ and stabilized by axion self-interactions~\cite{Hook:2017psm}, is discussed in Ref.~\cite{Gan:2025icr}.}.

Now, we provide a brief introduction to the experimental setup of the MICROSCOPE satellite. The MICROSCOPE satellite operates at an altitude of $h = 710 \, \km$. Therefore, we have
\bea
\frac{r}{R_\oplus} = \frac{h + R_\oplus}{R_\oplus} \simeq 1.1 \,,
\eea
which serves as one of the benchmark values for plotting the form factors discussed earlier. At this altitude, the gravitational acceleration is approximately $7.9\,\text{m}/\text{s}^2$. To test the EP violation, the MICROSCOPE satellite is equipped with two identical accelerometers, each containing a pair of test masses. One of the test masses is composed of Titanium alloy Ti/Al/V~$[90/6/4]$, where $90$, $6$, and $4$ denote the percentage of Ti, Al, and V components, respectively. The other test mass consists of Platinum alloy Pt/Rh~$[90/10]$, where $90$ and $10$ indicate the percentage of Pt and Rh components, respectively. Additionally, it is known that the Earth consists of $68\%$ $\text{SiO}_2$ and $32\%$ $\text{Fe}$. Using \Eq{dilaton_charge_num}, we have
\bea
\label{eq:dilaton_charge_MICROSCOPE}
Q_{\gamma, \ptalloy} - Q_{\gamma, \tialloy} = 1.94 \times 10^{-3}, \quad \quad Q_{\gamma,\oplus} = 1.90 \times 10^{-3}.
\eea
Here, $Q_{\gamma, \ptalloy} - Q_{\gamma, \tialloy}$ represents the dilaton charge difference between the Titanium and Platinum alloys, the materials used in the two test masses. $Q_{\gamma,\oplus}$ denotes the dilaton charge of the Earth.
\begin{figure}[t!]
\centering
\includegraphics[width=0.7\linewidth]{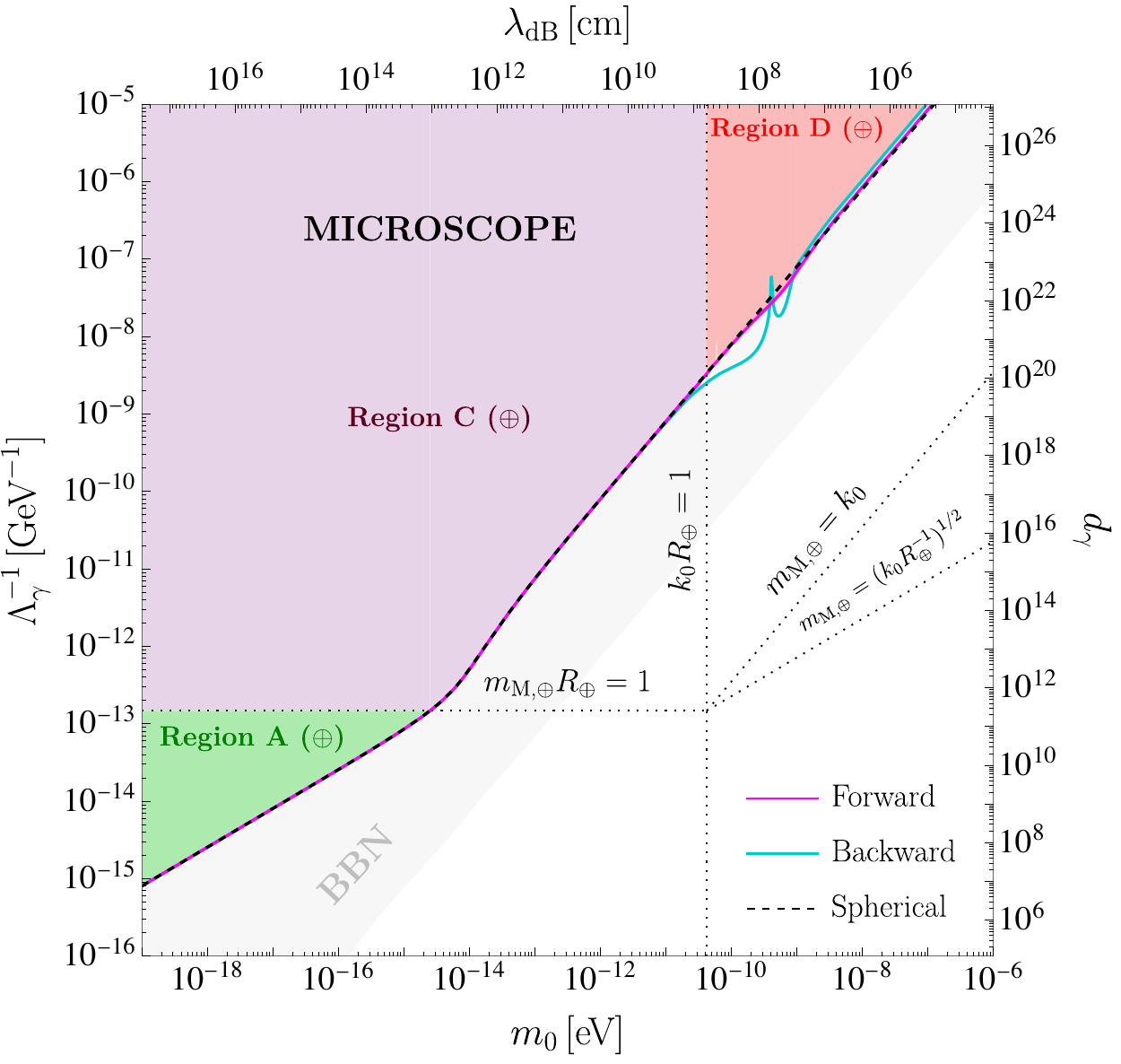}
\caption{MICROSCOPE constraint for the repulsive quadratic scalar-photon interaction. The magenta line represents the constraint based on the $\abs{\eta/\eta_\sph}$ ratio in the forward direction, while the cyan line represents the constraint imposed based on the $\abs{\eta/\eta_\sph}$ ratio in the backward direction. The black dashed line represents the previous constraint based on the spherically symmetric ansatz. In this plot, the conservative part of these two constraints is highlighted using shaded colors. The green region denotes perturbative Region A. The purple region represents Region C. Both Regions A and C correspond to the regime where the spherically symmetric ansatz is valid. The red region represents Region D, the non-perturbative region where the spherically symmetric ansatz is invalid, and all the partial waves up to $l_\text{max} \sim \text{few} \times k_0 R_\oplus$ contribute to the wave function. For comparison, the BBN constraint~\cite{Stadnik:2015kia,Sibiryakov:2020eir,Bouley:2022eer} is shown as a light gray shaded region.}
\label{fig:MICROSCOPE_constraint}
\end{figure}

As is well known, the experimental results of EP violation are expressed through the E$\ddot{\rm o}$tv$\ddot{\rm o}$s parameter, which is defined as
\bea
\eta(\testmass_1, \testmass_2) = 2 \frac{\abs{a_{\testmass_1,\sourcemass}-a_{\testmass_2,\sourcemass}}}{\abs{a_{\testmass_1,\sourcemass}+a_{\testmass_2,\sourcemass}}}.
\eea
Here, $a_{\testmass_1,\sourcemass}$ and $a_{\testmass_2,\sourcemass}$ are the accelerations of the test masses $\testmass_1$ and $\testmass_2$ induced by the background-induced force from the source $\sourcemass$. In the MICROSCOPE experiment, the source is the Earth, and the test masses are previously mentioned Platinum and Titanium alloys. The final result of MICROSCOPE~\cite{MICROSCOPE:2022doy}, released in 2022, reported the E$\ddot{\rm o}$tv$\ddot{\rm o}$s parameter as $\eta(\tialloy,\ptalloy)=[-1.5\pm 2.3\,({\rm stat})\pm 1.5\,({\rm syst})]\times 10^{-15}$. We compute the $1 \sigma$ uncertainty by doing the quadratic combination of both statistical and systematic uncertainties. Then we have $1\sigma$ uncertainty
\bea
\sigma_{\tialloy,\,\,\ptalloy} \simeq 2.7 \times 10^{-15},
\eea
which we use to impose the MICROSCOPE constraint.

Starting with the spherically symmetric ansatz discussed in \Subsec{bg_nonp_lowk}, we have
\bea
\label{eq:etovos_microscope_sph}
\eta_\text{sph} = \frac{\rho_\phi}{m_0^2} \frac{\mpl^2}{4\pi} \frac{Q_{\gamma, \oplus} (Q_{\gamma, \tialloy} - Q_{\gamma, \ptalloy})}{\Lambda_\gamma^4} J_+\left(\mMearth R_\oplus\right) \left[ 1 - \frac{1}{r} \, \frac{\mMearth^2 \mathcal{V}_\oplus}{4 \pi} J_+\left(\mMearth R_\oplus\right)\right],
\eea
which is the E$\ddot{\rm o}$tv$\ddot{\rm o}$s parameter for the previously mentioned test masses made of Platinum and Titanium alloys. In \Eq{etovos_microscope_sph}, we have
\bea
\frac{\mMearth^2 \mathcal{V}_\oplus}{4 \pi} = \frac{Q_{\gamma,\oplus} M_\oplus}{4 \pi \Lambda_\gamma^2}, \quad \quad \mMearth R_\oplus = \sqrt{\frac{3 Q_{\gamma,\oplus} M_\oplus}{4 \pi \Lambda_\gamma^2 R_\oplus}}. 
\eea
%
By performing the substitution $\Lambda_\gamma^2 \rightarrow \mpl^2/4\pi d_\gamma$, we can reproduce the same formula as shown in Refs.~\cite{Hees:2018fpg,Banerjee:2022sqg}.

Here, we compare the E$\ddot{\rm o}$tv$\ddot{\rm o}$s parameter obtained from the full partial wave analysis and phase space averaging with that derived from the spherically symmetric ansatz, as shown in \Eq{etovos_microscope_sph}. We have
\bea
\label{eq:etovos_microscope_ratio}
\frac{\eta}{\eta_\sph} 
\simeq \formfactorreduce - \left( 1 + \frac{R_\oplus}{2h}\right) \times k_0 r \frac{\partial \formfactorreduce}{\partial(k_0 r)}.
\eea
In the above formula, the approximate sign ``$\simeq$'' is valid for the hard sphere, which appropriately describes the region covered by the MICROSCOPE constraint. Here, we focus on the radial component of the background-induced force to simplify the analysis. We use the numerical values of the form factors shown in \Fig{nonp_formfactor_hardsph_finitesize} at $r=1.1R$ and then substitute them into \Eq{etovos_microscope_ratio}.  As a result, we obtain the ratio of the E$\ddot{\rm o}$tv$\ddot{\rm o}$s parameter, illustrated in \Fig{MICROSCOPE_force_ratio}. Moreover, we numerically compute the reduced form factor to the upper limit $k_0 R_\oplus = 500$ and find that it approaches a constant because of the balance between descreening enhancement and decoherence suppression. For this reason, we extrapolate the $\eta/\eta_\sph$ ratio over higher mass ranges to impose the MICROSCOPE constraint, as shown in \Fig{MICROSCOPE_constraint}.

In this plot, we use the magenta line to represent the constraint based on the E$\ddot{\rm o}$tv$\ddot{\rm o}$s parameter in the forward direction, and the cyan line to represent the constraint in the backward direction. The black dashed line represents the constraint based on the E$\ddot{\rm o}$tv$\ddot{\rm o}$s parameter computed using the spherically symmetric ansatz, as shown in \Eq{etovos_microscope_sph}. For the final MICROSCOPE constraint, we choose the conservative part of the constraints in forward and backward directions, and shade it accordingly. Utilizing the same color scheme as \Fig{phase_classify}, we use red to shade Region A for the Earth, green to shade Region C, and purple for Region D. The computation in Regions A and C with low incident momentum has already been presented in Refs.~\cite{Hees:2018fpg,Banerjee:2022sqg} using the spherically symmetric ansatz discussed in \Subsec{bg_nonp_lowk}. We extend their analysis to the high-momentum region where $kR_\oplus \gg 1$ using the full partial wave analysis and the phase space average. It is worth noting that even though the spherical symmetry breaks down, the slope of the constraint does not change. This occurs because the descreening enhancement and the decoherence suppression achieve a balance. Interestingly, from \Fig{MICROSCOPE_force_ratio}, we find that when $k_0 r > 10$, the background-induced force in the backward direction reverses direction. Therefore, when the satellite orbits around the Earth, the background-induced force is an AC force varying with the orbital period of the satellite~($\sim 1.6 \, \text{hours}$), rather than the DC centripetal force predicted by the spherically symmetric ansatz. For this reason, the specially designed satellite experiments can significantly enhance the sensitivity in this region. We leave this exploration for future work.

\section{Models for Varying Fine-Structure Constant}\label{sec:vary_alpha_UV}

For the quadratic scalar-photon interaction, our previous discussion is based on the effective field theory. In the following discussion, we introduce three types of UV models: (i) heavy $U(1)_Y$ charged fermions, (ii) heavy $U(1)_Y$ charged scalars, and (iii) the dark QCD axion. Using renormalization group analysis, we demonstrate that in a dense environment: Model (i) and (ii) can generate an effective mass squared of either positive sign (repulsive potential) or negative sign (attractive potential), depending on the values of the dark matter-heavy field couplings. Model (iii), however, leads to a negative effective mass squared~(attractive potential). 

\begin{figure}[t!]
\centering
\includegraphics[width=0.49\linewidth]{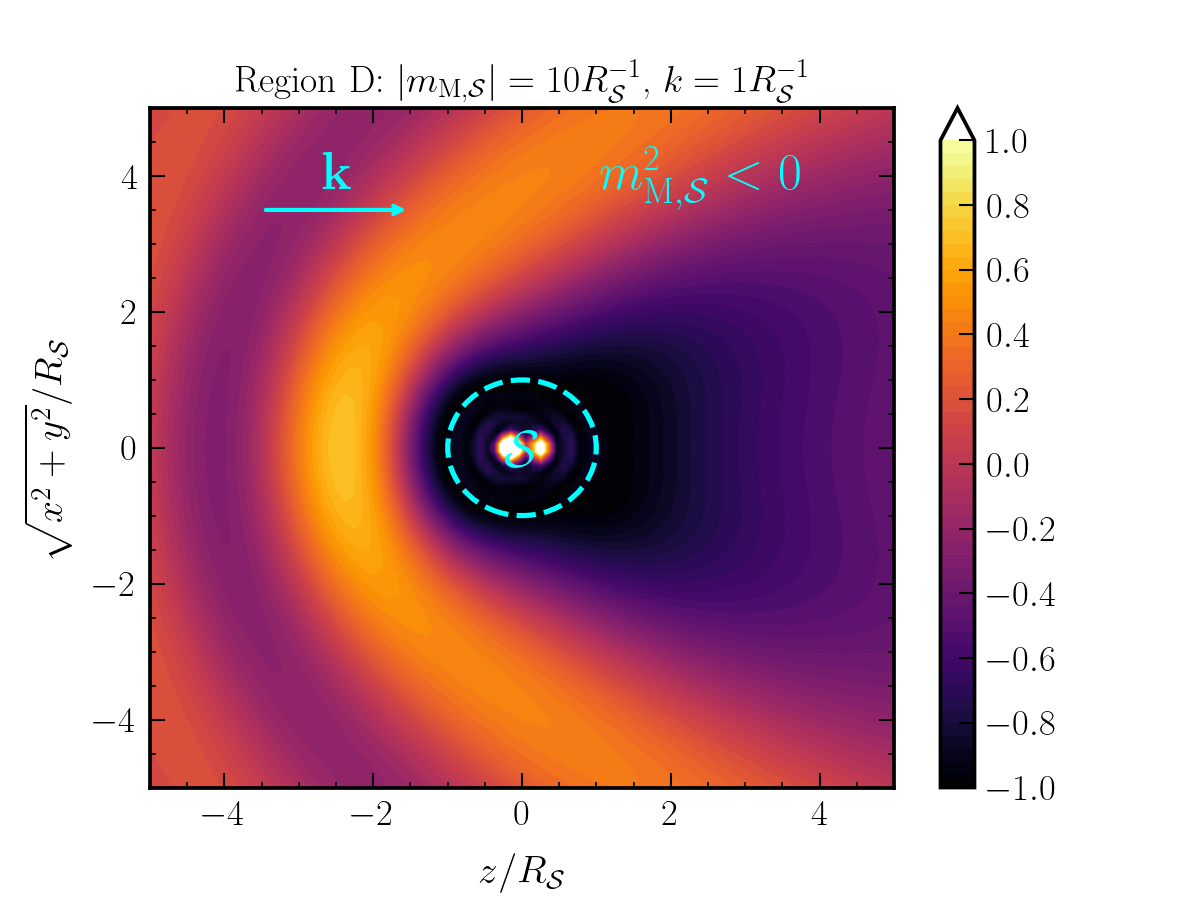}
\includegraphics[width=0.49\linewidth]{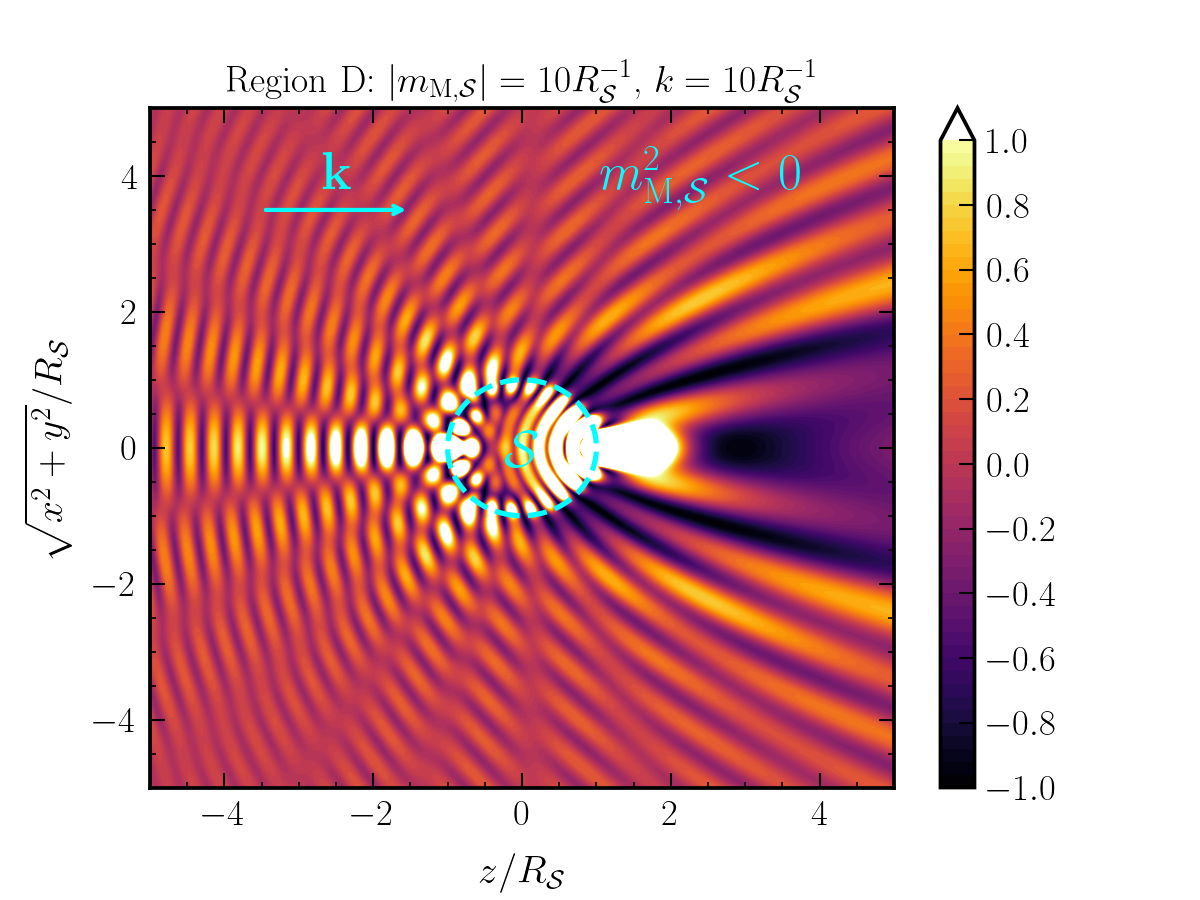}
\caption{The plots show $\abs{\psi}^2/\abs{\psi_0}^2-1$ for $\abs{\mMsource}=10\,R_\sourcemass^{-1}$ with different incident momentum $\veck$, illustrating resonance effects for an attractive potential~($\mMsource^2<0$). We set the scalar self-interaction to zero here. {\bf Left:} $k=1\,R_\sourcemass^{-1}$, corresponding to scattering pattern away from a resonance peak. Apart from a mild enhancement near the center, the overall scalar distribution is similar to that for the repulsive potential with the same $\abs{\mMsource}$, where the scalar field is screened inside the sphere. {\bf Right:} $k=10\,R_\sourcemass^{-1}$, corresponding to scattering pattern near the resonance condition $\sqrt{k^2+\mMsource^2}\,R_\sourcemass=(n+1/2)\,\pi$ with $n=4$. The scalar field is strongly enhanced inside $\sourcemass$ and redistributed outside $\sourcemass$. In the strongly resonant regime, the enhancement can reach orders of magnitude. This behavior is qualitatively different from the repulsive case shown in the left panel of \Fig{nonpeturb_penetrate}.}
\label{fig:nonpeturb_attractive}
\end{figure}

Before proceeding, we remind that we also present scattering patterns for the attractive potential in \Fig{nonpeturb_attractive}, to help readers build intuition for the differences between repulsive and attractive cases, even though our default setup throughout the paper assumes a repulsive potential. In practice, this extension is straightforward: in the partial-wave formalism summarized in \Appx{partial_wave_appx}, we simply apply the Wick rotation $\mM \rightarrow - i\,\mM$. When producing \Fig{nonpeturb_attractive}, we assume vanishing scalar self-interactions, so there is no stabilization mechanism or scalar phase transition.

\begin{figure}[h!]
\centering
\includegraphics[width=0.6\linewidth]{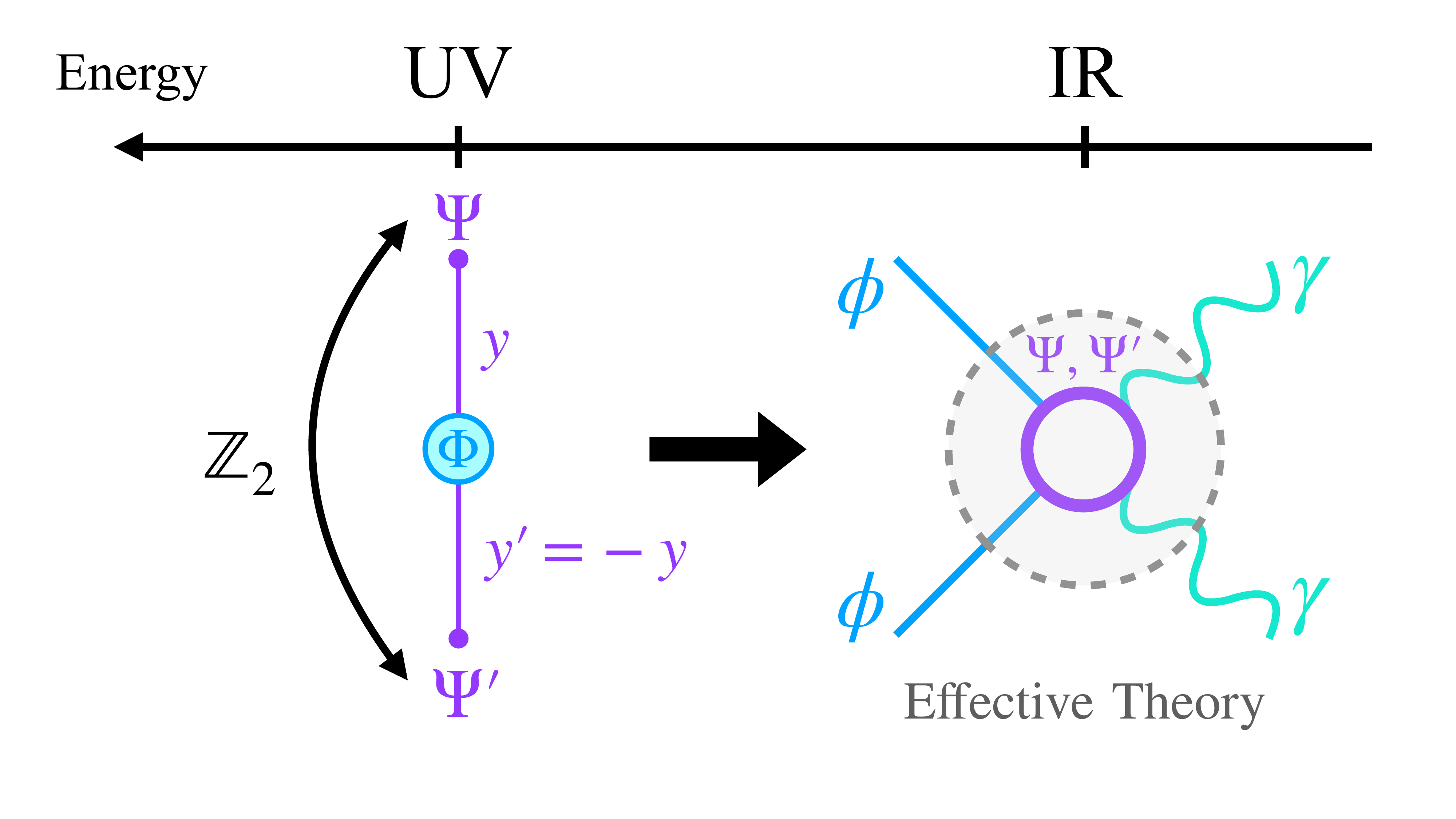}
\caption{The schematic diagram of the theory containing heavy $U(1)_Y$-charged fermions. The UV theory contains heavy fermionic messengers $\Psi$ and $\Psi'$ carrying the same hypercharge. $\Psi$ and $\Psi'$ are directly coupled to the ultralight scalar $\phi$ via the Yukawa couplings $y$ and $y'$. We impose the $\mathbb{Z}_2$ symmetry to protect mass and charge degeneracies between $\Psi$ and $\Psi'$. In the IR regime, the heavy messengers are integrated out, generating a quadratic scalar-photon coupling. This dimension-six operator is the lowest-order allowed operator in the effective theory.}
\label{fig:uv_model}
\end{figure}

\subsection{Heavy $U(1)_Y$-Charged Fermions}\label{sec:ChargedFermion}

In this section, we construct a quadratic scalar-photon coupling model based on two heavy $U(1)_Y$-charged fermions with an additional $\mathbb{Z}_2$ symmetry. A similar model was explored in Ref.~\cite{Gan:2023wnp} with different motivations. Here, we present a simplified version and perform a renormalization group analysis. In \Fig{uv_model}, we show the schematic diagram of this model, including both top-down and bottom-up perspectives. From a top-down perspective, this structure is generated naturally at the one-loop level with the introduction of a pair of vector-like fermions, $\Psi$ and $\Psi'$, which carry the same hypercharge under the $U(1)_Y$ gauge group and have the same mass satisfying $M_F \gtrsim \mathcal{O}({\rm TeV})$~\cite{Gan:2023wnp}. More specifically, we have
\begin{equation}
\Psi: (M_F^0, Q_F)\;,\quad \Psi': (M_F^0, Q_F).
\end{equation}

The Yukawa interactions in \Eq{lagF} provide radiative corrections to the mass of $\phi$. Some naturalness solutions require the ultralight real scalar to be a pseudo-Goldstone boson arising from spontaneous breaking of an approximate $U(1)$ symmetry~\cite{Brzeminski:2020uhm,Gan:2023wnp}. To this end, we embed $\phi$ into the angular component of a complex scalar $\Phi=\fdecayphi \exp(i\phi/\fdecayphi)/\sqrt2$, and introduce the Lagrangian,
\bea\label{eq:lagF}
{\cal L}\supset \left(\mathcal{Y}\Phi+\hc+M_F^0\right)\bar{\Psi}\Psi+\left(\mathcal{Y}^{\prime\dagger}\Phi+\hc+M_F^0\right)\bar{\Psi}'\Psi'
\eea
where, in general, $\mathcal{Y}$ and $\mathcal{Y}'$ are complex coupling parameters. Here and in what follows, we denote the fermion mass parameter in the Lagrangian by $M_{F}^0$. As the complex scalar $\Phi$ couples to the fermions, its nonzero vacuum expectation value in the $U(1)$ breaking phase contributes to the physical mass of $\Psi, \Psi'$. We therefore denote the tree-level fermion mass in the broken phase by $M_F$. Imposing the condition $\mathcal{Y}=\mathcal{Y}'=ye^{i c}$, where $c$ is the phase factor of the Yukawa coupling, we find that the Lagrangian given in \Eq{lagF} becomes invariant under the $\mathbb{Z}_2$ transformation:
\begin{equation}
\Phi\leftrightarrow \Phi^\dagger\;,\quad \Psi\leftrightarrow \Psi'.
\end{equation}
In the vacuum, the nonzero expectation value of the scalar field $\phi$ shifts the effective masses of the hypercharged fermions as follows,
\begin{equation} 
M_\Psi=M_F(1+\Delta_+)\;,\quad M_{\Psi'}=M_F(1+\Delta_-)\;, \quad M_F=M_F^0+m\cos(c)
\end{equation}
where $\Delta_\pm=y\fdecayphi[\cos(\phi/\fdecayphi\pm c)-\cos(c)]/\sqrt2 M_F$, with $|y\fdecayphi/M_F|\ll 1$, denotes the field-dependent mass correction. The oscillatory background of ultralight scalar dark matter induces time-dependent variations in the charged fermion masses at the UV scale. Consequently, in the IR limit, this effect leads to a $\mathbb{Z}_2$-even interaction between $\phi$ and the electromagnetic field strength. One straightforward approach to demonstrate this effect is via dimensional transmutation in the fine-structure constant at an energy scale $\mu\sim M_F$, given by
\begin{equation}\label{eq:alphaF}
\frac{1}{\alphaem(M_F^2)}=\beta_0\log\frac{\Lambda_{\rm QED}^2}{M_F^2(1+\Delta_+)(1+\Delta_-)}\;, 
\end{equation}
where $\Lambda_{\rm QED}$ is the QED Landau pole and $\beta_0=2 Q_F^2/3\pi$. As a result, $\Delta_\pm$ induces a variation in the fine-structure constant, $\delta\alphaem\simeq\alphaem^2\beta_0 (\Delta_++\Delta_-)$, leading to a quadratic coupling of the scalar field to the photon through the term $\delta\alphaem F_{\mu\nu}^2/4\alphaem$. This interaction predicts the quadratic coupling coefficient, which is
\bea
\mathcal{L}\supset-\frac{y\alphaem Q_F^2\cos(c)}{6\sqrt 2\pi M_F\fdecayphi}\phi^2 F_{\mu\nu}F^{\mu\nu}\;.
\eea
In this model, the sign of $\Lambda_\gamma^2$ is determined by the value of $c$. As previously discussed, $c$ can be chosen as an arbitrary real number because it serves as a phase factor.

\subsection{Heavy $U(1)_Y$-Charged Scalar}\label{sec:ChargedScalar}

Alternatively, the same $\mathbb{Z}_2$ symmetry can be embedded in the interactions involving $\phi$ and a complex scalar $\Phi$ heavier than TeV, which is assigned a mass and an SM $U(1)_Y$ charge as, $\Phi:  (M_S, Q_S)$, leading to the UV Lagrangian,
\begin{equation}
\mathcal{L}_{\rm UV}\supset -\left(M_S^2-\lambda\phi^2\right)\left|\Phi\right|^2\;. 
\end{equation}
We can apply the dimensional transmutation argument in \Eq{alphaF} to this scenario. In this case we need to take $M^2_F\rightarrow M^2_S$, $\beta_0=Q_S^2/6\pi$ and the quadratic coupling coefficient is given by,
\begin{equation}
 {\cal L}\supset-\frac{\lambda\alphaem Q_S^2}{48\pi M_S^2}\phi^2 F_{\mu\nu}F^{\mu\nu} 
\end{equation}
Similar to the fermionic UV model, the scalar model allows for either sign depending on the value of the quartic coupling $\lambda$. For instance, Chiral Perturbation Theory ($\chi$PT) suggests positive quadratic couplings of two QCD axions to two Standard Model charged pions~\cite{GrillidiCortona:2015jxo, DiLuzio:2020wdo}. However, due to the absence of $\mathbb{Z}_2$ symmetry in many QCD axion models, the linear axion-photon coupling is sizable and dominates experimental searches. In the next section, we will show that the isospin symmetry in the dark QCD sector can also realize the quadratic interaction.

\subsection{Dark QCD Axion}
Extended gauge and axion sectors commonly arise in string theory~\cite{Halverson:2018olu}. Inspired by QCD-like models~\cite{Bai:2010qg, Beadle:2023flm}, we consider an axion framework in the dark sector based on an $SU(N_c) \times U(1)'$ gauge symmetry. The model features two dark quarks, $u'$ and $d'$, which are charged under $U(1)'$ and transform in the fundamental representation of $SU(N_c)$. Assuming the dark quarks are much lighter than the dark confinement scale $\Lambda_c$, we impose approximate isospin symmetry, on the dark isospin doublet $q' = (u', d')^T$ and define its mass and charge matrices as,
\begin{equation}\label{eq:mQ} 
|\deltaiso|\ll 1,\quad x\in \mathbb{R}:\quad  \hat{m}_q' = m_q' \left(\mathbb{1} + \deltaiso \, \sigma^3\right), \quad \hat{Q}_q' = \frac{1}{2} \sigma^3 + x\,\mathbb{1}\;,
\end{equation}
where $\sigma^3$ is the third Pauli matrix. The dark $U(1)'$ gauge field kinetically mixes with the SM photon via the parameter $\epsilon$. The axion $a$ couples to the dark sector through shift-invariant interactions, including couplings to the dark QCD field strength $G_{\mu\nu}^{a\prime}$ and the dark photon field strength $F_{\mu\nu}'$,
\begin{equation}\label{eq:LDqcd2}
\mathcal{L}' \supset \epsilon F_{\mu\nu}F'^{\mu\nu} + \frac{a}{4\pi 
\fdecayaxion} \left( \alpha_s' N G^{a\prime}_{\mu\nu} \tilde{G}^{a\prime\mu\nu} + \alpha' E F_{\mu\nu}' \tilde{F}^{\prime\mu\nu} \right) + \frac{\partial_\mu a}{2 \fdecayaxion} \bar{q}'\gamma^\mu \gamma^5 \hat{C}_q^0 q' - \left( \bar{q}_L' \hat{m}{q'} q_R' + \hc \right)\;,
\end{equation}
where $G_{\mu\nu}^{a\prime}$ and $F_{\mu\nu}'$ are the field strengths of $SU(N_c)$ and $U(1)'$, with couplings $g_s' = \sqrt{4\pi \alpha_s'}$ and $e' = \sqrt{4\pi \alpha'}$, respectively.
The requirement for the axion coefficients can be described by the existence of a $2\times 2$ matrix $\hat{Q}_{A}$ which commutes with the mass matrix $ \hat{m}_{q^{\prime}}$ such that, $E=N_c\tr\big(\hat{Q}_A \hat{Q}_q^{\prime 2}\big)$, $N=\tr\big(\hat{Q}_A\big)/2$. The axion interactions respect a shift symmetry, ensuring that the $CP$-violating term $\theta' G^{\prime a}_{\mu\nu}\tilde{G}^{\prime a\mu\nu}$ can be absorbed by field redefinition. The kinetic mixing term introduces an effective electric charge for the dark quarks, $e\hat{Q}_{q'}^{\rm em} = \epsilon e' \hat{Q}_q'$, leading to their coupling with the SM photon. Both $a G\tilde{G}$ and $a F'\tilde{F}'$ terms can be simultaneously eliminated by the axion-dependent axial transformation, $q'\rightarrow \exp\big(ia\gamma_5 \hat{Q}_A/2\fdecayaxion\big)\,q'$. This transformation also modifies the coefficient of $ \bar{q}'\slashed{\partial} a\gamma^5 q'$ from $\hat{C}_q^0$ to $\hat{C}_q=\hat{C}_q^0-\hat{Q}_A$ in \Eq{LDqcd2} leading to,
\bea\label{eq:Dquark}
\mathcal{L}'&\supset\epsilon F_{\mu\nu}F'^{\mu\nu}-\left[\bar{q}_L^{\prime}\hat{m}_{q'}(a)q'_R+\hc\right]+\frac{\partial_\mu a}{2\fdecayaxion}\bar{q}'\gamma_\mu\gamma^5 \hat{C}_q q'\;.
\eea
At energies below $\Lambda_c$, chiral symmetry breaking occurs, leading to dark pions as the Goldstone bosons of the spontaneously broken $SU(2)_L \times SU(2)_R$ symmetry. The axion-dependent dark quark mass modifies the pion interactions, introducing an axion-dependent pion mass term that contributes to the quadratic axion-photon coupling. 
In the isospin symmetry limit, $\deltaiso=0$, employing the standard parametrization in two-flavor $\chi$PT~\cite{DiLuzio:2020wdo}, one finds,
\begin{equation}
\mathcal{L}'_\chi\supset m_{\pi'}^2 \fdecaypiprime^2\cos\left(\frac{a}{2\fdecayaxion}\right)\cos\frac{|\pi'|}{\fdecaypiprime}\;,\quad |\pi'|\equiv\sqrt{\pi^{\prime 
 2}_0+2\pi'_+\pi'_-}\;.
\end{equation}
where $\pi'_\pm$ are charged pions while $\pi_0'$ is neutral under $U(1)'$ with mass-square $m_{\pi'}^2$ and decay constant parameter $\fdecaypiprime$. The axion potential is minimized at $\langle a\rangle=2 n\pi$, $n=0,\pm 1, \pm 2\cdots$. Which means that the variation of the axion field value around its minima always reduce the axion dependent charged pion mass, $m_{\pi'}^2(a)=m_{\pi'}^2\cos(a/2\fdecayaxion)$, and therefore increase the energy scales ratio $\Lambda^2_{\rm QED}/m_{\pi'}^2(a)$. 
Following dimensional transmutation arguments, we find that the charged pions generate a negative contribution to $a^2 F^2$ at the one-loop level. More specifically, the quartic interaction $\lambda = m_{\pi'}^2 / 8 
\fdecayaxion^2 > 0$ leads to the axion-photon coupling,
\begin{equation}\label{eq:cFSUN}
\mathcal{L}' \supset -\frac{\epsilon^2\alpha'}{192\pi \fdecayaxion^2} a^2 F_{\mu\nu} F^{\mu\nu} + \deltaiso \frac{\epsilon\alpha'}{4\pi\fdecayaxion} \left(  a\,F_{\mu\nu} \tilde{F}'^{\mu\nu} + \epsilon \, a \, F_{\mu\nu} \tilde{F}^{\mu\nu} \right)\;. 
\end{equation}
Our result is consistent with Refs.~\cite{Kim:2023pvt,Bauer:2024yow,Bauer:2024hfv}, but differs in sign from Ref.~\cite{Beadle:2023flm}. We further point out that the approximate dark isospin symmetry suppresses linear axion-photon couplings, following
\bea
\label{eq:dark_isospin}
\text{Dark Isospin Symmetry}\,\,\,\,\,\deltaiso = 0 \quad \bm{\Longrightarrow} \quad a^2 F^2 \neq 0, \quad a F \widetilde{F} = 0. 
\eea
This implies that in certain axion models, even though the quadratic coupling is suppressed by $1/\fdecayaxion^2$, it can still dominate over the linear axion-photon coupling, which scales as $1/\fdecayaxion$ and provides the traditional observable~\cite{Sikivie:1983ip,ALPS:2009des,Redondo:2010dp,Sikivie:2013laa,Kahn:2016aff,DeRocco:2018jwe,Liu:2018icu,Berlin:2019ahk,Berlin:2020vrk,DMRadio:2022jfv}. This can be achieved through the cancellation of the linear operator under specific symmetries, such as the previously mentioned dark isospin symmetry, highlighting the significance of detecting quadratic axion couplings~\cite{Beadle:2023flm,Kim:2023pvt,Bauer:2024yow,Bauer:2024hfv}~\footnote{Even if the linear coupling is not canceled by an additional symmetry, a quadratic effect known as the ponderomotive effect can still arise. A comprehensive discussion of this phenomenon for different types of ultralight dark matter can be found in Ref.~\cite{Zhou:2025wax}. For the axion, the leading ponderomotive effect arises from the cross term between the linear axion–photon coupling and the derivative axion–fermion coupling.}.

In the UV regime, \Eq{Dquark} can be realized by introducing a neutral complex scalar $\Phi$ charged under a global $U(1)_\PQ$ symmetry. The corresponding Lagrangian is given by
\begin{equation}
\label{eq:UVDquark}
\mathcal{L} \supset -\Phi \bar{q}_L' \mathcal{Y} q_R' - \hc - \kappa \left( |\Phi|^2 - \frac{\fdecayaxion}{2} \right)^2 + \epsilon F_{\mu\nu} F'^{\mu\nu}\;, 
\end{equation}
where $\kappa > 0$. The spontaneous breaking of $U(1)_\PQ$ generates the axion field, $\Phi = \fdecayaxion \exp(ia/\fdecayaxion)/\sqrt{2}$, reducing \Eq{UVDquark} to \Eq{Dquark} with $\hat{C}_q=0$. The model described in \Eq{UVDquark} is analogous to the DFSZ-like scenario~\cite{Zhitnitsky:1980tq, Dine:1981rt}, where the light quarks are also charged under the PQ symmetry. Alternatively, we may consider a scenario in which $q'_{L}$ and $q'_R$ are neutral under the $U(1)_\PQ$, while the UV completion involves heavy dark quarks charged under $U(1)_\PQ$, corresponding to a KSVZ-like construction~\cite{Kim:1979if,Shifman:1979if}. Further details on this model are provided in~\Appx{uv_app}.

\section{Conclusion}
\label{sec:conclusion}

In this work, we systematically investigate the scattering force and background-induced force arising from the matter effect generated by a quadratic coupling between light scalar dark matter and Standard Model fields. By utilizing quantum mechanical scattering theory, we develop a unified framework to analyze both forces, perform a systematic classification of the parameter space~(as illustrated in \Fig{phase_classify}), and explore all regimes—crossing perturbative and non-perturbative, low-momentum and high-momentum regions. Moreover, we reconcile previous discussions based on different methodologies.

For the scattering force, we unify the perturbative and non-perturbative analyses scattered across previous works, identifying regions where decoherence effects become significant and where perturbative calculations break down. We also discuss experimental sensitivities based on acceleration tests, as well as theoretical sensitivities by comparing signals with thermal and quantum noise. In addition, we provide a comparison of the scattering force and the background-induced force and identify the regions where each force dominates.

For the background-induced force in the perturbative region, we not only reproduce the perturbative results derived from finite-density field theory using the Born approximation, but also extend the analysis of decoherence effects to scenarios involving sources of finite size. Furthermore, we conduct a thorough investigation of the non-perturbative region, where the screening effect becomes important. We revisit the previously employed spherically symmetric ansatz, which is suitable for the low-momentum near-field region, and also analyze its limitations in the high-momentum or far-field region. Additionally, we highlight the descreening effect, which serves as a transition from the spherically screening ansatz to the perturbative treatment. We discuss the experimental constraints, including inverse square law tests and short-range E$\ddot{\rm o}$t-Wash equivalence principle tests. Sensitivities associated with the background-induced force are evaluated using acceleration tests and by comparing predicted signals with thermal and quantum noise. Moreover, we go beyond the spherically symmetric ansatz and revisit the MICROSCOPE constraint in the high-momentum region, where the descreening and decoherence effects reach a balance.

Finally, we present UV models that give rise to the quadratic scalar-photon coupling, which induces variations in the fine-structure constant. We consider models involving heavy $U(1)_Y$-charged fermions, heavy $U(1)_Y$-charged scalars, and dark QCD axions, and clarify the signs of the effective mass-squared contributions from matter effects in each case through a renormalization group analysis.

\section*{Acknowledgement}

We thank Brian Batell, Yifan Chen, Sebastian Ellis, Christina Gao, Christophe Grojean, Yuval Grossman, Guilherme Guedes, Yikun Jiang, Hyungjin Kim, Zhifang Lin, Geraldine Servant, Konstantin Springmann, Ken Van Tilburg, Pham Ngoc Hoa Vuong, Lian-Tao Wang, Samuel Witte, Huangyu Xiao, Xingchen Xu, Yue Zhao, Kevin Zhou, Siyu Zhou, and Zilu Zhou for their helpful discussions. We thank Yifan Chen, Christina Gao, Hyungjin Kim, and Yue Zhao for the discussion on the experimental sensitivities. We thank Sebastian Ellis, Hyungjin Kim, and Pham Ngoc Hoa Vuong for the discussion on the models of the quadratic scalar interactions. We thank Yuval Grossman and Ken Van Tilburg for the discussion on the matter effect. We thank Siyu Zhou for the discussion on the perturbative finite-size effect. We thank the anonymous referee for helpful and constructive comments that improved the presentation of our results. XG thanks Rukeng Su for teaching him quantum mechanics and scattering theory during his undergraduate studies at Fudan University. This guidance has been invaluable for his subsequent studies and research.

The work of XG is supported by the Deutsche Forschungsgemeinschaft under
Germany’s Excellence Strategy - EXC 2121 “Quantum Universe” - 390833306. The work of DL is supported in part by the U.S. Department
of Energy under Grant No. DE-SC0007914 and in part by
the PITT PACC.  DL acknowledges support from The Leverhulme Trust, Grants RPG-2023-285 and funding from the French Programme d’investissements d’avenir through the Enigmass Labex. 
BY is supported in part by the NSF grant PHY-2309456. This work used the Maxwell computational resources operated at Deutsches Elektronen-Synchrotron DESY, Hamburg~(Germany).


\appendix

\section{Test Mass in Scalar Background}\label{appx:test_mass_bkg}

In this section, we derive the equation of motion of the test mass within the scalar background. We follow Ref.~\cite{Hees:2018fpg} but extend the discussion to a more generic situation in which the scalar field value varies inside the test mass. Because the test mass $\testmass$ moving inside the scalar background follows the geodesics, we can list the full Lagrangian of the test mass, which is
\bea
L_\testmass = - \int d^3 \delta \vecr_\testmass \, \rho_\testmass(\phi) \sqrt{g_{\mu\nu} \frac{dx^\mu_\testmass}{dt} \frac{dx^\nu_\testmass}{dt}} \stackrel{\text{NR}}{=} - \int d^3 \delta \vecr_\testmass \, \rho_\testmass(\phi)\Big[ 1 - \frac{\vecv^2_\testmass}{2} + \cdots \Big].
\eea
In the second equal sign, we take the NR limit of the test mass. $x_\testmass$ is the four-coordinate of the test mass, and $\vecv_\testmass$ is the three-velocity of the test mass. Throughout this paper, we impose a $\mathbb{Z}_2$ symmetry under the transformation $\phi \rightarrow -\phi$, which forbids linear terms in the Lagrangian. Therefore, after expanding $M_\testmass$ over $\phi$, we write the Lagrangian as
\bea
\label{eq:L_testmass_expand}
L_\testmass =  \int d^3 \delta \vecr_\testmass \Big[\rho_\testmass \frac{\vecv_\testmass^2}{2} - \frac{d \rho_\testmass}{d\phi^2}\bigg|_{\phi=0}\phi^2 + \cdots \Big], 
\eea
where
\bea
\frac{d \rho_\testmass}{d\phi^2}\bigg|_{\phi=0} = \frac{\mMtest^2}{2}.
\eea
Because the velocity of the test mass is negligible, we can neglect the cross-term between $\phi^2$ and the kinetic term of the test mass. From \Eq{L_testmass_expand}, we can find that the term denoting the variation of the test mass $\testmass$ sourced by the scalar field serves as the effective potential, $V_{\bg}$. Utilizing the Euler-Lagrange equation, we have
\bea
\vecF_{\bg} = - \vecnabla V_\bg,
\eea
where $V_\bg$ is the potential induced by the quadratic scalar-SM couplings. This potential is written as
\bea
V_\bg(t)  = \frac{\mMtest^2}{2} \int d^3 \delta \vecr_\testmass \, \phi^2(t,\delta \vecr_\testmass).
\eea
Here, we factor out the spatial gradient operator~$\vnabla$ from $V_\bg$, which applies to the arbitrary shape in the perturbative region when there is no strong backreaction. In the strongly coupled regime, where screening effects are significant to the test mass, this factorization is no longer valid. The background-induced force experienced by the test mass in this regime will be discussed in Ref.~\cite{Da:progress}. In the equation above, the effective potential $V_\bg$ induced by the scalar background contains a time-modulating component. However, within the mass range considered in this work, the scalar field oscillates much faster than the detector response time. As a result, we can directly average over time and treat the effective potential as time-independent. For the monochromatic incident wave, substituting the scalar's non-relativistic approximation $\phi = \Re[e^{-i \omega t} \psi]$ and doing time average over the background-induced potential, we have
\bea
\label{eq:V_bkg_timeave_appx}
\langle V_\bg\rangle_t 
& = \frac{\mMtest^2}{2} \int d^3 \delta \vecr _\testmass \, \frac{1}{2\pi/\omega} \int^{2\pi/\omega}_0 dt \, \abs{\psi(\delta \vecr_\testmass)}^2 \cos^2(\omega t - \arg \psi(\delta \vecr_\testmass))\\
& =  \frac{\mMtest^2}{4} \int d^3 \delta \vecr _\testmass \, \abs{\psi(\delta \vecr_\testmass)}^2.
\eea
From the above formula, we find that the background-induced potential is proportional to the probability amplitude of the scalar wave function. For the situation in which the incident scalar wave function is the plane wave, the $\abs{\psiinc}^2$ contribution is the constant of the spatial coordinate and thus does not induce the extra force on the test mass. For this reason, we drop the plane wave probability amplitude~(replace $\abs{\psi}^2$ with $\abs{\psi}^2 - \abs{\psi_0}^2$) in \Eq{V_bkg_timeave_appx} and write the background-induced potential as
\bea
\label{eq:V_bkg_timeave_appx_2}
\langle V_\bg\rangle_t = \frac{\mMtest^2}{2} \int d^3 \delta \vecr_\testmass  \left\{ \Re\left[ \psiinc^*(\delta \vecr_\testmass) \psisc(\delta \vecr_\testmass) \right] + \frac{\abs{\psi_\sct(\delta \vecr_\testmass)}^2}{2}\right\}.
\eea
In the spatial region where $\abs{\psisc} < \abs{\psiinc}$, we keep only $\Re\left[\psiinc^* \psisc\right]$, the interference term between the incident wave and the scattered wave, in the above formula. However, when the scattering process is non-perturbative, we have $\abs{\psisc} \sim \abs{\psiinc}$ in the scattering region. In this case, both terms in \Eq{V_bkg_timeave_appx_2} must be kept. Throughout the rest of the paper, we use \Eq{V_bkg_timeave_appx} to compute the additional force experienced by the test mass $\testmass$ moving in the scalar background. For notational simplicity, we omit the sign $\langle \cdots \rangle_t$. Unless specified, all background-induced potentials discussed in the rest of the paper are time-averaged.

\section{Forces in Quantum Field Theory}\label{appx:force_qft}

In this section, we provide a comprehensive derivation of scalar-mediated forces within the framework of quantum field theory for completeness. Using this formalism, we can simultaneously derive both the vacuum force and the background-induced force in the same framework. Such derivations help us to get the equivalent result as the method utilizing non-relativistic quantum mechanics in the weak-coupling region where the Born approximation is applicable, and the size of the source mass is small compared with the inverse of the incident momentum~(Region A, $\mM, k < R^{-1}$). Similar derivations can be found in Refs.~\cite{Notzold:1987ik,Horowitz:1993kw,Ferrer:1998ju,Ferrer:1999ad,Ferrer:2000hm,Ghosh:2022nzo,Blas:2022ovz,Ghosh:2024qai,VanTilburg:2024xib,Barbosa:2024pkl}, which studied the background-induced effects in neutrino physics and other new physics scenarios. 

Here, we focus on the scalar-induced forces acting on the test mass $\testmass$ due to the scalar potential generated by the source mass $\sourcemass$. As discussed in \Sec{matter_effect}, in an environment composed of ordinary matter, we can do the replacement $\langle F^2/4 \rangle \rightarrow -Q_{\gamma} \rho$. As a result, The vertex coupling between the test mass $\testmass$ and the scalar field is $Q_{\gamma,\testmass} M_\testmass/\Lambda_\gamma^2 = \mMtest^2 \mathcal{V}_\testmass$, and the vertex coupling between the source mass $\sourcemass$ and the scalar field is $Q_{\gamma,\sourcemass} M_\sourcemass/\Lambda_\gamma^2 = \mMsource^2 \mathcal{V}_\sourcemass$. Using these expressions, we can compute the one-loop scattering amplitude in a dense environment utilizing
\bea
\label{eq:amp}
i {\cal M} = (\mMtest^2 \mathcal{V}_\testmass)(\mMsource^2 \mathcal{V}_\sourcemass)\int \dbar^4 k \, D_{\rm mod}(k)D_{\rm mod}(k+q)\;,
\eea
where $q$ is the momentum transfer. To include the background effect, we introduce the modified propagator 
\bea
D_\text{mod}(k) \equiv D_\vac(k) + D_\bg(k),
\eea
which contains the vacuum propagator $D_\vac$ and the background propagator $D_\bg$. $D_\vac$ and $D_\bg$ are written as
\bea
D_\vac(k) = \frac{1}{k^2-m^2_0+i\epsilon}, \quad \quad D_\bg(k) = - 2\pi i \delta(k^2-m^2_0)f_\phi({\veck}),
\eea
where $f_\phi(\veck)$ is the dimensionless phase space distribution of the background particles. The $\delta$-function in $D_\bg$ indicates that the background scalars are on-shell.
\begin{figure}
\centering \includegraphics[width=0.57\columnwidth]{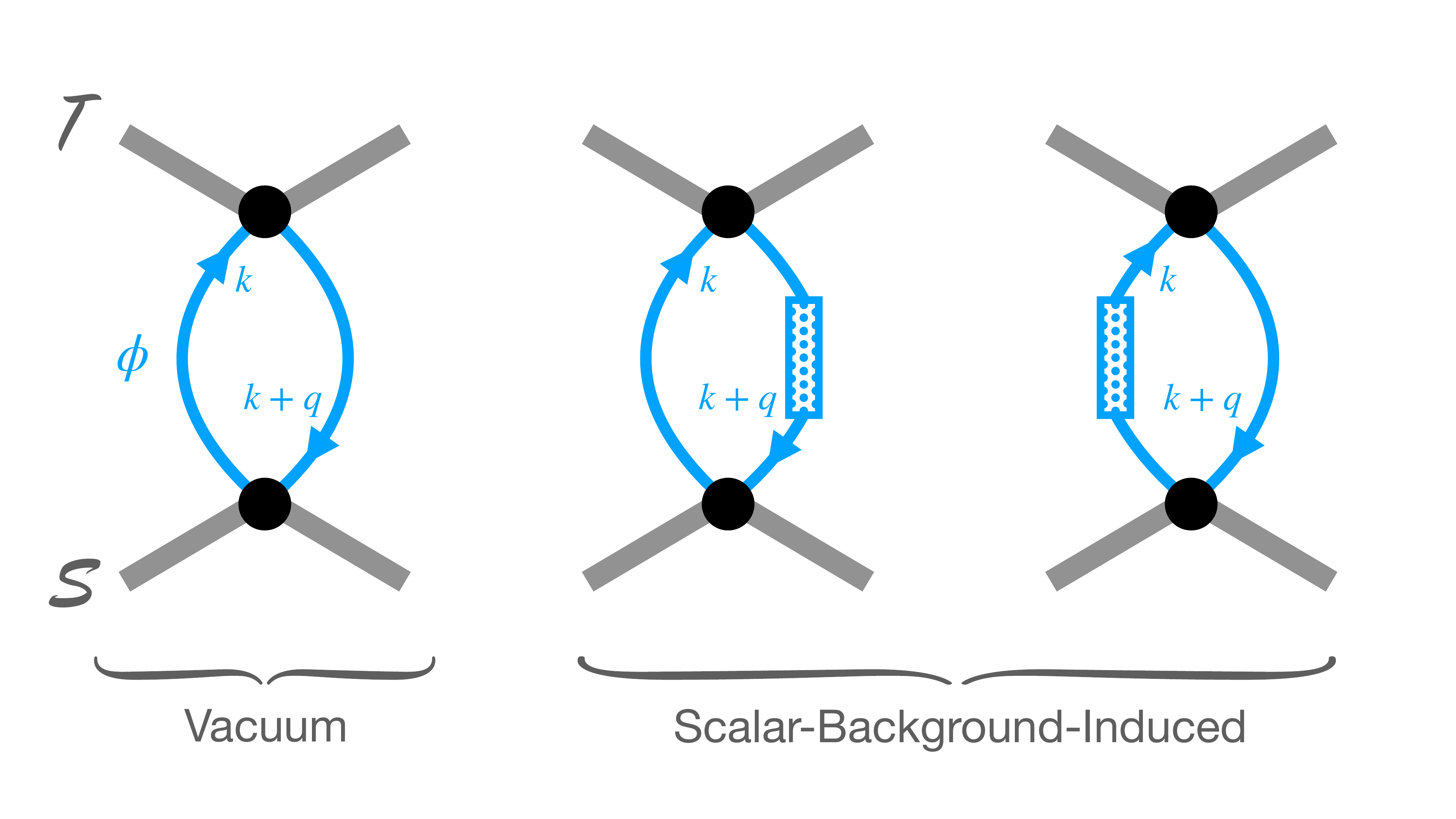}
\caption{The Feynman diagrams of the vacuum force~(left) and the scalar-background-induced force~(right) between $\testmass$ and $\sourcemass$. The thick gray lines represent the test mass $\testmass$ and the source mass $\sourcemass$. The black circles denote the interaction vertices between the test and source masses and the scalars. The smooth thin blue lines denote the scalar's vacuum propagators $D_\vac$. The thin blue lines attached by the dotted box denote the scalar's background propagators $D_\bg$.}
\label{fig:finite_size}
\end{figure}
In \Fig{finite_size}, we show the Feynman diagrams representing the vacuum force~(left panel) and background-induced force~(right panel). In this figure, the thick gray lines denote the macroscopic objects $\testmass$ and $\sourcemass$, respectively. The thin blue lines represent the scalar's vacuum propagators $D_\vac$. The thin blue lines with the dotted box attached represent the background correction $D_\bg$ in the scalar propagator, which arises from the coherent scattering between the background scalars and the ordinary matter. The black circles denote the interaction between $\testmass$ and $\sourcemass$ with the scalar. The advantage of this formalism is that it can describe both the vacuum effect (the quantum force) and the background effect in a unified, purely field-theoretical way. 

When both propagators in Eq.~(\ref{eq:amp}) take the $D_\vac$ term, it gives the amplitude from the exchange of two virtual scalars:
\begin{align}
i{\cal M}_\vac &= \frac{(\mMtest^2 \mathcal{V}_\testmass)(\mMsource^2 \mathcal{V}_\sourcemass)}{2}\int \dbar^4 k \, D_\vac(k)D_\vac(k+q) = i \frac{(\mMtest^2 \mathcal{V}_\testmass)(\mMsource^2 \mathcal{V}_\sourcemass)}{2} \times \frac{B_0(q^2;m_0,m_0)}{16\pi^2} ,\label{eq:ampvac}
\end{align}
where a symmetry factor of 1/2 has been included, and $B_0$ is the standard Passarino-Veltman function~\cite{Passarino:1978jh} simplified to be
\begin{align}
B_0(q^2;m_0,m_0) = \frac{\sqrt{q^2\left(q^2-4m^2_0\right)}}{q^2}\log\left[\frac{2m^2_0-q^2+\sqrt{q^2\left(q^2-4m^2_0\right)}}{2m^2_0}\right] + \cdots.
\end{align}
Here, ``$\cdots$'' denotes the terms that have no branch cuts in the $q^2$-plane, and thus do not contribute to the vacuum force. The vacuum force only depends on the imaginary part of ${\cal M}_\vac$, which, by definition, is connected to the discontinuity of the amplitude:
\bea
{\rm Disc}\left[i{\cal M}_\vac(q^2)\right] 
&\equiv i {\cal M}_\vac\left(q^2 + i\epsilon\right) - i{\cal M}_\vac\left(q^2-i\epsilon\right)\\
&= -2{\rm Im}\left[{\cal M}_\vac(q^2)\right]\\ 
&= -\Theta \left(q^2-4 m^2_0\right)\frac{(\mMtest^2 \mathcal{V}_\testmass)(\mMsource^2 \mathcal{V}_\sourcemass)}{2}\frac{1}{8 \pi  q^2 }\sqrt{q^2 \left(q^2-4 m^2_0\right)}\;,
\eea
where $\Theta$ is the step function. On the other hand, the background effect is captured by the ``crossing terms'' (namely, one of the propagators is on-shell while the other is off-shell):
\begin{align}
 i{\cal M}_\bg =  \frac{(\mMtest^2 \mathcal{V}_\testmass)(\mMsource^2 \mathcal{V}_\sourcemass)}{2} \int \dbar^4 k \left[D_\vac(k)D_\bg(k+q)+D_\bg(k)D_\vac(k+q)\right].\label{eq:ampbkg}
\end{align}
By integrating out the $k^0$ component first and using the non-relativistic approximation $q^\mu \simeq (0,\vecq)$, one obtains
\begin{align}
\label{eq:M_bg_appx}
{\cal M}_\bg =  (\mMtest^2 \mathcal{V}_\testmass)(\mMsource^2 \mathcal{V}_\sourcemass) \int  \frac{\dbar^3 \veck}{2\omega_\veck} f_\phi(\veck) \left(\frac{1}{\vecq^2+2\veck\cdot\vecq-i\epsilon}+\frac{1}{\vecq^2 - 2 \veck \cdot \vecq +i\epsilon}\right).
\end{align}
Since the source mass $\sourcemass$ is fixed while we measure the force experienced by the test mass $\testmass$, we express \Eq{M_bg_appx} in terms of the retarded Green's function. This formulation is also consistent with causality in quantum mechanical scattering theory, as shown in \Eq{psi_in_sca}, where the source mass generates only an outgoing wave without contributing to the incoming wave. 

The potential originating from scattering is extracted by matching the scattering amplitude in the momentum space to the position space using $V(\vecr) = - \int \dbar^3 \vecq \, e^{-i \vecq \cdot \vecr} \mathcal{M}(\vecq)$, where $\mathcal{M}(\vecq)$ can be amplitudes from both vacuum and background effects. For the vacuum amplitude, we use the dispersion technique~\cite{Feinberg:1989ps}
\begin{align}
\int  \dbar^3 \vecq \,e^{-i\vecq \cdot \vecr} {\cal M}_\vac(\vecq) = - \frac{1}{8\pi^{2}r}\int_{4m^2_0}^{\infty}\text{Disc}\left[i\mathcal{M}_\vac\left(t\right)\right] e^{-\sqrt{t}r}\text{d}t,
\end{align}
and then have
\begin{align}
\text{Vacuum Potential:}\quad V_\vac (\vecr) = -(\mMtest^2 \mathcal{V}_\testmass)(\mMsource^2 \mathcal{V}_\sourcemass) \frac{m_0}{32\pi^3 r^2} K_1\left(2m_0 r\right),
\end{align}
where $K_1$ is the modified Bessel function. In the near-field region where $r < m_0^{-1}$, the vacuum force follows the power law $V_\vac \propto 1/r^3$.
In the far-field region where $r > m_0^{-1}$, the vacuum force follows $V_\vac \propto e^{-2m_0 r}/r^{5/2}$, which means that the vacuum force is exponentially suppressed at large distances because of the nonzero mass of the mediator. These characteristic behaviors at short and long distances arise from the shift-symmetry-breaking operators in  \Eq{quad_phi_sm_general}, as shown by Refs.~\cite{Feinberg:1989ps,Ferrer:2000hm,Fichet:2017bng,Bauer:2023czj}.

On the other hand, the background-induced force is essentially a classical effect since there is only one off-shell propagator in the loop calculation. By performing the Fourier transform, we obtain

\begin{equation}
\text{Background-Induced Potential}:\quad V_\bg(\vecr) = -(\mMtest^2 \mathcal{V}_\testmass)(\mMsource^2 \mathcal{V}_\sourcemass) \frac{n_\phi}{4\pi r}\int \frac{\dbar^3 \veck}{\omega_\veck} \frac{f_\phi(\veck)}{n_\phi}\cos\left(kr-\veck\cdot\vecr\right).
\label{eq:bkgforce_appx}
\end{equation}
This is the relativistic form of the background-induced force, which is consistent with the previous work calculating the tree-level Compton scattering amplitude~\cite{VanTilburg:2024xib}. The major difference is that we calculate the closed loop containing both $D_\vac$ and $D_\bg$, while the Ref.~\cite{VanTilburg:2024xib} cuts the propagator $D_\bg$ and sets the scalars to be the external legs. Because we focus on the ultralight dark matter, we take the non-relativistic limit $\omega_\veck \simeq m_0$ and recover the scalar-background-induced potential derived in \Eq{Vbg_A}.

Before ending this section, we compare the effects of the vacuum and background-induced potentials. For the perturbative region which applies for most of the terrestrial experiments, we have
\bea
\abs{\frac{V_\vac}{V_\bg}} \sim \frac{m_0^3}{ \sigmak^{3} \, N_\phi} \times \frac{K_1(2m_0 r)}{m_0 r} \times \max\big[1,(\sigmak r)^2\big].
\eea
Here $N_\phi \sim n_\phi/\sigmak^3$ is the occupation number of the ultralight dark matter, and $\max\big[1,(\sigmak r)^2\big]$ arises from the finite-phase-space effect of $V_\bg$ as derived in \Eq{phase_space_factor}. For example, when $m_0 \sim 0.1\,\eV$, we have $N_\phi \sim 10^8$, which implies that $\abs{V_\vac/V_\bg} \ll 1$. In the non-perturbative region where the equivalence principle test on the MICROSCOPE satellite becomes relevant, there is an $\sim 1/(\mM R)^2$ suppression~(s-wave limit) on the background-induced force due to the screening effect as shown in \Eq{formfactor_sphscr_farfield_hard}. However, two factors counteract this suppression: (1) The enhancement of $V_\bg$ from the large occupation number~($N_\phi \sim 10^{44}$ for $m_0 \sim 10^{-10}\,\eV$) given that the MICROSCOPE experiment constrains a much lighter region. (2) The exponential suppression of $V_\vac$~($m_0 R_\oplus \sim 10^3$ for $m_0 \sim 10^{-10}\,\eV$). Therefore, $\abs{V_\vac/V_\bg} \ll 1$ remains valid for the MICROSCOPE experiment. Based on the discussion in this subsection, we conclude that the force from the vacuum potential is negligible compared to the background-induced potential in the macroscopic experiments discussed in this paper.

\section{Born Approximation}\label{appx:born_approx_appendix}

The Born approximation is the standard approach for dealing with the quantum mechanical scattering in the perturbative regime, where $|\psisc| < |\psi_0|$~\cite{Born:1926yhp}. In this section, we begin by introducing the solution to the Lippmann-Schwinger equation~\cite{Lippmann:1950zz}. We then derive the scattering amplitude, which describes the wave function's behavior in the asymptotic far-field limit. Next, we present analytical results for both the scattering cross section and the momentum-transfer cross section of a solid sphere. Finally, we prove that in the perturbative region where the finite-size effects become significant, the background-induced potential can be obtained by directly integrating the background-induced potential from point-to-point interaction. 

\subsection{Wave Function}\label{appx:wave_func_appendix}

In this subsection, we discuss the wave function in the perturbative limit. Substituting $\psi = \psiinc + \psisc$ into the Schr\"{o}dinger equation \Eq{Schrodinger_Eq}, we have the exact representation
\bea
\label{eq:lipp_sch_appx}
\psisc(\vecr) = \int_\mathcal{V} d^3 \delta \vecr \, G^{(\pm)}(\vecr,\delta \vecr) \, V_\eff(\delta \vecr) \, \psi(\delta \vecr),
\eea
which is known as the Lippmann-Schwinger equation~\cite{Lippmann:1950zz}. $G^{(\pm)}(\vecr,\delta\vecr)$ are the Green's functions represented as
\bea
G^{(\pm)}(\vecr,\delta\vecr) =\Big\langle \vecr \Big|\frac{1}{E - H_0 \pm i \epsilon}\Big| \delta \vecr \Big\rangle = - \frac{m_0}{2 \pi} \, \frac{\exp(\pm i \abs{\veck}\abs{\vecr-\delta \vecr})}{\abs{\vecr - \delta \vecr}},
\eea
where $(+)$ represents the retarded Green's function, and $(-)$ represents the advanced Green's function. $E = k^2/2m_0$ is the energy of the incident wave. $H_0 = - \vecnabla^2/2m_0$ is the free Hamiltonian. Because we focus on the outgoing scattered wave, we merely choose the retarded Green's function because of the causality. Substituting $\psi$ with zeroth order wave function $\psiinc$ on the right-hand-side of \Eq{lipp_sch_appx}, we have the first-order scattered wave function
\bea
\label{eq:born_approx_appx}
\psisc(\vecr;\veck) =  - \abs{\psi_0} \frac{m_0}{2 \pi} \int_\mathcal{V} d^3 \delta \vecr \frac{\exp(i \abs{\veck} \abs{\vecr - \delta \vecr})}{\abs{\vecr - \delta \vecr}} V_{\text{eff}}(\delta \vecr) \exp(i \veck \cdot \delta \vecr).
\eea
For the perturbativity condition to hold, we want $\abs{\psisc}_{r\sim R} \lesssim \abs{\psiinc}$. In the short-wavelength limit where $kR \lesssim 1$, the entire spherical volume contributes to the integration in \Eq{born_approx_appx}. So we have $\abs{\psisc}_{r \sim R} \sim \abs{\psi_0} \mM^2 R^3/R  \lesssim  \abs{\psi_0}$, which leads to $\mM  \lesssim R^{-1}$. In the long-wavelength limit where $kR \gtrsim 1$, the exponential term inside \Eq{born_approx_appx} is highly oscillatory, so only a shell of thickness $k^{-1}$ effectively contributes to the integration. This gives $\abs{\psisc}_{r\sim R} \sim \abs{\psi_0} \mM^2 R^2 k^{-1}/R \lesssim \abs{\psi_0}$, leading to $\mM \lesssim (kR^{-1})^{1/2}$. To summarize, we obtain
\bea
\label{eq:perturb_condition_appendix}
\text{Perturbativity Condition:\quad}
\left\{
\begin{aligned}
\text{Region A}: \quad  & kR \lesssim 1, \,\, \mM \lesssim R^{-1} \\
\text{Region B}: \quad & kR \gtrsim 1, \,\, \mM \lesssim ( k R^{-1})^{1/2}
\end{aligned}
\right.,
\eea
which can be utilized to divide the perturbative region of the parameter space shown in \Fig{phase_classify}. 

In the far-field region which satisfies $r > \max(k R^2, R)$, we can neglect the oscillating term, factor out the spherical wave $e^{ikr}/r$, and approximately write $\psisc$ as
\bea
\label{eq:psi_sc_born_far_appx}
\text{Far-Field}: \quad \psisc(\vecr;\veck) \simeq \underbrace{- \frac{m_0}{2 \pi} \int_\mathcal{V} d^3 \delta \vecr \, V_\eff(\delta \vecr) e^{-i \vecq \cdot \delta \vecr}}_{f(\hat{\vecr};\veck)} \times  \abs{\psi_0}  \frac{e^{ikr}}{r}.
\eea
Here we get the scattering amplitude $f(\hat{\vecr};\veck)$ listed in \Eq{f2}, which describes the asymptotic behavior of the wave function in the direction labeled by $\hat{\vecr}$. When discussing the system with the azimuthal symmetry, we can write $f(\hat{\vecr},\veck)$ as $f(\theta;k)$. For the solid sphere with a radius of $R$, we use the formula
\bea
\frac{1}{\mathcal{V}} \int_{\mathcal{V}} d^3 \delta \vecr \, e^{-i \vecq \cdot \delta \vecr} = \frac{ 3 (\sin(q R) - q R \cos(q R))  }{(qR)^3},
\eea
and get the analytical representation, which is
\bea
\label{eq:f_sphere_appx}
\text{Solid Sphere}:\quad f(\theta;k) = - \mM^2 R^3 \times \frac{\sin(qR)- qR \cos(qR)}{(qR)^3}
\simeq
\left\{
\begin{aligned}
& - \frac{\mM^2 \mathcal{V}}{4\pi} & \quad (qR < 1)\\
& \frac{\mM^2 \mathcal{V}}{4\pi} \times \frac{3\cos(qR)}{(qR)^2} & \quad (qR > 1)
\end{aligned}
\right..
\eea
Here, $\mathcal{V}=4\pi R^3/3$ is the volume of the solid sphere, $q = 2k \sin(\theta/2)$ is the magnitude of the momentum transfer. In the above formula, ``$=$'' denotes the exact integration, and ``$\simeq$'' denotes the scattering amplitude in different limits of $qR$. 

In the long-wavelength limit~($kR < 1$), we have:
\begin{itemize}
\item {All Direction:}
$f(\theta;k) \propto \mathcal{V}$. 
\end{itemize}

In the short-wavelength limit~($kR > 1$), we have:
\begin{itemize}
\item {Forward Direction~($\abs{\theta} < (kR)^{-1}$):} $f(\theta;k) \propto \mathcal{V}$. 
\item {Non-Forward Direction~($\abs{\theta} > (kR)^{-1}$):} 
$f(\theta;k) \propto \mathcal{V} \times \cos(qR)/(qR)^2$. 
\end{itemize}
The relation $f(\theta;k) \propto \mathcal{V} \propto \mathcal{N}$ represents the coherent enhancement of the scattering amplitude, where $\mathcal{N}$ is the number of particles in the scattered object. In the long-wavelength limit, this enhancement occurs in all directions. However, in the short-wavelength limit, coherent enhancement persists only in the forward direction, while in non-forward directions, it is suppressed due to finite-size decoherence effects.

\subsection{Cross Sections}\label{appx:sca_appendix}

To calculate the scattering cross section and the momentum transfer before or after the collision, one only needs to focus on the asymptotic behavior of the scattered wave. More specifically, we use the formulas of the cross section and momentum-transfer cross section
\bea
\label{eq:sigma_sigmat_appx}
\sigma(k) = \int d\Omega \abs{f(\theta;k)}^2, \quad \quad \sigmat(k) = \int d\Omega \abs{f(\theta;k)}^2 (1-\cos\theta).
\eea
To calculate the cross section of the solid sphere, we substitute \Eq{f_sphere_appx} into \Eq{sigma_sigmat_appx}. Then we have
\bea
\label{eq:sigma_appx}
\sigma(k) = \pi R^2 \times \left(\frac{2 \mM^4 R^2}{k^2}\right) \times F_\sigma(2kR),
\eea
where the dimensionless function $F_\sigma$ is defined as
\bea
\label{eq:F_sigma}
F_\sigma(x) 
& = \int^x_0 d\xi \, \frac{(\sin \xi - \xi \cos\xi)^2}{\xi^5} \\
& = \frac{\cos(2x)-1 + 2x \sin(2x) - 2x^2 + 2x^4}{8x^4}. 
\eea
The approximate form in different limits of $kR$ is
\bea
\sigma(k) \simeq \left\{
\begin{aligned}
& \frac{(\mM^2 \mathcal{V})^2}{4\pi} & \quad (kR < 1)\\
& \frac{(\mM^2 \mathcal{V})^2}{4\pi} \times \frac{9}{8}\frac{1}{(kR)^2} & \quad (kR > 1)
\end{aligned}
\right. .
\eea
In the region where $kR < 1$, we approximately have $\sigma/\pi R^2 \sim (\mM R)^4$, which gives the perturbativity bound $\mM < R^{-1}$. In the region where $kR > 1$, we have $\sigma/\pi R^2 \sim \mM^4 R^2/k^2$, which gives the perturbativity bound $\mM < (k R^{-1})^{1/2}$. In this region, because $\sigma(k) \sim f(\theta=0) \times (1/kR)^2$, we know that the major contribution comes from the wave function in the forward direction.

We also calculate the momentum-transfer cross section utilizing 
\beq
\label{eq:sigmat_AB_appx}
\begin{split}
\sigmat&=  \pi R^2   \times \frac{m_{\rm M}^4 }{ k^4 } \times F_{\text{T}}(2 k R).\\
\end{split}
\eeq
The dimensionless function $F_{\text{T}}$ is defined as
\bea
F_{\text{T}}(x) &= \int_0^{x} d\xi \, \frac{(\sin \xi - \xi \cos \xi)^2}{\xi^3} \\
&= \frac{-1+\cos 2x }{4x^2}  + \frac{\sin2x}{2x} + \frac{-1  - \Ci(2x)+ \gammaE + \log 2x}{2}.\\
\eea
Here $\Ci(x) = -\int_x^\infty d\xi \cos\xi/\xi$ is cosine integral function, and $\gammaE \simeq 0.57721$ is the Euler's constant. In long-wavelength and short-wavelength limits, the momentum-transfer cross section is approximately
\bea
\sigmat(k) \simeq 
\left\{ 
\begin{aligned}
& \frac{(\mM^2 \mathcal{V})^2}{4\pi} & \quad (kR < 1)\\
& \frac{(\mM^2 \mathcal{V})^2}{4\pi} \times \frac{9}{8} \frac{\log(4kR)}{(kR)^4}& \quad (kR > 1)
\end{aligned}
\right. ,
\eea
which appears in \Eq{region_a_sigmat} and \Eq{region_b_sigmat} in the main text. The result in the long-wavelength limit~($kR < 1$), is derived using the Taylor expansion $\Ci(2x) \simeq \gamma_E + \log(2x) - x^2 + x^4/6 +\mathcal{O}(x^6)$. The result in the short-wavelength limit~($kR > 1$) is derived using the asymptotic expansion $\Ci(2x) \simeq \sin(2x)/2x -\cos(2x)/4x^2$. Here, we find that $\sigmat$ increases much slower than $\sigma$ as $R$ increases. This is because when calculating $\sigmat$, the extra factor $(1-\cos\theta)$ suppresses the contribution of the wave function in the forward direction~($\abs{\theta} < (kR)^{-1}$).

\subsection{Background-Induced Force}\label{sec:born_bif}

In this subsection, we derive the general expression for the background-induced force under the first-order Born approximation. Combining \Eq{born_approx_appx} and \Eq{V_bkg_timeave_appx_2}, we have
\bea
\label{eq:Vbg_finitesize_full_appx}
V_\bg(\veck) & = - \frac{\rho_\phi}{m_0^2} \, \frac{(\mMtest^2 \mathcal{V}_\testmass) \, (\mMsource^2 \mathcal{V}_\sourcemass)}{4\pi r} \\
& \quad \times \frac{r}{\mathcal{V}_\testmass \mathcal{V}_\sourcemass}\int_{\mathcal{V}_\testmass} d^3 \delta \vecr_\testmass \int_{\mathcal{V}_\sourcemass} d^3 \delta \vecr_\sourcemass \frac{\cos(\abs{\veck}\abs{\vecr + \delta \vecr_\testmass - \delta \vecr_\sourcemass}-\veck \cdot (\vecr + \delta \vecr_\testmass - \delta \vecr_\sourcemass))}{\abs{\vecr + \delta \vecr_\testmass - \delta \vecr_\sourcemass}},
\eea
which is the full form of the background-induced potential in the perturbative region when both the test and source masses have a finite geometric shape. Here, $\vecr$ is the position vector pointing from $O_\sourcemass$, the reference point on the source mass, to $O_\testmass$, the reference point on the test mass. $\delta \vecr_\sourcemass$ denotes the position vector pointing from the $O_\sourcemass$ to a point within the source mass, and $\delta \vecr_\testmass$ is the position vector pointing from $O_\testmass$ to a point within the test mass. Thus, $\vecr + \delta \vecr_\testmass - \delta \vecr_\sourcemass$ gives the relative position vector from an infinitesimal volume element in the source to one in the test mass. The specific choices of $O_\sourcemass$ and $O_\testmass$ are flexible as long as they are fixed points attached to the respective experimental objects. For spherically symmetric objects, it is conventional and convenient to place these reference points at the centers of the spheres to simplify the computation. 

In \Eq{Vbg_finitesize_full_appx}, we prove strictly from quantum mechanics that in the perturbative limit, the total background-induced potential is the linear combination of the point-to-point potential between different volume elements. For this equation to hold, both the test and source masses must satisfy the perturbativity conditions, which are
\bea
\testmass \in \text{Region A}_\testmass \cup \text{Region B}_\testmass , \quad \sourcemass \in \text{Region A}_\sourcemass \cup \text{Region B}_\sourcemass. 
\eea
Given \Eq{Vbg_finitesize_full_appx}, we find an interesting phenomenon: When either the source mass or the test mass has a size larger than the de Broglie wavelength~($kR_\testmass > 1$ or $kR_\sourcemass > 1$), the cosine term in \Eq{Vbg_finitesize_full_appx} undergoes rapid oscillations during spatial integration. Such a spatial oscillation leads to the strong cancellation of the background-induced potential when the phase space configuration is monochromatic characterized by momentum $\veck$. When the phase space also has a finite width, things are more complicated. More detailed discussions are provided in \Appx{phase_space_finite_size} and in \Subsec{perturbative_bg}.

\section{Perturbative Decoherence Effects}\label{appx:phase_space_finite_size}

In this appendix, we discuss the decoherence effects coming from the finite phase space and the finite size in the perturbative region, where the Born approximation introduced in \Appx{born_approx_appendix} is valid. When the phase space distribution has a finite width, such as the boosted Maxwell-Boltzmann distribution listed in \Eq{f_dm}, it is necessary to average \Eq{Vbg_finitesize_full_appx} over the phase space. After performing this averaging, we have
\bea
\label{eq:form_factor_pert_full_appx}
\mathcal{F} = \frac{r}{\mathcal{V}_\testmass \mathcal{V}_\sourcemass}  \int_{\mathcal{V}_\testmass} d^3 \delta \vecr_\testmass \int_{\mathcal{V}_\sourcemass} d^3 \delta \vecr_\sourcemass \int \dbar^3 \veck \frac{f_\phi(\veck)}{n_\phi}  \frac{\cos(\abs{\veck}\abs{\vecr + \delta \vecr_\testmass - \delta \vecr_\sourcemass}-\veck \cdot (\vecr + \delta \vecr_\testmass - \delta \vecr_\sourcemass))}{\abs{\vecr + \delta \vecr_\testmass - \delta \vecr_\sourcemass}}.
\eea
To evaluate \Eq{form_factor_pert_full_appx}, one needs to perform a multidimensional integration over the source, test mass, and phase space. In this appendix, we consider several simplified scenarios where exact or approximate analytical results can be obtained. First, we discuss the background-induced force between two point-like masses using the phase space distribution in \Eq{f_dm}. In this case, the form factor reduces to a 2D integral, and we derive exact analytical expressions in the forward and backward directions. Next, we investigate the finite-size effect of a solid spherical source. To simplify the analysis, we assume a monochromatic phase-space distribution with the momentum $\veck$, i.e. $f_\phi(\vecp) = (2\pi)^3 n_\phi \delta^{(3)}(\vecp- \veck)$, which allows for exact analytical expressions in both the forward and backward directions. Finally, we discuss the finite-size effects of both the source and test masses in the far-field limit. The combined analysis incorporating finite-source and finite-phase-space effects in the perturbative region is presented in the main text, as shown in detail in \Subsec{perturbative_bg}.

\subsection{Finite-Phase-Space Effect}\label{appx:finite_phase_space_appx}

We begin with the situation in Region A, where both the source and test masses are point-like. In this region, the form factor is 
\bea
\label{eq:FA_appx}
\formfactor_\text{A}(\vecr)  = \int \dbar^3 \veck \, \frac{f_\phi(\veck)}{n_\phi}\cos(k r-\veck\cdot \vecr).
\eea
Substituting the boosted Maxwell-Boltzmann distribution in \Eq{f_dm} into $\formfactor_\text{A}$, we get
\begin{align}
\label{eq:F_phase_point_1_appx}
\formfactor_\text{A}(\vecr) 
& = \frac{1}{\left(2 \pi \sigmak^2\right)^{3/2}} \int_0^\pi d\theta_\veck \sin\theta_\veck \int_0^{2\pi}d\phi_\veck \int_0^\infty dk \, k^2 \,  e^{-\frac{k^2 + k_0^2}{2\sigmak^2}} e^{\frac{k k_0 \cos\theta_\veck}{\sigmak^2}} \cos\left(kr\left(1-\cos\theta \right)\right),
\end{align}
where the deflection angle $\theta$ is computed in \Eq{cos_theta}. For the above integration, we choose the spherical coordinate system shown in \Fig{frame}. Firstly, we expand the cosine term and integrate out $\phi_\veck$ by utilizing
\bea
\label{eq:F_phase_point_2_appx}
\int_0^{2\pi} \cos\left(\xi \cos\phi_\veck\right) \, d\phi_\veck = 2\pi J_0(\xi)\,, \quad \quad \int_0^{2\pi} \sin\left(\xi \cos\phi_\veck\right) \, d \phi_\veck = 0\,,
\eea
then we have
\begin{align}
\formfactor_\text{A}(\vecr) = \frac{2 \pi}{\left(2 \pi \sigmak^2\right)^{3/2}} \int_0^{\infty} dk \, k^2 e^{-\frac{k^2+k_0^2}{2\sigmak^2}}\int_0^\pi d \theta_\veck \sin \theta_\veck \, e^{\frac{k k_0 \cos\theta_\veck}{\sigmak^2}}\, J_0\left(kr\sin\theta_\veck \sin\theta_\vecr\right) \cos\left[kr\left(1-\cos\theta_\veck\cos\theta_\vecr\right)\right],\label{eq:FPS}
\end{align}
where $J_0$ is the Bessel function of the first kind. The above 2D integral can be computed numerically for arbitrary values of $\theta_\vecr$ and $r$, as shown in \Fig{FA_plt}. In particular, in the forward and backward directions, we have the analytical expressions
\bea
\label{eq:FA_forward_appx}
& \formfactor_\text{A}(\theta_\vecr = 0) = \frac{k_0^2 \left[ 1 + \erf\left(\frac{k_0}{\sqrt{2}\sigmak}\right) \right]}{2 \left( k_0^2 + \sigmak^4 \, r^2 \right)} + \frac{ e^{-2\sigmak^2 r^2} }{2\left(k_0^2 + \sigmak^4 \, r^2\right)} \Re\left[ e^{-2 i k_0 r} (k_0^2 - i k_0 \, \sigmak^2  \, r + 2 \sigmak^4 \, r^2) \,\, \erfc\left(\frac{ k_0 - 2 i \sigmak^2 \, r}{\sqrt{2} \sigmak}\right) \right]
\eea
and
\bea
\label{eq:FA_backward_appx}
\formfactor_\text{A}(\theta_\vecr = \pi) & = \frac{k_0^2 \, \erfc\left(\frac{k_0}{\sqrt{2}\sigmak}\right)}{2 (k_0^2 + \sigmak^4 \, r^2)} + \frac{e^{-2\sigmak^2 r^2}}{ 2(k_0^2 + \sigmak^4 \, r^2)} \Re\left[ e^{-2i k_0 r} (k_0^2 - i k_0 \, \sigmak^2 \, r + 2 \sigmak^4\, r^2) \erf\left(\frac{k_0 - 2i \sigmak^2 \, r}{\sqrt{2}\sigmak}\right)\right]\\
&  \quad + \frac{ e^{-2\sigmak^2 r^2} }{2(k_0^2 + \sigmak^4 \, r^2)} \left[ (k_0^2 + 2 \sigmak^4 \, r^2) \cos(2k_0 r) - k_0 \sigmak^2 \, r \, \sin(2k_0 r)\right].
\eea
Here $\erf(x)\equiv \frac{2}{\sqrt{\pi}}\int_0^x e^{-t^2} {\rm d}t$ is the error function and $\erfc(x)\equiv 1-\erf(x)$ is the complementary error function. In \Fig{FA_plt}, we present our numerical results, including the form factors in the forward and backward directions, computed using \Eq{FA_forward_appx} and \Eq{FA_backward_appx}. Additionally, we compute the form factor in an arbitrary direction by performing a 3D numerical integration using \Eq{F_phase_point_1_appx}. As an example, we choose $\theta_\vecr = \pi/2$, where the mean dark matter momentum $\veck_0$ is perpendicular to the position vector $\vecr$. The form factor in this direction is shown in gray.

In the short-distance limit, the forward and backward form factors listed in \Eq{FA_forward_appx} and \Eq{FA_backward_appx} can be approximated by
\bea
\label{eq:FA_forward_small_r}
k_0 r < 1: \quad \formfactor_\text{A}(\theta_\vecr = 0) \simeq 1 - \left[2 + \frac{k_0^2}{\sigmak^2} - \sqrt{\frac{2}{\pi}} e^{-\frac{k_0^2}{2 \sigmak^2}}\left( \frac{k_0}{\sigmak} + \frac{\sigmak}{k_0}\right) - \left(\frac{k_0^2}{\sigmak^2}  - \frac{\sigmak^2}{k_0^2} + 2 \right) \text{erf}\left(\frac{k_0}{\sqrt{2} \sigmak}\right) \right] \sigmak^2 \, r^2
\eea
and
\bea
\label{eq:FA_backward_small_r}
k_0 r < 1: \quad \formfactor_\text{A}(\theta_\vecr = \pi)\simeq 1 - \left[ 2 + \frac{k_0^2}{\sigmak^2}  + \sqrt{\frac{2}{\pi}} e^{-\frac{k_0^2}{2 \sigmak^2}}\left( \frac{k_0}{\sigmak} + \frac{\sigmak}{k_0}\right) + \left(\frac{k_0^2}{\sigmak^2}  - \frac{\sigmak^2}{k_0^2} + 2 \right) \text{erf}\left(\frac{k_0}{\sqrt{2} \sigmak}\right) \right] \sigmak^2 r^2. 
\eea
From \Eq{FA_forward_small_r} and \Eq{FA_backward_small_r}, we find that both form factors tend to one when the distance gets smaller. The reason is that the $\cos(kr-\veck \cdot \vecr)$ term in \Eq{F_phase_point_1_appx} is non-oscillatory when $r$ is small. In the large-distance limit, we can approximate \Eq{FA_forward_appx} and \Eq{FA_backward_appx} with
\bea
\label{eq:FA_forward_large_r}
k_0 r > 1: \quad \formfactor_\text{A}(\theta_\vecr = 0) \simeq \frac{k_0^2}{2k_0^2 + 2 \sigmak^4 r^2} \left[ 1 + \text{erf}\left(\frac{k_0}{\sqrt{2} \sigmak}\right) + \sqrt{\frac{2}{\pi}}\frac{\sigmak}{k_0}e^{-\frac{k_0^2}{2 \sigmak^2}} \right]
\eea
and
\bea
\label{eq:FA_backward_large_r}
k_0 r > 1: \quad \formfactor_\text{A}(\theta_\vecr = \pi) \simeq \frac{k_0^2}{2k_0^2 + 2 \sigmak^4 r^2} \left[ 1 - \text{erf}\left(\frac{k_0}{\sqrt{2} \sigmak}\right) - \sqrt{\frac{2}{\pi}}\frac{\sigmak}{k_0}e^{-\frac{k_0^2}{2 \sigmak^2}} \right].
\eea
We find that both form factors experience suppression due to the finite-phase-space effect when $k_0 r > 1$. This occurs because the term $\cos(kr-\veck \cdot \vecr)$ becomes highly oscillatory in this region. Moreover, from \Eq{FA_forward_large_r} and \Eq{FA_backward_large_r} we can find that the backward direction is suppressed more than the forward direction. This suppression arises because of two reasons. First, the boosted Maxwell-Boltzmann distribution gives more weight in the phase-space integration around $\veck \sim \veck_0$. Second, the term $\cos(kr - \veck \cdot \vecr)$ oscillates more rapidly when $\veck$ is anti-parallel to $\vecr$. Therefore, when $\vecr$ is anti-parallel to $\veck_0$, which means the test mass is located in the backward direction, the cancellation during the phase space integration becomes the strongest.

\subsection{Monochromatic Finite-Size Effect}\label{appx:finite_size_appx}

We now compute the finite-size effect of the experimental objects, assuming a monochromatic phase space distribution with momentum $\veck$.  For simplicity, we consider a source as a uniform sphere of radius  $R$ and begin with the case where the test mass is point-like. Under these conditions, the form factor in \Eq{form_factor_pert_full_appx} reduces to
\begin{align}
\formfactor_\sourcemass(\veck) \equiv \frac{r}{\mathcal{V}_\sourcemass}\int_\sourcemass
d^3\delta\vecr_\sourcemass \,\frac{\cos(\abs{\veck} \abs{\vecr-\delta\vecr_\sourcemass}-\veck\cdot\left(\vecr-\delta\vecr_\sourcemass\right))}{\left|\vecr - \delta\vecr_\sourcemass\right|}\;,
\end{align}
where $\mathcal{V}_\sourcemass=4\pi R^3_\sourcemass/3$ is the volume of the source, and the integral is performed over the that. For the forward and backward directions, we find the analytical expression of the forward and backward form factors, which are
\bea
\text{Forward:} \quad \formfactor_\sourcemass\left(\theta_\vecr = 0;\veck\right) &= 3\sqrt{\pi}\left(\frac{r}{kR^2_\sourcemass}\right)^{\frac{3}{2}}\left[\CFresnel \left(R_\sourcemass \sqrt{\frac{k}{\pi r}}\right)\sin\left(\frac{k R^2_\sourcemass}{2 r}\right)-\SFresnel \left(R_\sourcemass\sqrt{\frac{k}{\pi r}}\right)\cos\left(\frac{k R^2_\sourcemass}{2 r}\right)\right],
\eea
and
\bea
\text{Backward:} \quad \formfactor_\sourcemass\left(\theta_\vecr = \pi;\veck\right) 
&= \frac{3\sqrt{\pi}}{2}\left(\frac{r}{k R^2_\sourcemass}\right)^{\frac{3}{2}}\bigg[\frac{\sin\left(2 k r\right)\sin\left(2kR_\sourcemass\right)}{\sqrt{\pi k r}}\\
& \quad \quad +\left(\CFresnel\left(\xi_+\right)-\CFresnel\left(\xi_-\right)\right)\sin\left(\frac{kR^2_\sourcemass}{2r}\right) -\left(\SFresnel\left(\xi_+\right)-\SFresnel\left(\xi_-\right)\right)\cos\left(\frac{kR^2_\sourcemass}{2r}\right)\bigg].
\eea
In the above formulas, $\xi_{\pm}\equiv \left(2r \pm R_\sourcemass \right)\sqrt{k/(\pi r)}$.
$\CFresnel$ and $\SFresnel$ are Fresnel integrals defined as
\begin{align}
\CFresnel(z)\equiv \int_0^z \cos\left(\frac{\pi t^2}{2}\right) dt\;,\qquad    
\SFresnel(z)\equiv \int_0^z \sin\left(\frac{\pi t^2}{2}\right) dt\;.    
\end{align}
In the far-field region~($r > k R_\sourcemass^2$), they are reduced to
\bea
\label{eq:forward_formfactor_sphere_mono_appx}
\text{Forward:} \quad \formfactor_\sourcemass\left(\theta_\vecr = 0;\veck\right) & \simeq 1 \quad \quad \text{(Far-Field)},
\eea
and
\bea
\label{eq:backward_formfactor_sphere_mono_appx}
\text{Backward:} \quad \formfactor_\sourcemass\left(\theta_\vecr = \pi;\veck\right) & \simeq \frac{3(\sin\left(2kR_\sourcemass\right)-2kR_\sourcemass \cos\left(2kR_\sourcemass\right))}{(2kR_\sourcemass)^3} \times \cos\left(2kr\right) \quad \quad \text{(Far-Field)}.
\eea
The expansions of the form factors listed in \Eq{forward_formfactor_sphere_mono_appx} and \Eq{backward_formfactor_sphere_mono_appx} agree with the results derived from the Born approximation after the far-field expansion. 

Now, let us consider the case where both the source and test masses have finite sizes. Even though a complete investigation needs numerical evaluation, as shown in \Eq{form_factor_pert_full_appx}, we can obtain the analytical expression in the far field region, where $r > k R_\sourcemass^2$ and $r > k R_\testmass^2$. We express the cosine term in \Eq{Vbg_finitesize_full_appx} as the real part of an exponential function, apply a similar technique used to derive \Eq{psi_sc_born_far_appx}, and obtain
\bea
\label{eq:Vbg_far_appx}
V_\bg(\veck) \simeq \frac{\rho_\phi}{m_0^2} \, \mMtest^2 \mathcal{V}_\testmass  \, \abs{f_{\sourcemass \testmass}} \frac{\cos(kr - \veck \cdot \vecr +  \arg f_{\sourcemass \testmass})}{r} \quad \quad \text{(Far-Field)}.
\eea
As we have known in the previous text, $\vecq = \veck'-\veck$ is the momentum transfer vector, and it obeys $q = 2k \sin(\theta/2)$. $f_{\sourcemass \testmass}$ represents the effective scattering amplitude, accounting for the finite-size effect of both the test and source masses. It is given by
\bea
f_{\sourcemass \testmass} = - \frac{\mMsource^2 \mathcal{V}_\sourcemass}{4 \pi}  \times \underbrace{\frac{1}{\mathcal{V}_\testmass} \int_{\mathcal{V}_\testmass} d^3 \delta \vecr_\testmass \, e^{+i \vecq \cdot \delta \vecr_\testmass}}_{\text{Finite Test Mass}} \times 
\underbrace{\frac{1}{\mathcal{V}_\sourcemass}\int_{\mathcal{V}_\sourcemass} d^3 \delta \vecr_\sourcemass \, e^{-i \vecq \cdot \delta \vecr_\sourcemass}}_{\text{Finite Source Mass}}.
\eea
When the size of the test mass is negligible~($k R_\testmass < 1$), $f_{\sourcemass \testmass}$ in the above formula goes back to the standard quantum mechanical scattering amplitude. Therefore, we have the generic expression of the form factor in the far-field region, which is
\bea
\formfactor_{\sourcemass \testmass}(\veck) = \underbrace{\frac{1}{\mathcal{V}_\testmass} \int_{\mathcal{V}_\testmass} d^3 \delta \vecr_\testmass \, e^{+i \vecq \cdot \delta \vecr_\testmass} }_{\text{Finite Test Mass}}
\times \underbrace{\frac{1}{\mathcal{V}_\sourcemass} \int_{\mathcal{V}_\sourcemass} d^3 \delta \vecr_\sourcemass \, e^{-i \vecq \cdot \delta \vecr_\sourcemass}}_{\text{Finite Source Mass}} \times \cos(kr - \veck \cdot \vecr +  \arg f_{\sourcemass \testmass})  \quad \quad \text{(Far-Field)}.
\eea
From the above formula, we find that the finite-size effects of the source and the test masses can be factorized in the far-field limit. When the source and test masses are both spheres, we have
\bea
\formfactor_{\sourcemass \testmass}(\veck) = \underbrace{\frac{ 3 (\sin(q R_\testmass) - q R_\testmass \cos(q R_\testmass))  }{(qR_\testmass)^3}  }_{\text{Finite Test Mass}}
\times \underbrace{\frac{ 3 (\sin(q R_\sourcemass) - q R_\sourcemass \cos(q R_\sourcemass))  }{(qR_\sourcemass)^3} }_{\text{Finite Source Mass}} \times \cos(kr - \veck \cdot \vecr)  \quad \quad \text{(Far-Field)}.
\eea
From the above formula, we find that given the conditions that $k R_\sourcemass > 1$ and $k R_\testmass > 1$ and $\veck$ and $\vecr$ are not parallel, there is
\bea
\label{eq:FST_nonparallel_appx}
\mathcal{F}_{\sourcemass \testmass}(\veck) \simeq \underbrace{\frac{3\cos(qR_\testmass)}{(q R_\testmass)^2}}_{\text{Finite Test Mass}} \times \underbrace{\frac{3\cos(qR_\sourcemass)}{(q R_\sourcemass)^2}}_{\text{Finite Source Mass}} \times \cos(kr - \veck \cdot \vecr)  \quad \quad\quad \text{(Far-Field $+$ $\veck$ and $\vecr$ Not Parallel)}.
\eea
Therefore, the form factor has $(k R_\testmass)^{-2} \times (kR_\sourcemass)^{-2}$ suppression because of the finite size effect if the phase space distribution is monochromatic and $\veck$ is not parallel to $\vecr$. When $\veck$ and $\vecr$ are parallel~($\abs{\theta} < \min\left[(kR_\testmass)^{-1},\,(kR_\sourcemass)^{-1}\right]$),  the finite-size suppression in the far-field limit vanishes. More specifically, we have
\bea
\label{eq:FST_parallel_appx}
\formfactor_{\sourcemass \testmass}(\veck) \simeq \cos(kr - \veck \cdot \vecr) \quad \quad\quad \text{(Far-Field $+$ $\veck$ and $\vecr$ Parallel)}.
\eea

We stress that the above discussions are all based upon the assumption that the phase space is monochromatic, i.e., $f_\phi(\vecp) = (2\pi)^3 n_\phi \delta^{(3)}(\vecp- \veck)$. When considering the finite-size and finite-phase-space effects jointly, the suppressions from these two effects are not simply multiplicative. Specifically, when performing phase-space integration for point-to-point interactions, the oscillatory features of the background-induced potential are already significantly smoothed out due to the finite width of the phase space. Further integrating over the finite volumes of the source and test masses does not lead to significant additional cancellation. In fact, given the boosted Maxwell-Boltzmann distribution shown in \Eq{f_dm}, we have numerically verified in \Subsec{perturbative_bg} that in the perturbative region satisfying $\abs{r-R}/R \gtrsim \mathcal{O}(1)$, accounting for only the finite-phase-space suppression is sufficient to obtain a correct result with deviations of order $\mathcal{O}(10\%)$.

In the perturbative region, the finite-size effect becomes important only when $\abs{r-R}/R < 1$. It is worth noting that the analysis in this section cannot be directly applied to this case because the far-field condition is violated. The numerical approach for studying the $\abs{r-R}/R < 1$ case can be found in \Subsec{perturbative_bg}. In the non-perturbative region where the screening effect becomes important, the finite-size effect is relevant even when $\abs{r-R}/R \sim \mathcal{O}(1)$. However, the analysis based on the Born approximation is also invalid in that regime, as the perturbative condition breaks down when the screening effect becomes important. The detailed exploration of the non-perturbative region based on the partial wave analysis is shown in \Subsec{bg_nonp_highk}.

\section{Partial Wave Analysis}\label{appx:partial_wave_appx}

In this section, we introduce the partial wave analysis used in this paper. Unlike the Born approximation introduced in  \Appx{born_approx_appendix}, which is applicable only in the perturbative regions (Regions A and B), the partial wave analysis allows us to investigate both perturbative and non-perturbative regions, as outlined in \Fig{phase_classify}. In this section, we focus on the uniform solid sphere. The structure of this section is as follows. First, we introduce the decomposition of wave functions, the phase shift of the solid sphere, and the far-field limit of the wave function. Next, we introduce the computations of the cross section and the momentum-transfer cross section through the summation of the components of different partial waves. Finally, we compute the field configuration in the low-momentum limit. We reproduce the field configuration in \Subsec{bg_nonp_lowk}, which is derived from the ansatz assuming a spherical symmetry.

\subsection{Wave Functions}\label{appx:partial_wave_func_appx}

For the solid sphere with radius $R$, we have the incident wave function
\bea
\label{eq:inc_wave_appx}
\psiinc(\vecr;\veck) = \abs{\psi_0} e^{i\veck \cdot \vecr} = \abs{\psi_0} \,\sum^{\infty}_{l=0} (2l+1)\, i^l \, j_l(k r) \, P_l(\cos\theta),
\eea
and the scattered wave function
\bea
\label{eq:sca_wave_appx}
\psisc(\vecr;\veck) = \abs{\psi_0} \,\sum^{\infty}_{l=0} (2l+1)\, i^l \, A_l \, h_l(k r) \, P_l(\cos\theta),
\eea
where $\theta$ is the reflection angle as shown in \Fig{qm_sca} and \Fig{frame}. $h_l =  j_l+ i y_l$ are the spherical Hankel functions of the first kind, where $j_l$ are the spherical Bessel functions, and $y_l$ are the spherical Neumann functions. $A_l$ is the amplitude of the scattered wave, which can also be represented as
\bea
\label{eq:Al_Sl_deltal_appx}
A_l =  \frac{S_l - 1}{2} = \frac{e^{2i\delta_l}-1}{2},
\eea
where $S_l$ is the S-matrix for the $l$-th component of the partial wave, and $\delta_l$ is the corresponding phase shift of the $l$-th partial wave. To understand the physical meaning of the phase shift, we expand $j_l$, $y_l$, and $h_l$ in the far-field region as
\bea
\label{eq:sph_bessel_rinf}
j_l(kr) \stackrel{r\rightarrow\infty}{\longrightarrow}  \frac{\sin{(kr-\frac{l\pi}{2})}}{kr}, \quad \quad h_l(kr) \stackrel{r\rightarrow\infty}{\longrightarrow} \frac{\exp\left(i(kr-\frac{l\pi}{2})\right)}{ikr},
\eea
and have
\bea
\label{eq:psi_out_appx}
\psiout(\vecr;\veck) 
& = \psiinc(\vecr;\veck) + \psisc(\vecr;\veck) \\
& \stackrel{r\rightarrow\infty}{\longrightarrow} \abs{\psi_0} \sum_{l=0}^{\infty} (2l+1)\, i^l \, e^{i \delta_l} \, \frac{\sin\left(k r-\frac{l\pi}{2} + \delta_l \right)}{k r} \, P_l(\cos\theta). 
\eea
For the repulsive potential discussed in this paper, the phase shift obeys $\delta_l <0$, which means that the radial wave function is pushed outwards. In contrast, for the attractive potential, the phase shift obeys $\delta_l >0$, which means that the radial wave function is pulled inwards. We can also expand the scattered wave function \Eq{sca_wave_appx} in the far-field region and have
\bea
\label{eq:ftheta_partial}
f(\vecr;\veck) = \frac{1}{ik} \sum_{l=0}^\infty  (2l+1) \, A_l \, P_l(\cos\theta) = \frac{1}{k} \sum_{l=0}^\infty (2l+1) \, e^{i \delta_l} \, \sin\delta_l \, P_l(\cos\theta),
\eea
which is the scattering amplitude shown in \Eq{psi_in_sca}. 

\subsection{Phase Shifts of Solid Sphere}\label{appx:phase_shifts_appx}

From \Eq{sca_wave_appx} and \Eq{Al_Sl_deltal_appx}, we know that as long as the phase shifts are determined, the scattered wave function in \Eq{sca_wave_appx} is determined. Generally speaking, the phase shifts are determined by the potential shape and need to be solved numerically, as shown in Refs.~\cite{Buckley:2009in,Tulin:2013teo,Xu:2020qjk,Xu:2021lmg}. However, for the solid sphere potential~(as shown in \Eq{Veff_Eeff}) we are focusing on in this paper, we can determine the phase shift by matching the boundary condition between the wave function inside the sphere and the wave function outside the sphere. For a solid sphere with radius $R$ and e, the total wave function in the interior of the sphere is
\bea
\label{eq:inc_wave}
\psiint(\vecr;k) = \abs{\psi_0} \sum_{l=0}^{\infty} (2l+1) \, i^l \, P_{l}(\cos\theta) B_l \, j_l(k' r),
\eea
where $k' = \sqrt{k^2 - \mM^2}$ is the momentum inside the solid sphere. When $\mM > k$, which means the barrier is larger than the incident kinetic energy, we do the Wick rotation $k' = i \kappa'$, where $\kappa' = \sqrt{\mM^2 - k^2}$. $B_l$ is the amplitude of the interior wave function. By matching the boundary conditions at $r=R$, we find $A_l$ and $B_l$ satisfy
\bea
\label{eq:Al_Bl_appx}
A_l = - \frac{k j_l(k' R) j_{l+1}(k R) - k' j_l(k R) j_{l+1}(k' R) }{ k j_l(k' R) h_{l+1}(k R) - k' h_l(k R) j_{l+1}(k' R) }, \quad \quad B_l = \frac{k j_l(k R) h_{l+1}(k R) - k h_l(k R) j_{l+1}(k R) }{ k j_l(k' R) h_{l+1}(k R) - k' h_l(k R) j_{l+1}(k' R) }.
\eea
Therefore, the phase shift satisfies
\bea
\label{eq:tan_delta_l_solid_sphere_appx}
\tan \delta_l = \frac{A_l/i}{A_l+1} = \frac{k j_l(k' R) j_{l+1}(k R) - k' j_l(k R) j_{l+1}(k' R)}{ k j_l(k' R) y_{l+1}(k R) - k' y_l(k R) j_{l+1}(k' R) }. 
\eea
In the strongly coupled region that satisfies $\mM \gg k$, the potential can be treated as the hard sphere. In this situation, we use
\bea
\kappa' R \gg 1: \quad j_l(k' R) \simeq i^l \frac{e^{\kappa' R}}{2 \kappa' R}, 
\eea
and have
\bea
\label{eq:tan_delta_l_hard_sphere_appx}
\kappa' R \gg 1: \quad \tan \delta_l =  \frac{k j_{l+1}(k R) - i k' j_l(k R)}{k y_{l+1}(k R) - i k' y_l(k R)}\simeq \frac{j_l(k R)}{y_l(k R)} \quad \quad \quad \text{(Hard Sphere)}.
\eea
For the low-momentum hard-sphere scattering satisfying $k R < 1$, which is related to Region C as shown in \Fig{phase_classify}, we have the phase shift for the s-wave component, which is
\bea
\label{eq:phase_shift_swave}
k R < 1: \quad \delta_0 \simeq \tan\delta_0 \simeq - k R  \quad \quad \quad \text{(s-Wave)}.
\eea
For high-momentum scattering satisfying $kR \gtrsim 1$, corresponding to Region D or E, we sum the partial wave components up to $l_{\max} \sim kR$ to ensure sufficient resolution for capturing the excited higher-moment components of the partial waves. Particularly, throughout our numerical computation, we adjust the value of $l_{\max}$ to balance precision, computational speed, and numerical stability. Specifically, we set $l_{\max} = 0$ when $kR \leq 10^{-3}$, $l_{\max} = \min([2kR]+3,[kR]+300)$ when $kR > 10^{-3}$. Although a larger $l_{\max}$ provides higher resolution in principle, we avoid selecting an excessively large $l_{\max}$ to prevent numerical singularities. To ensure the reliability of our results, we varied the number of partial wave summation terms and confirmed the stability of our numerical outcomes following the above method for choosing $l_{\max}$ for the region satisfying $k R \lesssim 10^4$.

\subsection{Cross Sections}\label{appx:cross_sec_derivation_appx}

In this subsection, we list the formulas for computing the cross section $\sigma$ and the momentum-transfer cross-section $\sigmat$ for the elastic scattering, where the unitarity condition $\abs{S_l} = 1$ is satisfied, and the probability current is conserved. To calculate the cross section $\sigma$, we substitute \Eq{ftheta_partial} into \Eq{sigma}. Then, we utilize the orthogonality relation of the Legendre polynomials, which is 
\bea
\label{eq:Legendre_orthogonal_appx}
\int_{-1}^{+1} P_l(\cos{\theta}) P_{l'}(\cos{\theta}) \, d\cos{\theta}  = \frac{2}{2l+1}\delta_{ll'}.
\eea
Thus, we have
\bea
\label{eq:sigma_asym_appx}
\sigma = \frac{4\pi}{k^2} \sum_{l=0}^{\infty} (2l+1) \abs{A_l}^2 = \frac{4 \pi}{k^2} \sum_{l=0}^{\infty} (2l+1) \sin^2 \delta_l. 
\eea
Equivalently, we can write $\sigma$ in a more symmetric formalism as
\bea
\label{eq:sigma_sym_appx}
\sigma = \frac{4\pi}{k^2} \sum_{l=0}^{\infty} (l+1) (\abs{A_l}^2 + \abs{A_{l+1}}^2) =  \frac{4\pi}{k^2} \sum_{l=0}^{\infty} (l+1) (\sin^2\delta_l + \sin^2\delta_{l+1}). 
\eea
To derive \Eq{sigma_sym_appx} from \Eq{sigma_asym_appx}, we use the relation $\sum_{l=0}^\infty l \abs{A_l}^2 = \sum_{l=0}^\infty (l+1) \abs{A_{l+1}}^2$. 

To compute the momentum-transfer cross section, we substitute \Eq{ftheta_partial} into \Eq{sigma_eff}. Because there is a $\cos\theta$ term inside the integral in \Eq{sigma_eff}, we utilize the equation
\bea
\label{eq:Legendre_orthogonal_generic_appx}
\int_{-1}^{+1} P_l(\cos{\theta}) P_{l'}(\cos{\theta}) \cos{\theta} \, d \cos{\theta} = \frac{2 \max(l,l')}{(l+l') (l+l'+2) } \delta_{\abs{l-l'},1},
\eea
which is derived from \Eq{Legendre_orthogonal_appx} by using the Bonnet’s recursion formula
\bea
(2l+1)\,\cos\theta\,P_l(\cos\theta) = (l+1) \, P_{l+1}(\cos\theta) + l \, P_{l-1}(\cos\theta). 
\eea
Then we have
\bea
\label{eq:sigma_eff_sym_appx}
\sigmat = \frac{4\pi}{k^2} \sum_{l=0}^\infty [(2l+1) \abs{A_l}^2 - (l+1) A_l^* A_{l+1} - (l+1) A_l  A_{l+1}^*]. 
\eea
Similarly, we write the momentum-transfer cross section in a more symmetric form as
\bea
\label{eq:sigma_eff_A2_appx}
\sigmat = \frac{4 \pi}{k^2} \sum_{l=0}^\infty (l+1)(\abs{A_l}^2 + \abs{A_{l+1}}^2 -  A_l^* A_{l+1} - A_l  A_{l+1}^*).
\eea
In \Eq{sigma_eff_A2_appx}, the first two terms are exactly the same as the terms in \Eq{sigma_sym_appx}, which is the symmetric form of the scattering cross section. These terms come from the integral $\int d \Omega \abs{f(\theta;k)}^2$ in \Eq{sigma_eff}. The last two terms are conjugate mixing between $A_{l}$ and $A_{l+1}$. These terms come from the integral $-\int d \Omega \abs{f(\theta;k)}^2 \cos\theta$ in \Eq{sigma_eff}. 

Equivalently, we can write the momentum-transfer cross section as
\bea
\label{eq:sigma_eff_S_appx}
\sigmat = \frac{\pi}{k^2} \sum_{l=0}^\infty (l+1) ( 2 - S_l^* S_{l+1} - S^*_{l+1} S_l ).
\eea
To derive \Eq{sigma_eff_S_appx} from \Eq{sigma_eff_A2_appx}, we use $\abs{S_l}^2=1$, the unitarity condition for the elastic scattering. Using \Eq{Al_Sl_deltal_appx}, we can further simplify \Eq{sigma_eff_S_appx} to be
\bea
\label{eq:sigma_eff_delta_appx}
\sigmat =  \frac{\pi}{k^2} \sum_{l=0}^\infty (l+1) \sin^2(\delta_{l+1} - \delta_{l}). 
\eea

In the end, we summarize the mathematical formalisms of the scattering cross section and momentum-transfer cross section presented above for elastic scattering because they appear in various literature. \Eq{sigma_sym_appx} and \Eq{sigma_eff_sym_appx} are the scattering cross section and momentum-transfer cross section used in this work. \Eq{sigma_asym_appx} is the scattering cross section appearing in standard literature~\cite{Landau:1991wop,Sakurai:2011zz}. \Eq{sigma_eff_S_appx} is the momentum-transfer cross section appearing in Ref.~\cite{Day:2023mkb}. \Eq{sigma_eff_delta_appx} is the momentum-transfer cross section appearing in Ref.~\cite{bransden2006physics}.

\subsection{s-Wave Approximation}\label{appx:swave_approx_appx}

In this section, we reproduce the scalar profile of the spherically symmetric solution as discussed in \Subsec{bg_nonp_lowk} by taking the low-momentum limit of the scattered wave. We show that the spherically symmetric ansatz discussed in Refs.~\cite{Olive:2007aj,dePireySaintAlby:2017lwc,Berezhiani:2018oxf,Hees:2018fpg,Banerjee:2022sqg} comes from the s-wave approximation, which motivates us to further include the higher moment into the analysis as we have discussed in \Subsec{bg_nonp_highk}. Expanding the amplitude of the scattered wave and interior wave shown in \Eq{Al_Bl_appx} in the region where $kR  < 1$, we obtain the amplitude of the s-wave component of the scattered wave as
\bea
\label{eq:A0_lowk0_appx}
A_0 = -i \frac{\mM R - \tanh(m_\text{M} R)}{m_\text{M} R} \, kR - \left(\frac{\mM R - \tanh(\mM R)}{\mM R}\right)^2 \, (k R)^2+ \cdots,
\eea
and the amplitude of the s-wave component of the interior wave as
\bea
\label{eq:B0_lowk0_appx}
B_0 = \frac{1}{\cosh(m_\text{M} R)} + \frac{-i}{\cosh(m_\text{M} R)}\left(\frac{\mM R - \tanh(\mM R)}{\mM R}\right) (k R) + \cdots. 
\eea
In \Eq{A0_lowk0_appx} and \Eq{B0_lowk0_appx}, ``$\cdots$'' means the higher order terms in terms of $kR$.

The spherical Hankel and Bessel functions for the s-wave partial wave~($l=0$) are given by
\bea
h_0(k r) = - \frac{i}{k r} e^{ikr}, \quad \quad \quad \quad j_0(k' r) \simeq \frac{\sinh(\mM r)}{\mM r}.
\eea
Keeping only the leading-order term in $kR$ from \Eq{A0_lowk0_appx}, the scattered wave is written as
\bea
\label{eq:psisc_swave_appx}
\psisc(r) = \abs{\psi_0} A_0 \, h_0(kr) \simeq - \abs{\psi_0} \,\, \frac{\mM R - \tanh(\mM R)}{\mM} \frac{e^{ikr}}{r}.
\eea
The total wave outside the sphere is
\bea
\label{eq:psitot_swave_appx}
\psiout(r) =  \psiinc(r) + \psisc(r) \simeq \abs{\psi_0} e^{i \veck \cdot \vecr} - \abs{\psi_0}  \frac{\mM R - \tanh(\mM R)}{\mM} \frac{e^{ikr}}{r}. 
\eea
Keeping only the leading-order term in $kR$ from \Eq{B0_lowk0_appx}, the wave inside the sphere is given by
\bea
\psiint(r) \simeq \abs{\psi_0} \, \frac{\sinh(\mM r)}{\mM r \cosh(\mM R)}. 
\eea

In the near-field region where $r/R \sim \mathcal{O}(1)$, the $e^{i \veck \cdot \vecr}$ and $e^{ikr}$ terms in \Eq{psitot_swave_appx} are replaced by $1$ because $kr \sim kR \lesssim 1$, and \Eq{psitot_swave_appx} comes back to the spherically symmetric solution listed in \Eq{psi_bound}, which is explored in Refs.~\cite{Olive:2007aj,dePireySaintAlby:2017lwc,Berezhiani:2018oxf,Hees:2018fpg,Banerjee:2022sqg}. In the perturbative Region A where $\mM < R^{-1}$, \Eq{psitot_swave_appx} gives the background-induced potential as shown in \Eq{Vbg_A} by substituting \Eq{psitot_swave_appx} into \Eq{Vbg}, whose analytical formalism is given by Refs.~\cite{Horowitz:1993kw,Ferrer:1998ju,Ferrer:1999ad,Ferrer:2000hm,Ghosh:2022nzo,Blas:2022ovz,Ghosh:2024qai,VanTilburg:2024xib,Barbosa:2024pkl}.

In the long-wavelength region which is strongly coupled~($\mM > R^{-1}$, $kR \ll 1$), we substitute \Eq{psitot_swave_appx} into \Eq{Vbg}, and have the monochromatic background-induced potential given momentum $\veck$, which is
\bea
\label{eq:Vbg_swave_appx}
V_\bg(\veck) \simeq - \frac{\rho_\phi}{m_0^2} \frac{ (\mMsource^2 \mathcal{V}_\sourcemass) (\mMtest^2 \mathcal{V}_\testmass) }{4\pi r} \times \frac{3}{(\mM R)^2} \times \cos(kr -\veck \cdot \vecr).
\eea
Then we do the phase-space average over \Eq{Vbg_swave_appx} to take the decoherence effect into account and have
\bea
\label{eq:Vbg_formfactor_screen_appx}
\langle V_\bg\rangle_\veck \simeq - \frac{\rho_\phi}{m_0^2} \frac{ (\mMsource^2 \mathcal{V}_\sourcemass) (\mMtest^2 \mathcal{V}_\testmass) }{4\pi r} \times \frac{3}{(\mM R)^2} \times \formfactor_\text{A}. 
\eea
In \Eq{Vbg_swave_appx} and \Eq{Vbg_formfactor_screen_appx}, $3/(\mM R)^2$ is the form factor of the screening effect of the spherically symmetric ansatz in the hard sphere region, as shown in \Eq{formfactor_sphscr_farfield_hard} and \Eq{amplitudecalc_formfactor_sphscr_farfield_hard}. $\formfactor_\text{A}$ is the form factor describing the decoherence effect from the finite-phase-space, as shown in \Eq{phase_space_factor} and plotted in \Fig{FA_plt}. From the above discussion, we can find that even in Regions A and C, the spherical ansatz is incomplete because it only captures the information of the screening effect. More specifically, even if the s-wave approximation is justified, we can always find the far-field region satisfying $kr \gtrsim 1$ where the finite-phase-space induced decoherence suppression becomes important.

\section{Astrophysical Constraints}\label{appx:astro_constraint}

In this section, we estimate the energy loss constraint from the supernova, horizontal branch stars, and red giants following Refs.~\cite{Raffelt:1990yz,Cox:2024oeb}. The light $\phi$ particle can be on-shell produced in the stellar plasma. It will dissipate the energy of the hot environment if the produced light particle can escape freely. The dominant stellar energy loss contribution for the quadratic scalar-photon interaction comes from the $\gamma \gamma \rightarrow \phi \phi$ channel. Other channels, such as $e^- e^+ \rightarrow \phi \phi$, $e^- e^+ \gamma \rightarrow \phi \phi$, and $e \gamma \rightarrow e \phi \phi$, are either suppressed by the one-loop or the multi-body phase space. Here, the emission rate from $\gamma(k_1) \gamma(k_2) \rightarrow \phi(p_1) \phi(p_2)$ is
\bea
\label{eq:Gamma_AAphiphi}
\Gamma_{\gamma \gamma \rightarrow \phi \phi} = \frac{1}{4}\int\frac{\dbar^3 \veck_1}{2E_{\veck_1}}
\frac{\dbar^3 \veck_2}{2E_{\veck_2}}\frac{\dbar^3 \vecp_1}{2E_{\vecp_1}}\frac{\dbar^3 \vecp_2}{2E_{\vecp_2}}  f_{\gamma}(\veck_1) f_{\gamma}(\veck_2)(E_{\veck_1}+E_{\veck_2})(2\pi)^4\delta^4(k_{1}+k_{2}-p_{1}-p_{2})\sum_{\rm d.o.f}\left|\mathcal{M}_{\gamma\gamma\rightarrow \phi\phi}\right|^2\;,
\eea
where the factor of $1/4$ accounts for the symmetry factors from the real scalars ($1/2$) and the photon fields ($1/2$). $f_\gamma(\veck)\simeq\exp(-E_\veck/T)$ is the phase space density of the photons in the stellar's thermal bath. The unpolarized amplitude square and the polarization averaged cross section of $\gamma \gamma \rightarrow \phi \phi$ is
\bea
\label{eq:M_sigma_AAphiphi}
\sum_{\rm d.o.f}\left|\mathcal{M}_{\gamma\gamma\rightarrow \phi\phi}\right|^2=\frac{s^2}{2\Lambda^2_\gamma}
\;,\quad \sigma_{\gamma \gamma \rightarrow \phi \phi} = \frac{s}{128 \pi \Lambda_\gamma^4}\sqrt{1-\frac{ 4 m_0^2}{s}}\;,\quad s=(p_1+p_2)^2. 
\eea
Focusing on the $m_0 \ll E$ region and substituting \Eq{M_sigma_AAphiphi} into \Eq{Gamma_AAphiphi}, 
we have the $\phi$'s emission rate, which is written as
\bea
\Gamma_{\gamma \gamma \rightarrow \phi \phi} \simeq \frac{3}{2\pi^5} \frac{T^9}{\Lambda_{\gamma}^4}.
\eea
This result is consistent with the result of Ref.~\cite{Cox:2024oeb} but differs from that of Ref.~\cite{Olive:2007aj}. The discrepancy arises because Ref.~\cite{Olive:2007aj} oversimplifies the phase space integration when computing the emission rate in \Eq{Gamma_AAphiphi}. We define that $\Gamma_{\gamma \gamma \rightarrow \phi \phi} = \epsilon_\phi \, \rho_\text{core}$, where $\epsilon_\phi$ is the rate of energy loss to $\phi$ per unit stellar mass, and $\rho_\text{core}$ is the stellar core density. For SN1987A, we have $\rho_{\core,\SN} \simeq 3 \times 10^{14}\,\gram/\cm^3$, $\epsilon_{\phi,\SN} \lesssim 10^{19}\,\erg\, / \gram \cdot \second$, and the core temperature $T_{\SN} \simeq 30 \, \mev$~\cite{Cox:2024oeb}. For HB, we have $\rho_{\core,\HB} \simeq 10^{4}\,\gram/\cm^3$, $\epsilon_{\phi,\HB} \lesssim 10\,\erg\, / \gram \cdot \second$, and  $T_{\HB} \simeq 10^8\,\text{K} \simeq 8.6 \, \kev$~\cite{Cox:2024oeb}. For RG, we have $\rho_{\core,\RG} \simeq 10^{6}\,\gram/\cm^3$, $\epsilon_{\phi,\RG} \lesssim 10\,\erg\, / \gram \cdot \second$, and  $T_{\RG} \simeq 10^8\,\text{K} \simeq 8.6 \, \kev$~\cite{Raffelt:1990yz}. Finally, we derive the constraints for $\Lambda_\gamma$ which are
\bea
\text{SN1987A}: \,\, \Lambda_\gamma \gtrsim 1.8 \, \tev, \quad \quad \text{HB}: \,\, \Lambda_\gamma \gtrsim 2.5 \times 10^2 \, \gev, \quad \quad \text{RG}: \,\, \Lambda_\gamma \gtrsim 80 \, \gev.
\eea
Given that the constraint from SN1987A is stronger than the one from HB, we utilize the SN1987A constraint as the benchmark astrophysical constraint in the rest of the paper. One should note that $\phi$ has scattering with photons and other particles in the hot dense supernova environment. As $\Lambda_\gamma$ increases, its effective mean free path $\lambda_{\text{free},\phi}$ for energy exchanging scattering becomes smaller. When  $\lambda_{\text{free},\phi}$ becomes smaller than the supernova radius, $\phi$ is trapped inside, which effectively reduces the supernova energy loss constraint. We do not show it in the plots of this work for two reasons: 1.~The upper boundary of the SN1987A constraint caused by the trapping effect in the $m_0\text{-}\Lambda_\gamma^{-1}$ plane is $\sim 10^2$ times higher than the lower boundary~\cite{Cox:2024oeb}, which lies out of the region we are interested in. 2.~For HB and RG, the density and temperature of the core are much smaller than SN, which reduces the trapping effects and gives stronger constraints in the strongly-coupled regions. 

\section{Alternative Dark Axion UV Models}\label{appx:uv_app}
In this section, we present the KSVZ-like UV-complete model for~\Eq{LDqcd2}. 
For simplicity, we set $x=1/2$ in \Eq{mQ} and assume that the dark quarks $q'$ are invariant under $U(1)_\PQ$. In this scenario, we introduce a neutral scalar $\Phi$ and a pair of fermions $\Xi_{L, R}$. The charge assignments under $SU(N_c)\times U(1)'\times U(1)_\PQ$ for the aforementioned particles are given by
\bea
u'_{L,R}: (N_c, 1, 0)\;,\quad d'_{L,R}: (N_c, 0, 0)\;,\quad \Phi:(1, 0, -2)\;,\quad \Xi_L,\Xi^\dagger_R: (N_c, 1, 1)
\eea
The corresponding Lagrangian takes the form
\bea
-\mathcal{L}\supset \mathcal{Y}_\Xi\Phi \Xi_L^\dagger\Xi_R+\bar{q}_L'\hat{m}_{q'}q'_R+h.c.+\kappa \left(|\Phi|^2-\frac{\fdecayaxion}{2}\right)^2
\eea
In the ground state, the global $U(1)_\PQ$ symmetry is spontaneously broken by the nonzero vacuum expectation value of $\Phi$, which also generates the mass term in the UV model,
\bea
M_\Xi=\frac{{\cal Y}_\Xi \fdecayaxion}{\sqrt2 } e^{i\frac{a}{\fdecayaxion}}\;.
\eea
Since the fermion fields $\Xi$ are heavier than the dark confinement scale, $|M_\Xi| >\Lambda_c$, they can be integrated out. The resulting effective Lagrangian at energies below $|M_\Xi|$ reads,
\bea
-{\cal L}\supset \bar{q}_L'\hat{m}_{q'}q'_R+h.c.+\frac{a}{8\pi\fdecayaxion}\left(\alpha_s^{\prime}G^{a\prime}_{\mu\nu}\tilde{G}^{a\prime\mu\nu}+2 N_c\alpha' F_{\mu\nu}'\tilde{F}^{\prime \mu\nu}\right)
\eea
which reduces to \Eq{LDqcd2} with,
\bea
\hat{Q}_A=\frac12 \mathbb{1}\;,\quad E=N_c\;,\quad N=\frac12 \;,\quad \hat{C}_q^0=0\;.
\eea

\bibliography{References}

@article{Delaunay:2025pho,
    author = "Delaunay, C{\'e}dric and Geller, Michael and Heller-Algazi, Zamir and Perez, Gilad and Springmann, Konstantin",
    title = "{Natural Ultralight Dark Matter: The Quadratic Twin}",
    eprint = "2507.12514",
    archivePrefix = "arXiv",
    primaryClass = "hep-ph",
    month = "7",
    year = "2025"}

@article{Gan:2025icr,
    author = "Gan, Xucheng and Kim, Hyungjin and Mitridate, Andrea",
    title = "{Probing Quadratically Coupled Ultralight Dark Matter with Pulsar Timing Arrays}",
    eprint = "2510.13945",
    archivePrefix = "arXiv",
    primaryClass = "hep-ph",
    reportNumber = "DESY-25-134",
    month = "10",
    year = "2025"}

@article{Burrage:2024mxn,
    author = "Burrage, Clare and Elder, Benjamin and del Castillo, Yeray Garcia and Jaeckel, Joerg",
    title = "{Time-dependent density of quadratically coupled dark matter around ordinary matter objects}",
    eprint = "2410.23350",
    archivePrefix = "arXiv",
    primaryClass = "hep-ph",
    doi = "10.1103/PhysRevD.111.103526",
    journal = "Phys. Rev. D",
    volume = "111",
    number = "10",
    pages = "103526",
    year = "2025"}

@article{Burrage:2025grx,
    author = "Burrage, Clare and Macdonald, Angus and Ross, Michael P. and Rybka, Gray and Todarello, Elisa",
    title = "{Impact of cavities on the detection of quadratically coupled ultra-light dark matter}",
    eprint = "2507.16526",
    archivePrefix = "arXiv",
    primaryClass = "hep-ph",
    month = "7",
    year = "2025"}

@article{Horowitz:1993kw,
    author = "Horowitz, C. J. and Pantaleone, James T.",
    title = "{Long range forces from the cosmological neutrinos background}",
    eprint = "hep-ph/9306222",
    archivePrefix = "arXiv",
    reportNumber = "IUHET-249, IUNTC93-14",
    doi = "10.1016/0370-2693(93)90800-W",
    journal = "Phys. Lett. B",
    volume = "319",
    pages = "186--190",
    year = "1993"}

@article{Ferrer:1998ju,
    author = "Ferrer, F. and Grifols, J. A. and Nowakowski, M.",
    title = "{Long range forces induced by neutrinos at finite temperature}",
    eprint = "hep-ph/9806438",
    archivePrefix = "arXiv",
    reportNumber = "UAB-FT-448",
    doi = "10.1016/S0370-2693(98)01489-0",
    journal = "Phys. Lett. B",
    volume = "446",
    pages = "111--116",
    year = "1999"
}

@article{Ferrer:1999ad,
    author = "Ferrer, F. and Grifols, J. A. and Nowakowski, M.",
    title = "{Long range neutrino forces in the cosmic relic neutrino background}",
    eprint = "hep-ph/9906463",
    archivePrefix = "arXiv",
    reportNumber = "UAB-FT-468, IFT-P-053-99",
    doi = "10.1103/PhysRevD.61.057304",
    journal = "Phys. Rev. D",
    volume = "61",
    pages = "057304",
    year = "2000"
}

@article{Passarino:1978jh,
    author = "Passarino, G. and Veltman, M. J. G.",
    title = "{One Loop Corrections for e+ e- Annihilation Into mu+ mu- in the Weinberg Model}",
    reportNumber = "Print-79-0284 (UTRECHT)",
    doi = "10.1016/0550-3213(79)90234-7",
    journal = "Nucl. Phys. B",
    volume = "160",
    pages = "151--207",
    year = "1979"
}

@article{Botella:1986wy,
    author = "Botella, F. J. and Lim, C. S. and Marciano, W. J.",
    title = "{Radiative Corrections to Neutrino Indices of Refraction}",
    reportNumber = "PRINT-86-1162 (BNL)",
    doi = "10.1103/PhysRevD.35.896",
    journal = "Phys. Rev. D",
    volume = "35",
    pages = "896",
    year = "1987"
}

@article{Mirizzi:2009td,
    author = "Mirizzi, Alessandro and Pozzorini, Stefano and Raffelt, Georg G. and Serpico, Pasquale D.",
    title = "{Flavour-dependent radiative correction to neutrino-neutrino refraction}",
    eprint = "0907.3674",
    archivePrefix = "arXiv",
    primaryClass = "hep-ph",
    reportNumber = "CERN-PH-TH-2009-127, MPP-2009-81",
    doi = "10.1088/1126-6708/2009/10/020",
    journal = "JHEP",
    volume = "10",
    pages = "020",
    year = "2009"
}

@article{Huang:2023nqf,
    author = "Huang, Jihong and Zhou, Shun",
    title = "{Mikheyev-Smirnov-Wolfenstein matter potential at the one-loop level in the Standard Model}",
    eprint = "2307.04685",
    archivePrefix = "arXiv",
    primaryClass = "hep-ph",
    doi = "10.1103/PhysRevD.108.093010",
    journal = "Phys. Rev. D",
    volume = "108",
    number = "9",
    pages = "093010",
    year = "2023"
}

@article{Feinberg:1989ps,
    author = "Feinberg, G. and Sucher, J. and Au, C. K.",
    title = "{The Dispersion Theory of Dispersion Forces}",
    reportNumber = "CU-TP-437",
    doi = "10.1016/0370-1573(89)90111-7",
    journal = "Phys. Rept.",
    volume = "180",
    pages = "83",
    year = "1989"
}

@article{Notzold:1987ik,
    author = {N\"otzold, Dirk and Raffelt, Georg},
    title = "{Neutrino dispersion at finite temperature and density}",
    reportNumber = "MPI-PAE/PTh-87/87",
    doi = "10.1016/0550-3213(88)90113-7",
    journal = "Nucl. Phys. B",
    volume = "307",
    pages = "924--936",
    year = "1988"
}

@article{Fichet:2017bng,
    author = "Fichet, Sylvain",
    title = "{Quantum Forces from Dark Matter and Where to Find Them}",
    eprint = "1705.10331",
    archivePrefix = "arXiv",
    primaryClass = "hep-ph",
    doi = "10.1103/PhysRevLett.120.131801",
    journal = "Phys. Rev. Lett.",
    volume = "120",
    number = "13",
    pages = "131801",
    year = "2018"}

@article{Barbosa:2024pkl,
    author = "Barbosa, Sergio and Fichet, Sylvain",
    title = "{Background-induced forces from dark relics}",
    eprint = "2403.13894",
    archivePrefix = "arXiv",
    primaryClass = "hep-ph",
    doi = "10.1007/JHEP01(2025)021",
    journal = "JHEP",
    volume = "01",
    pages = "021",
    year = "2025"
}

@article{Grossman:2025cov,
    author = "Grossman, Yuval and Yu, Bingrong and Zhou, Siyu",
    title = "{Axion forces in axion backgrounds}",
    eprint = "2504.00104",
    archivePrefix = "arXiv",
    primaryClass = "hep-ph",
    month = "3",
    year = "2025",
    journal = ""}

@article{Cheng:2025fak,
    author = "Cheng, Yu and Ge, Shuailiang",
    title = "{Background-Enhanced Axion Force by Axion Dark Matter}",
    eprint = "2504.02702",
    archivePrefix = "arXiv",
    primaryClass = "hep-ph",
    month = "4",
    year = "2025",
    journal = ""}

@article{Ghosh:2024qai,
    author = "Ghosh, Mitrajyoti and Grossman, Yuval and Tangarife, Walter and Xu, Xun-Jie and Yu, Bingrong",
    title = "{The neutrino force in neutrino backgrounds: Spin dependence and parity-violating effects}",
    eprint = "2405.16801",
    archivePrefix = "arXiv",
    primaryClass = "hep-ph",
    doi = "10.1007/JHEP07(2024)107",
    journal = "JHEP",
    volume = "07",
    pages = "107",
    year = "2024"}

@article{Wolfenstein:1977ue,
    author = "Wolfenstein, L.",
    title = "{Neutrino Oscillations in Matter}",
    reportNumber = "COO-3066-102",
    doi = "10.1103/PhysRevD.17.2369",
    journal = "Phys. Rev. D",
    volume = "17",
    pages = "2369--2374",
    year = "1978"}

@article{Mikheyev:1985zog,
    author = "Mikheyev, S. P. and Smirnov, A. Yu.",
    title = "{Resonance Amplification of Oscillations in Matter and Spectroscopy of Solar Neutrinos}",
    journal = "Sov. J. Nucl. Phys.",
    volume = "42",
    pages = "913--917",
    year = "1985"
}

@article{Mikheyev:1986wj,
    author = "Mikheyev, S. P. and Smirnov, A. Yu.",
    title = "{Resonant amplification of neutrino oscillations in matter and solar neutrino spectroscopy}",
    doi = "10.1007/BF02508049",
    journal = "Nuovo Cim. C",
    volume = "9",
    pages = "17--26",
    year = "1986"
}

@article{Day:2023mkb,
    author = "Day, Hannah and Liu, Da and Luty, Markus A. and Zhao, Yue",
    title = "{Blowing in the dark matter wind}",
    eprint = "2312.13345",
    archivePrefix = "arXiv",
    primaryClass = "hep-ph",
    doi = "10.1007/JHEP07(2024)136",
    journal = "JHEP",
    volume = "07",
    pages = "136",
    year = "2024"}

@article{Luo:2024ocg,
    author = "Luo, Pengshun and Matsumoto, Shigeki and Sheng, Jie and Xing, Chuan-Yang and Zhu, Lin and Zhuge, Zhi-Jie",
    title = "{Detecting meV-Scale Dark Matter via Coherent Scattering with an Asymmetric Torsion Balance}",
    eprint = "2409.09950",
    archivePrefix = "arXiv",
    primaryClass = "hep-ph",
    month = "9",
    year = "2024",
    journal = ""}

@article{Gan:2023wnp,
    author = "Gan, Xucheng and Liu, Di",
    title = "{Cosmologically varying kinetic mixing}",
    eprint = "2302.03056",
    archivePrefix = "arXiv",
    primaryClass = "hep-ph",
    reportNumber = "DESY-23-013, LAPTH-003/23",
    doi = "10.1007/JHEP11(2023)031",
    journal = "JHEP",
    volume = "11",
    pages = "031",
    year = "2023"}

@article{Arakawa:2024lqr,
    author = "Arakawa, Jason and Zaheer, Muhammad H. and Eby, Joshua and Takhistov, Volodymyr and Safronova, Marianna S.",
    title = "{Bosenovae with quadratically-coupled scalars in quantum sensing experiments}",
    eprint = "2402.06736",
    archivePrefix = "arXiv",
    primaryClass = "hep-ph",
    reportNumber = "KEK-QUP-2024-0002, KEK-TH-2599, KEK-Cosmo-0338, IPMU24-0004",
    doi = "10.1007/JHEP08(2024)222",
    journal = "JHEP",
    volume = "08",
    pages = "222",
    year = "2024"}

@article{Brzeminski:2020uhm,
	archiveprefix = {arXiv},
	author = {Brzeminski, Dawid and Chacko, Zackaria and Dev, Abhish and Hook, Anson},
	doi = {10.1103/PhysRevD.104.075019},
	eprint = {2012.02787},
	journal = {Phys. Rev. D},
	number = {7},
	pages = {075019},
	primaryclass = {hep-ph},
	title = {{Time-varying fine structure constant from naturally ultralight dark matter}},
	volume = {104},
	year = {2021}}

@article{Batell:2021ofv,
    author = "Batell, Brian and Ghalsasi, Akshay",
    title = "{Thermal misalignment of scalar dark matter}",
    eprint = "2109.04476",
    archivePrefix = "arXiv",
    primaryClass = "hep-ph",
    reportNumber = "PITT-PACC-2119",
    doi = "10.1103/PhysRevD.107.L091701",
    journal = "Phys. Rev. D",
    volume = "107",
    number = "9",
    pages = "L091701",
    year = "2023"}

@article{Batell:2022qvr,
    author = "Batell, Brian and Ghalsasi, Akshay and Rai, Mudit",
    title = "{Dynamics of dark matter misalignment through the Higgs portal}",
    eprint = "2211.09132",
    archivePrefix = "arXiv",
    primaryClass = "hep-ph",
    reportNumber = "PITT-PACC-2213",
    doi = "10.1007/JHEP01(2024)038",
    journal = "JHEP",
    volume = "01",
    pages = "038",
    year = "2024"}

@article{Cyncynates:2024bxw,
    author = "Cyncynates, David and Simon, Olivier",
    title = "{Minimal targets for dilaton direct detection}",
    eprint = "2408.16816",
    archivePrefix = "arXiv",
    primaryClass = "hep-ph",
    month = "8",
    year = "2024",
    journal = ""}

@article{Cyncynates:2024ufu,
    author = "Cyncynates, David and Simon, Olivier",
    title = "{Scalar relics from the hot Big Bang}",
    eprint = "2410.22409",
    archivePrefix = "arXiv",
    primaryClass = "hep-ph",
    month = "10",
    year = "2024",
    journal = ""}

@article{Weiner:2005ac,
    author = "Weiner, Neal and Zurek, Kathryn M.",
    title = "{New matter effects and BBN constraints for mass varying neutrinos}",
    eprint = "hep-ph/0509201",
    archivePrefix = "arXiv",
    reportNumber = "INT-PUB-05-21",
    doi = "10.1103/PhysRevD.74.023517",
    journal = "Phys. Rev. D",
    volume = "74",
    pages = "023517",
    year = "2006"   }

@article{Ghalsasi:2016pcj,
    author = "Ghalsasi, Akshay and McKeen, David and Nelson, Ann E.",
    title = "{Probing nonstandard neutrino cosmology with terrestrial neutrino experiments}",
    eprint = "1609.06326",
    archivePrefix = "arXiv",
    primaryClass = "hep-ph",
    doi = "10.1103/PhysRevD.95.115039",
    journal = "Phys. Rev. D",
    volume = "95",
    number = "11",
    pages = "115039",
    year = "2017"   }

@article{Ellis:2019flb,
    author = "Ellis, Sebastian A. R. and Ipek, Seyda and White, Graham",
    title = "{Electroweak Baryogenesis from Temperature-Varying Couplings}",
    eprint = "1905.11994",
    archivePrefix = "arXiv",
    primaryClass = "hep-ph",
    doi = "10.1007/JHEP08(2019)002",
    journal = "JHEP",
    volume = "08",
    pages = "002",
    year = "2019" }

@article{Arvanitaki:2014faa,
	archiveprefix = {arXiv},
	author = {Arvanitaki, Asimina and Huang, Junwu and Van Tilburg, Ken},
	doi = {10.1103/PhysRevD.91.015015},
	eprint = {1405.2925},
	journal = {Phys. Rev. D},
	number = {1},
	pages = {015015},
	primaryclass = {hep-ph},
	title = {{Searching for dilaton dark matter with atomic clocks}},
	volume = {91},
	year = {2015}}

@article{VanTilburg:2015oza,
    author = "Van Tilburg, Ken and Leefer, Nathan and Bougas, Lykourgos and Budker, Dmitry",
    title = "{Search for ultralight scalar dark matter with atomic spectroscopy}",
    eprint = "1503.06886",
    archivePrefix = "arXiv",
    primaryClass = "physics.atom-ph",
    doi = "10.1103/PhysRevLett.115.011802",
    journal = "Phys. Rev. Lett.",
    volume = "115",
    number = "1",
    pages = "011802",
    year = "2015"}

@article{Stadnik:2016zkf,
    author = "Stadnik, Y. V. and Flambaum, V. V.",
    title = "{Improved limits on interactions of low-mass spin-0 dark matter from atomic clock spectroscopy}",
    eprint = "1605.04028",
    archivePrefix = "arXiv",
    primaryClass = "physics.atom-ph",
    doi = "10.1103/PhysRevA.94.022111",
    journal = "Phys. Rev. A",
    volume = "94",
    number = "2",
    pages = "022111",
    year = "2016"}

@article{kennedy2020precision,
  title={Precision metrology meets cosmology: improved constraints on ultralight dark matter from atom-cavity frequency comparisons},
  author={Kennedy, Colin J and Oelker, Eric and Robinson, John M and Bothwell, Tobias and Kedar, Dhruv and Milner, William R and Marti, G Edward and Derevianko, Andrei and Ye, Jun},
  journal={Physical Review Letters},
  volume={125},
  number={20},
  pages={201302},
  year={2020},
  publisher={APS}}

@article{Arvanitaki:2016fyj,
	archiveprefix = {arXiv},
	author = {Arvanitaki, Asimina and Graham, Peter W. and Hogan, Jason M. and Rajendran, Surjeet and Van Tilburg, Ken},
	doi = {10.1103/PhysRevD.97.075020},
	eprint = {1606.04541},
	journal = {Phys. Rev. D},
	number = {7},
	pages = {075020},
	primaryclass = {hep-ph},
	title = {{Search for light scalar dark matter with atomic gravitational wave detectors}},
	volume = {97},
	year = {2018}}

@article{Arvanitaki:2015iga,
    author = "Arvanitaki, Asimina and Dimopoulos, Savas and Van Tilburg, Ken",
    title = "{Sound of Dark Matter: Searching for Light Scalars with Resonant-Mass Detectors}",
    eprint = "1508.01798",
    archivePrefix = "arXiv",
    primaryClass = "hep-ph",
    doi = "10.1103/PhysRevLett.116.031102",
    journal = "Phys. Rev. Lett.",
    volume = "116",
    number = "3",
    pages = "031102",
    year = "2016"}

@article{Branca:2016rez,
    author = "Branca, Antonio and others",
    title = "{Search for an Ultralight Scalar Dark Matter Candidate with the AURIGA Detector}",
    eprint = "1607.07327",
    archivePrefix = "arXiv",
    primaryClass = "hep-ex",
    doi = "10.1103/PhysRevLett.118.021302",
    journal = "Phys. Rev. Lett.",
    volume = "118",
    number = "2",
    pages = "021302",
    year = "2017"}

@article{Manley:2019vxy,
    author = "Manley, Jack and Wilson, Dalziel and Stump, Russell and Grin, Daniel and Singh, Swati",
    title = "{Searching for Scalar Dark Matter with Compact Mechanical Resonators}",
    eprint = "1910.07574",
    archivePrefix = "arXiv",
    primaryClass = "astro-ph.IM",
    doi = "10.1103/PhysRevLett.124.151301",
    journal = "Phys. Rev. Lett.",
    volume = "124",
    number = "15",
    pages = "151301",
    year = "2020"}

@article{Hees:2016gop,
	archiveprefix = {arXiv},
	author = {Hees, A. and Gu\'ena, J. and Abgrall, M. and Bize, S. and Wolf, P.},
	doi = {10.1103/PhysRevLett.117.061301},
	eprint = {1604.08514},
	journal = {Phys. Rev. Lett.},
	number = {6},
	pages = {061301},
	primaryclass = {gr-qc},
	title = {{Searching for an oscillating massive scalar field as a dark matter candidate using atomic hyperfine frequency comparisons}},
	volume = {117},
	year = {2016}}

@article{Kalaydzhyan:2017jtv,
    author = "Kalaydzhyan, Tigran and Yu, Nan",
    title = "{Extracting dark matter signatures from atomic clock stability measurements}",
    eprint = "1705.05833",
    archivePrefix = "arXiv",
    primaryClass = "hep-ph",
    doi = "10.1103/PhysRevD.96.075007",
    journal = "Phys. Rev. D",
    volume = "96",
    number = "7",
    pages = "075007",
    year = "2017"}

@article{Hart:2019dxi,
    author = "Hart, Luke and Chluba, Jens",
    title = "{Updated fundamental constant constraints from Planck 2018 data and possible relations to the Hubble tension}",
    eprint = "1912.03986",
    archivePrefix = "arXiv",
    primaryClass = "astro-ph.CO",
    doi = "10.1093/mnras/staa412",
    journal = "Mon. Not. Roy. Astron. Soc.",
    volume = "493",
    number = "3",
    pages = "3255--3263",
    year = "2020"}

@article{Hannestad:1998xp,
    author = "Hannestad, Steen",
    title = "{Possible constraints on the time variation of the fine structure constant from cosmic microwave background data}",
    eprint = "astro-ph/9810102",
    archivePrefix = "arXiv",
    doi = "10.1103/PhysRevD.60.023515",
    journal = "Phys. Rev. D",
    volume = "60",
    pages = "023515",
    year = "1999"}

@article{Menegoni:2012tq,
    author = "Menegoni, Eloisa and Archidiacono, Maria and Calabrese, Erminia and Galli, Silvia and Martins, C. J. A. P. and Melchiorri, Alessandro",
    title = "{The Fine Structure Constant and the CMB Damping Scale}",
    eprint = "1202.1476",
    archivePrefix = "arXiv",
    primaryClass = "astro-ph.CO",
    doi = "10.1103/PhysRevD.85.107301",
    journal = "Phys. Rev. D",
    volume = "85",
    pages = "107301",
    year = "2012"}

@article{Stadnik:2015kia,
    author = "Stadnik, Y. V. and Flambaum, V. V.",
    title = "{Can dark matter induce cosmological evolution of the fundamental constants of Nature?}",
    eprint = "1503.08540",
    archivePrefix = "arXiv",
    primaryClass = "astro-ph.CO",
    doi = "10.1103/PhysRevLett.115.201301",
    journal = "Phys. Rev. Lett.",
    volume = "115",
    number = "20",
    pages = "201301",
    year = "2015"}

@article{Sibiryakov:2020eir,
    author = "Sibiryakov, Sergey and S\o{}rensen, Philip and Yu, Tien-Tien",
    title = "{BBN constraints on universally-coupled ultralight scalar dark matter}",
    eprint = "2006.04820",
    archivePrefix = "arXiv",
    primaryClass = "hep-ph",
    reportNumber = "DESY-19-234, CERN-TH-2020-091, INR-TH-2020-001",
    doi = "10.1007/JHEP12(2020)075",
    journal = "JHEP",
    volume = "12",
    pages = "075",
    year = "2020"}

@article{Bouley:2022eer,
    author = "Bouley, Thomas and S\o{}rensen, Philip and Yu, Tien-Tien",
    title = "{Constraints on ultralight scalar dark matter with quadratic couplings}",
    eprint = "2211.09826",
    archivePrefix = "arXiv",
    primaryClass = "hep-ph",
    reportNumber = "DESY-22-186",
    doi = "10.1007/JHEP03(2023)104",
    journal = "JHEP",
    volume = "03",
    pages = "104",
    year = "2023"}

@article{collaboration2021frequency,
  title={Frequency ratio measurements at 18-digit accuracy using an optical clock network},
  author={Collaboration, Boulder Atomic Clock Optical Network BACON},
  journal={Nature},
  volume={591},
  number={7851},
  pages={564--569},
  year={2021}}

@article{Barontini:2021mvu,
    author = "Barontini, G. and others",
    title = "{Measuring the stability of fundamental constants with a network of clocks}",
    eprint = "2112.10618",
    archivePrefix = "arXiv",
    primaryClass = "hep-ph",
    doi = "10.1140/epjqt/s40507-022-00130-5",
    journal = "EPJ Quant. Technol.",
    volume = "9",
    number = "1",
    pages = "12",
    year = "2022"}

@article{Kobayashi:2022vsf,
    author = "Kobayashi, Takumi and others",
    title = "{Search for Ultralight Dark Matter from Long-Term Frequency Comparisons of Optical and Microwave Atomic Clocks}",
    eprint = "2212.05721",
    archivePrefix = "arXiv",
    primaryClass = "physics.atom-ph",
    doi = "10.1103/PhysRevLett.129.241301",
    journal = "Phys. Rev. Lett.",
    volume = "129",
    number = "24",
    pages = "241301",
    year = "2022"}

@article{Filzinger:2023zrs,
    author = {Filzinger, M. and D\"orscher, S. and Lange, R. and Klose, J. and Steinel, M. and Benkler, E. and Peik, E. and Lisdat, C. and Huntemann, N.},
    title = "{Improved Limits on the Coupling of Ultralight Bosonic Dark Matter to Photons from Optical Atomic Clock Comparisons}",
    eprint = "2301.03433",
    archivePrefix = "arXiv",
    primaryClass = "physics.atom-ph",
    doi = "10.1103/PhysRevLett.130.253001",
    journal = "Phys. Rev. Lett.",
    volume = "130",
    number = "25",
    pages = "253001",
    year = "2023"}

@article{Hamaide:2022rwi,
    author = {Hamaide, Louis and M\"uller, Hendrik and Marsh, David J. E.},
    title = "{Searching for dilaton fields in the Lyman-\ensuremath{\alpha} forest}",
    eprint = "2210.03705",
    archivePrefix = "arXiv",
    primaryClass = "astro-ph.CO",
    doi = "10.1103/PhysRevD.106.123509",
    journal = "Phys. Rev. D",
    volume = "106",
    number = "12",
    pages = "123509",
    year = "2022"}

@article{AEDGE:2019nxb,
    author = "El-Neaj, Yousef Abou and others",
    collaboration = "AEDGE",
    title = "{AEDGE: Atomic Experiment for Dark Matter and Gravity Exploration in Space}",
    eprint = "1908.00802",
    archivePrefix = "arXiv",
    primaryClass = "gr-qc",
    reportNumber = "KCL-PH-TH/2019-65, CERN-TH-2019-126",
    doi = "10.1140/epjqt/s40507-020-0080-0",
    journal = "EPJ Quant. Technol.",
    volume = "7",
    pages = "6",
    year = "2020"}

@article{Badurina:2019hst,
    author = "Badurina, L. and others",
    title = "{AION: An Atom Interferometer Observatory and Network}",
    eprint = "1911.11755",
    archivePrefix = "arXiv",
    primaryClass = "astro-ph.CO",
    reportNumber = "AION-2019-001, CERN-TH-2019-199",
    doi = "10.1088/1475-7516/2020/05/011",
    journal = "JCAP",
    volume = "05",
    pages = "011",
    year = "2020"}

@article{MAGIS-100:2021etm,
    author = "Abe, Mahiro and others",
    collaboration = "MAGIS-100",
    title = "{Matter-wave Atomic Gradiometer Interferometric Sensor (MAGIS-100)}",
    eprint = "2104.02835",
    archivePrefix = "arXiv",
    primaryClass = "physics.atom-ph",
    reportNumber = "FERMILAB-PUB-21-031-AD-DI-FESS-QIS-T",
    doi = "10.1088/2058-9565/abf719",
    journal = "Quantum Sci. Technol.",
    volume = "6",
    number = "4",
    pages = "044003",
    year = "2021"}

@article{Zhao:2021din,
    author = "Zhao, Wei and Mei, Xitong and Gao, Dongfeng and Wang, Jin and Zhan, Mingsheng",
    title = "{Ultralight scalar dark matter detection with ZAIGA}",
    eprint = "2110.11564",
    archivePrefix = "arXiv",
    primaryClass = "physics.atom-ph",
    doi = "10.1142/S0218271822500377",
    journal = "Int. J. Mod. Phys. D",
    volume = "31",
    number = "05",
    pages = "2250037",
    year = "2022"}

@article{Piazza:2010ye,
    author = "Piazza, Federico and Pospelov, Maxim",
    title = "{Sub-eV scalar dark matter through the super-renormalizable Higgs portal}",
    eprint = "1003.2313",
    archivePrefix = "arXiv",
    primaryClass = "hep-ph",
    doi = "10.1103/PhysRevD.82.043533",
    journal = "Phys. Rev. D",
    volume = "82",
    pages = "043533",
    year = "2010"}

@article{Graham:2015ifn,
    author = "Graham, Peter W. and Kaplan, David E. and Mardon, Jeremy and Rajendran, Surjeet and Terrano, William A.",
    title = "{Dark Matter Direct Detection with Accelerometers}",
    eprint = "1512.06165",
    archivePrefix = "arXiv",
    primaryClass = "hep-ph",
    doi = "10.1103/PhysRevD.93.075029",
    journal = "Phys. Rev. D",
    volume = "93",
    number = "7",
    pages = "075029",
    year = "2016"}

@article{Bekenstein:1982eu,
    author = "Bekenstein, J. D.",
    title = "{Fine Structure Constant: Is It Really a Constant?}",
    doi = "10.1103/PhysRevD.25.1527",
    journal = "Phys. Rev. D",
    volume = "25",
    pages = "1527--1539",
    year = "1982"}

@article{Bekenstein:2002wz,
    author = "Bekenstein, J. D.",
    title = "{Fine structure constant variability, equivalence principle and cosmology}",
    eprint = "gr-qc/0208081",
    archivePrefix = "arXiv",
    doi = "10.1103/PhysRevD.66.123514",
    journal = "Phys. Rev. D",
    volume = "66",
    pages = "123514",
    year = "2002"}

@article{Damour:1996zw,
    author = "Damour, Thibault and Dyson, Freeman",
    title = "{The Oklo bound on the time variation of the fine structure constant revisited}",
    eprint = "hep-ph/9606486",
    archivePrefix = "arXiv",
    reportNumber = "IHES-P-96-40, IASSNS-HEP-96-62",
    doi = "10.1016/S0550-3213(96)00467-1",
    journal = "Nucl. Phys. B",
    volume = "480",
    pages = "37--54",
    year = "1996"}

@article{Damour:1994zq,
    author = "Damour, T. and Polyakov, Alexander M.",
    title = "{The String dilaton and a least coupling principle}",
    eprint = "hep-th/9401069",
    archivePrefix = "arXiv",
    doi = "10.1016/0550-3213(94)90143-0",
    journal = "Nucl. Phys. B",
    volume = "423",
    pages = "532--558",
    year = "1994"}

@article{Damour:1994ya,
    author = "Damour, T. and Polyakov, Alexander M.",
    title = "{String theory and gravity}",
    eprint = "gr-qc/9411069",
    archivePrefix = "arXiv",
    reportNumber = "IHES-P-94-1",
    doi = "10.1007/BF02106709",
    journal = "Gen. Rel. Grav.",
    volume = "26",
    pages = "1171--1176",
    year = "1994"}

@article{Damour:2002mi,
    author = "Damour, Thibault and Piazza, Federico and Veneziano, Gabriele",
    title = "{Runaway dilaton and equivalence principle violations}",
    eprint = "gr-qc/0204094",
    archivePrefix = "arXiv",
    reportNumber = "IHES-P-02-27, BICOCCA-FT-02-7, CERN-TH-2002-092",
    doi = "10.1103/PhysRevLett.89.081601",
    journal = "Phys. Rev. Lett.",
    volume = "89",
    pages = "081601",
    year = "2002"}

@article{Damour:2002nv,
    author = "Damour, T. and Piazza, F. and Veneziano, G.",
    title = "{Violations of the equivalence principle in a dilaton runaway scenario}",
    eprint = "hep-th/0205111",
    archivePrefix = "arXiv",
    reportNumber = "IHES-P-02-09, BICOCCA-FT-02-03, CERN-TH-2002-093",
    doi = "10.1103/PhysRevD.66.046007",
    journal = "Phys. Rev. D",
    volume = "66",
    pages = "046007",
    year = "2002"}

@article{Damour:2010rp,
    author = "Damour, Thibault and Donoghue, John F.",
    title = "{Equivalence Principle Violations and Couplings of a Light Dilaton}",
    eprint = "1007.2792",
    archivePrefix = "arXiv",
    primaryClass = "gr-qc",
    doi = "10.1103/PhysRevD.82.084033",
    journal = "Phys. Rev. D",
    volume = "82",
    pages = "084033",
    year = "2010"}

@article{Olive:2001vz,
    author = "Olive, Keith A. and Pospelov, Maxim",
    title = "{Evolution of the fine structure constant driven by dark matter and the cosmological constant}",
    eprint = "hep-ph/0110377",
    archivePrefix = "arXiv",
    reportNumber = "UMN-TH-2028-01, TPI-MINN-01-46, MCGILL-01-21",
    doi = "10.1103/PhysRevD.65.085044",
    journal = "Phys. Rev. D",
    volume = "65",
    pages = "085044",
    year = "2002"}

@article{Dvali:2001dd,
    author = "Dvali, G. R. and Zaldarriaga, Matias",
    title = "{Changing alpha with time: Implications for fifth force type experiments and quintessence}",
    eprint = "hep-ph/0108217",
    archivePrefix = "arXiv",
    reportNumber = "NYU-TH-01-08-07",
    doi = "10.1103/PhysRevLett.88.091303",
    journal = "Phys. Rev. Lett.",
    volume = "88",
    pages = "091303",
    year = "2002"}

@article{Wetterich:2002ic,
    author = "Wetterich, Christof",
    title = "{Probing quintessence with time variation of couplings}",
    eprint = "hep-ph/0203266",
    archivePrefix = "arXiv",
    reportNumber = "HD-THEP-02-11",
    doi = "10.1088/1475-7516/2003/10/002",
    journal = "JCAP",
    volume = "10",
    pages = "002",
    year = "2003"}

@article{Chacko:2002mf,
    author = "Chacko, Z. and Grojean, C. and Perelstein, M.",
    title = "{Fine structure constant variation from a late phase transition}",
    eprint = "hep-ph/0204142",
    archivePrefix = "arXiv",
    reportNumber = "SACLAY-T02-038",
    doi = "10.1016/S0370-2693(03)00766-4",
    journal = "Phys. Lett. B",
    volume = "565",
    pages = "169--175",
    year = "2003"}

@article{Armendariz-Picon:2002jez,
    author = "Armendariz-Picon, Christian",
    title = "{Predictions and observations in theories with varying couplings}",
    eprint = "astro-ph/0205187",
    archivePrefix = "arXiv",
    doi = "10.1103/PhysRevD.66.064008",
    journal = "Phys. Rev. D",
    volume = "66",
    pages = "064008",
    year = "2002"}

@article{Uzan:2002vq,
    author = "Uzan, Jean-Philippe",
    title = "{The Fundamental Constants and Their Variation: Observational Status and Theoretical Motivations}",
    eprint = "hep-ph/0205340",
    archivePrefix = "arXiv",
    doi = "10.1103/RevModPhys.75.403",
    journal = "Rev. Mod. Phys.",
    volume = "75",
    pages = "403",
    year = "2003"}

@article{Uzan:2010pm,
    author = "Uzan, Jean-Philippe",
    title = "{Varying Constants, Gravitation and Cosmology}",
    eprint = "1009.5514",
    archivePrefix = "arXiv",
    primaryClass = "astro-ph.CO",
    doi = "10.12942/lrr-2011-2",
    journal = "Living Rev. Rel.",
    volume = "14",
    pages = "2",
    year = "2011"}

@article{Uzan:2024ded,
    author = "Uzan, Jean-Philippe",
    title = "{Fundamental constants: from measurement to the universe, a window on gravitation and cosmology}",
    eprint = "2410.07281",
    archivePrefix = "arXiv",
    primaryClass = "astro-ph.CO",
    month = "10",
    year = "2024",
    journal = ""}

@article{Luo:2023cxo,
    author = "Luo, Xuheng and Mathur, Anubhav",
    title = "{Cosmic birefringence from CP-violating axion interactions}",
    eprint = "2311.03536",
    archivePrefix = "arXiv",
    primaryClass = "hep-ph",
    doi = "10.1007/JHEP08(2024)038",
    journal = "JHEP",
    volume = "08",
    pages = "038",
    year = "2024"}

@article{Kamionkowski:2024axz,
    author = "Kamionkowski, Marc and Mathur, Anubhav",
    title = "{Thermocoupled early dark energy}",
    eprint = "2411.09747",
    archivePrefix = "arXiv",
    primaryClass = "hep-ph",
    doi = "10.1103/PhysRevD.111.063551",
    journal = "Phys. Rev. D",
    volume = "111",
    number = "6",
    pages = "063551",
    year = "2025"}

@article{Preskill:1982cy,
    author = "Preskill, John and Wise, Mark B. and Wilczek, Frank",
    editor = "Srednicki, M. A.",
    title = "{Cosmology of the Invisible Axion}",
    reportNumber = "HUTP-82-A048, NSF-ITP-82-103",
    doi = "10.1016/0370-2693(83)90637-8",
    journal = "Phys. Lett. B",
    volume = "120",
    pages = "127--132",
    year = "1983"}

@article{Arias:2012az,
    author = "Arias, Paola and Cadamuro, Davide and Goodsell, Mark and Jaeckel, Joerg and Redondo, Javier and Ringwald, Andreas",
    title = "{WISPy Cold Dark Matter}",
    eprint = "1201.5902",
    archivePrefix = "arXiv",
    primaryClass = "hep-ph",
    reportNumber = "DESY-11-226, MPP-2011-140, CERN-PH-TH-2011-323, IPPP-11-80, DCPT-11-160",
    doi = "10.1088/1475-7516/2012/06/013",
    journal = "JCAP",
    volume = "06",
    pages = "013",
    year = "2012"}

@article{Schlamminger:2007ht,
    author = "Schlamminger, Stephan and Choi, K. -Y. and Wagner, T. A. and Gundlach, J. H. and Adelberger, E. G.",
    title = "{Test of the equivalence principle using a rotating torsion balance}",
    eprint = "0712.0607",
    archivePrefix = "arXiv",
    primaryClass = "gr-qc",
    doi = "10.1103/PhysRevLett.100.041101",
    journal = "Phys. Rev. Lett.",
    volume = "100",
    pages = "041101",
    year = "2008"}

@article{Wagner:2012ui,
    author = "Wagner, T. A. and Schlamminger, S. and Gundlach, J. H. and Adelberger, E. G.",
    title = "{Torsion-balance tests of the weak equivalence principle}",
    eprint = "1207.2442",
    archivePrefix = "arXiv",
    primaryClass = "gr-qc",
    doi = "10.1088/0264-9381/29/18/184002",
    journal = "Class. Quant. Grav.",
    volume = "29",
    pages = "184002",
    year = "2012"}

@article{Hees:2018fpg,
    author = "Hees, Aur\'elien and Minazzoli, Olivier and Savalle, Etienne and Stadnik, Yevgeny V. and Wolf, Peter",
    title = "{Violation of the equivalence principle from light scalar dark matter}",
    eprint = "1807.04512",
    archivePrefix = "arXiv",
    primaryClass = "gr-qc",
    doi = "10.1103/PhysRevD.98.064051",
    journal = "Phys. Rev. D",
    volume = "98",
    number = "6",
    pages = "064051",
    year = "2018"}

@article{Banerjee:2022sqg,
    author = "Banerjee, Abhishek and Perez, Gilad and Safronova, Marianna and Savoray, Inbar and Shalit, Aviv",
    title = "{The phenomenology of quadratically coupled ultra light dark matter}",
    eprint = "2211.05174",
    archivePrefix = "arXiv",
    primaryClass = "hep-ph",
    doi = "10.1007/JHEP10(2023)042",
    journal = "JHEP",
    volume = "10",
    pages = "042",
    year = "2023"}

@article{Berge:2017ovy,
    author = "Berg\'e, Joel and Brax, Philippe and M\'etris, Gilles and Pernot-Borr\`as, Martin and Touboul, Pierre and Uzan, Jean-Philippe",
    title = "{MICROSCOPE Mission: First Constraints on the Violation of the Weak Equivalence Principle by a Light Scalar Dilaton}",
    eprint = "1712.00483",
    archivePrefix = "arXiv",
    primaryClass = "gr-qc",
    doi = "10.1103/PhysRevLett.120.141101",
    journal = "Phys. Rev. Lett.",
    volume = "120",
    number = "14",
    pages = "141101",
    year = "2018"}

@article{Smith:1999cr,
    author = "Smith, G. L. and Hoyle, C. D. and Gundlach, J. H. and Adelberger, E. G. and Heckel, Blayne R. and Swanson, H. E.",
    title = "{Short range tests of the equivalence principle}",
    doi = "10.1103/PhysRevD.61.022001",
    journal = "Phys. Rev. D",
    volume = "61",
    pages = "022001",
    year = "2000"}

@book{Sakurai:2011zz,
    author = "Sakurai, Jun John and Napolitano, Jim",
    title = "{Modern Quantum Mechanics}",
    doi = "10.1017/9781108587280",
    isbn = "978-0-8053-8291-4, 978-1-108-52742-2, 978-1-108-58728-0",
    publisher = "Cambridge University Press",
    series = "Quantum physics, quantum information and quantum computation",
    month = "10",
    year = "2020"}

@book{Landau:1991wop,
    author = "Landau, Lev Davidovich and Lifshits, E. M.",
    title = "{Quantum Mechanics}: {Non-Relativistic Theory}",
    isbn = "978-0-7506-3539-4",
    publisher = "Butterworth-Heinemann",
    address = "Oxford",
    series = "Course of Theoretical Physics",
    volume = "v.3",
    year = "1991"}

@book{Bjorken:1965sts,
    author = "Bjorken, James D. and Drell, Sidney D.",
    title = "{Relativistic Quantum Mechanics}",
    isbn = "978-0-07-005493-6",
    publisher = "McGraw-Hill",
    address = "New York",
    series = "International Series In Pure and Applied Physics",
    year = "1965"}

@book{Greiner:1997xwk,
    author = "Greiner, Walter",
    title = "{Relativistic Quantum Mechanics. Wave Equations}",
    doi = "10.1007/978-3-662-03425-5",
    isbn = "978-3-662-03425-5, 978-3-662-02634-2, 978-3-540-67457-3, 978-3-662-04275-5",
    publisher = "Springer",
    address = "Berlin",
    year = "1997"}

@article{Damour:2010rm,
    author = "Damour, Thibault and Donoghue, John F.",
    title = "{Phenomenology of the Equivalence Principle with Light Scalars}",
    eprint = "1007.2790",
    archivePrefix = "arXiv",
    primaryClass = "gr-qc",
    doi = "10.1088/0264-9381/27/20/202001",
    journal = "Class. Quant. Grav.",
    volume = "27",
    pages = "202001",
    year = "2010"
}

@article{Ferrer:2000hm,
    author = "Ferrer, F. and Grifols, J. A.",
    title = "{Effects of Bose-Einstein condensation on forces among bodies sitting in a boson heat bath}",
    eprint = "hep-ph/0001185",
    archivePrefix = "arXiv",
    reportNumber = "UAB-FT-481",
    doi = "10.1103/PhysRevD.63.025020",
    journal = "Phys. Rev. D",
    volume = "63",
    pages = "025020",
    year = "2001"}

@article{VanTilburg:2024xib,
    author = "Van Tilburg, Ken",
    title = "{Wake forces in a background of quadratically coupled mediators}",
    doi = "10.1103/PhysRevD.109.096036",
    journal = "Phys. Rev. D",
    volume = "109",
    number = "9",
    pages = "096036",
    year = "2024"
}

@article{Ghosh:2022nzo,
    author = "Ghosh, Mitrajyoti and Grossman, Yuval and Tangarife, Walter and Xu, Xun-Jie and Yu, Bingrong",
    title = "{Neutrino forces in neutrino backgrounds}",
    eprint = "2209.07082",
    archivePrefix = "arXiv",
    primaryClass = "hep-ph",
    doi = "10.1007/JHEP02(2023)092",
    journal = "JHEP",
    volume = "02",
    pages = "092",
    year = "2023"}

@article{Blas:2022ovz,
    author = "Blas, Diego and Esteban, Ivan and Gonzalez-Garcia, M. C. and Salvado, Jordi",
    title = "{On neutrino-mediated potentials in a neutrino background}",
    eprint = "2212.03889",
    archivePrefix = "arXiv",
    primaryClass = "hep-ph",
    doi = "10.1007/JHEP04(2023)039",
    journal = "JHEP",
    volume = "04",
    pages = "039",
    year = "2023"}

@article{dePireySaintAlby:2017lwc,
    author = "de Pirey Saint Alby, Thibaut Arnoulx and Yunes, Nicol\'as",
    title = "{Cosmological Evolution and Solar System Consistency of Massive Scalar-Tensor Gravity}",
    eprint = "1703.06341",
    archivePrefix = "arXiv",
    primaryClass = "gr-qc",
    doi = "10.1103/PhysRevD.96.064040",
    journal = "Phys. Rev. D",
    volume = "96",
    number = "6",
    pages = "064040",
    year = "2017"}

@article{Berezhiani:2018oxf,
    author = "Berezhiani, Lasha and Khoury, Justin",
    title = "{Emergent long-range interactions in Bose-Einstein Condensates}",
    eprint = "1812.09332",
    archivePrefix = "arXiv",
    primaryClass = "hep-th",
    doi = "10.1103/PhysRevD.99.076003",
    journal = "Phys. Rev. D",
    volume = "99",
    number = "7",
    pages = "076003",
    year = "2019"}

@article{Fukuda:2018omk,
    author = "Fukuda, Hajime and Matsumoto, Shigeki and Yanagida, Tsutomu T.",
    title = "{Direct Detection of Ultralight Dark Matter via Astronomical Ephemeris}",
    eprint = "1801.02807",
    archivePrefix = "arXiv",
    primaryClass = "hep-ph",
    reportNumber = "IPMU18-0003",
    doi = "10.1016/j.physletb.2018.12.038",
    journal = "Phys. Lett. B",
    volume = "789",
    pages = "220--227",
    year = "2019"}

@article{Fukuda:2021drn,
    author = "Fukuda, Hajime and Shirai, Satoshi",
    title = "{Detection of QCD axion dark matter by coherent scattering}",
    eprint = "2112.13536",
    archivePrefix = "arXiv",
    primaryClass = "hep-ph",
    reportNumber = "IPMU21-0089",
    doi = "10.1103/PhysRevD.105.095030",
    journal = "Phys. Rev. D",
    volume = "105",
    number = "9",
    pages = "095030",
    year = "2022"}

@article{Will:2014kxa,
    author = "Will, Clifford M.",
    title = "{The Confrontation between General Relativity and Experiment}",
    eprint = "1403.7377",
    archivePrefix = "arXiv",
    primaryClass = "gr-qc",
    doi = "10.12942/lrr-2014-4",
    journal = "Living Rev. Rel.",
    volume = "17",
    pages = "4",
    year = "2014"}

@article{Murata:2014nra,
    author = "Murata, Jiro and Tanaka, Saki",
    title = "{A review of short-range gravity experiments in the LHC era}",
    eprint = "1408.3588",
    archivePrefix = "arXiv",
    primaryClass = "hep-ex",
    doi = "10.1088/0264-9381/32/3/033001",
    journal = "Class. Quant. Grav.",
    volume = "32",
    number = "3",
    pages = "033001",
    year = "2015"}

@article{Spero:1980zz,
    author = "Spero, R. and Hoskins, J. K. and Newman, R. and Pellam, J. and Schultz, J.",
    title = "{Test of the Gravitational Inverse-Square Law at Laboratory Distances}",
    doi = "10.1103/PhysRevLett.44.1645",
    journal = "Phys. Rev. Lett.",
    volume = "44",
    pages = "1645--1648",
    year = "1980"}

@article{Hoskins:1985tn,
    author = "Hoskins, J. K. and Newman, R. D. and Spero, R. and Schultz, J.",
    title = "{Experimental tests of the gravitational inverse square law for mass separations from 2-cm to 105-cm}",
    doi = "10.1103/PhysRevD.32.3084",
    journal = "Phys. Rev. D",
    volume = "32",
    pages = "3084--3095",
    year = "1985"}

@article{Hoyle:2000cv,
    author = "Hoyle, C. D. and Schmidt, U. and Heckel, Blayne R. and Adelberger, E. G. and Gundlach, J. H. and Kapner, D. J. and Swanson, H. E.",
    title = "{Submillimeter tests of the gravitational inverse square law: a search for 'large' extra dimensions}",
    eprint = "hep-ph/0011014",
    archivePrefix = "arXiv",
    doi = "10.1103/PhysRevLett.86.1418",
    journal = "Phys. Rev. Lett.",
    volume = "86",
    pages = "1418--1421",
    year = "2001"}

@article{Hoyle:2004cw,
    author = "Hoyle, C. D. and Kapner, D. J. and Heckel, Blayne R. and Adelberger, E. G. and Gundlach, J. H. and Schmidt, U. and Swanson, H. E.",
    title = "{Sub-millimeter tests of the gravitational inverse-square law}",
    eprint = "hep-ph/0405262",
    archivePrefix = "arXiv",
    doi = "10.1103/PhysRevD.70.042004",
    journal = "Phys. Rev. D",
    volume = "70",
    pages = "042004",
    year = "2004"}

@article{Kapner:2006si,
    author = "Kapner, D. J. and Cook, T. S. and Adelberger, E. G. and Gundlach, J. H. and Heckel, Blayne R. and Hoyle, C. D. and Swanson, H. E.",
    title = "{Tests of the gravitational inverse-square law below the dark-energy length scale}",
    eprint = "hep-ph/0611184",
    archivePrefix = "arXiv",
    doi = "10.1103/PhysRevLett.98.021101",
    journal = "Phys. Rev. Lett.",
    volume = "98",
    pages = "021101",
    year = "2007"}

@article{Tu:2007zz,
    author = "Tu, Liang-Cheng and Guan, Sheng-Guo and Luo, Jun and Shao, Cheng-Gang and Liu, Lin-Xia",
    title = "{Null Test of Newtonian Inverse-Square Law at Submillimeter Range with a Dual-Modulation Torsion Pendulum}",
    doi = "10.1103/PhysRevLett.98.201101",
    journal = "Phys. Rev. Lett.",
    volume = "98",
    pages = "201101",
    year = "2007"}

@article{Yang:2012zzb,
    author = "Yang, Shan-Qing and Zhan, Bi-Fu and Wang, Qing-Lan and Shao, Cheng-Gang and Tu, Liang-Cheng and Tan, Wen-Hai and Luo, Jun",
    title = "{Test of the Gravitational Inverse Square Law at Millimeter Ranges}",
    doi = "10.1103/PhysRevLett.108.081101",
    journal = "Phys. Rev. Lett.",
    volume = "108",
    pages = "081101",
    year = "2012"}

@article{Nobili:2012uj,
    author = "Nobili, A. M. and others",
    title = "{'Galileo Galilei' (GG): Space test of the weak equivalence principle to 10(-17) and laboratory demonstrations}",
    doi = "10.1088/0264-9381/29/18/184011",
    journal = "Class. Quant. Grav.",
    volume = "29",
    pages = "184011",
    year = "2012"}

@article{Touboul:2017grn,
    author = "Touboul, Pierre and others",
    title = "{MICROSCOPE Mission: First Results of a Space Test of the Equivalence Principle}",
    eprint = "1712.01176",
    archivePrefix = "arXiv",
    primaryClass = "astro-ph.IM",
    doi = "10.1103/PhysRevLett.119.231101",
    journal = "Phys. Rev. Lett.",
    volume = "119",
    number = "23",
    pages = "231101",
    year = "2017"}

@article{MICROSCOPE:2019jix,
    author = "Touboul, Pierre and others",
    collaboration = "MICROSCOPE",
    title = "{Space test of the Equivalence Principle: first results of the MICROSCOPE mission}",
    eprint = "1909.10598",
    archivePrefix = "arXiv",
    primaryClass = "gr-qc",
    doi = "10.1088/1361-6382/ab4707",
    journal = "Class. Quant. Grav.",
    volume = "36",
    number = "22",
    pages = "225006",
    year = "2019"}

@article{MICROSCOPE:2022doy,
    author = "Touboul, Pierre and others",
    collaboration = "MICROSCOPE",
    title = "{MICROSCOPE Mission: Final Results of the Test of the Equivalence Principle}",
    eprint = "2209.15487",
    archivePrefix = "arXiv",
    primaryClass = "gr-qc",
    doi = "10.1103/PhysRevLett.129.121102",
    journal = "Phys. Rev. Lett.",
    volume = "129",
    number = "12",
    pages = "121102",
    year = "2022"
}

@article{Buscaino:2015fya,
    author = "Buscaino, Brandon and DeBra, Daniel and Graham, Peter W. and Gratta, Giorgio and Wiser, Timothy D.",
    title = "{Testing long-distance modifications of gravity to 100 astronomical units}",
    eprint = "1508.06273",
    archivePrefix = "arXiv",
    primaryClass = "gr-qc",
    doi = "10.1103/PhysRevD.92.104048",
    journal = "Phys. Rev. D",
    volume = "92",
    number = "10",
    pages = "104048",
    year = "2015"
}

@article{Raffelt:1990yz,
    author = "Raffelt, Georg G.",
    title = "{Astrophysical methods to constrain axions and other novel particle phenomena}",
    reportNumber = "MPI-PAE-PTH-29-90",
    doi = "10.1016/0370-1573(90)90054-6",
    journal = "Phys. Rept.",
    volume = "198",
    pages = "1--113",
    year = "1990"}

@article{Olive:2007aj,
    author = "Olive, Keith A. and Pospelov, Maxim",
    title = "{Environmental dependence of masses and coupling constants}",
    eprint = "0709.3825",
    archivePrefix = "arXiv",
    primaryClass = "hep-ph",
    reportNumber = "UVIC-TH-07-12, UMN-TH-2618-07, FTPI-MINN-07-28",
    doi = "10.1103/PhysRevD.77.043524",
    journal = "Phys. Rev. D",
    volume = "77",
    pages = "043524",
    year = "2008"}

@article{Cox:2024oeb,
    author = "Cox, Peter and Dolan, Matthew J. and Hiskens, Frederick J.",
    title = "{Photonic Freeze-In}",
    eprint = "2412.17308",
    archivePrefix = "arXiv",
    primaryClass = "hep-ph",
    month = "12",
    year = "2024"}

@article{Beadle:2023flm,
    author = "Beadle, Carl and Ellis, Sebastian A. R. and Quevillon, J\'er\'emie and Hoa Vuong, Pham Ngoc",
    title = "{Quadratic coupling of the axion to photons}",
    eprint = "2307.10362",
    archivePrefix = "arXiv",
    primaryClass = "hep-ph",
    reportNumber = "CERN-TH-2023-136, DESY-23-097",
    doi = "10.1103/PhysRevD.110.035019",
    journal = "Phys. Rev. D",
    volume = "110",
    number = "3",
    pages = "035019",
    year = "2024"}

@article{Kim:2023pvt,
    author = "Kim, Hyungjin and Lenoci, Alessandro and Perez, Gilad and Ratzinger, Wolfram",
    title = "{Probing an ultralight QCD axion with electromagnetic quadratic interaction}",
    eprint = "2307.14962",
    archivePrefix = "arXiv",
    primaryClass = "hep-ph",
    reportNumber = "DESY-23-110",
    doi = "10.1103/PhysRevD.109.015030",
    journal = "Phys. Rev. D",
    volume = "109",
    number = "1",
    pages = "015030",
    year = "2024"}

@article{Bauer:2023czj,
    author = "Bauer, Martin and Rostagni, Guillaume",
    title = "{Fifth Forces from QCD Axions Scale Differently}",
    eprint = "2307.09516",
    archivePrefix = "arXiv",
    primaryClass = "hep-ph",
    reportNumber = "IPPP/23/35",
    doi = "10.1103/PhysRevLett.132.101802",
    journal = "Phys. Rev. Lett.",
    volume = "132",
    number = "10",
    pages = "101802",
    year = "2024"}

@article{Bauer:2024yow,
    author = "Bauer, Martin and Chakraborti, Sreemanti",
    title = "{On the Validity of Bounds on Light Axions for $f\lesssim10^{15}$ GeV}",
    eprint = "2408.06408",
    archivePrefix = "arXiv",
    primaryClass = "hep-ph",
    reportNumber = "IPPP/24/55",
    month = "8",
    year = "2024",
    journal = ""}

@article{Bauer:2024hfv,
    author = "Bauer, Martin and Chakraborti, Sreemanti and Rostagni, Guillaume",
    title = "{Axion bounds from quantum technology}",
    eprint = "2408.06412",
    archivePrefix = "arXiv",
    primaryClass = "hep-ph",
    reportNumber = "IPPP/24/53, IPPP/24/53",
    doi = "10.1007/JHEP05(2025)023",
    journal = "JHEP",
    volume = "05",
    pages = "023",
    year = "2025"}

@article{Banerjee:2025dlo,
    author = "Banerjee, Abhishek and Bloch, Itay M. and Bonnefoy, Quentin and Ellis, Sebastian A. R. and Perez, Gilad and Savoray, Inbar and Springmann, Konstantin and Stadnik, Yevgeny V.",
    title = "{Momentum and Matter Matter for Axion Dark Matter Matters on Earth}",
    eprint = "2502.04455",
    archivePrefix = "arXiv",
    primaryClass = "hep-ph",
    month = "2",
    year = "2025"}

@article{delCastillo:2025rbr,
    author = "del Castillo, Yeray Garcia and Hammett, Benjamin and Jaeckel, Joerg",
    title = "{Enhanced Axion-wind near Earth's Surface}",
    eprint = "2502.04456",
    archivePrefix = "arXiv",
    primaryClass = "hep-ph",
    month = "2",
    year = "2025"}

@article{Born:1926yhp,
    author = "Born, Max",
    title = {{Quantenmechanik der Sto\ss{}vorg\"ange}},
    doi = "10.1007/BF01397184",
    journal = "Z. Phys.",
    volume = "38",
    number = "11-12",
    pages = "803--827",
    year = "1926"}

@article{Weizsacker:1935bkz,
    author = "Weizsacker, C. F. V.",
    title = "{Zur Theorie der Kernmassen}",
    doi = "10.1007/BF01337700",
    journal = "Z. Phys.",
    volume = "96",
    pages = "431--458",
    year = "1935"}

@article{Bai:2010qg,
    author = "Bai, Yang and Hill, Richard J.",
    title = "{Weakly Interacting Stable Pions}",
    eprint = "1005.0008",
    archivePrefix = "arXiv",
    primaryClass = "hep-ph",
    reportNumber = "FERMILAB-PUB-10-001-T, EFI-PREPRINT-10-9",
    doi = "10.1103/PhysRevD.82.111701",
    journal = "Phys. Rev. D",
    volume = "82",
    pages = "111701",
    year = "2010"}

@article{Lippmann:1950zz,
    author = "Lippmann, B. A. and Schwinger, Julian",
    title = "{Variational Principles for Scattering Processes. I}",
    doi = "10.1103/PhysRev.79.469",
    journal = "Phys. Rev.",
    volume = "79",
    pages = "469--480",
    year = "1950"}

@article{GrillidiCortona:2015jxo,
    author = "Grilli di Cortona, Giovanni and Hardy, Edward and Pardo Vega, Javier and Villadoro, Giovanni",
    title = "{The QCD axion, precisely}",
    eprint = "1511.02867",
    archivePrefix = "arXiv",
    primaryClass = "hep-ph",
    doi = "10.1007/JHEP01(2016)034",
    journal = "JHEP",
    volume = "01",
    pages = "034",
    year = "2016"
}

@article{DiLuzio:2020wdo,
    author = "Di Luzio, Luca and Giannotti, Maurizio and Nardi, Enrico and Visinelli, Luca",
    title = "{The landscape of QCD axion models}",
    eprint = "2003.01100",
    archivePrefix = "arXiv",
    primaryClass = "hep-ph",
    reportNumber = "DESY 20-036, DESY-20-036",
    doi = "10.1016/j.physrep.2020.06.002",
    journal = "Phys. Rept.",
    volume = "870",
    pages = "1--117",
    year = "2020"
}

@article{Halverson:2018olu,
    author = "Halverson, James and Nelson, Brent D. and Ruehle, Fabian and Salinas, Gustavo",
    title = "{Dark Glueballs and their Ultralight Axions}",
    eprint = "1805.06011",
    archivePrefix = "arXiv",
    primaryClass = "hep-ph",
    doi = "10.1103/PhysRevD.98.043502",
    journal = "Phys. Rev. D",
    volume = "98",
    number = "4",
    pages = "043502",
    year = "2018"
}

@article{Sikivie:1983ip,
    author = "Sikivie, P.",
    editor = "Srednicki, M. A.",
    title = "{Experimental Tests of the Invisible Axion}",
    reportNumber = "PRINT-83-0597 (FLORIDA), UF-TP-83-13",
    doi = "10.1103/PhysRevLett.51.1415",
    journal = "Phys. Rev. Lett.",
    volume = "51",
    pages = "1415--1417",
    year = "1983",
    note = "[Erratum: Phys.Rev.Lett. 52, 695 (1984)]"}

@article{ALPS:2009des,
    author = "Ehret, Klaus and others",
    collaboration = "ALPS",
    title = "{Resonant laser power build-up in ALPS: A 'Light-shining-through-walls' experiment}",
    eprint = "0905.4159",
    archivePrefix = "arXiv",
    primaryClass = "physics.ins-det",
    reportNumber = "DESY-09-058",
    doi = "10.1016/j.nima.2009.10.102",
    journal = "Nucl. Instrum. Meth. A",
    volume = "612",
    pages = "83--96",
    year = "2009"}

@article{Redondo:2010dp,
    author = "Redondo, Javier and Ringwald, Andreas",
    title = "{Light shining through walls}",
    eprint = "1011.3741",
    archivePrefix = "arXiv",
    primaryClass = "hep-ph",
    reportNumber = "DESY-10-175, MPP-2010-149",
    doi = "10.1080/00107514.2011.563516",
    journal = "Contemp. Phys.",
    volume = "52",
    pages = "211--236",
    year = "2011"}

@article{Sikivie:2013laa,
    author = "Sikivie, P. and Sullivan, N. and Tanner, D. B.",
    title = "{Proposal for Axion Dark Matter Detection Using an LC Circuit}",
    eprint = "1310.8545",
    archivePrefix = "arXiv",
    primaryClass = "hep-ph",
    doi = "10.1103/PhysRevLett.112.131301",
    journal = "Phys. Rev. Lett.",
    volume = "112",
    number = "13",
    pages = "131301",
    year = "2014"}

@article{Kahn:2016aff,
    author = "Kahn, Yonatan and Safdi, Benjamin R. and Thaler, Jesse",
    title = "{Broadband and Resonant Approaches to Axion Dark Matter Detection}",
    eprint = "1602.01086",
    archivePrefix = "arXiv",
    primaryClass = "hep-ph",
    reportNumber = "MIT-CTP-4763, PUPT-2497",
    doi = "10.1103/PhysRevLett.117.141801",
    journal = "Phys. Rev. Lett.",
    volume = "117",
    number = "14",
    pages = "141801",
    year = "2016"}

@article{DeRocco:2018jwe,
    author = "DeRocco, William and Hook, Anson",
    title = "{Axion interferometry}",
    eprint = "1802.07273",
    archivePrefix = "arXiv",
    primaryClass = "hep-ph",
    doi = "10.1103/PhysRevD.98.035021",
    journal = "Phys. Rev. D",
    volume = "98",
    number = "3",
    pages = "035021",
    year = "2018"}

@article{Liu:2018icu,
    author = "Liu, Hongwan and Elwood, Brodi D. and Evans, Matthew and Thaler, Jesse",
    title = "{Searching for Axion Dark Matter with Birefringent Cavities}",
    eprint = "1809.01656",
    archivePrefix = "arXiv",
    primaryClass = "hep-ph",
    reportNumber = "MIT-CTP/5048",
    doi = "10.1103/PhysRevD.100.023548",
    journal = "Phys. Rev. D",
    volume = "100",
    number = "2",
    pages = "023548",
    year = "2019"}

@article{Berlin:2019ahk,
    author = "Berlin, Asher and D'Agnolo, Raffaele Tito and Ellis, Sebastian A. R. and Nantista, Christopher and Neilson, Jeffrey and Schuster, Philip and Tantawi, Sami and Toro, Natalia and Zhou, Kevin",
    title = "{Axion Dark Matter Detection by Superconducting Resonant Frequency Conversion}",
    eprint = "1912.11048",
    archivePrefix = "arXiv",
    primaryClass = "hep-ph",
    doi = "10.1007/JHEP07(2020)088",
    journal = "JHEP",
    volume = "07",
    number = "07",
    pages = "088",
    year = "2020"}

@article{Berlin:2020vrk,
    author = "Berlin, Asher and D'Agnolo, Raffaele Tito and Ellis, Sebastian A. R. and Zhou, Kevin",
    title = "{Heterodyne broadband detection of axion dark matter}",
    eprint = "2007.15656",
    archivePrefix = "arXiv",
    primaryClass = "hep-ph",
    doi = "10.1103/PhysRevD.104.L111701",
    journal = "Phys. Rev. D",
    volume = "104",
    number = "11",
    pages = "L111701",
    year = "2021"}

@article{DMRadio:2022jfv,
    author = "Brouwer, L. and others",
    collaboration = "DMRadio",
    title = "{Proposal for a definitive search for GUT-scale QCD axions}",
    eprint = "2203.11246",
    archivePrefix = "arXiv",
    primaryClass = "hep-ex",
    doi = "10.1103/PhysRevD.106.112003",
    journal = "Phys. Rev. D",
    volume = "106",
    number = "11",
    pages = "112003",
    year = "2022"}

@article{Zhou:2025wax,
    author = "Zhou, Kevin",
    title = "{Ponderomotive Effects of Ultralight Dark Matter}",
    eprint = "2502.01725",
    archivePrefix = "arXiv",
    primaryClass = "hep-ph",
    month = "2",
    year = "2025"}

@article{Hook:2017psm,
    author = "Hook, Anson and Huang, Junwu",
    title = "{Probing axions with neutron star inspirals and other stellar processes}",
    eprint = "1708.08464",
    archivePrefix = "arXiv",
    primaryClass = "hep-ph",
    doi = "10.1007/JHEP06(2018)036",
    journal = "JHEP",
    volume = "06",
    pages = "036",
    year = "2018"}

@article{Zhitnitsky:1980tq,
    author = "Zhitnitsky, A. R.",
    title = "{On Possible Suppression of the Axion Hadron Interactions. (In Russian)}",
    journal = "Sov. J. Nucl. Phys.",
    volume = "31",
    pages = "260",
    year = "1980"
}

@article{Dine:1981rt,
    author = "Dine, Michael and Fischler, Willy and Srednicki, Mark",
    title = "{A Simple Solution to the Strong CP Problem with a Harmless Axion}",
    reportNumber = "Print-81-0320 (IAS,PRINCETON)",
    doi = "10.1016/0370-2693(81)90590-6",
    journal = "Phys. Lett. B",
    volume = "104",
    pages = "199--202",
    year = "1981"
}

@article{Kim:1979if,
    author = "Kim, Jihn E.",
    title = "{Weak Interaction Singlet and Strong CP Invariance}",
    reportNumber = "UPR-0120T",
    doi = "10.1103/PhysRevLett.43.103",
    journal = "Phys. Rev. Lett.",
    volume = "43",
    pages = "103",
    year = "1979"
}

@article{Shifman:1979if,
    author = "Shifman, Mikhail A. and Vainshtein, A. I. and Zakharov, Valentin I.",
    title = "{Can Confinement Ensure Natural CP Invariance of Strong Interactions?}",
    reportNumber = "ITEP-64-1979",
    doi = "10.1016/0550-3213(80)90209-6",
    journal = "Nucl. Phys. B",
    volume = "166",
    pages = "493--506",
    year = "1980"
}

@article{Kubo:1966fyg,
    author = "Kubo, R.",
    title = "{The fluctuation-dissipation theorem}",
    doi = "10.1088/0034-4885/29/1/306",
    journal = "Rept. Prog. Phys.",
    volume = "29",
    number = "1",
    pages = "255",
    year = "1966"}

@article{Cagnoli:2000vr,
    author = "Cagnoli, G. and Gammaitoni, L. and Hough, J. and Kovalik, J. and McIntosh, S. and Punturo, M. and Rowan, S.",
    title = "{Very high Q measurements on a fused silica monolithic pendulum for use in enhanced gravity wave detectors}",
    doi = "10.1103/PhysRevLett.85.2442",
    journal = "Phys. Rev. Lett.",
    volume = "85",
    pages = "2442--2445",
    year = "2000"}

@article{Aspelmeyer:2013lha,
    author = "Aspelmeyer, Markus and Kippenberg, Tobias J. and Marquardt, Florian",
    title = "{Cavity Optomechanics}",
    eprint = "1303.0733",
    archivePrefix = "arXiv",
    primaryClass = "cond-mat.mes-hall",
    doi = "10.1103/RevModPhys.86.1391",
    journal = "Rev. Mod. Phys.",
    volume = "86",
    pages = "1391",
    year = "2014"}

@article{Smetana:2024xwi,
    author = "Smetana, Jiri and others",
    title = "{Sensitivity and control of a six-axis fused-silica seismometer}",
    eprint = "2405.13475",
    archivePrefix = "arXiv",
    primaryClass = "physics.ins-det",
    doi = "10.1103/PhysRevApplied.23.024013",
    journal = "Phys. Rev. Applied",
    volume = "23",
    number = "2",
    pages = "024013",
    year = "2025"}

@article{Gasser:1982ap,
    author = "Gasser, J. and Leutwyler, H.",
    title = "{Quark Masses}",
    reportNumber = "BUTP-6-1982-BERN",
    doi = "10.1016/0370-1573(82)90035-7",
    journal = "Phys. Rept.",
    volume = "87",
    pages = "77--169",
    year = "1982"}

@book{Blatt:1952ije,
    author = "Blatt, John Markus and Weisskopf, Victor Frederick",
    title = "{Theoretical nuclear physics}",
    doi = "10.1007/978-1-4612-9959-2",
    isbn = "978-0-471-08019-0",
    publisher = "Springer",
    address = "New York",
    year = "1952"}

@misc{kaplan1955nuclear,
  title={Nuclear physics},
  author={Kaplan, Irving and Teichmann, T},
  year={1955},
  publisher={American Institute of Physics}}

@article{friedel1952xiv,
  title={XIV. The distribution of electrons round impurities in monovalent metals},
  author={Friedel, J},
  journal={The London, Edinburgh, and Dublin Philosophical Magazine and Journal of Science},
  volume={43},
  number={337},
  pages={153--189},
  year={1952},
  publisher={Taylor \& Francis}}

@article{blandin1959knight,
  title={On the knight shift of alloys},
  author={Blandin, A and Daniel, E and Friedel, J},
  journal={Philosophical Magazine},
  volume={4},
  number={38},
  pages={180--182},
  year={1959},
  publisher={Taylor \& Francis}}

@article{blandin1960effets,
  title={Effets quadrupolaires dans la r{\'e}sonance magn{\'e}tique nucl{\'e}aire des alliages dilu{\'e}s},
  author={Blandin, Annie and Friedel, J},
  journal={Journal de Physique et le Radium},
  volume={21},
  number={10},
  pages={689--695},
  year={1960},
  publisher={Revue G{\'e}n{\'e}rale de l'Electricit{\'e}}}

@article{villain2016jacques,
  title={Jacques Friedel and the physics of metals and alloys},
  author={Villain, Jacques and Lavagna, Mireille and Bruno, Patrick},
  journal={Comptes Rendus Physique},
  volume={17},
  number={3-4},
  pages={276--290},
  year={2016},
  publisher={Elsevier}}

@article{Kohn:1965zz,
    author = "Kohn, W. and Luttinger, J. M.",
    title = "{New Mechanism for Superconductivity}",
    doi = "10.1103/PhysRevLett.15.524",
    journal = "Phys. Rev. Lett.",
    volume = "15",
    pages = "524--526",
    year = "1965"}

@article{bland2016galaxy,
  title={The galaxy in context: structural, kinematic, and integrated properties},
  author={Bland-Hawthorn, Joss and Gerhard, Ortwin},
  journal={Annual Review of Astronomy and Astrophysics},
  volume={54},
  number={1},
  pages={529--596},
  year={2016},
  publisher={Annual Reviews}}

@article{Baxter:2021pqo,
    author = "Baxter, D. and others",
    title = "{Recommended conventions for reporting results from direct dark matter searches}",
    eprint = "2105.00599",
    archivePrefix = "arXiv",
    primaryClass = "hep-ex",
    doi = "10.1140/epjc/s10052-021-09655-y",
    journal = "Eur. Phys. J. C",
    volume = "81",
    number = "10",
    pages = "907",
    year = "2021"}

@article{Buckley:2009in,
    author = "Buckley, Matthew R. and Fox, Patrick J.",
    title = "{Dark Matter Self-Interactions and Light Force Carriers}",
    eprint = "0911.3898",
    archivePrefix = "arXiv",
    primaryClass = "hep-ph",
    reportNumber = "FERMILAB-PUB-09-560-T",
    doi = "10.1103/PhysRevD.81.083522",
    journal = "Phys. Rev. D",
    volume = "81",
    pages = "083522",
    year = "2010"}

@article{Tulin:2013teo,
    author = "Tulin, Sean and Yu, Hai-Bo and Zurek, Kathryn M.",
    title = "{Beyond Collisionless Dark Matter: Particle Physics Dynamics for Dark Matter Halo Structure}",
    eprint = "1302.3898",
    archivePrefix = "arXiv",
    primaryClass = "hep-ph",
    doi = "10.1103/PhysRevD.87.115007",
    journal = "Phys. Rev. D",
    volume = "87",
    number = "11",
    pages = "115007",
    year = "2013"}

@article{Xu:2021lmg,
    author = "Xu, Xingchen and Farrar, Glennys R.",
    title = "{Constraints on GeV Dark Matter interaction with baryons, from a novel Dewar experiment}",
    eprint = "2112.00707",
    archivePrefix = "arXiv",
    primaryClass = "hep-ph",
    month = "12",
    year = "2021"}

@article{Xu:2020qjk,
    author = "Xu, Xingchen and Farrar, Glennys R.",
    title = "{Resonant scattering between dark matter and baryons: Revised direct detection and CMB limits}",
    eprint = "2101.00142",
    archivePrefix = "arXiv",
    primaryClass = "hep-ph",
    doi = "10.1103/PhysRevD.107.095028",
    journal = "Phys. Rev. D",
    volume = "107",
    number = "9",
    pages = "095028",
    year = "2023"}

@book{bransden2006physics,
  title={Physics of atoms and molecules},
  author={Bransden, Brian Harold and Joachain, Charles Jean},
  year={2006},
  publisher={Pearson Education India}}

@article{Gue:2025nxq,
    author = "Gu\'e, Jordan and Wolf, Peter and Hees, Aur\'elien",
    title = "{Search for QCD coupled axion dark matter with the MICROSCOPE space experiment}",
    eprint = "2504.00720",
    archivePrefix = "arXiv",
    primaryClass = "hep-ph",
    month = "4",
    year = "2025"}

@article{moritz2000geodetic,
  title={Geodetic reference system 1980},
  author={Moritz, Helmut},
  journal={Journal of geodesy},
  volume={74},
  number={1},
  pages={128--133},
  year={2000},
  publisher={Springer Science and Business Media LLC}
}

@article{IAUInter-DivisionA-GWorkingGrouponNominalUnitsforStellarPlanetaryAstronomy:2015fjh,
    author = "Mamajek, E. E. and others",
    collaboration = "IAU Inter-Division A-G Working Group on Nominal Units for Stellar \& Planetary Astronomy",
    title = "{IAU 2015 Resolution B3 on Recommended Nominal Conversion Constants for Selected Solar and Planetary Properties}",
    eprint = "1510.07674",
    archivePrefix = "arXiv",
    primaryClass = "astro-ph.SR",
    month = "10",
    year = "2015"}

@article{Tan:2016vwu,
    author = "Tan, Wen-Hai and Yang, Shan-Qing and Shao, Cheng-Gang and Li, Jia and Du, An-Bin and Zhan, Bi-Fu and Wang, Qing-Lan and Luo, Peng-Shun and Tu, Liang-Cheng and Luo, Jun",
    title = "{New Test of the Gravitational Inverse-Square Law at the Submillimeter Range with Dual Modulation and Compensation}",
    doi = "10.1103/PhysRevLett.116.131101",
    journal = "Phys. Rev. Lett.",
    volume = "116",
    number = "13",
    pages = "131101",
    year = "2016"}

@article{Webb:1998cq,
    author = "Webb, John K. and Flambaum, Victor V. and Churchill, Christopher W. and Drinkwater, Michael J. and Barrow, John D.",
    title = "{Evidence for time variation of the fine structure constant}",
    eprint = "astro-ph/9803165",
    archivePrefix = "arXiv",
    reportNumber = "AST-140398",
    doi = "10.1103/PhysRevLett.82.884",
    journal = "Phys. Rev. Lett.",
    volume = "82",
    pages = "884--887",
    year = "1999"}

@article{Petitjean:2009id,
    author = "Petitjean, Patrick and Srianand, Raghunathan and Chand, Hum and Ivanchik, Alexander and Noterdaeme, Pasquier and Gupta, Neeraj",
    title = "{Constraining fundamental constants of physics with quasar absorption line systems}",
    eprint = "0905.1516",
    archivePrefix = "arXiv",
    primaryClass = "astro-ph.CO",
    doi = "10.1007/s11214-009-9520-y",
    journal = "Space Sci. Rev.",
    volume = "148",
    pages = "289",
    year = "2009"}

@article{Srianand:2009gaa,
    author = "Srianand, R. and Petitjean, P. and Chand, H. and Noterdaeme, P. and Gupta, N.",
    editor = "Molaro, P. and Vangioni, E.",
    title = "{Probing the variation of fundamental constants using QSO absorption lines}",
    journal = "Mem. Soc. Ast. It.",
    volume = "80",
    number = "4",
    pages = "842--849",
    year = "2009"}

@article{Baryakhtar:2025uxs,
    author = "Baryakhtar, Masha and Simon, Olivier and Weiner, Zachary J.",
    title = "{Searching for coupled, hyperlight scalars across cosmic history}",
    eprint = "2502.04432",
    archivePrefix = "arXiv",
    primaryClass = "hep-ph",
    month = "2",
    year = "2025"}

@article{shlyakhter1976direct,
  title={Direct test of the constancy of fundamental nuclear constants},
  author={Shlyakhter, AI},
  journal={Nature},
  volume={264},
  number={5584},
  pages={340--340},
  year={1976},
  publisher={Nature Publishing Group UK London}}

@article{Gould:2006qxs,
    author = "Gould, C. R. and Sharapov, E. I. and Lamoreaux, S. K.",
    title = "{Time variability of alpha from realistic models of Oklo reactors}",
    eprint = "nucl-ex/0701019",
    archivePrefix = "arXiv",
    doi = "10.1103/PhysRevC.74.024607",
    journal = "Phys. Rev. C",
    volume = "74",
    pages = "024607",
    year = "2006"}

@article{Flambaum:2008hu,
    author = "Flambaum, V. V. and Wiringa, R. B.",
    title = "{Enhanced effect of quark mass variation in Th-229 and limits from Oklo data}",
    eprint = "0807.4943",
    archivePrefix = "arXiv",
    primaryClass = "nucl-th",
    doi = "10.1103/PhysRevC.79.034302",
    journal = "Phys. Rev. C",
    volume = "79",
    pages = "034302",
    year = "2009"}

@article{Meissner:1933ela,
    author = "Meissner, W. and Ochsenfeld, R.",
    title = {{Ein neuer Effekt bei Eintritt der Supraleitf\"ahigkeit}},
    doi = "10.1007/BF01504252",
    journal = "Naturwiss.",
    volume = "21",
    number = "44",
    pages = "787--788",
    year = "1933"}

@book{Jackson:1998nia,
    author = "Jackson, John David",
    title = "{Classical Electrodynamics}",
    isbn = "978-0-471-30932-1",
    publisher = "Wiley",
    year = "1998"}

@article{Arvanitaki:2023fij,
    author = "Arvanitaki, Asimina and Dimopoulos, Savas",
    title = "{A Diffraction Grating for the Cosmic Neutrino Background and Dark Matter}",
    eprint = "2303.04814",
    archivePrefix = "arXiv",
    primaryClass = "hep-ph",
    month = "3",
    year = "2023"}

@article{wheatley1968principles,
  title={Principles and methods of dilution refrigeration},
  author={Wheatley, John C and Vilches, OE and Abel, WR},
  journal={Physics Physique Fizika},
  volume={4},
  number={1},
  pages={1},
  year={1968},
  publisher={APS}}

@book{Pobell:2007egf,
    author = "Pobell, Frank",
    title = "{Matter and Methods at Low Temperatures}",
    doi = "10.1007/978-3-540-46360-3",
    isbn = "978-3-540-46356-6, 978-3-540-46360-3",
    publisher = "Springer",
    year = "2007"}

@article{burns1989optical,
  title={Optical binding},
  author={Burns, Michael M and Fournier, Jean-Marc and Golovchenko, Jene A},
  journal={Physical Review Letters},
  volume={63},
  number={12},
  pages={1233},
  year={1989},
  publisher={APS}}

@article{chen2015lateral,
  title={Lateral optical force on paired chiral nanoparticles in linearly polarized plane waves},
  author={Chen, Huajin and Jiang, Yikun and Wang, Neng and Lu, Wanli and Liu, Shiyang and Lin, Zhifang},
  journal={Optics letters},
  volume={40},
  number={23},
  pages={5530--5533},
  year={2015},
  publisher={Optical Society of America}}

@article{Rudolph:2024cey,
    author = "Rudolph, Henning and Deli\'c, Uro\v{s} and Hornberger, Klaus and Stickler, Benjamin A.",
    title = "{Quantum Optical Binding of Nanoscale Particles}",
    eprint = "2412.03204",
    archivePrefix = "arXiv",
    primaryClass = "quant-ph",
    doi = "10.1103/PhysRevLett.133.233603",
    journal = "Phys. Rev. Lett.",
    volume = "133",
    pages = "233603",
    year = "2024"}

@article{Da:progress,
    author = "Batell, Brian and Bhoonah, Amit and Ghalsasi, Akshay and Liu, Da and Luty, Markus A.",
    Note = {In Progress},
    journal = ""}

\end{document}